\documentclass[12pt]{report}
\usepackage{epsfig}
\usepackage{ODUthesis}

\begin{document}

\title{Inclusive and Exclusive Compton Processes in Quantum Chromodynamics}

\author{A. Psaker}
\principaladviser{Anatoly Radyushkin}
\member{Ian Balitsky}
\member{Charles Hyde-Wright}
\member{Mark Havey}
\member{John Adam}

\degrees{B.S. March 1997, University of Ljubljana\\
         M.S. May 2001, Old Dominion University}
\dept{Physics}

\submitdate{December 2005}

\abstract{In our work, we describe two types of Compton processes. 
As an example of an inclusive process, we consider the high-energy photoproduction of 
massive muon pairs off the nucleon. We analyze the process in the framework of the 
QCD parton model, in which the usual parton distributions emerge as a tool to 
describe the nucleon in terms of quark and gluonic degrees of freedom. 
To study its exclusive version, a new class of phenomenological functions is required, 
namely, generalized parton distributions. They can be considered as a generalization 
of the usual parton distributions measured in deeply inelastic lepton-nucleon scattering. 
Generalized parton distributions (GPDs) may be observed in hard exclusive reactions 
such as deeply virtual Compton scattering. We develop an extension of this particular 
process into the weak interaction sector. We also investigate a possible application 
of the GPD formalism to wide-angle real Compton scattering.}

\beforepreface

\prefacesection{Acknowledgments}
        
	I would like to express sincere and deep gratitude to my advisor Anatoly Radyushkin for 
	his generous support and guidance during my graduate student years. I would like to 
	extend my gratitude to all the members of my committee for their helpful comments and 
	suggestions. In particular, I'm grateful to Charles Hyde-Wright and Wally Melnitchouk 
	for various stimulating and valuable discussions. I would like to thank my colleagues 
	at the Physics Department of Old Dominion University and the Theory Group of Jefferson Lab. 
        I'm thankful to my family for their support, and to my wife Mateja for her patience and 
	continuous encouragement. This work was supported by the US Department of Energy 
	DE-FG02-97ER41028 and by the contract DE-AC05-84ER40150 under which the Southeastern 
	Universities Research Association (SURA) operates the Thomas Jefferson Accelerator 
	Facility. 
    
\afterpreface

\chapter{Introduction}

Strong interaction physics, the study of hadron structure in general,
is one of the most fascinating frontiers of modern science\nnfootnote{This dissertation 
follows the style of {\it The Physical Review}.}. The underlying theory is the universally 
accepted nonAbelian gauge field theory called quantum chromodynamics (QCD). 
In principle, QCD embraces all phenomena of hadronic physics. 
It postulates that hadrons (baryons and mesons) are composite objects made up of quarks, 
and that the color interaction between quarks is mediated by gluons as
the gauge bosons. The main difficulty of this elegant theory lies in the fact that it is 
formulated in terms of colored degrees of freedom (quarks and gluons) while the physical 
hadrons are colorless. How the hadrons are built out of quarks and gluons has yet to be 
answered, and it represents a challenging task.

Historically, QCD originated from the constituent quark model 
\cite{Zweig,Gell-Mann:1964nj}. In the 1960's, the development of high-energy accelerators 
made it possible to resolve the structure of hadrons for the first time. 
The studies of the hadron spectroscopy led to the concept of quarks
as the fundamental building blocks of hadrons, in particular, three quarks 
for baryons and a quark-antiquark pair for mesons. The natural step that followed, 
was the search for quarks in the experiments with the momentum transfer
large enough to look inside the hadrons. The first series of such
high-resolution experiments was performed at SLAC by probing the proton
structure. The process was known as deeply inelastic electron-proton
scattering (DIS). It is probably the most studied QCD process.
The data on DIS have strongly supported the parton model, a physical 
picture given by Feynman \cite{Bjorken:1969ja,Feynman:1969ej,Feynman}, 
in which the proton, when observed with high spatial resolution, 
is built out of almost-free and point-like constituents. They were called partons. 
Thus the dynamics of partons described by the QCD parton model should have 
the property of asymptotic freedom. In other words, the coupling constant of interaction 
between partons is vanishing at small distances (or large momentum transfers). 
In addition, the DIS experiments also suggested that charged partons should be spin-1/2 
particles that were later identified with quarks. Another important result from these 
experiments was the first evidence for the presence of electrically neutral gluons. 
It turns out that quarks carry only about a half of the total nucleon momentum.

The fact that the theory of quark dynamics should exhibit the desired property of 
the asymptotic freedom led physicists to consider the nonAbelian gauge 
field theory, originally constructed by Yang and Mills \cite{Yang:1954ek}, 
as the best candidate \cite{Gross:1973id,Politzer:1973fx,Politzer:1974fr}. 
In order to resolve several difficulties in the constituent quark model, 
a new quantum number, color, was introduced to quarks 
\cite{Bogolubov,Han:1965pf,Miyamoto}. By identifying color as the SU(3) symmetry 
of the nonAbelian gauge theory \cite{Fritzsch:2002jv,Fritzsch:1973pi}, QCD was 
finally established.

The description of all inclusive hard reactions in QCD is possible due to the so-called 
factorization. According to this property, the cross section of a particular QCD process 
splits (factorizes) into the hard (short-distance) and the soft (long-distance) parts. 
Introducing the factorization scale as a point of separation can only be possible 
by assuring the presence of a large invariant, such as the virtuality
of the probe or the momentum transfer 
\cite{Amati:1978wx,Amati:1978by,Libby:1978qf,Efremov:1978cu,Efremov:1980kz,
Efremov:1978xm,Ellis:1978ty,Mueller:1978xu}. 
The clear separation of scales is of the crucial importance. The asymptotically
free nature of QCD allows to discuss the short-distance interactions by means of 
perturbation theory. This is, however, not the case for the soft (nonperturbative) 
stages of interactions. They are expressed in terms of the hadronic matrix elements 
of well-defined quark-gluon operators, i.e. the QCD operators taken on the light-cone and 
sandwiched between the hadronic states. These matrix elements accumulate information 
about long-distance dynamics. They emerge as a result of the description 
of the hard hadronic processes using a powerful tool called the light-cone
operator product expansion (OPE) \cite{Wilson:1969zs,Brandt:1970kg}. 
Originally applied to DIS \cite{Gross:1973ju,Gross:1974cs,Georgi:1951sr}, 
the technique is based on the statement that, in the particular kinematical regime, 
the asymptotic behavior of the relevant scattering amplitudes is governed by the 
singularities on the light-cone which, in the framework of a light-cone expansion,
can be described in terms of contributions of definite twist \cite{Brandt:1970kg} 
(see \cite{Lazar:2002af} for a recent review). In other words, one performs a systematic 
expansion in the inverse powers of a characteristic momentum scale, and  extracts 
the leading and higher twist contributions, the latter containing information about 
quark-quark and quark-gluon correlations inside the hadron.

The hadronic matrix elements are process-independent nonperturbative 
objects measured with the help of different probes, such as photons and weak 
interaction bosons. For that reason, these objects are parametrized by universal functions, 
allowing to relate various light-cone dominated scattering processes to each other. 
The well-known example of such phenomenological functions is given by the usual parton 
distribution functions (PDFs). They parametrize forward hadronic matrix elements and, 
since these elements are related to inclusive cross sections, PDFs enter the description 
of hard inclusive reactions. Additional information on quark and gluon structure of hadrons 
is provided by measuring the nonforward hadronic matrix elements. They correspond to 
nondiagonal transitions between the hadronic states in momentum, flavor and spin spaces, 
and hence are described in terms of more general phenomenological functions known as 
the generalized parton distributions (GPDs) \cite{Muller:1998fv,Ji:1996ek,
Ji:1996nm,Radyushkin:1996nd,Radyushkin:1996ru,Radyushkin:1997ki}. These distributions
can be accessed in hard exclusive reactions \cite{Vanderhaeghen:1999xj,Goeke:2001tz}, 
such as deeply virtual Compton scattering (DVCS) \cite{Ji:1996ek,Ji:1996nm,
Radyushkin:1996nd,Radyushkin:1996ru,Radyushkin:1997ki} and deeply exclusive production 
of mesons (DMP) \cite{Radyushkin:1996ru,Radyushkin:1997ki,Collins:1996fb}. In particular, 
the DVCS process attracted a lot of attention, both theoretical and experimental.

The thesis is organized as follows. In Chapter II, we give a short review on the DIS process 
and discuss its relation to the forward virtual Compton scattering amplitude (VCA). 
In addition, we introduce parton distribution functions within the context of the parton model. 
The commonly used phenomenological functions, such as form factors, PDFs, distribution 
amplitudes (DAs) and GPDs are the subject of study in Chapter III. We list the definitions and 
discuss some of theoretical aspects, i.e. the basic properties of these functions and their relations 
to one another. Typically, GPDs can be measured in hard exclusive leptoproduction processes. 
DVCS, as the simplest process in this respect, is discussed in Chapter IV. 
The VCA is calculated at the leading twist within the framework of the nonlocal light-cone expansion 
of the product of currents in QCD string operators. Moreover, a simple model for the nucleon 
valence GPDs (i.e. those, which do not include the contribution from sea quarks) is introduced. 
In Chapter V, we apply the parton model to study the inclusive photoproduction of lepton pairs. 
Chapter VI is devoted to the exclusive version of the same process. It is related to time-like 
Compton scattering (TCS), that is the inverse of the DVCS process. With the help of a different set of 
GPDs, which cannot be accessible in the standard electromagnetic DVCS process, the extension into the 
weak interaction sector is made possible. This work is presented in Chapter VII. In Chapter VIII, 
we turn our attention to wide-angle real Compton scattering (WACS). Our approach is based on the 
handbag dominance, in other words, the light-cone singularities of the Compton scattering amplitude 
again play an important role \cite{Radyushkin:1998rt,Diehl:1998kh,Diehl:1999tr}. 
Here we use the formalism of double distributions (DDs) to describe nonforward matrix elements of 
light-cone operators. We study both the leading and next-to-leading twist contributions. 
Finally, we draw conclusions and outline future research plans in Chapter IX. 

\chapter{Deeply Inelastic Lepton Scattering}

\section{\ \,  Introduction}

We start with phenomenologically most important of the hard processes,
inclusive scattering of high-energy (charged and neutral) leptons
on hadrons (to be specific, we will consider here and throughout this thesis, except in Chapter VIII, 
a spin-1/2 hadronic target, e.g. a nucleon) with the exchange of vector bosons 
(photons, $W^{\pm}$ or $Z^{0}$). It played the key role in revealing the quark structure of hadrons, 
and it is, in addition to electron-positron annihilation into hadrons, the simplest process involving 
strongly interacting particles. Since leptons do not possess a resolvable internal structure, 
the reaction cross section depends solely on the internal structure of hadrons.
 
In Section II.2, we discuss the kinematics and the relation between the hadron tensor and 
the forward VCA. Both the electron and neutrino-induced DIS cross sections are given in terms of the 
relevant structure functions. The parton model is introduced in Section II.3. 
With the help of the parton model master formula, we calculate the DIS cross section, and 
express the structure functions in terms of PDFs.  

\section{\ \,  Kinematics}

\begin{figure}
\centerline{\epsfxsize=3in\epsffile{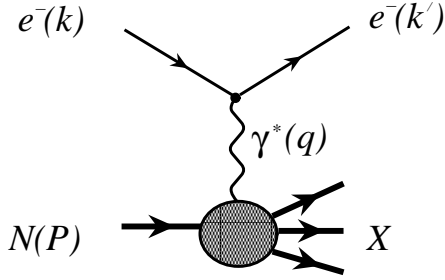}}
\caption{Deeply inelastic electron-nucleon scattering in the one-photon exchange approximation.}
\label{dis}
\end{figure}

Consider the electron scattering process off the nucleon target through a single photon exchange,
\begin{eqnarray}
e^{-}\left(k\right)+N\left(P\right) & \longrightarrow & e^{-}\left(k'\right)+X,
\label{eq:disprocess}
\end{eqnarray}
as illustrated in Fig. \ref{dis}. In other words, we consider only the quantum electrodynamic (or shortly QED) 
interaction between the electron and the nucleon, and further keep only its lowest order. We denote the nucleon 
four-momentum with \emph{P} and the initial and final momenta of electrons with
\emph{k} and $k'$, respectively. In the laboratory frame, in which the target is at rest, 
the nucleon four-momentum is $P=\left(M,\vec{0}\right)$, where \emph{M} stands for the nucleon mass. 
The electron four-momentum in the initial state is $k=\left(\omega,\vec{k}\right)$, and in the final state is 
$k'=\left(\omega',\vec{k}'\right)$. The momentum transfer to the target, carried by the virtual photon, 
is $q=k-k'=\left(\nu,\vec{q}\right)$. In this scattering process, due to the large \emph{q}, 
the nucleon breaks up and forms an infinite number of possible hadronic final
states labeled by \emph{X}, which remain unobserved. Thus only the outgoing electron is detected. One measures 
its energy $\omega'$ and the scattering angle $\theta$ relative to the incident beam of fixed energy $\omega$.

There are two characteristic Lorentz-invariant quantities of the process,
namely, the virtuality of the photon (since it is space-like, $q^{2}<0$,
one conventionally defines a positive quantity),
\begin{eqnarray}
q^{2} & \equiv & -Q^{2},
\label{eq:photonvirtuality}
\end{eqnarray}
which fixes the spatial resolution of the scattering process, $\lambda\sim1/\sqrt{Q^{2}}$,
and the invariant mass of the final hadronic state,
\begin{eqnarray}
W^{2} & \equiv & \left(P+q\right)^{2}=M^{2}+2\left(P\cdot q\right)-Q^{2}.
\label{eq:invariantmass}
\end{eqnarray}
In the laboratory frame, they are given by
\begin{eqnarray}
Q^{2} & = & 4\omega\omega'\sin^{2}\left(\theta/2\right)
\label{eq:photonvirtualitylabframe}
\end{eqnarray}
and 
\begin{eqnarray}
W^{2} & = & M^{2}+2M\nu-Q^{2}.
\label{eq:invariantmasssquaredlabframe}
\end{eqnarray}
Note that in Eq. (\ref{eq:photonvirtualitylabframe}) we have neglected the electron mass 
compared to its energy. Unlike elastic scattering, for which $W^{2}=M^{2}$ and the momentum transfer 
squared $Q^{2}$ is connected by the relation $\nu=Q^{2}/2M$ to the energy loss of the electron $\nu$, 
we now have two independent variables, $\nu$ and $Q^{2}$. The inclusive differential cross section 
(see, e.g. \cite{Feynman,Greiner1,Thomas:2001kw,Jaffe:1996zw}) factors into the electron tensor 
$L_{\mu\nu}^{\left(e\right)}$ and the hadron tensor $W^{\mu\nu}$, 
\begin{eqnarray}
\frac{d^{2}\sigma_{\left(eN\right)}}{d\Omega d\omega'} & = & \frac{\alpha^{2}}{Q^{4}}\frac{\omega'}
{\omega}L_{\mu\nu}^{\left(e\right)}W^{\mu\nu},
\label{eq:dissigma1}
\end{eqnarray}
where $d\Omega$ is the solid angle into which the electron is scattered
and $\alpha=e^{2}/4\pi\simeq1/137$ is the electromagnetic fine structure constant (in the following we 
use the Heaviside-Lorentz convention together with the natural choice of units, frequently used in 
particle physics, in which $\hbar=c=1$). The electron tensor can be computed exactly in QED. 
For unpolarized electron scattering, the expression for the spin-averaged electron tensor is
\begin{eqnarray}
L_{\mu\nu}^{\left(e\right)} & = & \frac{1}{2}\sum_{s,s'}\bar{u}(k,s)\gamma_{\mu}u(k',s')
\bar{u}(k',s')
\gamma_{\nu}u(k,s)\nonumber \\
& = & 2\left[k_{\mu}k'_{\nu}+k_{\nu}k'_{\mu}-g_{\mu\nu}\left(k\cdot k'\right)\right].
\label{eq:leptonictensordis}
\end{eqnarray}
The hadron tensor, on the other hand, describes the response of the nucleon and therefore, it 
should include all possible transitions of the nucleon from its ground state 
$\left|N\left(P,S\right)\right\rangle $ to any hadronic final state 
$\left|X\left(P'\right)\right\rangle $. It is worth noting that at present, the nonperturbative 
nature of strong interactions prevents us from calculating the hadron tensor directly within 
the framework of QCD. Formally, it can be written as
\begin{eqnarray}
W^{\mu\nu} & = & \frac{1}{4\pi M}\sum_{X}\left(2\pi\right)^{4}\delta^{\left(4\right)}
\left(P'-P-q\right)
\frac{1}{2}\sum_{S}\left\langle N\left(P,S\right)
\right|J_{EM}^{\mu}\left(0\right)\left|X\left(P'\right)\right\rangle \nonumber \\
&  & \times\left\langle X\left(P'\right)\right|J_{EM}^{\nu}\left(0\right)\left|N\left(P,S\right)
\right\rangle,
\label{eq:hadronictensordis}
\end{eqnarray}
where $J_{EM}^{\mu}$ is the quark electromagnetic current. Using the translation invariance,
\begin{eqnarray}
\left\langle N\left(P,S\right)\right|J_{EM}^{\mu}\left(z\right)\left|X\left(P'\right)
\right\rangle  & = & \left
\langle N\left(P,S\right)\right|J_{EM}^{\mu}\left(0\right)\left|X\left(P'\right)
\right\rangle e^{-i\left(P'-P\right)\cdot z},
\label{eq:translationinvariance}
\end{eqnarray}
and the completeness condition of the states $\left|X\left(P'\right)\right\rangle$,
\begin{eqnarray}
\sum_{X}\left|X\left(P'\right)\right\rangle \left\langle X\left(P'\right)\right| & = & \mathbf{1},
\label{eq:completeness}
\end{eqnarray}
the hadron tensor can be presented in a more compact form, namely,
\begin{eqnarray}
W^{\mu\nu} & = & \frac{1}{4\pi M}\int d^{4}z\; e^{i\left(q\cdot z\right)}\frac{1}{2}\sum_{S}\left
\langle N\left(P,S\right)
\right|J_{EM}^{\mu}\left(z\right)J_{EM}^{\nu}\left(0\right)\left|N\left(P,S\right)\right\rangle.
\label{eq:hadronictensorcompactform}
\end{eqnarray}
Adopting a simplifying notation,
\begin{eqnarray}
\left\langle N\left(P\right)\right|\mathcal{O}\left|N\left(P\right)\right
\rangle  & \equiv & 
\frac{1}{2}\sum_{S}\left\langle N\left(P,S\right)\right|\mathcal{O}\left|N\left(P,S\right)
\right\rangle,
\label{eq:simplifiednotation}
\end{eqnarray}
where $\mathcal{O}$ is an arbitrary operator we have
\begin{eqnarray}
W^{\mu\nu} & = & \frac{1}{4\pi M}\int d^{4}z\; e^{i\left(q\cdot z\right)}\left\langle N
\left(P\right)\right|J_{EM}^{\mu}
\left(z\right)J_{EM}^{\nu}\left(0\right)\left|N\left(P\right)\right\rangle.
\label{eq:hadronictensorcompactformfinal}
\end{eqnarray}
Interchanging the currents inside the matrix element leads to a delta
function $\delta^{\left(4\right)}\left(P'-P+q\right)$ which, in the
laboratory frame, requires that $E'=M-\nu<M$. Note that such a state
$\left|X\left(P'\right)\right\rangle $ does not exist since the nucleon
$\left|N\left(P,S\right)\right\rangle $ is the ground state baryon.
Hence the delta function cannot be satisfied and the expression with 
the interchanged currents vanishes. By adding this vanishing matrix element, 
one can also rewrite the hadron tensor as a current commutator,
\begin{eqnarray}
W^{\mu\nu} & = & \frac{1}{4\pi M}\int d^{4}z\; e^{i\left(q\cdot z\right)}
\left\langle N\left(P\right)
\right|\left[J_{EM}^{\mu}
\left(z\right),J_{EM}^{\nu}\left(0\right)\right]\left|N\left(P\right)\right\rangle.
\label{eq:hadronictensorascurrentcommutator}
\end{eqnarray}

\begin{figure}
\centerline{\epsfxsize=3in\epsffile{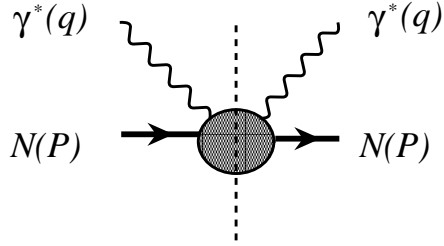}}
\caption[Forward virtual Compton scattering amplitude.] 
{Forward virtual Compton scattering amplitude. The dashed line denotes the cut, which gives 
the imaginary part, and it is further related to the hadron tensor through the optical theorem.}
\label{fvca}
\end{figure}

Next we discuss the relation between $W^{\mu\nu}$ and the object
known as the forward virtual Compton scattering amplitude (VCA). The latter,
shown in Fig. \ref{fvca}, is defined as a Fourier transform of the
correlation function of two electromagnetic currents,
\begin{eqnarray}
T^{\mu\nu} & = & i\int d^{4}z\; e^{i\left(q\cdot z\right)}\left\langle N
\left(P,S\right)\right|T\left\{ J_{EM}^{\mu}
\left(z\right)J_{EM}^{\nu}\left(0\right)\right\} \left|N\left(P,S\right)\right\rangle.
\label{eq:forwardvcsa}
\end{eqnarray}
The time-ordering symbol \emph{T} inside the matrix element instructs
us to place the operators into chronological order with the operator
having the later time argument to the left. The amplitude (\ref{eq:forwardvcsa}) is averaged
over the nucleon spin. It corresponds to the scattering amplitude for the Compton process off 
the proton in the forward direction, when contracted with the polarization vectors of the off-shell 
photon of momentum \emph{q}, $q^{2}\neq0$, 
\begin{eqnarray}
i\mathrm{T}\left(\gamma p\rightarrow\gamma p\right) & = & \left(-i\left|e\right|
\right)^{2}\epsilon_{\mu}^{*}
\left(q\right)\epsilon_{\nu}\left(q\right)\left(-iT^{\mu\nu}\right),
\label{eq:comptonscatteringoriginal}
\end{eqnarray}
or
\begin{eqnarray}
\mathrm{T}\left(\gamma p\rightarrow\gamma p\right) & = & e^{2}\epsilon_{\mu}^{*}\left(q\right)
\epsilon_{\nu}
\left(q\right)T^{\mu\nu}.
\label{eq:comptonscattering}
\end{eqnarray}
Here \emph{e} denotes the electric charge of the electron. With the help of the optical theorem, 
the hadron tensor can be written as
\begin{eqnarray}
W^{\mu\nu} & = & \frac{1}{2\pi M}\Im T^{\mu\nu}.
\label{eq:opticaltheorem}
\end{eqnarray}

Lastly, it is customary to express the hadron tensor, see 
Eqs. (\ref{eq:hadronictensorcompactformfinal}) and (\ref{eq:hadronictensorascurrentcommutator}), 
in terms of the so-called structure functions. The tensor expression is Lorentz-invariant, 
and it should also be invariant under parity and time reversal. Together with the electromagnetic
current conservation, $\partial_{\mu}J_{EM}^{\mu}=0$, implying the electromagnetic
gauge invariance,
\begin{eqnarray}
q_{\mu}W^{\mu\nu}=0 & \mathrm{and} & W^{\mu\nu}q_{\nu}=0,
\label{eq:gaugeinvariancedis}
\end{eqnarray}
one arrives at the most general form for $W^{\mu\nu}$ in the case of unpolarized scattering. 
It is determined only by two independent scalar structure functions (or alternatively the response 
functions), 
\begin{eqnarray}
W^{\mu\nu} & = & W_{1}\left(\nu,Q^{2}\right)\left[-g^{\mu\nu}+\frac{q^{\mu}q^{\nu}}{q^{2}}
\right]\nonumber \\
&  & +\frac{W_{2}\left(\nu,Q^{2}\right)}{M^{2}}\left[P^{\mu}-q^{\mu}
\frac{\left(P\cdot q\right)}{q^{2}}
\right]\left[P^{\nu}-q^{\nu}
\frac{\left(P\cdot q\right)}{q^{2}}\right].
\label{eq:hadronictensorwithw1w2}
\end{eqnarray}
The hadron tensor (\ref{eq:hadronictensorwithw1w2}) is symmetric with respect to 
Lorentz indices $\mu$ and $\nu$. In the case of polarized scattering the tensor has an 
extra term, which is antisymmetric with respect to $\mu$ and $\nu$ and contains two additional structure 
functions. We notice that both functions $W_{1}$ and $W_{2}$ depend on two invariant variables, 
$\nu=\left(P\cdot q\right)/M$ and $Q^{2}$. Substituting Eqs. (\ref{eq:leptonictensordis}) 
and (\ref{eq:hadronictensorwithw1w2}) into Eq. (\ref{eq:dissigma1}) and using 
Eq. (\ref{eq:photonvirtualitylabframe}), gives the differential cross section in the laboratory frame 
in the form
\begin{eqnarray}
\frac{d^{2}\sigma_{\left(eN\right)}}{d\Omega d\omega'} & = & 
\frac{\alpha^{2}}{4\omega^{2}\sin^{4}\left(\theta/2\right)}
\left[2W_{1}\left(\nu,Q^{2}\right)\sin^{2}\left(\theta/2\right)+W_{2}\left(\nu,Q^{2}\right)\cos^{2}
\left(\theta/2\right)\right].\nonumber \\
\label{eq:dissigma2}
\end{eqnarray}
Usually, $\nu$ is replaced by another Lorentz-invariant quantity, the Bjorken scaling variable,
\begin{eqnarray}
x_{B} & \equiv & \frac{Q^{2}}{2\left(P\cdot q\right)},
\label{eq:bjorkenx}
\end{eqnarray}
and further the dimensionfull \emph{W}-structure functions by their dimensionless
partners defined as
\begin{eqnarray}
F_{1}\left(x_{B},Q^{2}\right) & \equiv & MW_{1}\left(\nu,Q^{2}\right),\nonumber \\
F_{2}\left(x_{B},Q^{2}\right) & \equiv & \nu W_{2}\left(\nu,Q^{2}\right).
\label{eq:dimensionlessstructurefunctions}
\end{eqnarray}
Since the invariant mass of the unobserved hadronic final state is
always larger than the nucleon mass, it follows from Eqs. (\ref{eq:invariantmass})
and (\ref{eq:bjorkenx}) that
\begin{eqnarray}
x_{B} & = & \frac{Q^{2}}{W^{2}-M^{2}+Q^{2}}=\frac{1}{1+\left(W^{2}-M^{2}\right)/Q^{2}}
\in\left(0,1\right].
\label{eq:limitsonx}
\end{eqnarray}
In particular, the kinematical point $x_{B}=1$ corresponds to elastic
electron-nucleon scattering. In addition to $x_{B}$, there is another dimensionless
variable frequently used, the so-called inelasticity parameter,
\begin{eqnarray}
y & \equiv & \frac{\left(P\cdot q\right)}{\left(P\cdot k\right)}.
\label{eq:inelasticity}
\end{eqnarray}
It specifies, in the laboratory frame, the fraction of the electron
energy that is transferred to the nucleon, $y=1-\omega'/\omega$, and hence it is also 
kinematically constrained to the region $\left(0,1\right]$. The invariants $Q^{2}$, $x_{B}$ 
and \emph{y} are related to each other through
\begin{eqnarray}
Q^{2} & = & x_{B}y\left(s-M^{2}\right),
\label{eq:q^2xyrelation}
\end{eqnarray}
where $s\equiv\left(k+P\right)^{2}$ is the usual Mandelstam
invariant of the scattering process. The last formula is obtained by writing the Bjorken scaling 
variable in the laboratory frame, $x_{B}=Q^{2}/2M\nu$. Sometimes it is convenient to present 
the cross section in the invariant form. After some algebra, the cross section results, in terms 
of dimensionless structure functions, into
\begin{eqnarray}
\frac{d^{2}\sigma_{\left(eN\right)}}{dx_{B}dy} & = & \left(\frac{2\pi M\omega y}{\omega'}\right)
\frac{d^{2}\sigma_{\left(eN\right)}}{d\Omega d\omega'}\nonumber \\
& = & \frac{4\pi\alpha^{2}}{Q^{2}y}\left[y^{2}F_{1}\left(x_{B},Q^{2}\right)+\left(\frac{1-y}{x_{B}}-
\frac{M^{2}y}{s-M^{2}}
\right)F_{2}\left(x_{B},Q^{2}\right)\right].
\label{eq:finalresultinvariantdis}
\end{eqnarray}

For the sake of completeness, let us investigate the weak version
of inclusive lepton-nucleon scattering. In general, the weak boson
is used as a probe to study the nucleon structure. To be specific, we shall consider 
only the charged current case using muon neutrinos. The latter convert into muons by the emission of 
a $W^{+}$ boson. The reaction is
\begin{eqnarray}
\nu_{\mu}\left(k\right)+N\left(P\right) & \longrightarrow & \mu^{-}\left(k'\right)+X.
\label{eq:weakdisprocess}
\end{eqnarray}
Recall that the incident neutrino is always left-handed and for that
reason, there is no averaging over the initial spin. Moreover, having
the initial lepton (i.e. neutrino) state polarized yields an extra term in the lepton tensor that is 
antisymmetric in Lorentz indices $\mu$ and $\nu$,
\begin{eqnarray}
L_{\mu\nu}^{\left(\nu\right)} & = & 8\left[k_{\mu}k'_{\nu}+k_{\nu}k'_{\mu}-g_{\mu\nu}
\left(k\cdot k'\right)+i\epsilon_{\mu\nu\sigma\tau}k^{\sigma}k'^{\tau}\right].
\label{eq:leptonictensorneutrinodis}
\end{eqnarray}
In constructing the corresponding hadron tensor we have to respect that the parity in the weak 
interaction is no longer conserved. Thus, due to the presence of the parity-violating term 
$i\epsilon^{\mu\nu\alpha\beta}P_{\alpha}q_{\beta}$ in the hadron tensor, a third response 
function $W_{3}\left(\nu,Q^{2}\right)$ arises in the expression for the cross section. 
In analogy with Eq. (\ref{eq:dissigma1}), along with the factor replacement 
$\alpha/Q^{2}\rightarrow G_{F}/4\pi\sqrt{2}$, where 
$G_{F}\equiv g^{2}/4\sqrt{2}M_{W}^{2}\simeq1.166\cdot10^{-5}\;\mathrm{GeV}^{-2}$ 
is the so-called Fermi constant with \emph{g} being the weak interaction constant and 
$M_{W}$ the mass of the \emph{W}-boson (note that one chooses the coupling constant of 
the \emph{W}-boson to the nucleon as $g/2\sqrt{2}$, and that the factor $1/M_{W}^{2}$ 
in the definition of $G_{F}$ comes from the \emph{W}-boson propagator considered in the 
limit $Q^{2}\ll M_{W}^{2}$, in other words, one deals with the Fermi contact interaction), 
the laboratory differential cross section is
\begin{eqnarray}
\frac{d^{2}\sigma_{\left(\nu N\right)}}{d\Omega d\omega'} & = & \frac{G_{F}^{2}
\omega'^{2}}{2\pi^{2}}
\left[2W_{1}^{\left(\nu N\right)}\left(\nu,Q^{2}\right)\sin^{2}
\left(\theta/2\right)+W_{2}^{\left(\nu N\right)}\left(\nu,Q^{2}\right)\cos^{2}\left(\theta/2\right)
\right.\nonumber \\
&  & \left.+\frac{\omega+\omega'}{2M}W_{3}^{\left(\nu N\right)}\left(\nu,Q^{2}\right)\sin^{2}
\left(\theta/2\right)\right].
\label{eq:neutrinodissigma1}
\end{eqnarray}
Finally, in terms of the Lorentz invariant variables $x_{B}$ and \emph{y}, the invariant
cross section reads
\begin{eqnarray}
\frac{d^{2}\sigma_{\left(\nu N\right)}}{dx_{B}dy} & = & \frac{G_{F}^{2}\left(s-M^{2}\right)}{2\pi}
\left[x_{B}y^{2}F_{1}^{\left(\nu N\right)}
\left(x_{B},Q^{2}\right)+\left(1-y\right)F_{2}^{\left(\nu N\right)}
\left(x_{B},Q^{2}\right)\right.\nonumber \\
&  & \left.+y\left(1-\frac{y}{2}\right)F_{3}^{\left(\nu N\right)}\right].
\label{eq:finalresultinvarianneutrinodis}
\end{eqnarray}
Similarly to the electromagnetic case, the dimensionless structure functions in the weak interaction 
sector are defined as follows
\begin{eqnarray}
F_{1}^{\left(\nu N\right)}\left(x_{B},Q^{2}\right) & \equiv & MW_{1}^{\left(\nu N\right)}
\left(\nu,Q^{2}\right),\nonumber \\
F_{2}^{\left(\nu N\right)}\left(x_{B},Q^{2}\right) & \equiv & \nu W_{2}^{\left(\nu N\right)}
\left(\nu,Q^{2}\right),\nonumber \\
F_{3}^{\left(\nu N\right)}\left(x_{B},Q^{2}\right) & \equiv & \nu W_{3}^{\left(\nu N\right)}
\left(\nu,Q^{2}\right).
\label{eq:dimensionlessstructurefunctionsneutrino}
\end{eqnarray}

\section{\ \,  Bjorken Scaling and the Parton Model}

The nucleon structure with the size of the order $\lambda<0.2\;\mathrm{fm}$
can only be resolved at sufficiently large $Q^{2}\geq1\;\mathrm{GeV}^{2}$.
Having large $Q^{2}$ also implies large values of $\nu$, more precisely
$\nu^{2}\gg Q^{2}$, since the ratio $x_{B}=Q^{2}/2M\nu$ is finite and
bounded between 0 and 1. Accordingly, a massive hadronic state \emph{X}
is produced with the invariant mass equal to
\begin{eqnarray}
W^{2} & = & Q^{2}\left(1-x_{B}\right)/x_{B}+M^{2}\gg M^{2}.
\label{eq:invariantmasssquaredindisregion}
\end{eqnarray}
It lies well above the resonance region, $W\geq2\;\mathrm{GeV}$. In this kinematical regime,
known as the deeply inelastic region, the structure functions $F\left(x_{B},Q^{2}\right)$ extracted 
from the measured inelastic cross sections do not depend significantly on $Q^{2}$ but rather 
only on $x_{B}$. This interesting feature is termed Bjorken 
scaling \cite{Bjorken:1968dy,Bloom:1969kc,Breidenbach:1969kd,Friedman:1972sy}. It can be stated in 
the following way: in the Bjorken limit, where both $Q^{2}\rightarrow\infty$ and $\nu\rightarrow\infty$ 
whereas $x_{B}$ is kept fixed, one experimentally observes
\begin{eqnarray}
F\left(x_{B},Q^{2}\right) & \longrightarrow & F\left(x_{B}\right).
\label{eq:scaling}
\end{eqnarray}

The fact that, for sufficiently large values of $Q^{2}$, the structure
functions are independent of $Q^{2}$ implies that the nucleon, or any
other hadron, is made of point-like constituents. Recall that a finite size object 
must have a form factor and hence introduce some dependence on $Q^{2}$. 
Furthermore, since the structure functions are Lorentz invariant, one can study the
scattering process in any reference frame. However, the description is considerably
simplified, if we look at the nucleon in a very fast moving system 
with its momentum approaching to infinity along the \emph{z}-direction,
i.e. in the so-called infinite momentum frame (the physics of the process is, of course, 
independent of this choice). Then the transverse momenta, the rest masses of the 
nucleon constituents and, for consistency, the nucleon mass can be neglected and accordingly, 
the nucleon momentum is $P^{\mu}=\left(P_{z},\vec{0}_{\bot},P_{z}\right)$. In other words, 
the structure of the nucleon is described only in terms of the longitudinal momenta of its constituents. 
This is the basis of the parton model of Feynman \cite{Feynman:1969ej}, which gives the clearest 
physical interpretation of scaling. In the parton picture, the nucleon is viewed as a collection 
of noninteracting, point-like constituents, the partons. The interaction of the electron with the 
nucleon can be then viewed as the incoherent sum of interactions between the 
electron and the individual partons. This approximation is valid as long as 
the duration of the electron-parton interaction, which is regarded as elastic scattering, 
is so short that the interaction between the partons themselves can be safely neglected. 
Thus we consider the scattering process in the impulse approximation by picking up only the 
lowest-order electromagnetic contribution, and neglect all the QCD corrections associated with the 
exchange or emission of gluons.

In the infinite momentum frame, a given parton is characterized by
the longitudinal momentum fraction $x\in\left[0,1\right]$ of
the total nucleon momentum $P$,
\begin{eqnarray}
p & = & xP.
\label{eq:partonlongitudinalmomentum}
\end{eqnarray}
In addition, for each parton species \emph{a}, we define the parton
distribution function (PDF) $f_{a/N}\left(x\right)$. It describes
the probability of finding a parton of type \emph{a} at the longitudinal
momentum fraction \emph{x} inside the target nucleon \emph{N}. Note that 
these functions cannot be computed using QCD perturbation theory.
All fractions \emph{x} have to add up to 1 and hence the
normalization
\begin{eqnarray}
\sum_{a}\int_{0}^{1}dx\; xf_{a/N}\left(x\right) & = & 1
\label{eq:normalization}
\end{eqnarray}
holds. The scattered parton has the momentum $p'=p+q$, given by the
four-momentum conservation, where $q$ is the momentum of the virtual photon. 
Since the parton is on its mass shell, it follows that
\begin{eqnarray}
p'{}^{2} & = & \left(xP+q\right)^{2}\nonumber \\
& = & x^{2}M^{2}+2x\left(P\cdot q\right)-Q^{2}\nonumber \\
& \approx & 0.
\label{eq:scatteredpartonmass}
\end{eqnarray}
Neglecting $x^{2}M^{2}$ compared to $Q^{2}$ and $\nu$ (recall that $\left(P\cdot q\right)=M\nu$) yields
\begin{eqnarray}
x & = & \frac{Q^{2}}{2\left(P\cdot q\right)}\equiv x_{B}.
\label{eq:momentumfraction}
\end{eqnarray}
The longitudinal momentum fraction is found to be identical to the
Bjorken scaling variable. It means that the parton must have the 
fraction $x_{B}$ of the nucleon momentum in order to absorb the virtual photon. 
It should be emphasized, however, that the variable $x_{B}$ has this simple meaning only 
in the infinite momentum frame.

\begin{figure}
\centerline{\epsfxsize=3in\epsffile{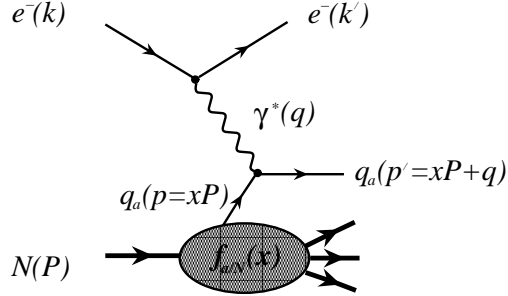}}
\caption{Kinematics of deeply inelastic electron-nucleon scattering in the parton model.}
\label{dispartonmodel}
\end{figure}

Due to the fact that the partons are point-like and noninteracting, the DIS cross section for 
nucleon-electron scattering in the parton picture is simply given by the incoherent sum over 
all the contributing partons (i.e. the partonic cross sections for elastic scattering of an 
electron from the individual partons), weighted with the proper distribution functions. 
The fundamental relation, known as the parton model master formula, reads
\begin{eqnarray}
\sigma\left[e^{-}\left(k\right)N\left(P\right)\rightarrow e^{-}\left(k'\right)X\right] & = & 
\sum_{a}\int_{0}^{1}dx\; f_{a/N}\left(x\right)\nonumber \\
&  & \times\sigma\left[e^{-}\left(k\right)q_{a}\left(xP\right)\rightarrow e^{-}
\left(k'\right)q_{a}\left(p'\right)\right],
\label{eq:totalcrosssectiondeepinelasticscattering}
\end{eqnarray}
where the sum runs over quarks and antiquarks, in order words, over all charged partons because 
they are the ones that interact with the virtual photon. Schematically, see  Fig. \ref{dispartonmodel}, 
the hadronic interaction is broken into a PDF part, represented by the blob, and hard (parton-electron) 
scattering part. According to the factorization, PDFs do not interfere with the hard scattering part. 
For that reason, they are universal in a sense that they are same for all inclusive hard scattering processes, 
not only for electromagnetic DIS \cite{Sterman:1995fz}.

Let us work out the explicit leading-order formula for the DIS cross
section using Eq. (\ref{eq:totalcrosssectiondeepinelasticscattering}). First we calculate,
at the most elementary level, the cross section for electron scattering
from a single parton. For $M^{2}\ll -u,s,Q^{2}$, where $u\equiv\left(k'-P\right)^{2}$
is the Mandelstam variable, one has
\begin{eqnarray}
Q^{2}=x_{B}ys & \mathrm{and} & y=\frac{s+u}{s},
\label{eq:Qsquaredandy}
\end{eqnarray}
and the invariant form of the partonic double differential cross section
is (see, e.g. \cite{Greiner1,Peskin:1995ev})
\begin{eqnarray}
\frac{d^{2}\sigma\left(e^{-}q_{a}\rightarrow e^{-}q_{a}\right)}{dx_{B}dy} & = & 
\frac{2\pi\alpha^{2}Q_{a}^{2}s}{Q^{4}}\left[1+\left(1-y\right)^{2}\right]x_{B}
\delta\left(x-x_{B}\right).
\label{eq:partoniccrosssection}
\end{eqnarray}
The electric charge of the parton of type \emph{a} in units of $\left|e\right|$ is denoted by $Q_{a}$. 
Substituting Eq. (\ref{eq:partoniccrosssection}) into Eq. (\ref{eq:totalcrosssectiondeepinelasticscattering}) 
results in
\begin{eqnarray}
\frac{d^{2}\sigma\left(e^{-}N\rightarrow e^{-}X\right)}{dx_{B}dy} & = & \frac{2\pi\alpha^{2}s}{Q^{4}}
\left[1+\left(1-y\right)^{2}\right]\sum_{a}Q_{a}^{2}x_{B}f_{a/N}\left(x_{B}\right).
\label{eq:partonmodeldis}
\end{eqnarray}
Comparing now Eqs. (\ref{eq:partonmodeldis}) and (\ref{eq:finalresultinvariantdis}),
under the assumption that $M^{2}\ll s$, gives
\begin{eqnarray}
2\left[x_{B}y^{2}F_{1}\left(x_{B},Q^{2}\right)+\left(1-y\right)F_{2}\left(x_{B},Q^{2}\right)
\right] & = & \left[1+\left(1-y\right)^{2}\right]\sum_{a}Q_{a}^{2}x_{B}f_{a/N}\left(x_{B}
\right).\nonumber \\
\label{eq:relationbetweenformulas}
\end{eqnarray}
We notice that the right-hand side of Eq. (\ref{eq:relationbetweenformulas}) depends only 
on $x_{B}$ and \emph{y}. Therefore,
\begin{eqnarray}
x_{B}y^{2}F_{1}\left(x_{B},Q^{2}\right)+\left(1-y\right)F_{2}
\left(x_{B},Q^{2}\right) & = & \mathrm{const.}\;\;\;\;\;
\mathrm{for}\;\mathrm{all}\; Q^{2}
\label{eq:constant}
\end{eqnarray}
or, 
\begin{eqnarray}
F_{1}\left(x_{B},Q^{2}\right) & = & F_{1}\left(x_{B}\right),\nonumber \\
F_{2}\left(x_{B},Q^{2}\right) & = & F_{2}\left(x_{B}\right).
\label{eq:asymptoticvalues}
\end{eqnarray}
Hence the structure functions in the parton model exhibit the scaling
property, i.e. they do not depend on $Q^{2}$. Conversely, Bjorken scaling suggests the 
asymptotically free quark dynamics, the property that the quark interaction gets weaker at short distances. 
In addition, by comparing powers of \emph{y} in Eq. (\ref{eq:relationbetweenformulas}), we can express 
the \emph{F}-structure functions in terms of PDFs, namely,
\begin{eqnarray}
F_{1}\left(x_{B}\right) & = & \frac{1}{2}\sum_{a}Q_{a}^{2}f_{a/N}\left(x_{B}\right),\nonumber \\
F_{2}\left(x_{B}\right) & = & \sum_{a}Q_{a}^{2}x_{B}f_{a/N}\left(x_{B}\right).
\label{eq:f1andf2expressions}
\end{eqnarray}
Finally, from Eq. (\ref{eq:f1andf2expressions}) one can read off the following simple relation specific 
to the scattering of electrons from massless fermions
\begin{eqnarray}
F_{2}\left(x_{B}\right) & = & 2x_{B}F_{1}\left(x_{B}\right).
\label{eq:callangrossrelation}
\end{eqnarray}
It is known as the Callan-Gross relation \cite{Callan:1969uq}, and it is an important 
evidence that the partons detected in DIS are indeed the spin-1/2 quarks of hadron 
spectroscopy \cite{Sterman:1995fz}. It is worth noting that a more descriptive notation is 
commonly used, in which $f_{a/N}\left(x\right)$ is replaced by the flavor labels $u_{N}\left(x\right), 
d_{N}\left(x\right),s_{N}\left(x\right),\bar{u}_{N}\left(x\right)$, etc. Then the proton structure 
function $F_{1}$ is
\begin{eqnarray}
F_{1p}\left(x\right) & = & \frac{1}{2}\left\{ \frac{4}{9}\left[u_{p}\left(x\right)+\bar{u}_{p}
\left(x\right)\right]+\frac{1}{9}\left[d_{p}\left(x\right)+\bar{d}_{p}\left(x\right)\right]+
\frac{1}{9}\left[s_{p}\left(x\right)+\bar{s}_{p}\left(x\right)\right]\right\},\nonumber \\ 
\label{eq:f1expressionwithflavorindices}
\end{eqnarray}
and similarly for $F_{2p}\left(x\right)$. 

In summary, the parton model serves both to explain scaling and to
identify partons as quarks. However, it is important to add that ultimately,
the success of this simple model must be justified in quantum field theory, 
in particular within the framework of the QCD operator product expansion (OPE). 
It turns out that, in the Bjorken regime, the dominant contribution to $W^{\mu\nu}$, 
see Eq. (\ref{eq:hadronictensorcompactformfinal}), comes from the integration region 
$0\leq z^{2}\leq\mathrm{const.}/Q^{2}$. Accordingly, the DIS cross section is dominated 
by the light-cone region, $z^{2}\rightarrow0$, of the space-time integration 
in Eq. (\ref{eq:hadronictensorcompactformfinal}) \cite{Muta:1987mz}. In fact, the scattering amplitudes 
of other hard inclusive and exclusive processes are also governed by the product of currents 
$J^{\mu}\left(z\right)J^{\nu}\left(0\right)$ near the light-cone. Thus OPE is a powerful tool which, 
not only recovers the parton model predictions but also allows to analyze, in a systematic way, 
the terms in the light-cone expansion contributing to a given power of $1/Q^{2}$.

\chapter{Phenomenological Functions}

\section{\ \,  Introduction}

The fundamental particles, from which hadrons are built, are known.
They are quarks and gluons. The interactions between them are described
by the QCD Lagrangian, which is also established. Unfortunately, knowing
the first principles is not sufficient. We still need to understand
how QCD works, in other words, how to translate the information obtained
from experiments on the hadronic level into the language of quark
and gluon fields. One may, for example, consider projecting these
fields onto hadronic states. The resulting matrix elements can be
interpreted as hadronic wave functions \cite{Radyushkin:2002qt,Radyushkin:2004sr}. 
An alternative approach to describe the hadronic structure is to use different phenomenological
functions. The well-known examples are form factors, usual parton distributions  
functions and distribution amplitudes. Since they have been around for a long time, they are termed 
the old phenomenological functions. We discuss them separately in Sections III.2, III.3 and III.4. 
On the other hand, the concept of generalized parton distributions \cite{Muller:1998fv,Ji:1996ek,
Ji:1996nm,Radyushkin:1996nd,Radyushkin:1996ru,Radyushkin:1997ki} (for reviews, see 
\cite{Radyushkin:2000uy,Goeke:2001tz,Diehl:2003ny} and recently \cite{Belitsky:2005qn}) is new. 
These new phenomenological functions are hybrids of the old ones and therefore, provide a unified 
and more detailed description of the hadronic structure. Different species of generalized 
parton distributions are presented in Section III.5 together with some of their general properties.

\section{\ \,  Form Factors}

Form factors are defined through the matrix elements of electromagnetic
and weak (neutral and charged) currents between the hadronic states.
In particular, the matrix element of the electromagnetic current (for
the weak currents, see Section VII.3),
\begin{eqnarray}
J_{EM}^{\mu}\left(0\right) & = & \sum_{f}Q_{f}\bar{\psi}_{f}\left(0\right)
\gamma^{\mu}\psi_{f}\left(0\right),
\label{eq:emcurrent}
\end{eqnarray}
where $Q_{f}$ is the electric charge (in units of $\left|e\right|$)
of the quark of flavor \emph{f}, between the nucleon states $N\left(p_{1},s_{1}\right)$
and $N\left(p_{2},s_{2}\right)$ is parametrized in terms of two independent
nucleon electromagnetic form factors known as the Dirac and Pauli
form factors. Namely,
\begin{eqnarray}
\left\langle N\left(p_{2},s_{2}\right)\right|J_{EM}^{\mu}\left(0\right)
\left|N\left(p_{1},s_{1}\right)\right\rangle  & = & \bar{u}\left(p_{2},s_{2}\right)
\left[\gamma^{\mu}F_{1}\left(t\right)-i\sigma^{\mu\nu}\frac{r_{\nu}}{2M}F_{2}
\left(t\right)\right] \nonumber \\
&  & \times u\left(p_{1},s_{1}\right),
\label{eq:formfactors1}
\end{eqnarray}
where $\bar{u}\left(p_{2},s_{2}\right)$ and $u\left(p_{1},s_{1}\right)$ are the Dirac spinors, 
$r=p_{1}-p_{2}$ is the overall momentum transfer and the invariant $t=r^{2}$. The elastic 
form factors should not be confused with the dimensionless structure functions 
$F_{1}\left(x_{B},Q^{2}\right)$ and $F_{2}\left(x_{B},Q^{2}\right)$ introduced in the preceding 
chapter. Note also that for elastic scattering, we have $Q^{2}=-t$. The nucleon mass \emph{M} in 
Eq. (\ref{eq:formfactors1}) is introduced only for dimensional convenience. Similarly to the flavor 
decomposition of the electromagnetic current given by Eq. (\ref{eq:emcurrent}), 
the Dirac and Pauli form factors can also be expressed in terms of their flavor components,
\begin{eqnarray}
F_{1}\left(t\right) & = & \sum_{f}Q_{f}F_{1f}\left(t\right),\nonumber \\
F_{2}\left(t\right) & = & \sum_{f}Q_{f}F_{2f}\left(t\right).
\label{eq:flavordecomposition}
\end{eqnarray}
Their limiting values at $t=0$ are known. The Dirac form factor gives
the total electric charge of the nucleon \emph{N}, $F_{1}\left(t=0\right)=Q_{N}$,
and the Pauli form factor gives its anomalous magnetic moment, 
$F_{1}\left(t=0\right)=\kappa_{N}$. For the proton, we have $\kappa_{p}=1.793$, and for the neutron 
$\kappa_{n}=-1.913$. The $Q^{2}$-dependence of form factors is a clear evidence for the extended structure 
of the nucleon, in particular the charge and current distributions. 

Note that writing the matrix element $\left\langle N\left(p_{2},s_{2}
\right)\right|J_{EM}^{\mu}\left(0\right)\left|N\left(p_{1},s_{1}
\right)\right\rangle$ in  the most general form, 
one might expect to include, in addition to terms $\gamma^{\mu}$
and $\sigma^{\mu\nu}r_{\nu}$, also terms like $\left(p_{1}+p_{2}\right)^{\mu}$,
$r^{\mu}$ and $\sigma^{\mu\nu}\left(p_{1}+p_{2}\right)_{\nu}$. However,
with the help of the Gordon identity,
\begin{eqnarray}
\bar{u}\left(p_{2},s_{2}\right)\gamma^{\mu}u\left(p_{1},s_{1}\right) & = & \frac{1}{2M}\bar{u}
\left(p_{2},s_{2}\right)\left[\left(p_{1}+p_{2}\right)^{\mu}-i
\sigma^{\mu\nu}r_{\nu}\right]u\left(p_{1},s_{1}\right),
\label{eq:gordonidentity}
\end{eqnarray}
one can express the terms $\left(p_{1}+p_{2}\right)^{\mu}$ as the
linear combination of terms $\gamma^{\mu}$ and $\sigma^{\mu\nu}r_{\nu}$.
Moreover, the electromagnetic current conservation, $r_{\mu}J_{EM}^{\mu}=0$,
implies that the term $r^{\mu}$ should vanish and hence the most
general form reduces to the expression (\ref{eq:formfactors1}).

Instead of $F_{1}$ and $F_{2}$, we often introduce the so-called Sachs electric and magnetic 
form factors, $G_{E}$ and $G_{M}$, respectively. Then the matrix element takes the form
\begin{eqnarray}
\left\langle N\left(p_{2},s_{2}\right)\right|J_{EM}^{\mu}\left(0\right)\left|N
\left(p_{1},s_{1}\right)\right\rangle  & = & \bar{u}\left(p_{2},s_{2}\right)
\left[\gamma^{\mu}\frac{G_{E}\left(t\right)-\left(t/4M^{2}\right)G_{M}
\left(t\right)}{1-t/4M^{2}}\right.\nonumber \\
&  & \left.-i\sigma^{\mu\nu}\frac{r_{\nu}}{2M}\frac{G_{M}\left(t\right)-G_{E}
\left(t\right)}{1-t/4M^{2}}\right]u\left(p_{1},s_{1}\right),\nonumber \\
\label{eq:emcurrentandgegm}
\end{eqnarray}
where
\begin{eqnarray}
G_{E}\left(t\right) & = & F_{1}\left(t\right)+\frac{t}{4M^{2}}F_{2}
\left(t\right),
\label{eq:sachelectricff}
\end{eqnarray}
and
\begin{eqnarray}
G_{M}\left(t\right) & = & F_{1}\left(t\right)+F_{2}\left(t\right).
\label{eq:sachmagneticff}
\end{eqnarray}
These form factors are normalized at $Q^{2}=0$ in the following way
\begin{eqnarray}
G_{Ep}\left(t=0\right) & = & 1,\nonumber \\
G_{Mp}\left(t=0\right) & = & \mu_{p}=2.793
\label{eq:protonsachffs}
\end{eqnarray}
for the proton and
\begin{eqnarray}
G_{En}\left(t=0\right) & = & 0,\nonumber \\
G_{Mn}\left(t=0\right) & = & \mu_{n}=-1.913
\label{eq:neutronsachffs}
\end{eqnarray}
for the neutron. Their magnetic moments are given in terms of the nuclear magneton, 
$\mu=\left|e\right|/2M_{p}=5.051\cdot10^{-27}\;\mathrm{A}\mathrm{m}^{2}$ in the SI units.

\begin{figure}
\centerline{\epsfxsize=3in\epsffile{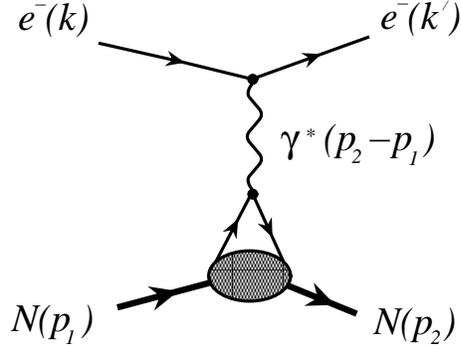}}
\caption{Elastic electron-nucleon scattering in the one-photon exchange approximation.}
\label{ffs}
\end{figure}

The nucleon electromagnetic form factors can be measured through elastic
electron-nucleon scattering,
\begin{eqnarray}
e^{-}\left(k\right)+N\left(p_{1}\right) & \longrightarrow & e^{-}
\left(k'\right)+N\left(p_{2}\right).
\label{eq:elasticscatteringreaction}
\end{eqnarray}
The process is shown in the one-photon exchange approximation in Fig. \ref{ffs}. 

\section{\ \,  Parton Distribution Functions}

Parton distribution functions (PDFs) are defined through the forward matrix 
elements of the light-like correlation functions, i.e. the quark and gluon fields 
separated by light-like distances. For the unpolarized case we have
\begin{eqnarray}
\left\langle N\left(P,S\right)\right|\bar{\psi}_{f}\left(-z/2\right)\gamma^{\mu}\psi_{f}
\left(z/2\right)\left|N\left(P,S\right)\right\rangle _{z^{2}=0}=
\bar{u}\left(P,S\right)\gamma^{\mu}u\left(P,S\right)\nonumber \\
\times\int_{0}^{1}dx\;
\left[e^{ix\left(P\cdot z\right)}f_{N}\left(x\right)-e^{-ix\left(P\cdot z\right)}
\bar{f}_{N}\left(x\right)\right],
\label{eq:unpolPDFs}
\end{eqnarray}
and for the polarized one
\begin{eqnarray}
\left\langle N\left(P,S\right)\right|\bar{\psi}_{f}\left(-z/2\right)\gamma^{\mu}
\gamma_{5}\psi_{f}\left(z/2\right)\left|N\left(P,S\right)\right
\rangle _{z^{2}=0}=\bar{u}\left(P,S\right)\gamma^{\mu}\gamma_{5}u\left(P,S\right)\nonumber \\
\times\int_{0}^{1}dx\;\left[e^{ix\left(P\cdot z\right)}
\Delta f_{N}\left(x\right)+e^{-ix\left(P\cdot z\right)}\Delta\bar{f}_{N}\left(x\right)\right].
\label{eq:polPDFs}
\end{eqnarray}
To make connection with GPDs, which are usually discussed in the
region $-1\leq x\leq1$ (the variable $x$ runs from 0 to 1 for quarks and from $-1$ to 0 for 
antiquarks), it is convenient to introduce new distribution functions,
\begin{equation}
\tilde{f}_{N}\left(x\right)=\left\{ \begin{array}{cc}
f_{N}\left(x\right) & x>0\\
-\bar{f}_{N}\left(-x\right) & x<0\\
\end{array}\right.
\label{eq:newfunctionunpol}
\end{equation}
and
\begin{equation}
\Delta\tilde{f}_{N}\left(x\right)=\left\{ \begin{array}{cc}
\Delta f_{N}\left(x\right) & x>0\\
\Delta\bar{f}_{N}\left(-x\right) & x<0\\
\end{array}\right.
\label{eq:newfunctionpol}
\end{equation}
and alternatively, write the integrals over \emph{x} in Eqs. (\ref{eq:unpolPDFs})
and (\ref{eq:polPDFs}) as
\begin{eqnarray}
\int_{0}^{1}dx\;\left[e^{ix\left(P\cdot z\right)}f_{N}
\left(x\right)-e^{-ix\left(P\cdot z\right)}\bar{f}_{N}\left(x\right)
\right] & = & \int_{-1}^{1}dx\; e^{ix\left(P\cdot z\right)}
\tilde{f}_{N}\left(x\right),\nonumber \\
\int_{0}^{1}dx\;\left[e^{ix\left(P\cdot z\right)}
\Delta f_{N}\left(x\right)+e^{-ix\left(P\cdot z\right)}\Delta\bar{f}_{N}
\left(x\right)\right] & = & \int_{-1}^{1}dx\; e^{ix\left(P\cdot z\right)}
\Delta\tilde{f}_{N}\left(x\right).
\label{eq:newintegrals}
\end{eqnarray}
Furthermore, we observe that the definition of PDFs has the form
of the plane-wave decomposition. Thus it allows to give the momentum-space interpretation. 
For example, $f_{N}\left(x\right)\left(\bar{f}_{N}\left(x\right)\right)$ is the probability to 
find the quark (antiquark) of flavor \emph{f} carrying the momentum $xP$ inside a fast-moving 
nucleon \emph{N} having the momentum $P$.

Parton distribution functions have been intensively studied in hard
inclusive processes for the last three decades. The classic example
in this respect is the deeply inelastic scattering process shown
in Fig. \ref{dis}. Its structure functions are directly expressed
in terms of PDFs. By substituting Eq. (\ref{eq:opticaltheorem}) into the 
expression (\ref{eq:dissigma1}), it is easy to see that the DIS cross section is given, 
via the optical theorem, by the imaginary part of the forward virtual Compton scattering amplitude.

\begin{figure}
\centerline{\epsfxsize=6in\epsffile{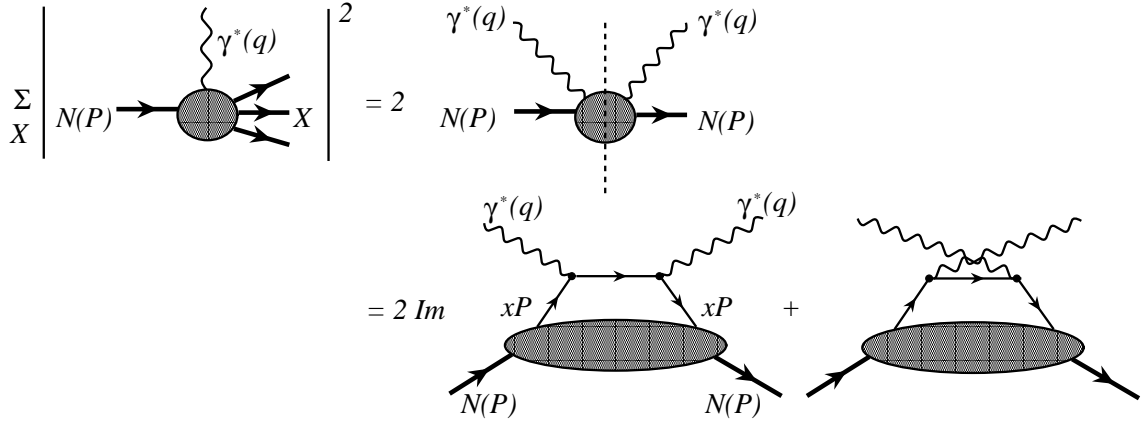}}
\caption[Deeply inelastic electron-nucleon scattering cross section as the imaginary part of the forward 
virtual Compton scattering amplitude.]
{Deeply inelastic electron-nucleon scattering cross section as the imaginary part of the forward 
virtual Compton scattering amplitude. The summation over \emph{X} reflects the inclusive nature of 
the nucleon structure description by parton distribution functions.}
\label{disandoptical}
\end{figure}

Let us consider this amplitude in the specific kinematics known as the Bjorken limit. 
Here the invariant momentum transfer to the nucleon system is sufficiently 
large, $-q^{2}\equiv Q^{2}\rightarrow\infty$, together with large total 
center-of-mass energy of the virtual photon-nucleon system, 
$s\equiv\left(P+q\right)^{2}\rightarrow\infty$, while the 
Bjorken ratio $x_{B}\equiv Q^{2}/2\left(P\cdot q\right)$ is finite. 
In the deeply inelastic region discussed in Section II.3, we have, for instance, 
$Q^{2}\geq1\;\mathrm{GeV}^{2}$ and $W^{2}\equiv s\geq4\;\mathrm{GeV}^{2}$. 
Hence in this particular regime, the behavior of the forward VCA is dominated by short distances, 
i.e. when the separation between the two point-like photon-quark 
vertices (note that the photons couple to the quarks of the nucleon) in the amplitude 
is light-like. As a result, QCD factorization works and the amplitude factorizes into 
a convolution of a perturbatively calculable hard scattering process at the level 
of quarks and gluons, and process independent matrix elements, containing the soft 
nonperturbative information about the nucleon structure. These matrix elements are parametrized 
in terms of PDFs. Schematically, factorization allows to write the leading-order 
amplitude in the form of the so-called handbag diagrams. In these diagrams, two photons 
couple to the same quark line, as illustrated in Fig. \ref{disandoptical}. It is worth noting at this 
point that throughout this thesis we work only to the leading order in the strong coupling 
$\alpha_{s}$. Perturbative corrections produce logarithmic dependence of PDFs on the 
scale $Q^{2}$, in other words, they define the evolution of parton distributions, 
which can be calculated in QCD \cite{Gribov:1972ri,Altarelli:1977zs,Dokshitzer:1977sg}. 

Now taking the imaginary part of the forward VCA generates the delta function,
\begin{eqnarray}
\Im\frac{1}{\left(xP\pm q\right)^{2}+i\epsilon} & = & -\frac{\pi}{2\left(P\cdot q\right)}
\delta\left(x\mp x_{B}\right),
\label{eq:imaginarypartofpropagator}
\end{eqnarray}
which selects two points, $x=\pm x_{B}$, after the integration over
the momentum fraction \emph{x}. Thus in the DIS process we measure 
parton distribution functions $\tilde{f}_{N}\left(x\right)$ and 
$\Delta\tilde{f}_{N}\left(x\right)$ at $x=x_{B}$ corresponding to the quark 
PDFs, and at $x=-x_{B}$ for those of antiquarks. Unlike the form factors, one deals, in the 
case of the parton distribution functions, with a light-like separation instead of a point 
vertex, and also the initial and final nucleon momenta are equal.

In summary, the unpolarized DIS experiments have mapped out the quark and gluon distributions 
in the nucleon while the polarized DIS experiments have shown that quarks carry a small 
fraction of the nucleon spin. As a result, new investigations to understand the nucleon spin 
became necessary.

\section{\ \,  Distribution Amplitudes}

Distribution amplitudes (DAs) (sometimes also referred to as the hadronic wave functions) 
\cite{Radyushkin:1977gp,Efremov:1978rn,Efremov:1979qk,Lepage:1979zb,Lepage:1980fj} describe hadrons 
in hard exclusive scattering processes and therefore, in addition to usual parton distribution 
functions, provide complementary information about the hadronic structure. 
They are defined through the vacuum-to-hadron matrix elements 
$\left\langle 0\right|...\left|P\right\rangle $ of light-cone operators. 
For example, in the pion case we write
\begin{eqnarray}
\left\langle 0\right|\bar{\psi}_{d}\left(-z/2\right)\gamma^{\mu}\gamma_{5}\psi_{u}
\left(z/2\right)\left|\pi^{+}\left(P\right)\right\rangle _{z^{2}=0} & = & iP^{\mu}f_{\pi}
\int_{-1}^{1}d\alpha\; e^{i\alpha\left(P\cdot z\right)/2}\varphi_{\pi^{+}}\left(\alpha\right).
\nonumber \\
\label{eq:pionda}
\end{eqnarray}
The fractions of the pion momentum carried by the quarks are $\left(1\pm\alpha\right)/2$.
One can interpret $\varphi_{\pi^{+}}\left(\alpha\right)$ as the probability
amplitude to find a positive fast-moving pion $\pi^{+}$ in a quark-antiquark
state $\bar{u}d$, with the longitudinal pion momentum $P$ shared
in fractions $\left(1+\alpha\right)/2$ and $\left(1-\alpha\right)/2$.
The distribution $\varphi_{\pi^{+}}\left(\alpha\right)$ is an even
function in the relative fraction $\alpha$ \cite{Radyushkin:2000uy}.

\begin{figure}
\centerline{\epsfxsize=6in\epsffile{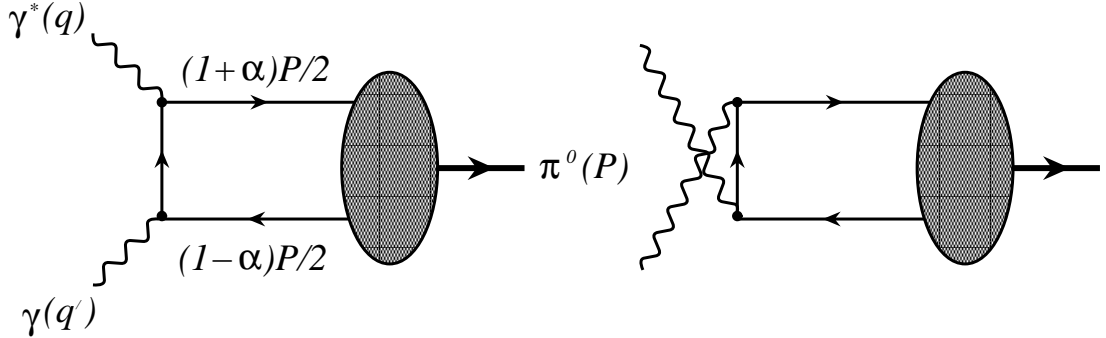}}
\caption[The handbag contribution to the process $\gamma^{*}\gamma\rightarrow\pi^{0}$.]
{The handbag contribution to the process $\gamma^{*}\gamma\rightarrow\pi^{0}$. 
The light-cone dominance is secured by the high virtuality of the incoming photon $\gamma^{*}$.}
\label{photonphotonpi}
\end{figure}

The simplest and cleanest process, in which the pion DA can be accessed, 
is the transition \cite{Lepage:1980fj,Musatov:1997pu}
\begin{eqnarray}
\gamma^{*}\left(q\right)+\gamma\left(q'\right) & \longrightarrow & \pi^{0}\left(P\right).
\label{eq:pionreaction}
\end{eqnarray}
For large virtuality $Q^{2}$, the leading-order contribution 
to the amplitude is given by the handbag diagrams depicted in Fig.
\ref{photonphotonpi}.

\section{\ \,  Generalized Parton Distributions}

The key idea of the GPD approach is the so-called hybridization. As hybrids of 
form factors, PDFs and DAs, generalized parton distributions provide the most complete 
information about the hadronic structure. Parton distributions parametrize the longitudinal 
momentum distributions (in an infinite momentum frame) of partons in the hadron while 
the Fourier transforms of form factors in impact parameter space (impact parameter measures 
the transverse distance of the struck parton from the hadron center) describe the transverse 
coordinate distributions of the hadron's constituents \cite{Burkardt:2000za,Burkardt:2002hr}. 
Generalized parton distributions, on the other hand, encapsulate at the same time the 
longitudinal momentum and transverse coordinate distributions, 
thereby providing a much more detailed and comprehensive, three-dimensional snapshot of 
the substructure of the hadron. In addition, the universality of GPDs allows to 
develop a unified description of wide variety of different hard processes, both inclusive 
and exclusive. 

Generalized parton distributions can be, in general, divided into two groups, namely, 
the skewed parton distributions (SPDs) and the double distributions (DDs). 

\subsection{\ \,  Skewed Parton Distributions} 

\begin{figure}
\centerline{\epsfxsize=6in\epsffile{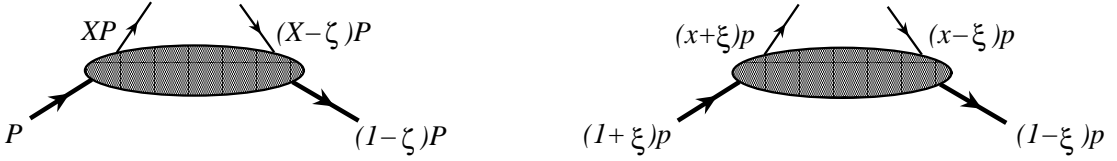}}
\caption{Comparison between the descriptions in terms of nonforward and off-forward parton 
distribution functions.}
\label{schemes}
\end{figure}

There are two implementations of the SPD formalism, as illustrated
in Fig. \ref{schemes}. The nonforward parton distributions (NFPDs) 
\cite{Radyushkin:1997ki} have the advantage of using the variables similar
to those of the usual PDFs. The distributions depend on $X$, the
fraction of the plus component of the light-cone momentum $P^{+}$ of the hadron 
carried by the parton (the plus and minus light-cone components are defined by 
$a^{\pm}\equiv\left(a^{0}\pm a^{3}\right)/\sqrt{2}$ for any Lorentz four-vector \emph{a}), 
or alternatively, the longitudinal momentum fraction with respect to the initial 
hadron momentum \emph{P}; on the skewness parameter $\zeta$ specifying the difference 
between the initial and final hadron plus momenta, $r^{+}=\zeta P^{+}$; and on the 
invariant momentum transfer $t=r^{2}$. For instance, the nonforward parton distribution 
$F_{\zeta}^{f}\left(X,t\right)$ is the probability amplitude that the initial fast-moving hadron, 
having longitudinal momentum $P^{+}$, emits a parton of flavor \emph{f} carrying the momentum 
$XP^{+}$ while the final hadron, having longitudinal momentum $\left(1-\zeta\right)P^{+}$, 
absorbs a parton of flavor \emph{f} carrying the momentum $\left(X-\zeta\right)P^{+}$. 
In this particular scheme, the initial and final hadron momenta are not treated symmetrically. 
The off-forward parton distributions (OFPDs) \cite{Ji:1996ek,Ji:1996nm}, 
on the other hand, use symmetric variables expressed 
in terms of the average hadron momentum, $p=\left(p_{1}+p_{2}\right)/2$, with $p_{1}$ 
being the momentum of the initial hadron and $p_{2}$ the momentum of the final one. 
In the symmetric scheme the hadron longitudinal (the plus component) 
momenta are $\left(1\pm\xi\right)p^{+}$ and accordingly, those of the active partons 
become $\left(x\pm\xi\right)p^{+}$. Similarly to NFPDs, the off-forward
distributions are defined for each quark flavor, and are the functions of three variables, 
namely, the light-cone momentum fraction $x$, the skewness, $\xi\equiv r^{+}/2p^{+}$, 
(here the skewness is introduced as the coefficient of proportionality between the light-cone 
plus components of the momentum transfer and the average hadron momentum, and like $\zeta$ 
varies between 0 and 1) and the invariant \emph{t}. In fact, one should bear in mind that both 
the nonforward and off-forward parton distributions also depend weakly (i.e. logarithmically) on 
the probing scale $Q^{2}$.

The two pairs of variables $\left(X,\zeta\right)$ and $\left(x,\xi\right)$
are related to each other. The conversions between them are 
\begin{eqnarray}
X & = & \frac{x+\xi}{1+\xi}
\label{eq:xandX}
\end{eqnarray}
and
\begin{eqnarray}
\zeta & = & \frac{2\xi}{1+\xi}.
\label{eq:zetaandksi}
\end{eqnarray}
We emphasize that these variables solely characterize the plus, or the longitudinal, components of 
the momenta of the partons involved, however, their transverse momenta are related to the \emph{t}-dependence 
of GPDs. The latter is driven by the \emph{t}-dependence of the corresponding elastic form factors. 
Thus it is possible to access simultaneously the longitudinal momentum and transverse position of the 
parton in the infinite momentum frame \cite{Burkardt:2000za,Burkardt:2002hr}. Moreover, 
by removing the parton with the light-cone momentum 
fraction $x+\xi$ from the hadron and replacing it, at some later point on the light-cone, 
with the parton of the momentum fraction $x-\xi$, one can say that SPDs (or GPDs, in general) measure 
the coherence between two different parton momentum states of the hadron whereas usual PDFs yield 
only the probability that a parton carries a fraction \emph{x} of the hadron momentum. 
In addition to the parton momentum as well as the spin correlations, we can even consider the 
matrix elements corresponding to different hadrons in the initial and final states, e.g. 
the proton-to-neutron transition accessible through the exclusive charged pion electroproduction, 
the proton-to-$\Lambda$ transition in the kaon electroproduction, the nucleon-to-delta transition, 
and hence study flavor nondiagonal GPDs.

Let us focus on the off-forward parton distributions. At the leading, twist-2 level, the hadron 
structure information can be parametrized in terms of two unpolarized OFPDs, $H_{f}\left(x,\xi,t\right)$ 
and $E_{f}\left(x,\xi,t\right)$, and two polarized OFPDs, $\tilde{H}_{f}\left(x,\xi,t\right)$ and 
$\tilde{E}_{f}\left(x,\xi,t\right)$. Since $-1\leq x\leq1$ in Fig. \ref{schemes}, the momentum fractions 
$x\pm\xi$ of the active partons can be either positive or negative. Positive and negative momentum 
fractions correspond to quarks and antiquarks, respectively. Therefore, each OFPD has three distinct 
regions. When $\xi\leq x\leq1$, both partons represent quarks while for $-1\leq x\leq-\xi$, they are both 
antiquarks. In these two regions, the distributions are just a generalization of the usual PDFs. 
In the central region, $-\xi\leq x\leq\xi$, which is often referred to as the mesonic region, 
the parton with a positive momentum $\left(x+\xi\right)p^{+}$ is going out from the blob and represents 
a quark. The returning parton has a negative momentum and therefore, should be treated as an outgoing 
antiquark with the momentum $\left(\xi-x\right)p^{+}$. The total momentum of the quark-antiquark pair, 
$r^{+}=2\xi p^{+}$, is then shared in fractions 
$\left(x+\xi\right)p^{+}=\left(1+x/\xi\right)r^{+}/2$ and 
$\left(\xi-x\right)p^{+}=\left(1-x/\xi\right)r^{+}/2$. In this 
region of \emph{x}, which is not present in deeply inelastic scattering,
OFPDs behave like meson distribution amplitudes with $\alpha=x/\xi$.

Clearly, in the nonforward kinematics, SPDs uncover much richer information about hadronic 
structure, which is not accessible in the DIS process. This new information can be extracted with the 
study of hard exclusive processes, such as deeply exclusive photon or meson electroproduction 
(one refers to the former as deeply virtual Compton scattering or shortly DVCS), which turns out to be a much 
more difficult task due to the small cross sections. Nevertheless, high-energy and high-luminosity 
electron accelerators combined with large acceptance spectrometers give a unique opportunity to perform 
precision studies of such reactions.

\begin{figure}
\centerline{\epsfxsize=6in\epsffile{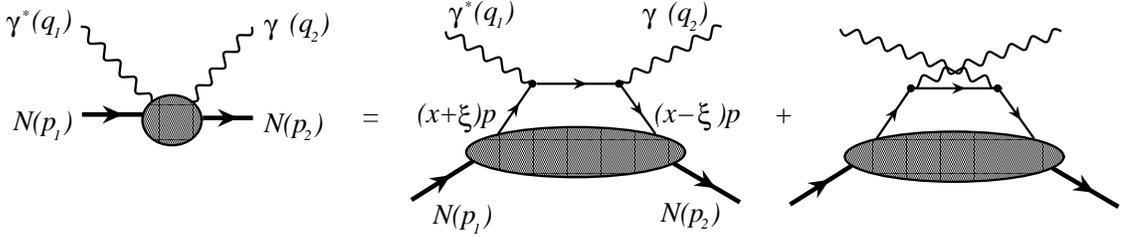}}
\caption[Handbag contribution to the nonforward virtual Compton scattering amplitude.]
{Handbag contribution to the nonforward virtual Compton scattering amplitude. The hard quark propagator 
(this particular quark propagates along the light-cone) in the \emph{s}- and \emph{u}-channel diagrams 
is convoluted with generalized parton distributions.}
\label{nonfvca}
\end{figure}

The factorization into short and long distance dynamics is more general.
Having large space-like virtuality of the initial photon is sufficient
for QCD factorization to work \cite{Ji:1996nm,Radyushkin:1997ki,Ji:1998xh,Collins:1998be}. 
In particular, in the DVCS process, which will be studied 
in detail in Chapter IV, the initial photon is highly virtual, 
$-q_{1}^{2}\rightarrow\infty$, while the final photon is on shell, 
$q_{2}^{2}=0$. In the leading-twist handbag approximation (DVCS is a handbag-dominated 
process for $-q_{1}^{2}$ as low as $2\;\mathrm{GeV}^{2}$) illustrated in Fig. \ref{nonfvca}, 
the hard short-distance part of the so-called nonforward virtual Compton scattering amplitude 
factorizes from the nonperturbative long-distance part. The latter is represented by the lower 
blob and contains now the nonforward matrix elements 
$\left\langle N\left(p_{2}\right)\right|...\left|N\left(p_{1}\right)\right\rangle $ 
of the same quark and gluon operators as in the forward case. These
matrix elements, describing the nucleon structure, are parametrized
in terms of GPDs. Even though the parton picture of DVCS, see Fig. \ref{nonfvca}, 
looks similar to that of DIS, see Fig. \ref{disandoptical},
there are three crucial differences. In DVCS, one deals with the skewed
kinematics, in which the plus momenta of the initial and final hadrons as well as partons 
are not equal. Furthermore, the invariant momentum transfer \emph{t} in DVCS is small but not zero. 
These two extra degrees of freedom, $\zeta$ (or $\xi$) and \emph{t}, make the dynamics of DVCS rich 
and diverse. Finally, in DVCS (or in any other hard exclusive process) the virtual Compton 
scattering amplitude described by GPDs appears at the amplitude level whereas in DIS 
(or in any other inclusive process) the amplitude described by PDFs enters through the optical 
theorem at the level of the cross section.

Skewed parton distributions have interesting properties linking them
to usual PDFs and form factors. In the forward limit, $p_{1}=p_{2}$
and $r=0$, $\xi=0$, $t=0$, they reduce to PDFs obtained from the
DIS process. In particular, the OFPDs $H_{f}$ and $\tilde{H}_{f}$
coincide with the quark density distribution $f_{N}\left(x\right)$
and the quark helicity distribution $\Delta f_{N}\left(x\right)$
given by Eqs. (\ref{eq:newfunctionunpol}) and (\ref{eq:newfunctionpol}). We write the so-called 
reduction formulas,
\begin{equation}
H_{f}\left(x,0,0\right)=\left\{ \begin{array}{cc}
f_{N}\left(x\right) & x>0\\
-\bar{f}_{N}\left(-x\right) & x<0\\
\end{array}\right.
\label{eq:forwardlimit1}
\end{equation}
and
\begin{equation}
\tilde{H}_{f}\left(x,0,0\right)=\left\{ \begin{array}{cc}
\Delta f_{N}\left(x\right) & x>0\\
\Delta\bar{f}_{N}\left(-x\right) & x<0,\\
\end{array}\right.
\label{eq:forwardlimit2}
\end{equation}
while the OFPDs $E_{f}$ and $\tilde{E}_{f}$ have no connections to PDFs. 
They are always accompanied with the momentum transfer \emph{r} and therefore, simply not visible in DIS. 
One can say that $E_{f}$ and $\tilde{E}_{f}$ encode completely new information on the hadron structure, 
which is not accessible in inclusive measurements. Even though they have no analogue in the forward 
limit like $H_{f}$ and $\tilde{H}_{f}$, their limits do exist. 

In the local limit, $z=0$, off-forward parton distributions reduce to the standard vector 
and axial vector form factors. In other words, the first moments of OFPDs, 
obtained by integrating the distributions over \emph{x}, are equal to the nucleon elastic 
form factors (i.e. the Dirac, Pauli, axial and pseudoscalar form factors). 
For any $\xi$ one has the following relations for a particular quark flavor 
\begin{eqnarray}
\int_{-1}^{1}dx\; H_{f}\left(x,\xi,t\right) & = & F_{1f}
\left(t\right),\nonumber \\
\int_{-1}^{1}dx\; E_{f}\left(x,\xi,t\right) & = & F_{2f}
\left(t\right),\nonumber \\
\int_{-1}^{1}dx\;\tilde{H}_{f}\left(x,\xi,t\right) & = & g_{Af}
\left(t\right),\nonumber \\
\int_{-1}^{1}dx\;\tilde{E}_{f}\left(x,\xi,t\right) & = & g_{Pf}
\left(t\right).
\label{eq:sumrules}
\end{eqnarray}
We call these relations the sum rules. It is important to note that
these sum rules are model and $\xi$-independent (the dependence on $\xi$ drops out 
after integration over \emph{x}). 

Generalized parton distributions are also relevant for the nucleon spin structure and have 
received considerable attention in recent years in connection with the so-called proton spin puzzle 
(for reviews, see, e.g. \cite{Lampe:1998eu,Filippone:2001ux,Bass:2004xa}). 
Namely, certain low moments of GPDs can be related to the total angular momentum carried by quarks 
and gluons (or generically, partons) in the nucleon \cite{Ji:1996ek}. In particular, 
in terms of the off-forward distributions, the second moment of the unpolarized OFPDs at $t=0$ 
gives the quark total angular momentum,
\begin{eqnarray}
J_{q} & = & \frac{1}{2}\sum_{f}\int_{-1}^{1}dx\; x
\left[H_{f}\left(x,\xi,t=0\right)+E_{f}\left(x,\xi,t=0\right)\right].
\label{eq:quarkangularmomentum}
\end{eqnarray}
The above equation is independent of $\xi$ (again the $\xi$-dependence of $H_{f}$ and $E_{f}$ 
is removed by integration over \emph{x}). The quark angular momentum, on the other hand, 
decomposes into the quark intrinsic spin $\Delta\Sigma$ and the quark orbital angular momentum 
$L_{q}$,
\begin{eqnarray}
J_{q} & = & \frac{1}{2}\Delta\Sigma+L_{q},
\label{eq:quarkangularmomentumdecomposition}
\end{eqnarray}
where $\Delta\Sigma$ is measured through the polarized DIS process.
Substituting Eq. (\ref{eq:quarkangularmomentum}) into 
Eq. (\ref{eq:quarkangularmomentumdecomposition}) we can determine 
$L_{q}$. Moreover, the total spin of the nucleon comes from quarks and gluons,
\begin{eqnarray}
\frac{1}{2} & = & J_{q}+J_{g},
\label{eq:totalnucleonspin}
\end{eqnarray}
where $J_{g}$ is the total angular momentum carried by gluons. Thus we can also extract the gluon 
contribution to the nucleon spin. Note that $\Delta\Sigma/2$ accounts only for approximately $30$ \% 
of the nucleon spin. In summary, by measuring GPDs we obtain information about the 
angular momentum distributions of quarks and gluons in the nucleon. 
Therefore, the DVCS process, combined with measurements of the 
quark helicity distributions from inclusive deeply inelastic scattering, can unravel the orbital 
angular momentum carried by partons, on which little or no information is currently available.

\subsection{\ \,  Double Distributions}

\begin{figure}
\centerline{\epsfxsize=3in\epsffile{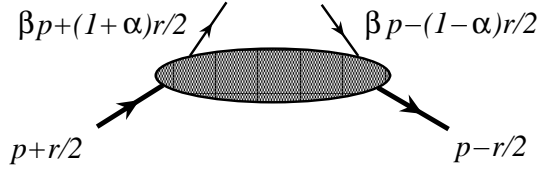}}
\caption{Description in terms of double distributions.}
\label{DDscheme}
\end{figure}

Two approaches are used to model generalized parton distributions:
\begin{itemize}
\item A direct calculation of GPDs in specific dynamical models, such as
the bag model \cite{Ji:1997gm}, the chiral soliton model \cite{Petrov:1998kf}, 
the light-cone formalism \cite{Diehl:1998kh}, etc.
\item A phenomenological construction \cite{Radyushkin:1998es,Mankiewicz:1997uy,Musatov:1999xp} 
based on reduction formulas relating GPDs to PDFs 
$f_{N}\left(x\right)$ and $\Delta f_{N}\left(x\right)$ and form factors 
$F_{1}\left(t\right)$, $F_{2}\left(t\right)$, $g_{A}\left(t\right)$ and 
$g_{P}\left(t\right)$.
\end{itemize}
The most convenient way to construct models in the second approach, 
and further study the interplay between \emph{x}, $\xi$ and \emph{t} dependencies of GPDs 
is performed using the formalism of double distributions (DDs) 
$f\left(\beta,\alpha,t\right)$. Here we only consider the so-called $\alpha$-DDs 
corresponding to the symmetric description with respect to the initial and 
final hadron momenta. In the parton picture of double distributions shown in Fig. \ref{DDscheme}, 
the active parton momentum, $k^{+}=\beta p^{+}+\left(1+\alpha\right)r^{+}/2$, is represented 
as the sum of two components $\beta p^{+}$ and $\left(1+\alpha\right)r^{+}/2$.
The former specifies the momentum flow in the \emph{s}-channel due to the plus component of the 
average hadron momentum $p^{+}$, and the latter specifies the momentum flow in the \emph{t}-channel 
due to the plus component of the momentum transfer $r^{+}$. Despite the proportionality between 
$r^{+}$ and $p^{+}$, they correspond to the momentum fluxes in two different channels. Their 
superposition is the main feature of double distributions. In addition, it is important to note 
another characteristic feature, i.e. the absence of the $\xi$-dependence in $f\left(\beta,\alpha,t\right)$.

Thus double distributions are hybrids, which look like usual parton distributions with respect to 
the variable $\beta$, and like distribution amplitudes with respect to $\alpha$. 
Therefore, when modeling DDs we usually represent a double distribution in the factorized form as the 
product of a usual PDF in the $\beta$-direction and a distribution amplitude, which drives 
the $\alpha$-profile. The connection between the DD variables $\alpha$ and $\beta$ and the OFPD 
variables \emph{x} and $\xi$ can be established through the formula $r^{+}=2\xi p^{+}$, namely, 
$x=\beta+\xi\alpha$.
\begin{figure}
\centerline{\epsfxsize=4in\epsffile{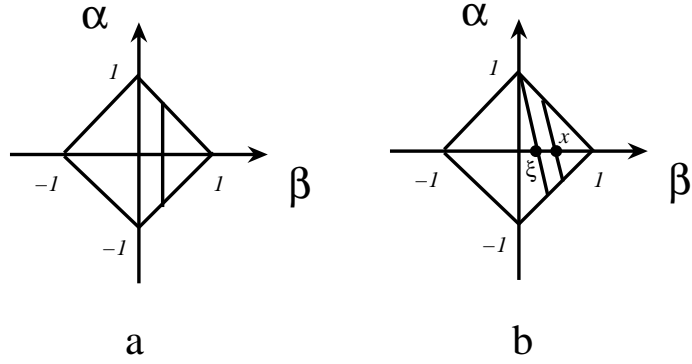}}
\caption{Conversion from a double distribution to an off-forward parton distribution function with a zero 
and nonzero skewness. }
\label{scan}
\end{figure}

The support area of double distributions is the rhombus $\left|\alpha\right|+\left|\beta\right|\leq1$. 
Moreover, due to the hermiticity properties and time-reversal invariance of the nonforward matrix 
elements, they are even functions in $\alpha$, 
$f\left(\beta,\alpha,t\right)=f\left(\beta,-\alpha,t\right)$. This property is termed as the Munich 
symmetry \cite{Mankiewicz:1997uy}. Now, in order to obtain the usual PDFs from DDs (recall that the 
former correspond to the forward limit, $\xi=0$ and $t=0$, of OFPDs), one should simply integrate them along 
the vertical lines $\beta=x$, see the rhombus (a) in Fig. \ref{scan}. On the other hand, to get OFPDs 
$H\left(x,\xi,t\right)$ (we skip the flavor index for convenience) with a nonzero skewness 
(e.g. for $\xi=1$, they behave like meson distribution amplitudes), one should integrate DDs 
$f\left(\beta,\alpha,t\right)$ along the parallel lines $\alpha=\left(x-\beta\right)/\xi$ with a 
$\xi$-dependent slope,
\begin{eqnarray}
H\left(x,\xi,t\right) & = & \int_{-1}^{1}d\beta
\int_{-1+\left|\beta\right|}^{1-\left|\beta\right|}d\alpha\;\delta
\left(\beta+\xi\alpha-x\right)f\left(\beta,\alpha,t\right),
\label{eq:relationofpdsdds}
\end{eqnarray}
as illustrated in the rhombus (b) in Fig. \ref{scan}. We call this process of integration or scanning, 
the DD-tomography. To summarize, double distributions are primary objects producing SPDs after appropriate 
integration.

\chapter{Deeply Virtual Compton Scattering}

\section{\ \,  Introduction}

\begin{figure}
\centerline{\epsfxsize=6in\epsffile{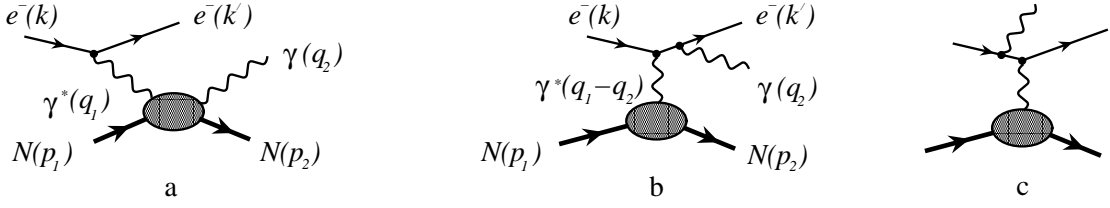}}
\caption{DVCS (a) and Bethe-Heitler (b and c) diagrams contributing to 
electroproduction of a real photon.}
\label{DVCSdiagrams}
\end{figure}

In recent years, significant effort was made to access GPDs through
the measurement of hard exclusive leptoproduction processes, such as deeply virtual Compton scattering 
or deeply exclusive meson production. The simplest process in this respect is DVCS. It can be accessed 
through the following reaction
\begin{eqnarray}
e^{-}\left(k\right)+N\left(p_{1}\right) & \longrightarrow & e^{-}\left(k'\right)+N
\left(p_{2}\right)+\gamma\left(q_{2}\right),
\label{eq:DVCSreaction}
\end{eqnarray}
and it is illustrated in Fig. \ref{DVCSdiagrams}. There are three relevant
diagrams. The nucleon blob with two photon legs, see the diagram (a),
represents the virtual Compton scattering amplitude, which will eventually 
become the subject of our study. This diagram is referred to as the DVCS diagram or 
the Compton contribution. Unfortunately, the final real photon 
can be emitted not only by the nucleon, but also by the electron. The latter is 
presented by the remaining two diagrams. They are referred to as the Bethe-Heitler 
contribution. Here lower part, the nucleon blob, stands for the electromagnetic form factor while 
the upper part can be exactly calculated in QED. Despite this disadvantage 
(in measuring of the VCA by extracting it from the cross section, the pure DVCS process is 
always in competition with the Bethe-Heitler process), in addition to a small cross section, deeply 
virtual Compton scattering is still regarded to be the cleanest tool to access the underlying GPDs. 

At this point, we focus only on the Compton part. Hence an electron (or muon) scatters off 
a nucleon via the exchange (in the leading-order QED) of a space-like photon with virtuality 
$q_{1}^{2}=\left(k-k'\right)^{2}<0$, producing an intact nucleon (with altered momentum) 
and a real photon in the final state. Since we turn a virtual photon into a real one there is 
always a nonzero momentum transfer. At the quark level, in leading twist, the electromagnetic current 
couples to different quark species with strength proportional to the squares of the quark charges, 
selecting specific linear combinations of GPDs. Flavor-specific GPDs can be reconstructed by considering 
DVCS from different hadrons (protons and neutrons, for instance), and using isospin or flavor symmetry 
to relate GPDs in the proton to those in the neutron.

\begin{figure}
\centerline{\epsfxsize=4in\epsffile{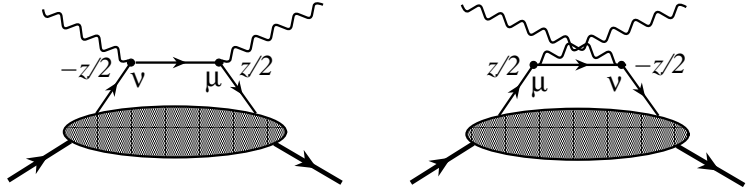}}
\caption{Handbag diagrams (\emph{s}- and \emph{u}-channel) in the virtual Compton scattering amplitude.}
\label{handbagdiagramsDVCS}
\end{figure}

In the Bjorken regime (recall that in addition to the large invariant mass of 
the photon-nucleon system, $\left(p_{1}+q_{1}\right)^{2}\rightarrow\infty$, 
the initial photon is also highly virtual, nevertheless, their ratio 
$x_{B}\equiv-q_{1}^{2}/2\left(p_{1}\cdot q_{1}\right)$ is finite), the VCA is 
dominated by light-like distances. The dominant light-cone singularities, 
which generate the leading power contributions in $1/\left|q_{1}^{2}\right|$ to 
the amplitude, are represented by two handbag diagrams shown in Fig. \ref{handbagdiagramsDVCS}, 
in which the (hard) quark propagator is convoluted with the soft function parametrized in terms of GPDs. 
In addition, keeping the momentum transfer squared to the nucleon, $t\equiv\left(p_{1}-p_{2}\right)^{2}$, 
as small as possible, one arrives at the DVCS kinematics, $s>-q_{1}^{2}\gg-t$. 
This particular kinematics implies, on one hand, that $-q_{1}^2$ should be large enough 
to ensure scaling regime for the amplitude and, on the other hand, it implies small \emph{t}.  

One of the methods to study the virtual Compton scattering amplitude in the DVCS kinematics is the 
approach based on the nonlocal light-cone expansion of the product of currents in QCD string operators 
in coordinate space \cite{Balitsky:1987bk}. It will be employed in the present work.

A detailed derivation of the leading-twist (and to the lowest order in $\alpha_{s}$) VCA is provided 
in Section IV.2. In Section IV.3, we discuss the kinematics, common to all DVCS-like reactions. 
After introducing the simple model for nucleon GPDs, see Section IV.4, both the Compton and Bethe-Heitler 
cross sections are estimated in Section IV.5.

\section{\ \,  Virtual Compton Scattering Amplitude}

We begin with an analysis of some general aspects of the VCA. In the
most general nonforward case, the amplitude is given by a Fourier 
transform of the correlation function of two electroweak currents. 
In particular, for the standard virtual Compton process on the nucleon, where both currents are 
electromagnetic one has \cite{Radyushkin:2000uy}
\begin{eqnarray}
T_{EM}^{\mu\nu} & = & i\int d^{4}x\int d^{4}y\; e^{-i
\left(q_{1}\cdot x\right)+i\left(q_{2}\cdot y\right)}\left
\langle N\left(p_{2},s_{2}\right)
\right|T\left\{ J_{EM}^{\mu}\left(y\right)J_{EM}^{\nu}\left(x\right)
\right\} \left|N\left(p_{1},s_{1}\right)
\right\rangle .\nonumber \\
\label{eq:standardVCA1}
\end{eqnarray}
Due to the current conservation, the amplitude is transverse with respect to the incoming virtual 
and outgoing real photon momenta, 
\begin{eqnarray}
T_{EM}^{\mu\nu}q_{1\nu}=0 & \mathrm{and} & q_{2\mu}T_{EM}^{\mu\nu}=0.
\label{eq:currentconservationVCA}
\end{eqnarray}
It will be convenient in the analysis to use symmetric coordinates,
defined by introducing center and relative coordinates of the points
\emph{x} and \emph{y}, $X\equiv\left(x+y\right)/2$ and $z\equiv y-x$.
Accordingly, the amplitude takes the form
\begin{eqnarray}
T_{EM}^{\mu\nu} & = & i\int d^{4}X
\int d^{4}z\; e^{-i\left(q_{1}-q_{2}\right)\cdot X+i\left(q_{1}+q_{2}\right)\cdot z/2}\nonumber \\
&  & \times\left\langle N\left(p_{2},s_{2}\right)\right|T\left\{ J_{EM}^{\mu}
\left(X+z/2\right)J_{EM}^{\nu}\left(X-z/2\right)\right\} \left|N\left(p_{1},s_{1}\right)\right\rangle .
\label{eq:standardVCA2}
\end{eqnarray}
Furthermore, in order to treat the initial and final nucleons in a symmetric manner, 
we introduce, as independent momentum variables, the averages of the photon and nucleon momenta, 
$q\equiv\left(q_{1}+q_{2}\right)/2$ and $p\equiv\left(p_{1}+p_{2}\right)/2$, and the overall momentum 
transfer to the nucleon, $r\equiv p_{1}-p_{2}=q_{2}-q_{1}$. Then we can write 
\begin{eqnarray}
q^{2}=q_{1}^{2}/2-t/4 & \mathrm{with} & t=r^{2}.
\label{eq:qsquaredstandardVCA}
\end{eqnarray}
From the on-mass-shell condition, $p_{1}^{2}=p_{2}^{2}=M^{2}$, it
follows that 
\begin{eqnarray}
p^{2}=M^{2}-t/4 & \mathrm{and} & \left(p\cdot r\right)=0,
\label{eq:frommassshellconditionsstandardVCA}
\end{eqnarray}
where \emph{M} denotes the mass of the nucleon. After performing
the translation in \emph{x} in Eq. (\ref{eq:standardVCA2}), 
$\left\langle p_{2}\right|J^{\mu}\left(x\right)\left|p_{1}\right\rangle 
=\left\langle p_{2}\right|J^{\mu}\left(0\right)\left|p_{1}\right\rangle e^{-i\left(p_{1}-p_{2}
\right)\cdot x}$, and integrating over the center coordinate, one finds 
\begin{eqnarray}
T_{EM}^{\mu\nu} & = & \left(2\pi\right)^{4}\delta^{\left(4\right)}\left(p_{1}+q_{1}-p_{2}-q_{2}\right)
\mathsf{\mathcal{T}}_{EM}^{\mu\nu},
\label{eq:standardVCA3}
\end{eqnarray}
where
\begin{eqnarray}
\mathsf{\mathcal{T}}_{EM}^{\mu\nu} & = & i\int d^{4}z\; e^{i\left(q\cdot z\right)}
\left\langle N\left(p-r/2,s_{2}\right)
\right|T\left\{ J_{EM}^{\mu}\left(z/2\right)J_{EM}^{\nu}
\left(-z/2\right)\right\} 
\left|N\left(p+r/2,s_{1}\right)\right\rangle .\nonumber \\
\label{eq:reducedVCAamplitude}
\end{eqnarray}
We call the amplitude (\ref{eq:reducedVCAamplitude}) the reduced virtual Compton scattering amplitude. 
It appears in the invariant matrix element (or alternatively the T-matrix) of the standard 
electromagnetic DVCS process.

Having introduced the reduced VCA we demonstrate its calculation at
the twist-2 level in the DVCS kinematics, which amounts to neglecting contributions of the 
order $M^{2}/q^{2}$ and $t/q^{2}$ in the hard part of the amplitude, and keeping the \emph{t}-dependence 
only in the soft part. It was noticed that a straightforward use of the twist-2 result for $t\neq0$ 
leads to inconsistencies, e.g. the VCA is not electromagnetic gauge invariant \cite{Vanderhaeghen:1999xj}. 
The invariance is restored through inclusion of the twist-3 corrections to the amplitude. 
They are power suppressed in $q^{2}$, and have been calculated by several groups 
\cite{Anikin:2000em,Penttinen:2000dg,Belitsky:2000vx,Radyushkin:2000jy,
Radyushkin:2000ap,Kivel:2000cn,Kivel:2000fg,Kivel:2000rb,Belitsky:2000vk,Belitsky:2001hz,Belitsky:2001yp} 
using different approaches.

Since the final state photon is on shell, $q_{2}^{2}=0$,
it follows in this particular kinematics that 
$\left(r\cdot q_{1}\right)\simeq-q_{1}^{2}/2=x_{B}\left(p_{1}\cdot q_{1}\right)$. 
Hence the momentum transfer should have a large component in the direction
of the average nucleon momentum, 
\begin{eqnarray}
r & = & 2\eta p+\Delta,
\label{eq:momentumtransferVCA}
\end{eqnarray}
characterized by the skewness parameter, 
\begin{eqnarray}
\eta & \equiv & \frac{\left(r\cdot q\right)}{2\left(p\cdot q\right)}.
\label{eq:etaparameter}
\end{eqnarray}
In the DVCS kinematics, the remainder $\Delta$ in Eq. (\ref{eq:momentumtransferVCA})
is transverse to both \emph{p} and \emph{q}, 
$\left(\Delta\cdot p\right)=-2\eta p^{2}\rightarrow0$ and 
$\left(\Delta\cdot q\right)=-t/4\rightarrow0$ \cite{Radyushkin:2000jy}. 
In the language of the parton model, the component $2\eta p$ of the momentum transfer 
would be identified with the longitudinal component $r_{\Vert}$ while $\Delta$ would be a transverse 
component $r_{\bot}$ \cite{Radyushkin:2000ap}. Moreover, $\eta$ coincides with another 
scaling variable $\xi$ defined as \cite{Ji:1996ek}
\begin{eqnarray}
\xi & \equiv & -\frac{q^{2}}{2\left(p\cdot q\right)}.
\label{eq:ksiparameter}
\end{eqnarray}
For that reason, $\xi$ in DVCS is frequently also referred to
as skewness. It is easy to verify that $\xi=x_{B}/\left(2-x_{B}\right)$
in the limit $t/q^{2}\rightarrow0$. In DIS, on the other hand, the asymptotic region, which is also 
characterized by the large virtuality of the incoming photon, corresponds to the situation in 
which $q_{1}=q_{2}$, $p_{1}=p_{2}$, $r=0$ and accordingly, $\eta=0$ and $\xi=x_{B}$.

Let us now turn to the formal light-cone expansion of the time-ordered
product $T\left\{ J_{EM}^{\mu}\left(z/2\right)J_{EM}^{\nu}\left(-z/2\right)\right\}$
in the coordinate representation. The expansion is performed in terms
of QCD string operators, as discussed in Ref. \cite{Balitsky:1987bk}. The string 
operators have gauge links along the straight line between the fields 
(the gauge link connecting the two space-time points corresponds to the 
summation over twist-0 longitudinal gluons, and it disappears, e.g. in the 
Fock-Schwinger gauge, $z^{\alpha}A_{\alpha}\left(z\right)=0$), however, for brevity we 
will not write them explicitly. The leading light-cone singularity, $z^{2}\rightarrow0$, 
is given by the sum of two (\emph{s}- and \emph{u}-channel) handbag 
diagrams shown in Fig. \ref{handbagdiagramsDVCS}. The hard part of each of the 
diagrams begins at zeroth order in $\alpha_{s}$ with the purely tree level diagrams, 
in which the virtual and real photons interact with the (massless) quarks. We have
\begin{eqnarray}
iT\left\{ J_{EM}^{\mu}\left(z/2\right)J_{EM}^{\nu}\left(-z/2\right)\right\}  & = & i
\sum_{f}Q_{f}^{2}\left\{ \bar{\psi}_{f}\left(z/2\right)\gamma^{\mu}i\not\! S\left(z\right)
\gamma^{\nu}\psi_{f}\left(-z/2\right)\right.\nonumber \\
&  & \left.+\bar{\psi}_{f}\left(-z/2\right)\gamma^{\nu}i\not\! S\left(-z\right)\gamma^{\mu}
\psi_{f}\left(z/2\right)\right\},
\label{eq:DVCSexpansion1}
\end{eqnarray}
where $Q_{f}$ denotes the electric charge of the quark with flavor
\emph{f} in units of $\left|e\right|$. Note that the vertices, in
fact, contribute the factor $\left(-i\left|e\right|Q_{f}\right)^{2}$
rather than $Q_{f}^{2}$. For convenience, we do not include an extra factor of 
$-e^{2}$ in the expression for the VCA, however, it will be included
in the T-matrix. The free quark propagator between the initial and
final quark fields in the coordinate representation is given by 
\begin{eqnarray}
\not\! S\left(z\right)=\frac{\not\! z}{2\pi^{2}\left(z^{2}-i0\right)^{2}} & = & \int
\frac{d^{4}l}{\left(2\pi\right)^{4}}\; e^{-i\left(l\cdot z\right)}\frac{\not l}{l^{2}+i0}.
\label{eq:freequarkpropagator}
\end{eqnarray}
Using the $\gamma$-matrix formula,
\begin{eqnarray}
\gamma^{\mu}\gamma^{\rho}\gamma^{\nu} & = & \left(s^{\mu\rho\nu\eta}+i\epsilon^{\mu\rho\nu\eta}
\gamma_{5}\right)\gamma_{\eta},
\label{eq:gammamatrixformula}
\end{eqnarray}
where $s^{\mu\rho\nu\eta}\equiv g^{\mu\rho}g^{\nu\eta}+g^{\mu\eta}g^{\rho\nu}-g^{\mu\nu}g^{\rho\eta}$
is the symmetric and $\epsilon^{\mu\rho\nu\eta}$ is the antisymmetric tensor in the Lorentz indices 
$\mu$ and $\nu$, we express the original bilocal quark operators in Eq. (\ref{eq:DVCSexpansion1}) 
in terms of operators with only one Lorentz index,
\begin{eqnarray}
iT\left\{ J_{EM}^{\mu}\left(z/2\right)J_{EM}^{\nu}\left(-z/2\right)\right\} =\nonumber \\
\frac{z_{\rho}}{2\pi^{2}z^{4}}\sum_{f}Q_{f}^{2}\left\{ s^{\mu\rho\nu\eta}\left[\bar{\psi}_{f}
\left(-z/2\right)\gamma_{\eta}\psi_{f}\left(z/2\right)-\bar{\psi}_{f}\left(z/2\right)
\gamma_{\eta}\psi_{f}
\left(-z/2\right)\right]\right.\nonumber \\
\left.+i\epsilon^{\mu\rho\nu\eta}\left[\bar{\psi}_{f}\left(-z/2\right)\gamma_{\eta}
\gamma_{5}\psi_{f}\left(z/2\right)+\bar{\psi}_{f}\left(z/2\right)\gamma_{\eta}\gamma_{5}\psi_{f}
\left(-z/2\right)\right]\right\}.
\label{eq:DVCSexpansion2}
\end{eqnarray}
It is customary to write these new QCD bilocal operators (the vector and axial vector string operators) 
as
\begin{eqnarray}
\mathcal{O}_{\eta}^{f}\left(z\left|0\right.\right) & = & \left[\bar{\psi}_{f}\left(-z/2\right)
\gamma_{\eta}\psi_{f}\left(z/2\right)-\bar{\psi}_{f}\left(z/2\right)\gamma_{\eta}\psi_{f}\left(-z/2
\right)\right],\nonumber \\
\mathcal{O}_{5\eta}^{f}\left(z\left|0\right.\right) & = & \left[\bar{\psi}_{f}\left(-z/2\right)
\gamma_{\eta}\gamma_{5}\psi_{f}\left(z/2\right)+\bar{\psi}_{f}\left(z/2\right)\gamma_{\eta}\gamma_{5}
\psi_{f}\left(-z/2\right)\right],
\label{eq:stringoperatorsDVCS}
\end{eqnarray}
and accordingly, Eq. (\ref{eq:DVCSexpansion2}) turns into
\begin{eqnarray}
iT\left\{ J_{EM}^{\mu}\left(z/2\right)J_{EM}^{\nu}\left(-z/2\right)\right\}  & = & 
\frac{z_{\rho}}{2\pi^{2}z^{4}}\sum_{f}Q_{f}^{2}\left\{ s^{\mu\rho\nu\eta}
\mathcal{O}_{\eta}^{f}\left(z\left|0\right.\right)+i\epsilon^{\mu\rho\nu\eta}\mathcal{O}_{5\eta}^{f}
\left(z\left|0\right.\right)\right\} .\nonumber \\
\label{eq:DVCSexpansion3}
\end{eqnarray}

The string operators in Eq. (\ref{eq:stringoperatorsDVCS}) do not
have a definite twist. To isolate their twist-2 part one uses a Taylor
series expansion of $\mathcal{O}_{\eta}^{f}\left(z\left|0\right.\right)$
and $\mathcal{O}_{5\eta}^{f}\left(z\left|0\right.\right)$ in the
relative coordinate \emph{z}. This gives local operators, 
$\bar{\psi}_{f}\left(0\right)\gamma_{\eta}D_{\mu_{1}}...
D_{\mu_{n}}\psi_{f}\left(0\right)$ and $\bar{\psi}_{f}\left(0\right)\gamma_{\eta}\gamma_{5}D_{\mu_{1}}...
D_{\mu_{n}}\psi_{f}\left(0\right)$, where $D_{\mu}$ is the covariant derivative, 
which are not symmetric in their indices. To get the twist-2 contribution, one should keep only the 
totally symmetric traceless parts of the coefficients in the expansion. 
As it was shown in Ref. \cite{Balitsky:1987bk}, the totally symmetric parts can be carried out by the 
following operation
\begin{eqnarray}
\left[\mathcal{O}_{\eta}^{f}\left(z\left|0\right.\right)\right]_{sym} & = & \partial_{\eta}
\int_{0}^{1}d\beta\nonumber \\
&  & \times\left[\bar{\psi}_{f}\left(-\beta z/2\right)\not\! z\psi_{f}
\left(\beta z/2\right)-\bar{\psi}_{f}\left(\beta z/2\right)\not\! z\psi_{f}
\left(-\beta z/2\right)\right],\nonumber \\
\left[\mathcal{O}_{5\eta}^{f}\left(z\left|0\right.\right)\right]_{sym} & = & \partial_{\eta}
\int_{0}^{1}d\beta\nonumber \\
&  & \times\left[\bar{\psi}_{f}\left(-\beta z/2\right)\not\! z\gamma_{5}\psi_{f}
\left(\beta z/2\right)+\bar{\psi}_{f}\left(\beta z/2\right)\not\! z\gamma_{5}\psi_{f}
\left(-\beta z/2\right)\right],\nonumber \\
\label{eq:symmetricpartDVCS}
\end{eqnarray}
where $\partial_{\eta}\equiv\partial/\partial z^{\eta}$ is the derivative with respect to the 
relative coordinate. It becomes clear now why the term string is used: the argument of $\psi_{f}$ and 
$\bar{\psi}_{f}$ takes all the values on the string from $-z/2$ to $z/2$. The subtraction of traces 
can be achieved by imposing the harmonic condition on the contracted vector and axial vector string 
operators,
\begin{eqnarray}
\mathcal{O}^{f+}\left(z\right) & \equiv & z^{\eta}\mathcal{O}_{\eta}^{f}\left(z\left|0\right.\right)=
\left[\bar{\psi}_{f}\left(-z/2\right)\not\! z\psi_{f}\left(z/2\right)+
\left(z\rightarrow-z\right)\right],\nonumber \\
\mathcal{O}_{5}^{f-}\left(z\right) & \equiv & z^{\eta}\mathcal{O}_{5\eta}^{f}\left(z\left|0\right.
\right)=\left[\bar{\psi}_{f}\left(-z/2\right)\not\! z\gamma_{5}\psi_{f}\left(z/2\right)-
\left(z\rightarrow-z\right)\right],
\label{eq:contractedoperatorsDVCS}
\end{eqnarray}
which appear on the right-hand-side of Eq. (\ref{eq:symmetricpartDVCS}). In other words, these two 
operators should satisfy d'Alembert equation with respect
to \emph{z}
\begin{eqnarray}
\partial^{2}\left[\mathcal{O}^{f+}\left(z\right)\right]_{twist-2} & = & 0,
\label{eq:d'Alembert}
\end{eqnarray}
and similarly for the twist-2 part of $\mathcal{O}_{5}^{f-}\left(z\right)$. Note that we have assigned 
an extra superscript $\pm$ to the operators (\ref{eq:contractedoperatorsDVCS}) because they possess 
the symmetry with respect to the change $z\rightarrow-z$.

To compute the amplitude (\ref{eq:reducedVCAamplitude}), the contracted twist-2 operators have to be 
sandwiched between the initial and final nucleon states, and ultimately integrated over \emph{z}. 
Thus we need to construct a parametrization for these nonforward nucleon matrix elements. 
The most convenient way is a decomposition into plane waves, i.e. a spectral representation, 
where the relevant spectral functions correspond to GPDs (to be specific, here and in the following, 
we use the off-forward parton distributions). In principle, we need to provide a parametrization 
valid everywhere in \emph{z} since the coordinate \emph{z} runs over the whole four-dimensional space. 
However, it turns out that the inclusion of the $z^{2}$ terms in the matrix elements generate 
$M^{2}/q^{2}$ and $t/q^{2}$ corrections to the amplitude (such corrections are analogous to 
the well-known target mass corrections in DIS \cite{Nachtmann:1973mr,Georgi:1976ve}) and hence 
will be neglected. It is, therefore, sufficient to provide a parametrization only on 
the light-cone \cite{Radyushkin:2000ap}, namely,
\begin{eqnarray}
\left\langle N\left(p_{2},s_{2}\right)\right|\mathcal{O}^{f+}\left(z\right)
\left|N\left(p_{1},s_{1}\right)\right\rangle_{z^2=0}  & = & \bar{u}
\left(p_{2},s_{2}\right)\not\! zu\left(p_{1},s_{1}\right)\nonumber\\
&  & \times\int_{-1}^{1}dx\; e^{ix\left(p\cdot z\right)}H_{f}^{+}\left(x,\xi,t\right)\nonumber \\
&  & +\bar{u}\left(p_{2},s_{2}\right)
\frac{\left(\not\! z\not\! r-\not\! r\not\! z\right)}{4M}u\left(p_{1},s_{1}\right)\nonumber \\
&  & \times\int_{-1}^{1}dx\; e^{ix\left(p\cdot z\right)}E_{f}^{+}\left(x,\xi,t\right),\nonumber \\
\left\langle N\left(p_{2},s_{2}\right)\right|\mathcal{O}_{5}^{f-}
\left(z\right)\left|N\left(p_{1},s_{1}
\right)\right\rangle_{z^2=0}  & = & \bar{u}
\left(p_{2},s_{2}\right)\not\! z\gamma_{5}u\left(p_{1},s_{1}\right)\nonumber \\
&  &\times\int_{-1}^{1}dx\; e^{ix\left(p\cdot z\right)}\tilde{H}_{f}^{+}\left(x,\xi,t\right)\nonumber \\
&  & -\bar{u}\left(p_{2},s_{2}\right)\frac{\left(r\cdot z\right)}{2M}
\gamma_{5}u\left(p_{1},s_{1}\right)\nonumber \\
&  & \times\int_{-1}^{1}dx\; e^{ix\left(p\cdot z\right)}\tilde{E}_{f}^{+}\left(x,\xi,t\right).
\label{eq:parametrizationDVCS}
\end{eqnarray}
The flavor dependent OFPDs in the parametrization (\ref{eq:parametrizationDVCS}) refer to the 
corresponding quark flavor \emph{f} in the nucleon \emph{N}. Apart from the scale $Q^2\equiv-q^2$, 
there are three variables necessary to specify OFPDs, namely, the usual light-cone momentum 
fraction \emph{x}, the invariant momentum transfer \emph{t} to the target and the skewness 
parameter $\xi$, which corresponds to the light-cone momentum fraction transferred to the target and 
characterizes the momentum asymmetry. Recall that the variables \emph{x} and $\xi$ only define 
the longitudinal momenta of the partons involved, that is their plus components. Schematically, 
the parton going out of the parent nucleon in Fig. \ref{handbagdiagramsDVCS} carries the fraction 
$\left(x+\xi\right)$ of the average nucleon momentum \emph{p} while the momentum of the returning parton 
is $\left(x-\xi\right)p$. 

The OFPDs $H_{f}\left(x,\xi,t\right)$, $E_{f}\left(x,\xi,t\right)$, $\tilde{H}_{f}\left(x,\xi,t\right)$ 
and $\tilde{E}_{f}\left(x,\xi,t\right)$, introduced in Section III.5, parametrize the matrix elements of 
operators $\bar{\psi}_{f}\left(-z/2\right)\not\! z\psi_{f}\left(z/2\right)$ and 
$\bar{\psi}_{f}\left(-z/2\right)\not\! z\gamma_{5}\psi_{f}\left(z/2\right)$, do not have symmetry 
with respect to the change $x\rightarrow-x$. In particular, the function $H_{f}\left(x\right)$ in 
the forward limit, see Eq. (\ref{eq:forwardlimit1}), corresponds for positive $x$ to the quark 
distribution while for negative $x$, it corresponds to the minus antiquark distribution. 
On the other hand, the plus distributions, introduced through the matrix elements of (anti)symmetrized 
operators given by Eq. (\ref{eq:contractedoperatorsDVCS}), do have well defined 
symmetry properties with respect the scaling variable \emph{x}. Simply by transforming 
$z\rightarrow-z$ and $x\rightarrow-x$ in Eq. (\ref{eq:parametrizationDVCS}), we can establish 
the following
\begin{eqnarray}
H_{f}^{+}\left(x\right) & = & - H_{f}^{+}\left(-x\right),\nonumber \\
E_{f}^{+}\left(x\right) & = & - E_{f}^{+}\left(-x\right),\nonumber \\
\tilde{H}_{f}^{+}\left(x\right) & = & \tilde{H}_{f}^{+}\left(-x\right),\nonumber \\
\tilde{E}_{f}^{+}\left(x\right) & = & \tilde{E}_{f}^{+}\left(-x\right).
\label{eq:symmetrypropertiesplusGPDs}
\end{eqnarray}
It is easy to notice that the plus OFPDs are determined by the sum of quark and antiquark distributions. 
Since the quark distribution includes both the valence and sea contributions, and further the antiquark 
distribution only the sea contribution, the plus distribution turns into the sum of the valence and 
twice the sea quark distributions.  

After substitution of Eq. (\ref{eq:parametrizationDVCS}) into the right-hand
side of Eq. (\ref{eq:symmetricpartDVCS}), we first take the derivative 
with respect to \emph{z}. Then we perform integration by parts over
the parameter $\beta$ and keep only the surface terms with the arguments
$\bar{\psi}_{f}\left(\pm z/2\right)$ and $\psi_{f}\left(\pm z/2\right)$.
Finally, the integral over \emph{z} is carried out with the help of
the inversion formula for $\not\! S\left(z\right)$, 
\begin{eqnarray}
\int d^{4}z\; e^{i\left(l\cdot z\right)}
\frac{z_{\rho}}{2\pi^{2}\left(z^{2}-i0\right)^{2}} & = & \frac{l_{\rho}}{\left(l^{2}+i0\right)}.
\label{eq:integraloverzDVCS}
\end{eqnarray}
Here the momentum \emph{l} is given by $l=\left(xp+q\right)$, so
that $l^{2}$ in the denominator of Eq. (\ref{eq:integraloverzDVCS})
becomes $l^{2}=2\left(p\cdot q\right)\left(x-\xi\right)$. The expression
for the reduced VCA reads
\begin{eqnarray}
\mathsf{\mathcal{T}}_{EMtwist-2}^{\mu\nu}=\frac{1}{2\left(p\cdot q\right)}\sum_{f}Q_{f}^{2}
\int_{-1}^{1}\frac{dx}{\left(x-\xi+i0\right)}\nonumber \\
\times\Bigg\lbrace H_{f}^{+}\left(x,\xi,t\right)\left[l^{\mu}\bar{u}\left(p_{2},s_{2}
\right)\gamma^{\nu}u\left(p_{1},s_{1}\right)\right.\nonumber \\
\left.+l^{\nu}\bar{u}\left(p_{2},s_{2}\right)\gamma^{\mu}u\left(p_{1},s_{1}\right)-g^{\mu\nu}
\bar{u}\left(p_{2},s_{2}\right)\not lu\left(p_{1},s_{1}\right)\right]\nonumber \\
+E_{f}^{+}\left(x,\xi,t\right)\left[l^{\mu}\bar{u}\left(p_{2},s_{2}\right)
\frac{\gamma^{\nu}\not\! r-\not\! r\gamma^{\nu}}{4M}u\left(p_{1},s_{1}\right)\right.\nonumber \\
\left.+l^{\nu}\bar{u}\left(p_{2},s_{2}\right)\frac{\gamma^{\mu}\not\! r-\not\! r\gamma^{\mu}}{4M}u
\left(p_{1},s_{1}\right)-g^{\mu\nu}\bar{u}\left(p_{2},s_{2}\right)
\frac{\left(\not l\not\! r-\not\! r\not l\right)}{4M}u\left(p_{1},s_{1}\right)\right]\nonumber \\
-i\epsilon^{\mu\nu\rho\eta}\left[\tilde{H}_{f}^{+}\left(x,\xi,t\right)l_{\rho}
\bar{u}\left(p_{2},s_{2}\right)\gamma_{\eta}\gamma_{5}u\left(p_{1},s_{1}\right)\right.\nonumber \\
\left.-\tilde{E}_{f}^{+}\left(x,\xi,t\right)l_{\rho}\bar{u}\left(p_{2},s_{2}\right)
\frac{r_{\eta}}{2M}\gamma_{5}u\left(p_{1},s_{1}\right)\right]\Bigg\rbrace,
\label{eq:reducedVCAresult1}
\end{eqnarray}
and can be further simplified as follows. Let us perform the light-cone
decomposition of $\gamma$-matrix, 
\begin{eqnarray}
\gamma^{\mu} & = & a^{\mu}\not\! n_{1}+b^{\mu}\not\! n_{2}+\gamma_{\bot}^{\mu}.
\label{eq:lightconedecomposition1}
\end{eqnarray}
The four-vectors $n_{1}$ and $n_{2}$ in Eq. (\ref{eq:lightconedecomposition1}) 
are light-like, $n_{1}^{2}=n_{2}^{2}=0$, and satisfy the condition
$\left(n_{1}\cdot n_{2}\right)=1$. Identifying $n_{1}\rightarrow p$,
$n_{2}\rightarrow q_{2}$ and neglecting the transverse component 
$\gamma_{\bot}^{\mu}$, since it corresponds to the higher-twist contributions,
Eq. (\ref{eq:lightconedecomposition1}) takes the form 
\begin{eqnarray}
\gamma^{\mu} & = & \frac{1}{\left(p\cdot q_{2}\right)}
\left(q_{2}^{\mu}\not\! p+p^{\mu}\not\! q_{2}\right).
\label{eq:lightconedecomposition2}
\end{eqnarray}
Using the above decomposition, the Dirac equation, 
$\not\! p_{1}u\left(p_{1},s_{1}\right)=\not\! p_{2}u\left(p_{2},s_{2}\right)=0$ 
(recall that we neglect the nucleon mass), and the fact that $H_{f}^{+}$ and $E_{f}^{+}$ are odd 
functions in \emph{x}, the reduced VCA in the leading twist transforms into
\begin{eqnarray}
\mathsf{\mathcal{T}}_{EMtwist-2}^{\mu\nu}=\frac{1}{2\left(p\cdot q\right)}\sum_{f}Q_{f}^{2}
\int_{-1}^{1}\frac{dx}{\left(x-\xi+i0\right)}\nonumber \\
\times\Bigg\lbrace H_{f}^{+}\left(x,\xi,t\right)\left[\frac{1}{\left(p\cdot q_{2}\right)}
\left(p^{\mu}q_{2}^{\nu}+p^{\nu}q_{2}^{\mu}\right)-g^{\mu\nu}\right]\bar{u}\left(p_{2},s_{2}
\right)\not\! q_{2}u\left(p_{1},s_{1}\right)\nonumber \\
+E_{f}^{+}\left(x,\xi,t\right)\left[\frac{1}{\left(p\cdot q_{2}\right)}
\left(p^{\mu}q_{2}^{\nu}+p^{\nu}q_{2}^{\mu}\right)-g^{\mu\nu}\right]\bar{u}\left(p_{2},s_{2}\right)
\frac{\left(\not\! q_{2}\not\! r-\not\! r\not\! q_{2}\right)}{4M}u\left(p_{1},s_{1}\right)\nonumber \\
-\tilde{H}_{f}^{+}\left(x,\xi,t\right)\left[\frac{1}{\left(p\cdot q_{2}\right)}i
\epsilon^{\mu\nu\rho\eta}q_{2\rho}p_{\eta}\right]\bar{u}\left(p_{2},s_{2}\right)\not\! q_{2}
\gamma_{5}u\left(p_{1},s_{1}\right)\nonumber \\
+\tilde{E}_{f}^{+}\left(x,\xi,t\right)\left[\frac{1}{\left(p\cdot q_{2}\right)}i
\epsilon^{\mu\nu\rho\eta}q_{2\rho}p_{\eta}\right]\frac{\left(q_{2}\cdot r\right)}{2M}
\bar{u}\left(p_{2},s_{2}\right)\gamma_{5}u\left(p_{1},s_{1}\right)\Bigg\rbrace.
\label{eq:reducedVCAresult2}
\end{eqnarray}
The amplitude has both real and imaginary parts. Since OFPDs are real functions, the imaginary part of 
the amplitude comes only from the singularity of the expression $1/\left(x-\xi+i0\right)$. The latter 
can be calculated by applying the formula
\begin{eqnarray}
\frac{1}{\left(x\pm i0\right)} & \equiv & P\left(\frac{1}{x}\right)
\mp i\pi\delta\left(x\right),
\label{eq:principalvalueformula}
\end{eqnarray}
where \emph{P} denotes the principal value. Thus the real part of $\mathsf{\mathcal{T}}_{EM}^{\mu\nu}$ 
is obtained using the principal value prescription. The imaginary part, on the other hand, generates 
the delta function $\delta\left(x-\xi\right)$, which assures taking OFPDs at the specific point, $x=\xi$. 
In other words, the imaginary part of the amplitude is directly proportional to OFPDs evaluated along 
the line $x=\xi$. Moreover, from the tensor structure of the VCA in Eq. (\ref{eq:reducedVCAresult2}), 
it is easy to see that the amplitude exactly satisfies the electromagnetic gauge invariance with respect 
to the final real photon,
\begin{eqnarray}
q_{2\mu}\mathsf{\mathcal{T}}_{EMtwist-2}^{\mu\nu} & = & 0.
\label{eq:gaugeinvariancetwist2finalphoton}
\end{eqnarray}
The gauge invariant condition with respect to the initial virtual photon that has the momentum 
$q_{1\nu}=q_{2\nu}-r_{\nu}$ will only be satisfied if $\mathsf{\mathcal{T}}_{EMtwist-2}^{\mu\nu}r_{\nu}=0$. 
However, for the part of $\mathsf{\mathcal{T}}_{EMtwist-2}^{\mu\nu}$, which is symmetric with respect to 
$\mu\leftrightarrow\nu$, we find with the help of Eqs. (\ref{eq:momentumtransferVCA}) and 
(\ref{eq:etaparameter}),
\begin{eqnarray}
\mathsf{\mathcal{T}}_{EMtwist-2\left(sym\right)}^{\mu\nu}r_{\nu} & \sim & r^{\mu}-
\frac{\left(r\cdot q_{2}\right)}{\left(p\cdot q_{2}\right)}p^{\mu}\nonumber \\
& = & \Delta^{\mu}-\mathcal{O}\left(r^{2}\right)p^{\mu}.
\label{eq:gaugeinvarianctwist2initialphoton}
\end{eqnarray}
Since the components of $\Delta$ are all of the order $\sqrt{\left|t\right|}$, 
see Ref. \cite{Radyushkin:2000ap}, it follows that the leading-twist VCA is gauge invariant to accuracy 
$\mathcal{O}\left(\sqrt{\left|t\right|}\;\right)$. This violation of electromagnetic gauge invariance 
is a higher-twist (i.e. twist-3) level effect. 

\section{\ \,  Kinematics}

In the generalized DVCS process,
\begin{eqnarray}
l_{1}\left(k\right)+N\left(p_{1}\right) & \longrightarrow & l_{2}\left(k'\right)+N'
\left(p_{2}\right)+\gamma\left(q_{2}\right),
\label{eq:reactionformulageneralDVCS}
\end{eqnarray}
there are a lepton $l_{1}$ with the four-momentum \emph{k} and a nucleon $N\left(p_{1}\right)$ in the 
initial state, and  a lepton $l_{2}\left(k'\right)$, a nucleon $N'\left(p_{2}\right)$ and a real photon  
$\gamma\left(q_{2}\right)$ in a final state. The process has two contributions that are represented by two 
types of diagrams, see Fig. \ref{DVCSdiagrams}. The DVCS diagram corresponds to the emission of the real 
photon from the nucleon blob. In the Bethe-Heitler diagram, the real photon is emitted from a lepton leg. 
We denote the four-momenta as $k=\left(\omega,\vec{k}\right)$, 
$p_{1}=\left(M,\vec{0}\right)$, $k'=\left(\omega',\vec{k}'\right)$, 
$p_{2}=\left(E_{2},\vec{p}_{2}\right)$ and $q_{2}=\left(\nu_{2},\vec{q}_{2}\right)$. The differential 
cross section for lepton scattering off a nucleon, to produce a final state with a lepton, 
nucleon and a real photon is
\begin{eqnarray}
d\sigma & = & \frac{1}{2s}\left|\mathrm{T}\right|^{2}\frac{1}{\left(2\pi\right)^{5}}\;
\delta^{\left(4\right)}\left(k+p_{1}-k'-p_{2}-q_{2}\right)\frac{d^{3}k'}{2\omega'}
\frac{d^{3}p_{2}}{2E_{2}}\frac{d^{3}q_{2}}{2\nu_{2}},
\label{eq:generaldiffcrosssectionDVCS}
\end{eqnarray}
where T represents the scattering amplitude. It contains both the Compton and Bethe-Heitler contributions,
\begin{eqnarray}
\mathrm{T} & = & \mathrm{T}_{C}+\mathrm{T}_{BH}.
\label{eq:totalamplitudeDVCS}
\end{eqnarray}
In the laboratory frame, as the target rest frame, the total lepton-nucleon
center-of-mass energy squared is $s\equiv\left(k+p_{1}\right)^{2}=2\omega M+M^{2}$. 
Note that in the following, we neglect both lepton masses. Integrating 
Eq. (\ref{eq:generaldiffcrosssectionDVCS}) over the photon momentum gives
\begin{eqnarray}
d\sigma & = & \frac{1}{2s}\left|\mathrm{T}\right|^{2}\frac{1}{\left(2\pi\right)^{5}}\;
\delta\left[\left(k+p_{1}-k'-p_{2}\right)^{2}\right]\frac{\omega'd\omega'd\Omega'}{2}
\frac{d^{3}p_{2}}{2E_{2}}.
\label{eq:diffcrosssection1DVCS}
\end{eqnarray}
The delta function in the cross section (\ref{eq:diffcrosssection1DVCS}) provides the constraint 
$\hat{s}+M^{2}-2\left[\left(\nu_{1}+M\right)E_{2}-\vec{q}_{1}\cdot\vec{p}_{2}\right]=0$, 
where the invariant $\hat{s}\equiv\left(p_{1}+q_{1}\right)^{2}$ and
$q_{1}=k-k'=\left(\nu_{1},\vec{q}_{1}\right)$ is the four-momentum
of the virtual photon. Furthermore, we choose the coordinate system
so that the \emph{z}-axis is in the direction of the incident lepton
in the plane formed by the lepton momenta. After integrating Eq. (\ref{eq:diffcrosssection1DVCS}) 
over the magnitude of $\vec{p}_{2}$, the differential cross section reads 
\begin{eqnarray}
d\sigma & = & \frac{1}{16s\left(2\pi\right)^{5}}
\frac{\omega'\left|\vec{p}_{2}\right|^{2}}{\left|\left(\nu_{1}+M\right)
\left|\vec{p}_{2}\right|-\left|\vec{q}_{1}\right|E_{2}\cos\phi_{12}\right|}
\left|\mathrm{T}\right|^{2}d\omega'd\Omega'd\Omega_{2},
\label{eq:diffcrosssection2DVCS}
\end{eqnarray}
where $\phi_{12}$ is the angle between vectors $\vec{q}_{1}$ and
$\vec{p}_{2}$.
\begin{figure}
\centerline{\epsfxsize=5in\epsffile{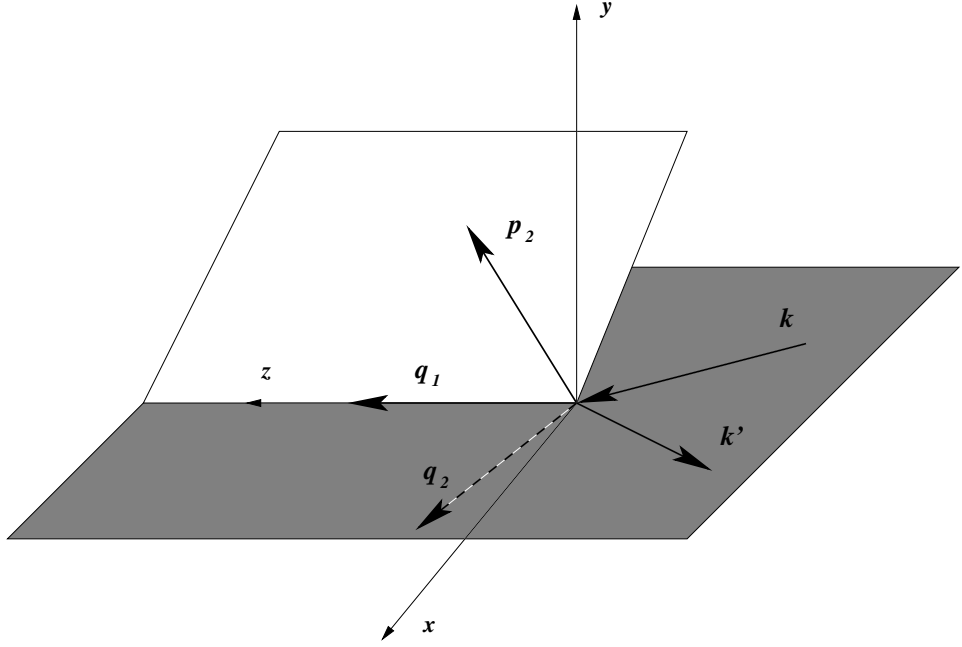}}
\caption{Kinematics of the generalized DVCS process in the target rest frame.}
\label{kinematics}
\end{figure}

Alternatively, one can choose the coordinate system depicted in Fig. 
\ref{kinematics}, in which the virtual photon four-momentum has no 
transverse components, $q_{1}=\left(\nu_{1},0,0,\left|\vec{q}_{1}\right|\right)$,
and the incoming and outgoing lepton four-momenta are 
$k=\omega\left(1,\sin\phi,0,\cos\phi\right)$ 
and $k'=\omega'\left(1,\sin\phi',0,\cos\phi'\right)$, respectively. 
In this reference frame, the azimuthal angle of the recoiled nucleon corresponds to 
the angle $\varphi$ between the lepton and nucleon scattering planes. 
Using now the delta function $\delta\left[\left(k+p_{1}-k'-p_{2}\right)^{2}\right]$ 
in Eq. (\ref{eq:diffcrosssection1DVCS}) to integrate over the polar angle $\phi_{2}$ 
of the outgoing nucleon, we find 
\begin{eqnarray}
\cos\phi_{2} & = & \frac{2E_{2}\left(\omega-\omega'\right)-q_{1}^{2}\left(1-1/x_{B}
\right)-t}{2\sqrt{E_{2}^{2}-M^{2}}\;\sqrt{\left(\omega-\omega'\right)^{2}-q_{1}^{2}}},
\label{eq:outgoinghadronapolarangle}
\end{eqnarray}
and the cross section assumes the form 
\begin{eqnarray}
d\sigma & = & \frac{1}{16s}\left|\mathrm{T}\right|^{2}\frac{1}{\left(2\pi\right)^{4}}
\frac{\omega'}{\sqrt{\left(\omega-\omega'\right)^{2}-q_{1}^{2}}}d\omega'dE_{2}d
\left(\cos\phi'\right)d\varphi,
\label{eq:diffcrosssection3DVCS}
\end{eqnarray}
with the angle $\phi'$ denoting the polar angle of the scattered
lepton. 

Instead of the kinematical variables $\omega'$, $E_{2}$
and $\cos\phi'$, it is convenient to express the differential cross section in terms 
of the invariant variables, e.g. $\left(y,t,x_{B}\right)$ or $\left(Q_{1}^{2},t,x_{B}\right)$, 
where $Q_{1}^{2}\equiv-q_{1}^{2}$ and 
$y\equiv\left(p_{1}\cdot q_{1}\right)/\left(p_{1}\cdot k\right)$. In the laboratory frame, these 
invariants are given by
\begin{eqnarray}
Q_{1}^{2} & = & 2M\omega yx_{B},\nonumber \\
y & = & \frac{\omega-\omega'}{\omega},\nonumber \\
t & = & 2M^{2}-2ME_{2}.
\label{eq:invariantsinlabframe}
\end{eqnarray}
The last invariant can also be written as 
$t\equiv\left(q_{1}-q_{2}\right)^{2}=-Q_{1}^{2}-2\nu_{2}\left(\omega y-
\sqrt{\omega^{2}y^{2}+Q_{1}^{2}}\;\cos\theta_{\gamma\gamma}\right)$, 
where the energy of the outgoing real photon is given by the energy conservation, 
$\nu_{2}=M+\omega y-E_{2}$. By combining both expressions for \emph{t}, we can express 
it as a function of the angle $\theta_{\gamma\gamma}$ between the incoming virtual and 
outgoing real photon, namely,
\begin{eqnarray}
t & =2M^{2}-M & \frac{2M^{2}+Q_{1}^{2}+2\left(M+\omega y\right)\left(\omega y-
\sqrt{\omega^{2}y^{2}+Q_{1}^{2}}\;\cos\theta_{\gamma\gamma}\right)}{M+\omega y-
\sqrt{\omega^{2}y^{2}+Q_{1}^{2}}\;\cos\theta_{\gamma\gamma}}.\nonumber \\
\label{eq:invariantt}
\end{eqnarray}
Moreover, from the three-momentum conservation, $\vec{q}_{1}=\vec{k}-\vec{k}'$,
and alternative expression for the invariant $Q_{1}^{2}$ in the laboratory
frame, $Q_{1}^{2}=-\left(k-k'\right)^{2}=2\omega\omega'\left(1-\cos\phi_{lep}\right)$,
we get for the polar angles of the incoming and scattered leptons,
\begin{eqnarray}
\cos\phi & = & \frac{1}{\omega}\left(\sqrt{\omega^{2}y^{2}+Q_{1}^{2}}+
\omega\left(1-y\right)\cos\phi'\right)
\label{eq:leptonpolarangles1}
\end{eqnarray}
and
\begin{eqnarray}
\cos\phi' & = & \frac{2\omega^{2}y\left(1-y\right)-Q_{1}^{2}}{2
\omega\left(1-y\right)\sqrt{\omega^{2}y^{2}+Q_{1}^{2}}},
\label{eq:leptonpolarangles2}
\end{eqnarray}
respectively. The Jacobian is then equal to
\begin{eqnarray}
J & \equiv & \left|
\frac{\partial\left(\omega',E_{2},\cos\phi'\right)}{\partial\left(y,t,x_{B}\right)}
\right|=\frac{\omega^{2}y^{2}\left(\omega+Mx_{B}\right)}{2
\left(1-y\right)\left(\omega^{2}y^{2}+Q_{1}^{2}\right)^{3/2}}
\label{eq:Jacobian}
\end{eqnarray}
and accordingly, the differential cross section turns into 
\begin{eqnarray}
\frac{d^{4}\sigma}{dx_{B}dydtd\varphi} & = & 
\frac{d^{4}\sigma}{d\left(\cos\phi'\right)d\omega'dE_{2}d\varphi}\; J\nonumber \\
& = & \frac{1}{32s}\frac{1}{\left(2\pi\right)^{4}}
\frac{1+x_{B}\left(M/\omega\right)}{\left[y+2x_{B}\left(M/\omega\right)\right]^{2}}
\left|T\right|^{2}.
\label{eq:generaldiffcross}
\end{eqnarray}
Note that using the invariant $Q_{1}^{2}$ instead of \emph{y} yields 
\begin{eqnarray}
\frac{d^{4}\sigma}{dx_{B}dQ_{1}^{2}dtd\varphi} & = & \frac{1}{64s}\frac{1}{\left(2\pi\right)^{4}}
\frac{1+x_{B}\left(M/\omega\right)}{M\omega x_{B}\left[y+2x_{B}\left(M/\omega\right)\right]^{2}}
\left|T\right|^{2}.
\label{eq:generaldiffcrosswithQ1squared}
\end{eqnarray}

\begin{figure}
\centerline{\epsfxsize=5in\epsffile{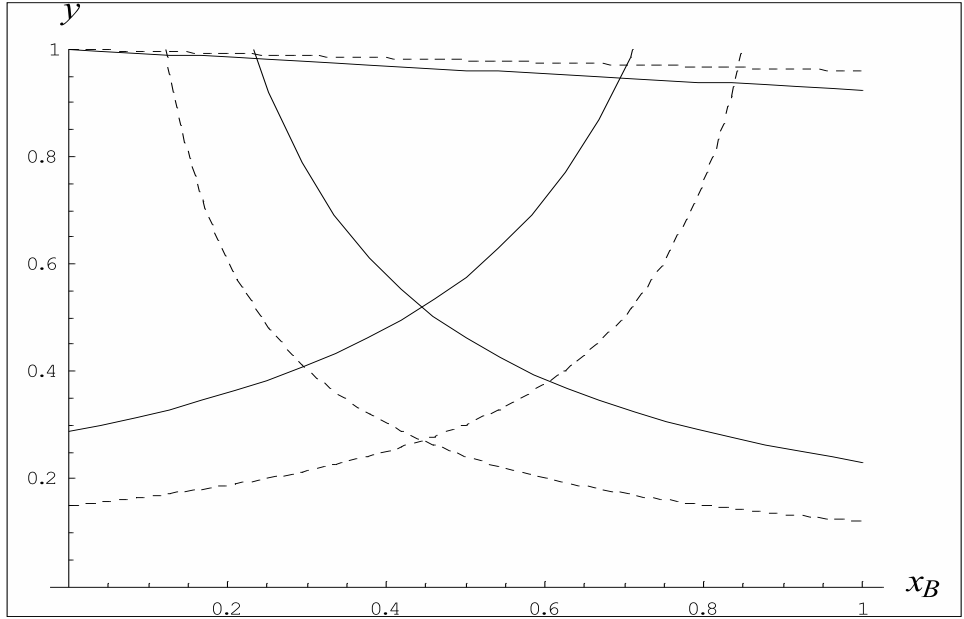}}
\caption{Kinematically allowed region for $\hat{s}\geq4\;\mathrm{GeV}^{2}$ and 
$Q_{1}^{2}\geq2.5\;\mathrm{GeV}^{2}$ with $\omega=5.75\;\mathrm{GeV}$ (solid line) 
and $\omega=11\;\mathrm{GeV}$ (dashed line) lepton beam.}
\label{xyplane}
\end{figure}

In the following, we determine the kinematically allowed region for the generalized DVCS process. 
In other words, one needs to find the upper and lower limits on $x_{B}$, \emph{y} and \emph{t}. 
We require the following constraints: 
\begin{itemize}
\item The energy of the incoming lepton beam is fixed. In particular, we consider two examples 
that are both relevant to Jefferson Lab, namely, $\omega=5.75\;\mathrm{GeV}$ and 
$\omega=11\;\mathrm{GeV}$. 
\item The invariant mass of the virtual photon-nucleon system should
be above the resonance region, 
$\hat{s}\equiv\left(p_{1}+q_{1}\right)^{2}\geq\hat{s}_{min}=4\;\mathrm{GeV}^{2}$. 
\item The virtuality of the incoming photon has to be large enough to secure
the light-cone dominance, $Q_{1}^{2}\geq Q_{1min}^{2}=2.5\;\mathrm{GeV}^{2}$. 
\item The momentum transfer squared to the nucleon should be kept as small
as possible, $t/q^{2}\ll1$. 
\end{itemize}
The second constraint, $M^{2}-Q_{1}^{2}\left(1-1/x_{B}\right)\geq\hat{s}_{min}$,
together with $Q_{1}^{2}=x_{B}y\left(s-M^{2}\right)$ implies that 
\begin{eqnarray}
y_{min} & = & \frac{\hat{s}_{min}-M^{2}}{\left(s-M^{2}\right)\left(1-x_{B}\right)}.
\label{eq:ymin}
\end{eqnarray}
On the other hand, the scaling variable \emph{y} reaches its maximum value,
\begin{eqnarray}
y_{max} & = & \left(1+\frac{M^{2}x_{B}}{s-M^{2}}\right)^{-1},
\label{eq:ymax}
\end{eqnarray}
when the incoming lepton is aligned along the \emph{z}-axis at the angle $\phi_{lep}=180^{0}$. 
If one plots the region in the $x_{B}y$ plane, as illustrated in Fig. \ref{xyplane}, 
then Eqs. (\ref{eq:ymin}), (\ref{eq:ymax}) and $y=Q_{1min}^{2}/x_{B}\left(s-M^{2}\right)$ 
(the latter comes from the third constraint) correspond to its boundaries. 
Next, both lower and upper limits of the invariant \emph{t} can be found if we go in the virtual 
photon-nucleon center-of-mass frame. In this particular frame, at the scattering angles between 
the initial and final nucleons equal to $0^{0}$ and $180^{0}$, the invariant momentum attains, 
up to relative corrections of the order $x_{B}M^{2}/Q_{1}^{2}$, its kinematical limits,
\begin{eqnarray}
t_{min} & = & \frac{-M^{2}x_{B}^{2}}{1-x_{B}\left(1-M^{2}/Q_{1}^{2}\right)},\nonumber \\
t_{max} & = & \frac{M^{2}x_{B}^{2}-2M^{2}x_{B}+Q_{1}^{2}\left(x_{B}-1\right)/x_{B}}{1-x_{B}
\left(1-M^{2}/Q_{1}^{2}\right)}.
\label{eq:tlimits}
\end{eqnarray}
They are presented, within the kinematically allowed regions, in Table
\ref{tabletlimits}. Since we require small \emph{t}, the upper limit becomes
irrelevant. In addition, having small \emph{t}  yields a low-energy nucleon with 
$E_{2}=M\left(1-t/2M^{2}\right)$ and a high-energy real photon in the final state. 

\begin{table}
\caption{Lower and upper limits of the invariant momentum transfer $t$ within the kinematically 
allowed region illustrated in Fig. \ref{xyplane} for two different lepton beam energies $\omega$.}
\label{tabletlimits}
\vspace{12pt}
\begin{center}
\begin{tabular}{lcc}\hline\hline\\
$\omega\left[\mathrm{GeV}\right]$ & 
$-t_{min}\left[\mathrm{GeV}^{2}\right]$ & 
$-t_{max}\left[\mathrm{GeV}^{2}\right]$ \\\\
\hline
 & & \\
5.75 & 0.058 & 10.019  \\
 & & \\
11 & 0.014 & 19.853  \\\\ \hline\hline
\end{tabular}
\end{center}
\end{table}
\begin{figure}
\centerline{\epsfxsize=5in\epsffile{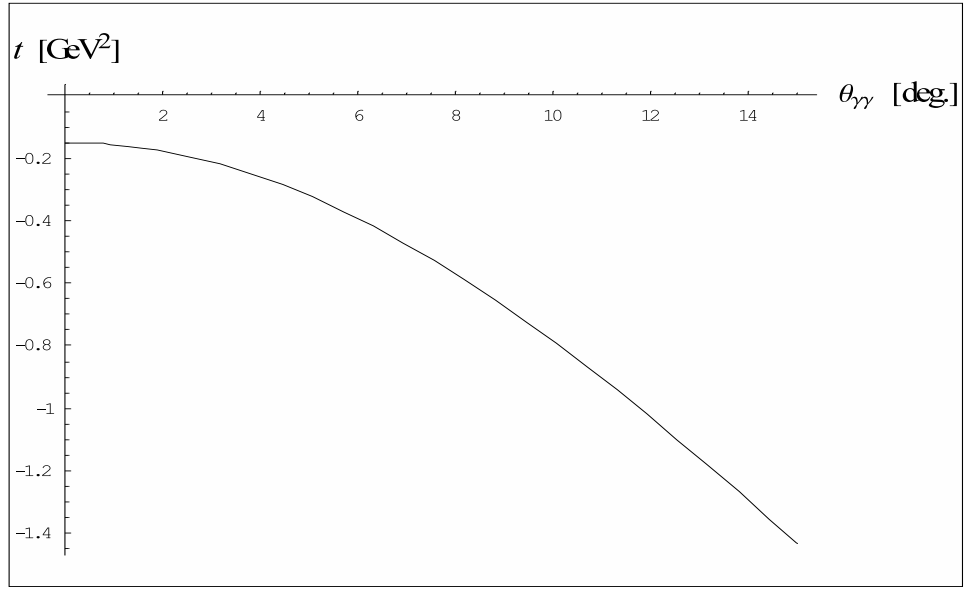}}
\caption{Invariant momentum transfer $t$ plotted as a function of the angle 
$\theta_{\gamma\gamma}$ between the incoming virtual and outgoing real photon in the 
target rest frame for $Q_{1}^{2}=2.5\;\mathrm{GeV}^{2}$ and $x_{B}=0.35$.}
\label{t}
\end{figure}
\begin{table}
\caption{Polar angles $\phi$ and $\phi'$ of the incoming and scattered leptons, respectively, 
in the target rest frame for $Q_{1}^{2}=2.5\;\mathrm{GeV}^{2}$ and $x_{B}=0.35$ with two different 
lepton beam energies $\omega$.}
\label{tableangles}
\vspace{12pt}
\begin{center}
\begin{tabular}{lcc}\hline\hline\\
$\omega\left[\mathrm{GeV}\right]$ & 
$\phi\left[\mathrm{deg}.\right]$ & 
$\phi'\left[\mathrm{deg}.\right]$ \\\\
\hline
 & & \\
5.75 & 12.6 & 39.9  \\
 & & \\
11 & 18 & 28.2  \\\\ \hline\hline
\end{tabular}
\end{center}
\end{table}

Finally, for each lepton energy we pick one kinematical point within the allowed region 
in the $x_{B}y$ plane. In particular, we take $Q_{1}^{2}=2.5\;\mathrm{GeV}^{2}$ for the virtuality 
of the initial photon and $x_{B}=0.35$. In Fig. \ref{t}, the invariant \emph{t} is plotted against 
the angle $\theta_{\gamma\gamma}$. It is customary to present the plot for the values of \emph{t} 
up to $-1\;\mathrm{GeV}^{2}$, which happens at $\theta_{\gamma\gamma}\simeq12^{0}$, even though we 
require that, in principle, $-t$ should be much smaller than $Q_{1}^{2}$. The values of \emph{t} 
vary from $-0.15\;\mathrm{GeV}^{2}$ to $-1.433\;\mathrm{GeV}^{2}$ for angles between 
$0\leq\theta_{\gamma\gamma}\leq15^{0}$. Note that by fixing $Q_{1}^{2}$ and $x_{B}$, 
the product $\omega y=Q_{1}^2/2Mx_{B}$, which appears in Eq. (\ref{eq:invariantt}) is also fixed and 
accordingly, we end up with the same $\theta_{\gamma\gamma}$-dependence for both beam energies. 
For convenience, we set the angle between the lepton and nucleon scattering planes to 
$\varphi=0$, and arrive at the so-called in-plane kinematics. The polar angles of both incoming 
and scattered leptons are then fixed and given in Table \ref{tableangles}.

\section{\ \,  Toy Model}

Our simple model has the following properties: 
\begin{itemize}
\item We assume that the sea quark contribution is negligible. For that
reason, the plus OFPDs in the parametrization (\ref{eq:parametrizationDVCS}) are equal to 
the valence OFPDs, $H_{f}^{+}=H_{f}^{val}$ and $\tilde{H}_{f}^{+}=\tilde{H}_{f}^{val}$, 
with the quark flavor $f=u,\; d$, and similarly for $E_{f}^{+}$ and $\tilde{E}_{f}^{+}$
distributions.
\item We assume that for all distributions the \emph{t}-dependence factorize from the 
dependence on other two scaling variables. The dependence of GPDs on the invariant \emph{t} is 
then characterized by the corresponding form factors. 
\item The $\xi$-dependence of OFPDs appears only in the $\tilde{E}_{f}^{+}$ 
distribution. 
\end{itemize}
The parametrization of the unpolarized quark OFPDs is taken from Ref. \cite{Guichon:1998xv}. 
Namely,
\begin{eqnarray}
H_{u}^{val}\left(x,\xi,t\right) & = & u_{N}^{val}\left(x\right)
F_{1u}\left(t\right)/2,\nonumber \\
H_{d}^{val}\left(x,\xi,t\right) & = & d_{N}^{val}
\left(x\right)F_{1d}\left(t\right),\nonumber \\
E_{u}^{val}\left(x,\xi,t\right) & = & u_{N}^{val}
\left(x\right)F_{2u}\left(t\right)/2,\nonumber \\
E_{d}^{val}\left(x,\xi,t\right) & = & d_{N}^{val}
\left(x\right)F_{2d}\left(t\right).
\label{eq:HandE}
\end{eqnarray}
In particular, for the proton target the unpolarized valence quark distributions 
in the proton are given by \cite{Radyushkin:1998rt}
\begin{eqnarray}
u_{p}^{val}\left(x\right) & = & 1.89x^{-0.4}
\left(1-x\right)^{3.5}\left(1+6x\right),\nonumber \\
d_{p}^{val}\left(x\right) & = & 0.54x^{-0.6}\left(1-x\right)^{4.2}\left(1+8x\right).
\label{eq:unpolvalencedistributions}
\end{eqnarray}
They closely reproduce corresponding GRV parametrizations \cite{Gluck:1994uf} at a low normalization 
point, $-q_{1}^{2}\simeq1\;\mathrm{GeV}^{2}$ \cite{Musatov:1999gv}. 

The \emph{u}- and \emph{d}-quark form factors in Eq. (\ref{eq:HandE}) can be extracted 
from the proton and neutron Dirac and Pauli form factors according to 
$F_{1p\left(n\right)}=Q_{u}F_{1u\left(d\right)}+Q_{d}F_{1d\left(u\right)}$ 
and $F_{2p\left(n\right)}=Q_{u}F_{2u\left(d\right)}+Q_{d}F_{2d\left(u\right)}$. 
Furthermore, the proton and neutron form factors are related to the Sachs electric and 
magnetic form factors, see Eqs. (\ref{eq:sachelectricff}) and (\ref{eq:sachmagneticff}), through
\begin{eqnarray}
F_{1p\left(n\right)}\left(t\right) & = & \left[G_{Ep\left(n\right)}\left(t\right)-
\frac{t}{4M^{2}}G_{Mp\left(n\right)}\left(t\right)\right]
\left(1-\frac{t}{4M^{2}}\right)^{-1},\nonumber \\
F_{2p\left(n\right)}\left(t\right) & = & \left[G_{Mp\left(n\right)}
\left(t\right)-G_{Ep\left(n\right)}
\left(t\right)\right]\left(1-\frac{t}{4M^{2}}\right)^{-1},
\label{eq:f1andf2factors}
\end{eqnarray}
where the nucleon mass is $M\simeq0.94\;\mathrm{GeV}$. In the region of small \emph{t}, both Sachs 
form factors are well described by a dipole fit, 
\begin{eqnarray}
G_{Ep}\left(t\right)=\frac{G_{Mp}\left(t\right)}{1+\kappa_{p}}=
\frac{G_{Mn}\left(t\right)}{\kappa_{n}}=\left(1-\frac{t}{\Lambda^{2}}
\right)^{-2} & \mathrm{and} & G_{En}\left(t\right)=0,
\label{eq:Sachsformfactors}
\end{eqnarray}
with the parameter $\Lambda^2=0.71\;\mathrm{GeV}^2$ and $\kappa_{p}=1.793$
and $\kappa_{n}=-1.913$ are the proton and neutron anomalous magnetic
moments, respectively. In the polarized case, we take for the valence 
distributions \cite{Belitsky:2001ns}
\begin{eqnarray}
\tilde{H}_{u}^{val}\left(x,\xi,t\right) & = & \Delta u_{p}^{val}
\left(x\right)\left(1-\frac{t}{m_{A}^{2}}\right)^{-2},\nonumber \\
\tilde{H}_{d}^{val}\left(x,\xi,t\right) & = & \Delta d_{p}^{val}\left(x\right)
\left(1-\frac{t}{m_{A}^{2}}\right)^{-2},
\label{eq:Htilda}
\end{eqnarray}
with the mass parameter $m_{A}=1.03\;\mathrm{GeV}$. The \emph{t}-dependence in Eq. (\ref{eq:Htilda}) 
corresponds to the ratio $g_{A}\left(t\right)/g_{A}\left(t=0\right)$. The polarized valence quark 
distributions in the proton can be expressed in terms of the unpolarized distributions in the following 
way \cite{Goshtasbpour:1995eh}
\begin{eqnarray}
\Delta u_{p}^{val} & = & \cos\theta_{D}\left(u_{p}^{val}-
\frac{2}{3}d_{p}^{val}\right),\nonumber \\
\Delta d_{p}^{val} & = & \cos\theta_{D}\left(-\frac{1}{3}d_{p}^{val}\right),
\label{eq:polvalencedistributions}
\end{eqnarray}
where $\cos\theta_{D}=\left[1+\mathrm{H}_{0}\left(1-x^{2}\right)/\sqrt{x}\;\right]^{-1}$
with $\mathrm{H}_{0}=0.06$. Finally, for the $\widetilde{E}_{f}$
distribution we accept the pion pole dominated ansatz,
\begin{eqnarray}
\tilde{E}_{u}^{val}\left(x,\xi,t\right) & = & \frac{1}{2}F_{\pi}
\left(t\right)\frac{\theta\left(\left|x\right|<\xi\right)}{2\xi}\;
\phi_{\pi}\left(\frac{x+\xi}{2\xi}\right),\nonumber \\
\tilde{E}_{d}^{val}\left(x,\xi,t\right) & = & -\tilde{E}_{u}^{val}
\left(x,\xi,t\right).
\label{eq:Etilda}
\end{eqnarray}
The function $F_{\pi}\left(t\right)$ is taken in the form valid for $-t\ll M^{2}$ \cite{Penttinen:1999th}, 
\begin{eqnarray}
F_{\pi}\left(t\right) & = & 4g_{A}\left(t=0\right)M^{2}
\left[\frac{1}{\left(m_{\pi}^{2}-t\right)/\mathrm{GeV^{2}}}-
\frac{1.7}{\left(1-t/2\;\mathrm{GeV^{2}}\right)^{2}}\right],
\label{eq:Fpi}
\end{eqnarray}
where $m_{\pi}\simeq0.14\;\mathrm{GeV}$ denotes the pion mass and 
$g_{A}\left(t=0\right)=1.267$. For the pion distribution amplitude in Eq. (\ref{eq:Etilda}) we 
choose, for simplicity, its asymptotic form 
\begin{eqnarray}
\phi_{\pi}\left(u\right) & = & 6u\left(1-u\right).
\label{eq:Fipi}
\end{eqnarray}

\section{\ \,  Cross Section}

Having defined the kinematics and described the simple model, in the
following we compute separately the unpolarized cross sections for
the Compton and Bethe-Heitler contributions to the DVCS process on
a proton target using an electron beam. In fact, since the scattering amplitude consists 
of two parts, see Eq. (\ref{eq:totalamplitudeDVCS}), its modulus squared 
(and, accordingly the cross section) is given by the sum of three terms, namely, 
the Compton term, the Bethe-Heitler term and the interference term. Here we only consider 
the first two terms. Their cross sections are plotted for the in-plane kinematics 
against the angle $\theta_{\gamma\gamma}$ by taking the same kinematical point, 
$Q_{1}^{2}=2.5\;\mathrm{GeV}^{2}$ and $x_{B}=0.35$, however, for two beam energies, 
$\omega=5.75\;\mathrm{GeV}$ and $11\;\mathrm{GeV}$.

\begin{figure}
\centerline{\epsfxsize=5in\epsffile{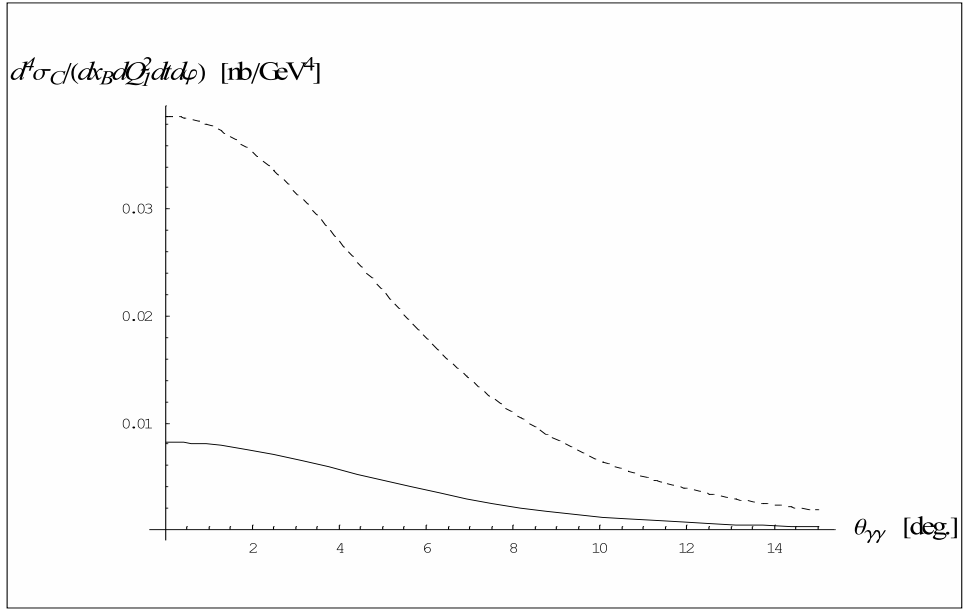}}
\caption{Compton cross section $\sigma_{C}$ plotted as a function of the angle 
$\theta_{\gamma\gamma}$ between the incoming virtual and outgoing real photon in the target rest frame 
for $Q_{1}^{2}=2.5\;\mathrm{GeV}^{2}$ and $x_{B}=0.35$ with $\omega=5.75\;\mathrm{GeV}$ (solid line) and 
$\omega=11\;\mathrm{GeV}$ (dashed line) electron beam.}
\label{standardCompton}
\end{figure}

With the help of the momentum-space Feynman rules for QED we can immediately
write down the T-matrix for the pure Compton process. Denoting the
polarization vector of the final real photon by $\epsilon_{\mu}^{*}\left(q_{2}\right)$, 
we have 
\begin{eqnarray}
i\mathrm{T}_{C} & = & \bar{u}\left(k'\right)\left(i\left|e\right|
\gamma^{\lambda}\right)u\left(k\right)
\left(\frac{-ig_{\nu\lambda}}{q_{1}^{2}}\right)\left(-e^{2}\right)\epsilon_{\mu}^{*}\left(q_{2}
\right)\left(-i\mathcal{T}_{EM}^{\mu\nu}\right),
\label{eq:Tmatrixcompton1}
\end{eqnarray}
or
\begin{eqnarray}
\mathrm{T}_{C} & = & \frac{\left|e\right|^{3}}{q_{1}^{2}}\bar{u}\left(k'\right)
\gamma_{\nu}u\left(k\right)\epsilon_{\mu}^{*}\left(q_{2}\right)\mathcal{T}_{EM}^{\mu\nu},
\label{eq:Tmatrixcompton2}
\end{eqnarray}
where $\mathcal{T}_{EM}^{\mu\nu}$ is given by the expression (\ref{eq:reducedVCAresult2}). 
Then one should average the square of Eq. (\ref{eq:Tmatrixcompton2}) over the initial 
proton and electron spins, and further sum it over the final proton and electron spins 
and photon polarizations. Note that this particular summation is performed using the 
Feynman gauge prescription, i.e. one can replace
\begin{eqnarray}
\sum_{\gamma\; polar.}\epsilon_{\mu}^{*}\left(q_{2}\right)\epsilon_{\alpha}
\left(q_{2}\right) & \longrightarrow & -g_{\mu\alpha}
\label{eq:feynmangauge}
\end{eqnarray}
by virtue of the Ward identity. As a result, we get a factorized expression,
\begin{eqnarray}
\overline{\left|\mathrm{T}_{C}\right|^{2}} & = & 
\frac{\left(4\pi\alpha\right)^{3}}{q_{1}^{4}}L_{\nu\beta}^{C}H_{C}^{\nu\beta},
\label{eq:comptonamplitudesquaredDVCS}
\end{eqnarray}
in terms of the electron and hadron tensors. Neglecting the electron mass, the electron tensor 
simply reads
\begin{eqnarray}
L_{\nu\beta}^{C} & = & 2\left[k_{\nu}k'_{\beta}+k_{\beta}k'_{\nu}-g_{\nu\beta}
\left(k\cdot k'\right)\right].
\label{eq:leptonictensorDVCS}
\end{eqnarray}
The hadron tensor requires some algebra, however, if we define the convolution integrals of OFPDs 
as a new set of functions,
\begin{eqnarray}
\mathcal{H}^{+}\left(\xi,t\right) & \equiv & \sum_{f}Q_{f}^{2}\int_{-1}^{1}
\frac{dx}{\left(x-\xi+i0\right)}H_{f}^{+}\left(x,\xi,t\right),\nonumber \\
\mathcal{E}^{+}\left(\xi,t\right) & \equiv & \sum_{f}Q_{f}^{2}\int_{-1}^{1}
\frac{dx}{\left(x-\xi+i0\right)}E_{f}^{+}\left(x,\xi,t\right),\nonumber \\
\tilde{\mathcal{H}}^{+}\left(\xi,t\right) & \equiv & \sum_{f}Q_{f}^{2}\int_{-1}^{1}
\frac{dx}{\left(x-\xi+i0\right)}\tilde{H}_{f}^{+}\left(x,\xi,t\right),\nonumber \\
\tilde{\mathcal{E}}^{+}\left(\xi,t\right) & \equiv & \sum_{f}Q_{f}^{2}
\int_{-1}^{1}\frac{dx}{\left(x-\xi+i0\right)}\tilde{E}_{f}^{+}\left(x,\xi,t\right),
\label{eq:integralsofGPDsDVCS}
\end{eqnarray}
and, in addition, neglect terms $\mathcal{O}\left(t/q_{1}^{2}\right)$ and 
$\mathcal{O}\left(M^{2}/q_{1}^{2}\right)$, then $H_{C}^{\nu\beta}$ can be written 
in a compact form as
\begin{eqnarray}
H_{C}^{\nu\beta} & = & -\frac{1}{2}\mathcal{T}_{EM}^{\mu\nu}\left(\mathcal{T}_{\mu EM}^{\beta}
\right)^{*}\nonumber \\
& = & - \left[\left(1-\xi^{2}\right)\left(\left|\mathcal{H}^{+}\right|^{2}+\left|
\tilde{\mathcal{H}}^{+}\right|^{2}\right)-\left(\xi^{2}+\frac{t}{4M^{2}}\right)\left|
\mathcal{E}^{+}\right|^{2}\right.\nonumber \\
&  & \left.-\xi^{2}\frac{t}{4M^{2}}\left|\tilde{\mathcal{E}}^{+}\right|^{2}-2\xi^{2}
\Re\left(\mathcal{H}^{+*}\mathcal{E}^{+}+\tilde{\mathcal{H}}^{+*}\tilde{\mathcal{E}}^{+}
\right)\right]\nonumber \\
&  & \times\left[g^{\nu\beta}-\frac{1}{\left(p\cdot q_{2}\right)}
\left(p^{\nu}q_{2}^{\beta}+p^{\beta}q_{2}^{\nu}\right)+
\frac{M^{2}}{\left(p\cdot q_{2}\right)^{2}}\left(1-\frac{t}{4M^{2}}\right)q_{2}^{\nu}q_{2}^{\beta}
\right].
\label{eq:hadronictensorDVCS}
\end{eqnarray}
The initial overall factor of $1/2$ in Eq. (\ref{eq:hadronictensorDVCS}) comes from averaging 
over the initial proton spin. Substituting now Eq. (\ref{eq:comptonamplitudesquaredDVCS}) into 
Eq. (\ref{eq:generaldiffcrosswithQ1squared}) gives the unpolarized differential cross section 
for the Compton contribution to the standard electromagnetic DVCS process 
\begin{eqnarray}
\frac{d^{4}\sigma_{C}}{dx_{B}dQ_{1}^{2}dtd\varphi} & = & \frac{\alpha^{3}}{16\pi Q_{1}^{4}}
\frac{1+x_{B}\left(M/\omega\right)}{M^{2}\omega^{2}\left[2+\left(M/\omega\right)\right]x_{B}
\left[y+2x_{B}\left(M/\omega\right)\right]^{2}}L_{\nu\beta}^{C}H_{C}^{\nu\beta}.\nonumber \\
\label{eq:diffcrosssectionComptonDVCS}
\end{eqnarray}
Its dependence on the angle $\theta_{\gamma\gamma}$ is shown in Fig. \ref{standardCompton}.

\begin{figure}
\centerline{\epsfxsize=5in\epsffile{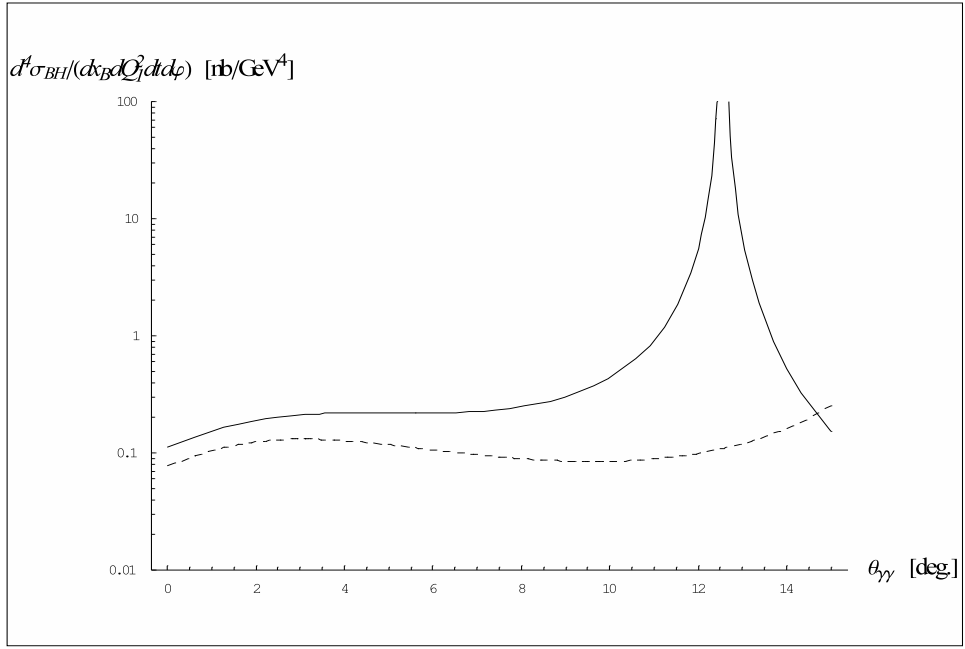}}
\caption{Bethe-Heitler cross section $\sigma_{BH}$ plotted as a function of the angle 
$\theta_{\gamma\gamma}$ between the incoming virtual and outgoing real photon in the target rest frame 
for $Q_{1}^{2}=2.5\;\mathrm{GeV}^{2}$ and $x_{B}=0.35$ with $\omega=5.75\;\mathrm{GeV}$ (solid line) and 
$\omega=11\;\mathrm{GeV}$ (dashed line) electron beam.}
\label{standardBH1}
\end{figure}

\begin{figure}
\centerline{\epsfxsize=5in\epsffile{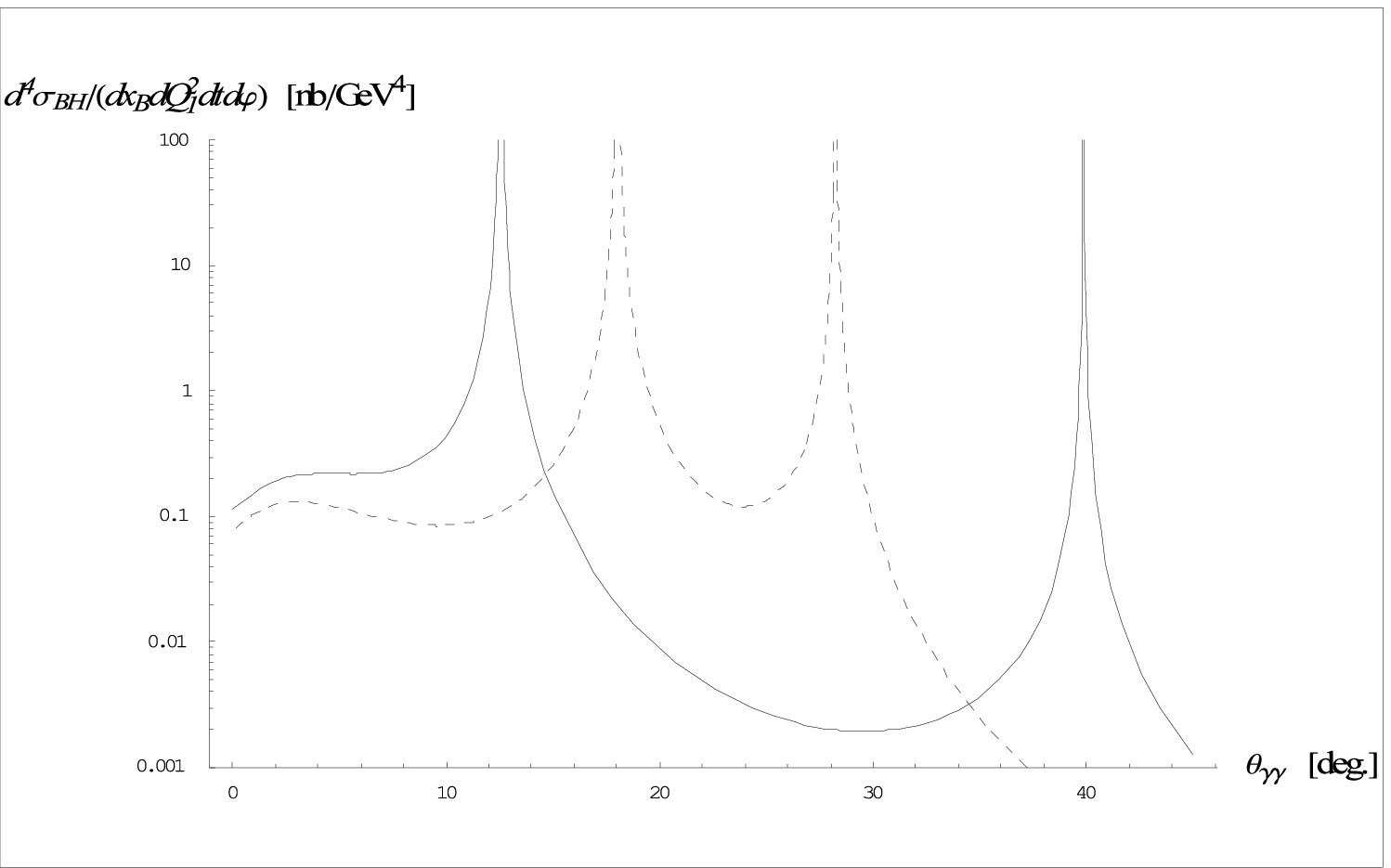}}
\caption{Bethe-Heitler cross section $\sigma_{BH}$ plotted as a function of the angle 
$\theta_{\gamma\gamma}$ between the incoming virtual and outgoing real photon in the target rest frame 
for $Q_{1}^{2}=2.5\;\mathrm{GeV}^{2}$ and $x_{B}=0.35$ with $\omega=5.75\;\mathrm{GeV}$ (solid line) and 
$\omega=11\;\mathrm{GeV}$ (dashed line) electron beam.}
\label{standardBH2}
\end{figure}

\begin{figure}
\centerline{\epsfxsize=5in\epsffile{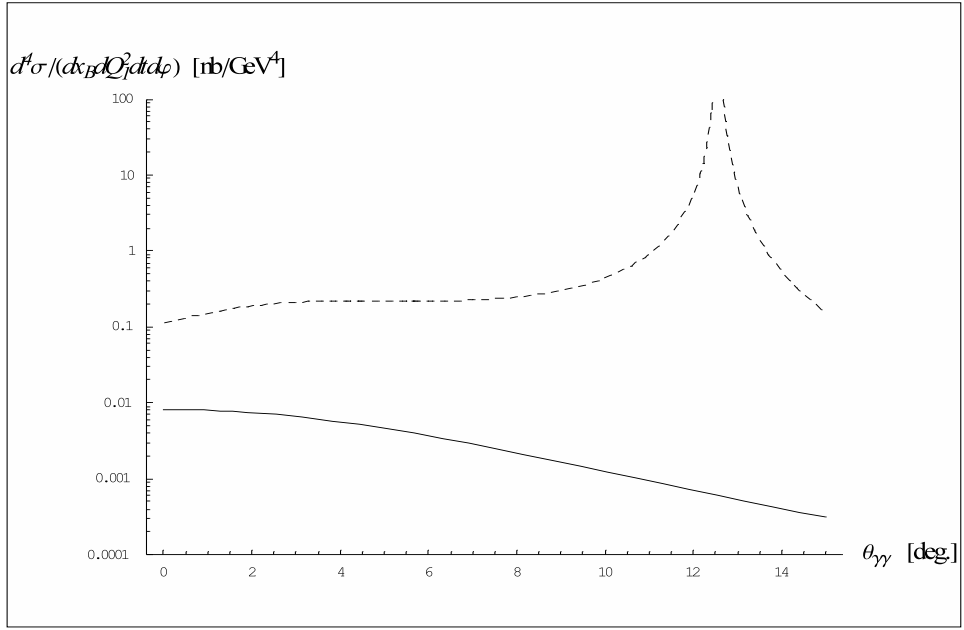}}
\caption{Compton cross section $\sigma_{C}$ (solid line) and Bethe-Heitler cross section $\sigma_{BH}$ 
(dashed line) plotted as a function of the angle $\theta_{\gamma\gamma}$ between the incoming virtual and 
outgoing real photon in the target rest frame for $Q_{1}^{2}=2.5\;\mathrm{GeV}^{2}$ and $x_{B}=0.35$ with 
$\omega=5.75\;\mathrm{GeV}$ electron beam.}
\label{togetherDVCS}
\end{figure}

Apart from the Compton part, the Bethe-Heitler contribution can be
fully calculable in QED with the knowledge on the nucleon form factors. The amplitude, 
which in contrast to $\mathrm{T}_{C}$ is purely real, emerges from two Bethe-Heitler diagrams 
shown in Fig. \ref{DVCSdiagrams},
\begin{eqnarray}
i\mathrm{T}_{BH} & = & \bar{u}\left(k'\right)\left(i\left|e\right|\gamma^{\mu}\right)
\epsilon_{\mu}^{*}\left(q_{2}\right)
\frac{i\left(\not\! k'+\not\! q_{2}\right)}{\left(k'+q_{2}\right)^{2}}\left(i\left|e\right|
\gamma^{\nu}\right)u\left(k\right)\left(\frac{-ig_{\nu\lambda}}{\left(q_{1}-q_{2}\right)^{2}}
\right)\nonumber \\
&  & \times\left(-i\left|e\right|\right)\left\langle p\left(p_{2},s_{2}\right)
\right|J_{EM}^{\lambda}\left(0\right)\left|p\left(p_{1},s_{1}\right)\right\rangle \nonumber \\
&  & +\bar{u}\left(k'\right)\left(i\left|e\right|\gamma^{\nu}\right)
\frac{i\left(\not\! k-\not\! q_{2}\right)}{\left(k-q_{2}\right)^{2}}\left(i\left|e\right|
\gamma^{\mu}\right)\epsilon_{\mu}^{*}\left(q_{2}\right)u\left(k\right)
\left(\frac{-ig_{\nu\lambda}}{\left(q_{1}-q_{2}\right)^{2}}\right)\nonumber \\
&  & \times\left(-i\left|e\right|\right)\left\langle p\left(p_{2},s_{2}\right)
\right|J_{EM}^{\lambda}\left(0\right)\left|p\left(p_{1},s_{1}\right)\right\rangle.
\label{eq:TmatrixBH1}
\end{eqnarray}
Using the relations, $\bar{u}\left(k'\right)\gamma^{\mu}\not\! k'=2\bar{u}\left(k'\right)k'^{\mu}$ 
and $\not\! k\gamma^{\mu}u\left(k\right)=2k^{\mu}u\left(p\right)$, we simplify the numerators of 
both fermion propagators in Eq. (\ref{eq:TmatrixBH1}), and obtain for the T-matrix
\begin{eqnarray}
\mathrm{T}_{BH} & = & \frac{\left|e\right|^{3}}{t}\epsilon_{\mu}^{*}\left(q_{2}\right)\bar{u}
\left(k'\right)
\left[\frac{\gamma^{\mu}\not\! q_{2}\gamma^{\nu}+2k'^{\mu}\gamma^{\nu}}{2\left(k'\cdot q_{2}
\right)}+
\frac{-\gamma^{\nu}\not\! q_{2}\gamma^{\mu}+2\gamma^{\nu}k^{\mu}}{-2\left(k\cdot q_{2}\right)}
\right]u\left(k\right)\nonumber \\
&  & \times\left\langle p\left(p_{2},s_{2}\right)\right|J_{\nu}^{EM}\left(0\right)
\left|p\left(p_{1},s_{1}
\right)\right\rangle.
\label{eq:TmatrixBH2}
\end{eqnarray}
The matrix element of the proton transition current, see Section III.2, is parametrized 
in terms of the usual Dirac and Pauli proton form factors,
\begin{eqnarray}
\left\langle p\left(p_{2},s_{2}\right)\right|J_{\nu}^{EM}\left(0\right)\left|p\left(p_{1},s_{1}\right)
\right\rangle  & = & \bar{u}\left(p_{2},s_{2}\right)\left[F_{1p}\left(t\right)\gamma_{\nu}-F_{2p}
\left(t\right)\frac{i\sigma_{\nu\lambda}r^{\lambda}}{2M}\right]u\left(p_{1},s_{1}\right).\nonumber \\
\label{eq:protonmatrixelementemdvcs}
\end{eqnarray}
The spin-averaged square of Eq. (\ref{eq:TmatrixBH2}) is again written
in terms of two tensors,
\begin{eqnarray}
\overline{\left|\mathrm{T}_{BH}\right|^{2}} & = & 
\frac{\left(4\pi\alpha\right)^{3}}{t^{2}}L_{BH}^{\nu\beta}H_{\nu\beta}^{BH}.
\label{eq:BHamplitudesquaredDVCS}
\end{eqnarray}
Here the hadron tensor reduces to a simple expression, namely,
\begin{eqnarray}
H_{\nu\beta}^{BH} & = & \frac{1}{2}\sum_{s_{1},s_{2}}\left\langle p\left(p_{2},s_{2}\right)
\right|J_{\nu}^{EM}\left(0\right)\left|p\left(p_{1},s_{1}\right)\right\rangle \left\langle p
\left(p_{2},s_{2}\right)\right|J_{\beta}^{EM}\left(0\right)\left|p\left(p_{1},s_{1}\right)\right
\rangle ^{*}\nonumber \\
& = & t\left[g_{\nu\beta}-\frac{r_{\nu}r_{\beta}}{t}\right]\left[F_{1p}\left(t\right)+F_{2p}
\left(t\right)\right]^{2}\nonumber \\
&  & +4\left[p_{1\nu}+\frac{r_{\nu}}{2}\right]\left[p_{1\beta}+\frac{r_{\beta}}{2}\right]
\left[F_{1p}^{2}\left(t\right)-\frac{t}{4M^{2}}F_{2p}^{2}\left(t\right)\right]
\label{eq:BHhadronictensorDVCS}
\end{eqnarray}
and the lepton tensor is
\begin{eqnarray}
L_{BH}^{\nu\beta} & = & -\frac{1}{2}\mathrm{Tr}\left\{ \not\! k'
\left[\frac{\gamma^{\mu}\not\! q_{2}\gamma^{\nu}+2k'^{\mu}\gamma^{\nu}}{2\left(k'\cdot q_{2}\right)}+
\frac{\gamma^{\nu}\not\! q_{2}\gamma^{\mu}-2\gamma^{\nu}k^{\mu}}{2\left(k\cdot q_{2}\right)}
\right]\right.\nonumber \\
&  & \left.\times\not\! k\left[\frac{\gamma^{\beta}\not\! q_{2}\gamma_{\mu}+2\gamma^{\beta}k'_{\mu}}{2
\left(k'\cdot q_{2}\right)}+\frac{\gamma_{\mu}\not\! q_{2}\gamma^{\beta}-2k_{\mu}\gamma^{\beta}}{2
\left(k\cdot q_{2}\right)}\right]\right\} \nonumber \\
& = & \frac{2}{\left(k'\cdot q_{2}\right)}\left[k^{\nu}q_{2}^{\beta}+k^{\beta}q_{2}^{\nu}-g^{\nu\beta}
\left(k\cdot q_{2}\right)\right]+\frac{2}{\left(k\cdot q_{2}\right)}
\left[k'^{\nu}q_{2}^{\beta}+k'^{\beta}q_{2}^{\nu}-g^{\nu\beta}
\left(k'\cdot q_{2}\right)\right]\nonumber \\
&  & +\frac{2}{\left(k'\cdot q_{2}\right)\left(k\cdot q_{2}\right)}\left[\left(k\cdot q_{2}\right)
\left[k^{\nu}k'^{\beta}+k^{\beta}k'^{\nu}+2k'^{\nu}k'^{\beta}\right]\right.\nonumber \\
&  & -\left(k'\cdot q_{2}\right)
\left[k^{\nu}k'^{\beta}+k^{\beta}k'^{\nu}+2k^{\nu}k^{\beta}\right]\nonumber \\
&  & +\left(k\cdot k'\right)
\left[k^{\nu}q_{2}^{\beta}+k^{\beta}q_{2}^{\nu}-k'^{\nu}q_{2}^{\beta}-k'^{\beta}q_{2}^{\nu}+2k^{
\nu}k'^{\beta}+2k^{\beta}k'^{\nu}\right]\nonumber \\
&  & \left.+2g^{\nu\beta}\left(k\cdot k'\right)\left[\left(k'\cdot q_{2}\right)-\left(k\cdot q_{2}
\right)-\left(k\cdot k'\right)\right]\right].
\label{eq:BHleptonictensorDVCS}
\end{eqnarray}
Finally, the unpolarized differential cross section reads
\begin{eqnarray}
\frac{d^{4}\sigma_{BH}}{dx_{B}dQ_{1}^{2}dtd\varphi} & = & \frac{\alpha^{3}}{16\pi t^{2}}
\frac{1+x_{B}\left(M/\omega\right)}{M^{2}\omega^{2}\left[2+\left(M/\omega\right)
\right]x_{B}\left[y+2x_{B}\left(M/
\omega\right)\right]^{2}}L_{BH}^{\nu\beta}H_{\nu\beta}^{BH}.\nonumber \\
\label{eq:diffcrosssectionfinalBHemdvcs}
\end{eqnarray}
The Bethe-Heitler contribution is illustrated on a logarithmic scale in 
Figs. \ref{standardBH1} and \ref{standardBH2}. We plot the latter for a wider range in 
$\theta_{\gamma\gamma}$ to be able to see both poles at the angles $\phi$ and $\phi'$, 
given in Table \ref{tableangles}. The poles corresponds to the situation, when the outgoing real photon is 
collinear either with the incoming or scattered electron. In Fig. \ref{togetherDVCS}, a logarithmic plot 
of both contributions together is shown for $\omega=5.75\;\mathrm{GeV}$. The comparison of the results shows 
that the Bethe-Heitler signal is well above the Compton one. This is easy to see just by comparing the 
factorized expressions (\ref{eq:comptonamplitudesquaredDVCS}) and (\ref{eq:BHamplitudesquaredDVCS}), and 
recalling that $t\ll q_{1}^{2}$. The presence of $1/t^{2}$ in the Bethe-Heitler part enhances its 
contribution with respect to the Compton part, which is proportional to $1/q_{1}^{4}$. 

One way to minimize the contamination with the dominating Bethe-Heitler process is to 
find the kinematical regions, where the Bethe-Heitler contribution is suppressed, or 
at least comparable with the Compton contribution. Another way is to exploit the interference between 
the two processes. This approach is based on the fact that $\mathrm{T}_{C}$ 
has both real and imaginary parts while $\mathrm{T}_{BH}$ is purely real. By incorporating the 
interference terms between the Compton and Bethe-Heitler amplitudes, we can disentangle $\Re\mathrm{T}_{C}$ 
and $\Im\mathrm{T}_{C}$. In particular, using the positron beam in addition to the electron beam, 
one can measure the so-called beam-charge asymmetry, which is sensitive to the real part of $\mathrm{T}_{C}$. 
Furthermore, measuring the single-spin (or alternatively the beam-spin asymmetry) by considering electrons 
with opposite helicities gives access to the imaginary part of $\mathrm{T}_{C}$. 
First experiments of this kind were performed at Jefferson Lab \cite{Stepanyan:2001sm} 
and at Hermes \cite{Airapetian:2001yk}. As a result, we can project out independently both parts of the 
amplitude and accordingly, probe different linear combinations of OFPDs. 

\chapter{Inclusive Photoproduction of Lepton Pairs}

\section{\ \,  Introduction}

\begin{figure}
\centerline{\epsfxsize=3in\epsffile{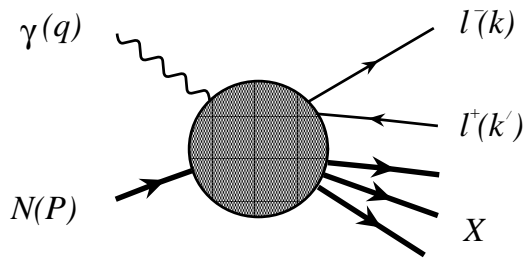}}
\caption{Inclusive photoproduction of lepton pairs.}
\label{inclusivephotoproductionfigure}
\end{figure}

By the inclusive photoproduction of lepton pairs, see Ref. \cite{Psaker:2004qf}, we refer to the reaction, 
in which a high-energy real photon $\gamma$ (with the four-momentum \emph{q}) scatters inelastically 
from a nucleon $N\left(P\right)$ emitting a pair of leptons (electrons or muons) with momenta $k$ and $k'$,
\begin{eqnarray}
\gamma\left(q\right)+N\left(P\right) & \longrightarrow & l^{-}\left(k\right)+l^{+}\left(k'\right)+X.
\label{eq:inclusivereaction}
\end{eqnarray}
The process is shown in Fig. \ref{inclusivephotoproductionfigure}, where again \emph{X} labels a system 
of hadrons produced through inelastic processes. The reaction (\ref{eq:inclusivereaction}) is a crossed 
channel to inclusive virtual Compton scattering,
\begin{eqnarray}
e^{-}+N & \longrightarrow & e^{-}+\gamma+X,
\label{eq:inclusiveVCS}
\end{eqnarray}
which was originally studied in the parton model in Ref. \cite{Brodsky:1972yx}. 

\begin{figure}
\centerline{\epsfxsize=6in\epsffile{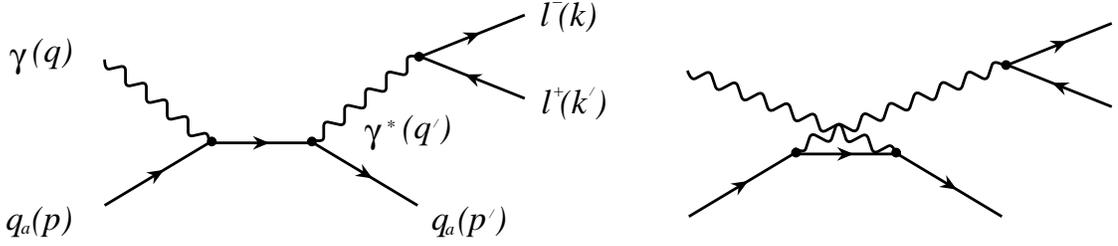}}
\caption{Compton contribution to the inclusive photoproduction of lepton pairs in the parton model.}
\label{inclusivecompton}
\end{figure}

\begin{figure}
\centerline{\epsfxsize=4in\epsffile{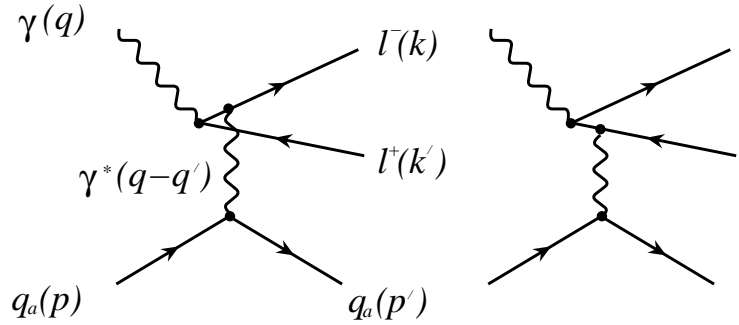}}
\caption{Bethe-Heitler contribution to the inclusive photoproduction of lepton pairs in the parton model.}
\label{inclusivebetheheitler}
\end{figure}

In the framework of the QCD parton model introduced in Chapter II,
the elementary photon-parton scattering subprocess can be viewed in
two different ways. The incident photon can either scatter off a parton or split into a lepton pair. 
Thus we have two contributions to the process at the amplitude level. According to the first scenario, 
known as the Compton contribution, a heavy time-like photon, $\gamma^{*}\left(q'\right)$ with $q'^{2}>0$, 
is produced and decays eventually into a pair of leptons, as shown in Fig. \ref{inclusivecompton}. 
We discuss this particular subprocess (see also \cite{Bjorken:1969uu}) in Section V.2. 
The second scenario is illustrated illustrated in Fig. \ref{inclusivebetheheitler} and corresponds to 
the Bethe-Heitler mechanism. It will be studied in Section V.3. The real and virtual photons can 
be interchanged, and hence both the Compton and Bethe-Heitler contributions consist of two Feynman diagrams.

\section{\ \,  Compton Subprocess}

The invariant matrix element for the Compton part comes from two Feynman diagrams 
shown in Fig. \ref{inclusivecompton}. Neglecting the parton masses by taking $p^{2}=p'^{2}=0$, we have
\begin{eqnarray}
i\mathrm{T}_{C} & = & \bar{u}\left(k\right)\left(i\left|e\right|
\gamma^{\lambda}\right)v\left(k'\right)
\left(\frac{-ig_{\mu\lambda}}{q'^{2}}\right)\bar{u}\left(p'\right)\left(-iQ_{a}
\left|e\right|\gamma^{\mu}\right)
\frac{i\left(\not\! p+\not\! q\right)}{\left(p+q\right)^{2}}\nonumber \\
&  & \times\left(-iQ_{a}\left|e\right|\gamma^{\nu}\right)
\epsilon_{\nu}\left(q\right)u
\left(p\right)\nonumber \\
&  & +\bar{u}\left(k\right)\left(i\left|e\right|
\gamma^{\lambda}\right)v\left(k'\right)
\left(\frac{-ig_{\mu\lambda}}{q'^{2}}\right)\bar{u}
\left(p'\right)\left(-iQ_{a}\left|e\right|
\gamma^{\nu}\right)
\epsilon_{\nu}\left(q\right)
\frac{i\left(\not\! p-\not\! q'\right)}{\left(p-q'\right)^{2}}\nonumber \\
&  & \times\left(-iQ_{a}\left|e\right|\gamma^{\mu}\right)u\left(p\right),
\label{eq:comptonamplitude1}
\end{eqnarray}
where $\epsilon_{\nu}\left(q\right)$ with  $q^{2}=0$ denotes the polarization vector of the initial 
real photon. By simplifying the numerator of each propagator in Eq. (\ref{eq:comptonamplitude1}), we 
obtain for the T-matrix
\begin{eqnarray}
\mathrm{T}_{C} & = & -\frac{\left|e\right|^{3}Q_{a}^{2}}{q'^{2}}\bar{u}\left(k\right)\gamma_{\mu}v
\left(k'\right)
\epsilon_{\nu}\left(q\right)\bar{u}\left(p'\right)\left[
\frac{\gamma^{\mu}\not\! q\gamma^{\nu}+2\gamma^{\mu}p^{\nu}}{2\left(p\cdot q\right)}
\right.\nonumber \\
&  & \left.+
\frac{-\gamma^{\nu}\not\! q'\gamma^{\mu}+2\gamma^{\nu}p^{\mu}}{-2\left(p'\cdot q\right)}
\right]u\left(p\right).
\label{eq:comptonamplitude2}
\end{eqnarray}

We will assume that both the nucleon target and the photon beam are unpolarized. 
Then the square of Eq. (\ref{eq:comptonamplitude2}) has to be averaged over the initial parton 
and photon polarizations, and further summed over the final parton and lepton polarizations. 
It can be written in a factorized form in the following way
\begin{eqnarray}
\overline{\left|\mathrm{T}_{C}\right|^{2}} & = & 
\frac{\left(4\pi\alpha\right)^{3}Q_{a}^{4}}{2q'^{4}}L_{\mu\rho}^{C}H_{C}^{\mu\rho}.
\label{eq:Tamplitudesquaredcompton}
\end{eqnarray}
The lepton tensor is
\begin{eqnarray}
L_{\mu\rho}^{C} & = & 4\left[k_{\mu}k'_{\rho}+k_{\rho}k'_{\mu}-g_{\mu\rho}
\left[\left(k\cdot k'\right)+m^{2}\right]\right],
\label{eq:leptonictensorcompton}
\end{eqnarray}
where \emph{m} denotes the lepton mass. The hadron tensor, on the other hand, is more complicated. 
In the Feynman gauge, it reads
\begin{eqnarray}
H_{C}^{\mu\rho} & = & 2\left[\frac{A_{C}}{\left(p\cdot q\right)^{2}}+
\frac{B_{C}}{\left(p'\cdot q\right)^{2}}+
\frac{C_{C}}{\left(p\cdot q\right)\left(p'\cdot q\right)}\right],\nonumber \\
A_{C} & = & \left(p\cdot q\right)\left[p'^{\mu}q^{\rho}+p'^{\rho}q^{\mu}-g^{\mu\rho}
\left(p'\cdot q\right)\right],\nonumber \\
B_{C} & = & \left(p\cdot q'\right)\left[p'^{\mu}p^{\rho}+p'^{\rho}p^{\mu}\right]+\left(p'
\cdot q'\right)
\left[p^{\mu}q'^{\rho}+p^{\rho}q'^{\mu}-2p^{\mu}p^{\rho}\right]\nonumber \\
&  & +\left(p\cdot p'\right)\left[2p^{\mu}p^{\rho}-p^{\mu}q'^{\rho}-p^{\rho}q'^{\mu}\right]+
\frac{q'^{2}}{2}
\left[-p'^{\mu}p^{\rho}-p'^{\rho}p^{\mu}+g^{\mu\rho}(p\cdot p')\right]\nonumber \\
&  & -g^{\mu\rho}\left(p\cdot q'\right)\left(p'\cdot q'\right),\nonumber \\
C_{C} & = & \left(p\cdot q\right)
\left[p'^{\mu}q'^{\rho}+p'^{\rho}q'^{\mu}-p'^{\mu}p^{\rho}-p'^{\rho}p^{\mu}\right]+
\left(p'\cdot q\right)\left[2p^{\mu}p^{\rho}-p^{\mu}q'^{\rho}-p^{\rho}q'^{\mu}\right]\nonumber \\
&  & +\left(p\cdot p'\right)
\left[-p^{\mu}q^{\rho}-p^{\rho}q^{\mu}-p^{\mu}q'^{\rho}-p^{\rho}q'^{\mu}\right]\nonumber \\
&  & +g^{\mu\rho}\left[2\left(p\cdot p'\right)\left(p\cdot q'\right)+\left(q\cdot q'\right)
\left(p\cdot p'\right)+
\left(p\cdot q'\right)\left(p'\cdot q\right)-\left(p\cdot q\right)\left(p'\cdot q'
\right)\right].\nonumber \\
\label{eq:hadronictensorcompton}
\end{eqnarray}
At this point, it is convenient to introduce the Mandelstam variables for the scattering 
subprocess at the parton level, namely,
\begin{eqnarray}
\hat{s} & = & \left(p+q\right)^{2}=\left(p'+q'\right)^{2},\nonumber \\
\hat{t} & = & \left(p'-p\right)^{2}=\left(q'-q\right)^{2},\nonumber \\
\hat{u} & = & \left(q'-p\right)^{2}=\left(p'-q\right)^{2}.
\label{eq:mandelstamvariables}
\end{eqnarray}
The four-momentum conservation implies that $\hat{s}+\hat{t}+\hat{u}=q'^{2}$.
With the help of Eq. (\ref{eq:mandelstamvariables}) one can immediately verify the gauge 
invariance of the hadron tensor,
\begin{eqnarray}
q'_{\mu}q'_{\rho}H_{C}^{\mu\rho} & = & 2\left[\frac{\hat{s}q'^{2}-q'^{4}-\hat{s}
\hat{t}+q'^{2}\hat{t}+q'^{2}\hat{u}}
{\hat{s}}+
\frac{\left(q'^{2}-\hat{u}\right)\left(\hat{u}\hat{t}+q'^{2}\hat{t}\right)-q'^{4}
\hat{t}}{\hat{u}^{2}}
\right.\nonumber \\
&  & \left.-\frac{\hat{s}\hat{u}\left(\hat{s}-q'^{2}\right)+\hat{u}^{2}\left(q'^{2}-
\hat{u}\right)-\hat{u}\hat{t}
\left(q'^{2}-\hat{u}\right)}{\hat{s}\hat{u}}\right]\nonumber \\
& = & 2\left[-\hat{t}-\hat{t}+2\hat{t}\right]\nonumber \\
& = & 0.
\label{eq:gaugeinvariance}
\end{eqnarray}
In addition, by setting $q'^{2}=0$ in Eq. (\ref{eq:hadronictensorcompton}), in other words, 
the outgoing photon now becomes real, and also replacing the lepton tensor with the metric tensor 
$-g_{\mu\rho}$, we easily recover the high-energy limit of the usual Compton scattering process, 
\begin{eqnarray}
\overline{\left|\mathrm{T}_{C}\right|^{2}} & \sim & -\left[\frac{\hat{u}}{\hat{s}}+
\frac{\hat{s}}{\hat{u}}\right].
\label{eq:usualcompton}
\end{eqnarray}

To calculate the unpolarized scattering cross section, we should integrate 
Eq. (\ref{eq:Tamplitudesquaredcompton}) over the Lorentz-invariant phase space defined as
\begin{eqnarray}
d\Pi & = & (2\pi)^{4}\;\delta^{(4)}\left(p+q-p'-k-k'\right)\left[\frac{d^{3}p'}{(2\pi)^{3}2E_{p'}}
\right]\left[
\frac{d^{3}k}{(2\pi)^{3}2\omega}\right]\left[\frac{d^{3}k'}{(2\pi)^{3}2\omega'}\right]
\nonumber \\
\label{eq:phasespace}
\end{eqnarray}
in a specific frame of reference, and divide the result of integration by the flux factor. 
The latter is written in the invariant form as $2\hat{s}$. Since the brute-force contraction 
of tensors $L_{\mu\rho}^{C}$ and $H_{C}^{\mu\rho}$ is somewhat tedious, we use a little trick instead. 
One notices that $L_{\mu\rho}^{C}$ depends only on the momenta of the final leptons. For that reason, 
we can first integrate the lepton tensor over $\vec{k}$ and $\vec{k}'$, then contract it with 
$H_{C}^{\mu\rho}$, and finally perform integration over the remaining momentum $\vec{p\,}'$. 
Switching to the symmetric momentum variables (i.e. the lepton pair four-momentum, $q'=k+k'$, and the 
relative four-momentum, $\kappa=\left(k-k'\right)/2$) and, with the help of the formula 
\begin{eqnarray}
\frac{d^{3}k}{2\omega} & = & \int d^{4}k\;\delta^{+}\left(k^{2}-m^{2}\right),
\label{eq:4dimintegral}
\end{eqnarray}
where $\delta^{+}\left(k^{2}-m^{2}\right)\equiv\delta\left(k^{2}-m^{2}\right)\theta\left(k_{0}\right)$,
writing the three-dimensional Lorentz-invariant volume in the momentum space as the four-dimensional 
integral, we find
\begin{eqnarray}
\int\frac{d^{3}k}{2\omega}\int\frac{d^{3}k'}{2\omega'}\;\delta^{\left(4\right)}\left(p+q-p'-k-k'
\right)L_{\mu\rho}^{C} & =\nonumber \\
\int d^{4}q'\;\delta^{(4)}\left(p+q-p'-q'\right)\Bigg\lbrace\int d^{4}\kappa\;\delta^{+}
\left[\left(q'/2+
\kappa\right)^{2}-m^{2}
\right]\nonumber \\\times\delta^{+}
\left[\left(q'/2-\kappa\right)^{2}-m^{2}\right]L_{\mu\rho}^{C}\Bigg\rbrace.
\label{eq:symmetricvariablesintegral}
\end{eqnarray}
The lepton tensor is now equal to
\begin{eqnarray}
L_{\mu\rho}^{C} & = & 4\left[\frac{1}{2}q'_{\mu}q'_{\rho}-2\kappa_{\mu}\kappa_{\rho}-g_{\mu\rho}
\frac{q'^{2}}{2}\right].
\label{eq:comptonleptonictensorsymmetric}
\end{eqnarray}
The integral over the relative momentum on the right-hand-side of
Eq. (\ref{eq:symmetricvariablesintegral}) has a tensor structure,
and it can be constructed out of the tensors $g_{\mu\rho}$ and $q'_{\mu}q'_{\rho}$
and scalar functions of the invariant mass of the lepton pair
$q'^{2}$. Moreover, the Ward identity implies that this integral must assume the form
\begin{equation}
\int d^{4}\kappa\;\delta^{+}\left[\left(q'/2+\kappa\right)^{2}-m^{2}\right]\delta^{+}\left[
\left(q'/2-\kappa\right)^{2}-m^{2}
\right]L_{\mu\rho}^{C}=\Phi\left(q'^{2}\right)\left[q'^{2}g_{\mu\rho}-q'_{\mu}q'_{\rho}\right],
\label{eq:newform}
\end{equation}
where the scalar function $\Phi\left(q'^{2}\right)$ is regular (i.e. it has no pole) at $q'^{2}=0$. 
Next we calculate $\Phi\left(q'^{2}\right)$. It is a function of a Lorentz scalar and thus frame independent. 
As a reference frame, it is convenient to choose the center-of-mass frame of the lepton pair. 
In this particular frame, the symmetric momenta are simply given by $q'=\left(\sqrt{q'^{2}},\vec{0}\right)$ 
and $\kappa=\left(0,\vec{\kappa}\right)$. After contracting both sides of Eq. (\ref{eq:newform}) with 
$g^{\mu\rho}$, then using 
\begin{eqnarray}
\int dx\int dy\;\delta(x)\delta(y) & = & 2\int dx\int dy\;\delta(x+y)\delta(x-y),
\label{eq:deltafunctionformula}
\end{eqnarray}
and finally integrating over $\kappa$ (the angular integration simply gives $4\pi$), we get for the scalar 
function
\begin{eqnarray}
\Phi\left(q'^{2}\right) & = & -\frac{4}{3q'^{2}}\int d^{4}\kappa\;\delta^{+}\left[\left(q'/2+
\kappa\right)^{2}-m^{2}
\right]
\delta^{+}\left[\left(q'/2-\kappa\right)^{2}-m^{2}\right]\left[3q'^{2}/2+2\kappa^{2}
\right]\nonumber \\
& = & -\frac{16\pi}{3q'^{2}}\int d\left(\left|\vec{\kappa}\right|^{2}\right)
\int d\kappa_{0}\;\delta^{+}
\left[2\sqrt{q'^{2}}\;
\kappa_{0}\right]\delta^{+}\left[q'^{2}/2+2\kappa_{0}^{2}-2\left|\vec{\kappa}\right|^{2}-2m^{2}
\right]\nonumber \\
&  & \times\left|\vec{\kappa}\right|\left[3q'^{2}/2+2\kappa_{0}^{2}-2\left|\vec{\kappa}\right|^{2}
\right]\nonumber \\
& = & -\frac{2\pi}{3}\frac{q'^{2}+2m^{2}}{q'^{2}}\sqrt{1-\frac{4m^{2}}{q'^{2}}}.
\label{eq:scalarfunction}
\end{eqnarray}
Note that the scalar function $\Phi\left(q'^{2}\right)<0$. For the sake of completeness, it is worth 
showing that generalizing the calculation of $\Phi\left(q'^{2}\right)<0$  to any reference frame 
indeed yields the same result. For instance, integration over $\kappa_{0}$ gives 
\begin{eqnarray}
\Phi\left(q'^{2}\right) & \sim & \int d\left(\cos\vartheta\right)
\int d\left(\left|\vec{\kappa}\right|^{2}\right)\;
\left|\vec{\kappa}\right|\delta^{+}\left[q'^{2}/2-2\left|\vec{\kappa}\right|^{2}\left(1-
\left|\vec{q\,}'
\right|^{2}\cos^{2}\vartheta/q_{0}'^{2}\right)-2m^{2}\right]\nonumber \\
&  & \times\left[3q'^{2}/2-2\left|\vec{\kappa}\right|^{2}\left(1-\left|\vec{q\,}'\right|^{2}\cos^{2}
\vartheta/q_{0}'^{2}\right)\right],
\label{eq:generalexpressionforphi1}
\end{eqnarray}
where $\vartheta$ is the angle between vectors $\vec{\kappa}$ and $\vec{q}\,'$ in the lepton scattering plane. 
Furthermore, integrating over $\left|\vec{\kappa}\right|^{2}$ together with the substitution 
$z=\left|\vec{q\,}'\right|\cos\vartheta/q_{0}'$ produces the integral of the type 
\begin{equation}
\int_{-\left|\vec{q\,}'\right|/q_{0}'}^{\left|\vec{q\,}'\right|/q_{0}'}
\frac{dz}{\left(1-z^{2}\right)^{3/2}},
\label{eq:generalintegralcompton}
\end{equation}
which leads to the same result for $\Phi\left(q'^{2}\right)$ given by Eq. (\ref{eq:scalarfunction}).

Thus the double integral (\ref{eq:symmetricvariablesintegral}) reduces now to
\begin{eqnarray}
\int\frac{d^{3}k}{2\omega}\int\frac{d^{3}k'}{2\omega'}\;\delta^{\left(4\right)}\left(p+q-p'-k-k'
\right)L_{\mu\rho}^{C} & = & 
\int d^{4}q'\;
\delta^{(4)}\left(p+q-p'-q'\right)\nonumber \\
&  & \times\Phi\left(q'^{2}\right)\left[q'^{2}g_{\mu\rho}-q'_{\mu}q'_{\rho}\right].
\label{eq:symmetricvariablesintegralnew}
\end{eqnarray}
Contracting the tensor in the integrand with $H_{C}^{\mu\rho}$ can be easily carried out. In terms of 
the subprocess Mandelstam variables, we obtain
\begin{eqnarray}
\left[q'^{2}g_{\mu\rho}-q'_{\mu}q'_{\rho}\right]H_{C}^{\mu\rho} & = & 4q'^{2}
\left[\frac{\hat{u}}{\hat{s}}+
\frac{\hat{s}}{\hat{u}}+
\frac{2q'^{2}\hat{t}}{\hat{s}\hat{u}}\right].
\label{eq:comptoncontraction}
\end{eqnarray}
Note that the second term on the left-hand side of Eq. (\ref{eq:comptoncontraction}) gives no contribution 
due to the condition (\ref{eq:gaugeinvariance}). After combining the result of contraction with 
Eqs. (\ref{eq:Tamplitudesquaredcompton}) and (\ref{eq:phasespace}) and using Eq. (\ref{eq:4dimintegral}), 
this time for the momentum $p'$, the unpolarized cross section for the Compton subprocess reads
\begin{eqnarray}
\sigma_{C} & = & \int d^{4}q'\int d^{4}p'\;\delta^{(4)}\left(p+q-p'-q'\right)\delta^{+}
\left(p'^{2}\right)\frac{1}{2\hat{s}}
\frac{1}{\left(2\pi\right)^{5}}\frac{\left(4\pi\alpha\right)^{3}Q_{a}^{4}}{2q'^{4}}\nonumber \\
&  & \times\Phi\left(q'^{2}\right)4q'^{2}\left[\frac{\hat{u}}{\hat{s}}+\frac{\hat{s}}{\hat{u}}+
\frac{2q'^{2}\hat{t}}{\hat{s}\hat{u}}\right].
\label{eq:comptoncrosssection1}
\end{eqnarray}
The delta function $\delta^{(4)}\left(p+q-p'-q'\right)$ can be trivially
integrated over $p'$. The final integral over $q'$ is performed in the photon-parton
center-of-mass frame. Since we want the cross section $\sigma_{C}$ to
be differential in the invariant mass of the lepton pair,
an extra substitution
\begin{eqnarray}
\int d^{4}q' & = & \int d^{4}q'\int dM_{pair}^{2}\;\delta\left(q'^{2}-M_{pair}^{2}\right)
\label{eq:substitution}
\end{eqnarray}
is being made. Then
\begin{eqnarray}
\sigma_{C} & = & \pi\int dM_{pair}^{2}\int d\left(\cos\theta_{cm}\right)\int d\left(\left|
\vec{q\,}'\right|^{2}\right)\int dq'_{0}\;
\delta^{+}\left(M_{pair}^{2}+\hat{s}-2\sqrt{\hat{s}}\; q'_{0}\right)\nonumber \\
&  & \times\delta\left(q_{0}'^{2}-\left|\vec{q\,}'\right|^{2}-M_{pair}^{2}\right)\left|\vec{q\,}'
\right|\frac{1}{\hat{s}}
\frac{1}{\left(2\pi\right)^{5}}\frac{\left(4\pi\alpha\right)^{3}Q_{a}^{4}}{M_{pair}^{2}}\nonumber \\
&  & \times\Phi\left(M_{pair}^{2}\right)\left[\frac{\hat{u}}{\hat{s}}+\frac{\hat{s}}{\hat{u}}+
\frac{2M_{pair}^{2}\hat{t}}{\hat{s}\hat{u}}\right],
\label{eq:comptoncrosssection2}
\end{eqnarray}
where $\theta_{cm}$ is the angle between the directions of the incoming
real and outgoing virtual photons in the photon-parton center-of-mass 
frame. The center-of-mass energy $E_{cm}$ is equally distributed between the photon and 
the parton and hence the invariant $\hat{s}=E_{cm}^{2}$. In the differential form, after integrations 
over $q'_{0}$ and $\left|\vec{q\,}'\right|^{2}$, we have
\begin{eqnarray}
\frac{d^{2}\sigma_{C}}{dM_{pair}^{2}d\left(\cos\theta_{cm}\right)} & = & \pi\left(
\frac{\hat{s}-M_{pair}^{2}}{4\hat{s}}\right)
\frac{1}{\hat{s}}\frac{1}{\left(2\pi\right)^{5}}\frac{\left(4\pi\alpha
\right)^{3}Q_{a}^{4}}{M_{pair}^{2}}\nonumber \\
&  & \times\Phi\left(M_{pair}^{2}\right)\left[\frac{\hat{u}}{\hat{s}}+\frac{\hat{s}}{\hat{u}}+
\frac{2M_{pair}^{2}\hat{t}}{\hat{s}\hat{u}}\right].
\label{eq:comptoncrosssection3}
\end{eqnarray}
Finally, replacing $\cos\theta_{cm}$ by the invariant $\hat{t}$
through the relation 
$\hat{t}=M_{pair}^{2}-\sqrt{\hat{s}}\left(q'_{0}-\sqrt{q_{0}'^{2}-M_{pair}^{2}}\cos\theta_{cm}\right)$, 
where $q'_{0}=\left(\hat{s}+M_{pair}^{2}\right)/2\sqrt{\hat{s}}$, we arrive at the partonic cross section
\begin{eqnarray}
\frac{d^{2}\sigma_{C}}{dM_{pair}^{2}d\hat{t}} & = & -\left(\frac{2\alpha^{3}Q_{a}^{4}}{3}
\right)\left(1+\frac{2m^{2}}{M_{pair}^{2}}\right)
\sqrt{1-\frac{4m^{2}}{M_{pair}^{2}}}\;\frac{1}{\hat{s}^{2}M_{pair}^{2}}\nonumber \\
&  & \times\left[\frac{\hat{s}^{2}+\hat{u}^{2}+2M_{pair}^{2}\hat{t}}{\hat{s}\hat{u}}\right].
\label{eq:comptoncrosssection4}\end{eqnarray}

We calculate now the cross section for photon-nucleon inelastic scattering
with the help of the parton model master formula, see Section II.3. The process cross
section is obtained by summing Eq. (\ref{eq:comptoncrosssection4})
over all types \emph{a} of charged partons and all possible longitudinal momentum
fractions \emph{x}, where the single summands must be weighted with
the proper PDFs $f_{a/N}\left(x\right)$. Namely,
\begin{eqnarray}
\sigma\left[\gamma\left(q\right)N\left(P\right)\rightarrow l^{-}\left(k\right)l^{+}
\left(k'\right)X\right] & =\nonumber \\
\sum_{a}\int_{0}^{1}dx\; f_{a/N}\left(x\right)\sigma\left[\gamma\left(q\right)q_{a}
\left(xP\right)\rightarrow l^{-}
\left(k\right)l^{+}\left(k'\right)q_{a}\left(p'\right)\right].
\label{eq:partonmodelformulaphotoproduction}
\end{eqnarray}
In addition, the subprocess Mandelstam variables are expressed in
terms of the process variables. The invariant $\hat{t}$ is identical
to $t\equiv\left(q'-q\right)^{2}$ . Neglecting the nucleon mass,
the invariant $\hat{s}$ is simply given by $\hat{s}=xs$, where 
$s\equiv\left(P+q\right)^{2}$. Accordingly, the Mandelstam invariant 
$\hat{u}=M_{pair}^{2}-t-xs$. Recall that by virtue of the vanishing mass of the scattered parton, 
$p'^{2}=\left(xP+q-q'\right)^{2}=0$, the longitudinal momentum fraction coincides with the Bjorken 
scaling variable,
\begin{eqnarray}
x & = & -\frac{t}{2\left[P\cdot\left(q-q'\right)\right]}\equiv x_{B}.
\label{eq:momentumfractionandbjorkenvariable}
\end{eqnarray}
Substituting Eq. (\ref{eq:comptoncrosssection4}) into Eq. (\ref{eq:partonmodelformulaphotoproduction}) 
ultimately yields the Compton differential cross section for the inclusive photoproduction of lepton pairs 
in the parton model
\begin{eqnarray}
\frac{d^{3}\sigma_{C}\left(\gamma N\rightarrow l^{-}l^{+}X\right)}{dM_{pair}^{2}dtdx_{B}} & = & 
\sum_{a}\int_{0}^{1}dx\;
\delta\left(x-x_{B}\right)f_{a/N}\left(x\right)\left\{ \frac{2\alpha^{3}Q_{a}^{4}}{3}
\left(1+\frac{2m^{2}}{M_{pair}^{2}}\right)\right.\nonumber \\
&  & \times\sqrt{1-\frac{4m^{2}}{M_{pair}^{2}}}\;
\frac{1}{\left(xs\right)^{2}M_{pair}^{2}}\nonumber \\
&  & \left.\times\left[\frac{\left(xs\right)^{2}+\left(M_{pair}^{2}-t-xs\right)^{2}+2M_{pair}^{2}
t}{xs\left(xs+t-M_{pair}^{2}\right)}\right]\right\} \nonumber \\
& = & \frac{2\alpha^{3}}{3}\left(1+\frac{2m^{2}}{M_{pair}^{2}}
\right)\sqrt{1-\frac{4m^{2}}{M_{pair}^{2}}}\;
\frac{1}{\left(x_{B}s\right)^{2}M_{pair}^{2}}\nonumber \\
&  & \times\left[
\frac{\left(x_{B}s\right)^{2}+\left(M_{pair}^{2}-t-x_{B}s\right)^{2}+2M_{pair}^{2}
t}{x_{B}s\left(x_{B}s+t-M_{pair}^{2}\right)}\right]\nonumber \\
&  & \times\sum_{a}Q_{a}^{4}f_{a/N}\left(x_{B}\right).
\label{eq:finalcomptonprocess}
\end{eqnarray}

\section{\ \,  Bethe-Heitler Subprocess}

The Bethe-Heitler amplitude is calculated from the Feynman diagrams shown in 
Fig. \ref{inclusivebetheheitler}. Adopting the notation of the preceding subsection, we write
\begin{eqnarray}
i\mathrm{T}_{BH} & = & \bar{u}\left(p'\right)\left(-iQ_{a}\left|e\right|
\gamma^{\lambda}\right)u\left(p\right)
\left(\frac{-ig_{\mu\lambda}}{\left(q-q'\right)^{2}}\right)\bar{u}\left(k\right)
\left(i\left|e\right|\gamma^{\mu}\right)
\frac{i\left(\not\! q-\not\! k'+m\right)}{\left(q-k'\right)^{2}-m^{2}}\nonumber \\
&  & \times\left(i\left|e\right|\gamma^{\nu}\right)\epsilon_{\nu}\left(q\right)v
\left(k'\right)\nonumber \\
&  & +\bar{u}\left(p'\right)\left(-iQ_{a}\left|e\right|\gamma^{\lambda}\right)u\left(p\right)
\left(\frac{-ig_{\mu\lambda}}{\left(q-q'\right)^{2}}\right)\bar{u}\left(k\right)\left(i\left|e
\right|\gamma^{\nu}\right)
\epsilon_{\nu}\left(q\right)
\frac{i\left(\not\! k-\not\! q+m\right)}{\left(k-q\right)^{2}-m^{2}}\nonumber \\
&  & \times\left(i\left|e\right|\gamma^{\mu}\right)v\left(k'\right)
\label{eq:bhamplitude1}
\end{eqnarray}
or
\begin{eqnarray}
\mathrm{T}_{BH} & = & \frac{\left|e\right|^{3}Q_{a}}{t}\bar{u}\left(p'\right)
\gamma_{\mu}u\left(p\right)
\epsilon_{\nu}\left(q\right)\bar{u}\left(k\right)\left[
\frac{\gamma^{\mu}\not\! q\gamma^{\nu}-2\gamma^{\mu}k'^{\nu}}{-2\left(k'\cdot q\right)}
\right.\nonumber \\
&  & \left.+\frac{-\gamma^{\nu}\not\! q\gamma^{\mu}+2k^{\nu}\gamma^{\mu}}{-2\left(k\cdot q\right)}
\right]v\left(k'\right).
\label{eq:bhamplitude2}
\end{eqnarray}
Again, after averaging $\left|\mathrm{T}_{BH}\right|^{2}$ over the polarization 
states of the initial particles and summing over the polarization states of 
the final parton, we have for the spin-averaged square of the amplitude
\begin{eqnarray}
\overline{\left|\mathrm{T}_{BH}\right|^{2}} & = & 
\frac{\left(4\pi\alpha\right)^{3}Q_{a}^{2}}{2t^{2}}L_{\mu\rho}^{BH}H_{BH}^{\mu\rho}.
\label{eq:Tamplitudesquaredbh}
\end{eqnarray}
A straightforward calculation of traces gives
\begin{eqnarray}
L_{BH}^{\mu\rho} & = & 4\left[\frac{A_{BH}}{\left(k\cdot q\right)^{2}}+
\frac{B_{BH}}{\left(k'\cdot q\right)^{2}}+
\frac{C_{BH}}{\left(k\cdot q\right)\left(k'\cdot q\right)}\right],\nonumber \\
A_{BH} & = & \left(k\cdot q\right)\left[k'^{\mu}q^{\rho}+k'^{\rho}q^{\mu}-g^{\mu\rho}
\left(k'\cdot q\right)\right]-m^{2}
\left[k^{\mu}k'^{\rho}+k^{\rho}k'^{\mu}-g^{\mu\rho}\left(k\cdot k'\right)\right]\nonumber \\
&  & +m^{2}\left[k'^{\mu}q^{\rho}+k'^{\rho}q^{\mu}-g^{\mu\rho}\left(k'\cdot q\right)\right]-m^{2}
\left(k\cdot q\right)g^{\mu\rho}+m^{4}g^{\mu\rho},\nonumber \\
B_{BH} & = & \left(k'\cdot q\right)\left[k^{\mu}q^{\rho}+k^{\rho}q^{\mu}-g^{\mu\rho}
\left(k\cdot q\right)\right]-m^{2}
\left[k^{\mu}k'^{\rho}+k^{\rho}k'^{\mu}-g^{\mu\rho}\left(k\cdot k'\right)\right]\nonumber \\
&  & +m^{2}\left[k^{\mu}q^{\rho}+k^{\rho}q^{\mu}-g^{\mu\rho}\left(k\cdot q\right)\right]-m^{2}
\left(k'\cdot q
\right)g^{\mu\rho}+m^{4}g^{\mu\rho},\nonumber \\
C_{BH} & = & 2\left(k\cdot k'\right)
\left[k^{\mu}k'^{\rho}+k^{\rho}k'^{\mu}-g^{\mu\rho}\left(k\cdot k'
\right)\right]-\left(k\cdot q\right)
\left[k^{\mu}k'^{\rho}+k^{\rho}k'^{\mu}-2k'^{\mu}k'^{\rho}\right]\nonumber \\
&  & -\left(k'\cdot q\right)\left[k^{\mu}k'^{\rho}+k^{\rho}k'^{\mu}-2k^{\mu}k^{\rho}\right]-
\left(k\cdot k'\right)
\left[k^{\mu}q^{\rho}+k^{\rho}q^{\mu}+k'^{\mu}q^{\rho}+k'^{\rho}q^{\mu}\right]\nonumber \\
&  & +2g^{\mu\rho}\left(k\cdot k'\right)\left[\left(k\cdot q\right)+\left(k'\cdot q\right)-m^{2}
\right]-2m^{2}q^{\mu}q^{\rho}
\label{eq:leptonictensorbh}
\end{eqnarray}
for the lepton tensor and
\begin{eqnarray}
H_{\mu\rho}^{BH} & = & 2\left[p_{\mu}p'_{\rho}+p_{\rho}p'_{\mu}-g_{\mu\rho}\left(p\cdot p'
\right)\right]
\label{eq:hadronictensorbh}
\end{eqnarray}
for the hadron tensor.

Recall now that the expression (\ref{eq:Tamplitudesquaredbh}) has to be integrated
over the final-state momenta constrained by a four-momentum delta
function, see Eq. (\ref{eq:phasespace}). However, before contracting
tensors, we perform integration of $L_{BH}^{\mu\rho}$ over both lepton three-momenta. 
One defines this particular integral as
\begin{eqnarray}
\mathcal{L}_{BH}^{\mu\rho} & \equiv & \int\frac{d^{3}k}{2\omega}\int\frac{d^{3}k'}{2\omega'}\;
\delta^{(4)}\left(q'-k-k'\right)L_{BH}^{\mu\rho},
\label{eq:integratedleptonictensorbh1}
\end{eqnarray}
which, with the help of the formula (\ref{eq:4dimintegral}), turns
into
\begin{eqnarray}
\mathcal{L}_{BH}^{\mu\rho} & \equiv & \int d^{4}k\int d^{4}k'\;\delta^{(4)}
\left(q'-k-k'\right)\delta^{+}
\left(k^{2}-m^{2}\right)\delta^{+}\left(k'^{2}-m^{2}\right)L_{BH}^{\mu\rho}.\nonumber \\
\label{eq:integratedleptonictensorbh2}
\end{eqnarray}
We evaluate $\mathcal{L}_{BH}^{\mu\rho}$ in the center-of-mass frame
of the lepton pair. Unfortunately, due to presence of a large number
of terms in the lepton tensor, the calculation requires more algebra.
Let us demonstrate the method by calculating only the integral of
the first term in the tensor (\ref{eq:leptonictensorbh}), namely,
\begin{eqnarray}
\mathcal{L}_{BH\left(1\right)}^{\mu\rho} & = & \int d^{4}k\int d^{4}k'\;
\delta^{(4)}\left(q'-k-k'\right)\delta^{+}
\left(k^{2}-m^{2}\right)\delta^{+}\left(k'^{2}-m^{2}\right)
\left[\frac{4k'^{\mu}q^{\rho}}{\left(k\cdot q\right)}\right].\nonumber \\
\label{eq:firstterm1}
\end{eqnarray}
Integration over $k'$ leads to
\begin{eqnarray}
\mathcal{L}_{BH\left(1\right)}^{\mu\rho} & = & 4q'^{\mu}q^{\rho}\int d^{4}k\;
\delta^{+}\left(k^{2}-m^{2}
\right)\delta^{+}
\left[\left(q'-k\right)^{2}-m^{2}\right]\left[\frac{1}{\left(k\cdot q\right)}
\right]\nonumber \\
&  & -4q^{\rho}\int d^{4}k\;\delta^{+}\left(k^{2}-m^{2}\right)\delta^{+}
\left[\left(q'-k\right)^{2}-m^{2}\right]
\left[\frac{k^{\mu}}{\left(k\cdot q\right)}\right].
\label{eq:firstterm2}
\end{eqnarray}
For simplicity, we choose the incident photon momentum along the \emph{z}-axis so that 
$q=\left(q_{0},0,0,q_{0}\right)$. The four-momentum of the outgoing lepton is 
$k=\left(k_{0},\vec{k}_{\bot},k_{z}\right)$. Accordingly, the calculation of the first 
integral in Eq. (\ref{eq:firstterm2}) (we will call it the function $I_{0}$ for later convenience) 
goes as follows
\begin{eqnarray}
I_{0} & = & \pi\int dk_{z}\int d\left(\left|\vec{k}_{\bot}\right|^{2}\right)\int dk_{0}\;
\delta^{+}\left(k_{0}^{2}-\left|\vec{k}_{\bot}
\right|^{2}-k_{z}^{2}-m^{2}\right)\delta^{+}\left(q'^{2}-2\sqrt{q'^{2}}\; k_{0}
\right)\nonumber \\
&  & \times\frac{1}{q_{0}\left(k_{0}-k_{z}\right)}\nonumber \\
& = & -\frac{\pi}{2\sqrt{q'^{2}}\; q_{0}}
\int_{-\sqrt{q'^{2}/4-m^{2}}}^{\sqrt{q'^{2}/4-m^{2}}}
\frac{dk_{z}}{k_{z}-\sqrt{q'^{2}}/2}\nonumber \\
& = & \frac{\pi}{2\left(q\cdot q'\right)}\ln\left[
\frac{1+\sqrt{1-4m^{2}/q'^{2}}}{1-\sqrt{1-4m^{2}/q'^{2}}}\right].
\label{eq:inot}
\end{eqnarray}
Note that we have expressed the initial photon energy in the invariant
form, $q_{0}=\left(q\cdot q'\right)/\sqrt{q'^{2}}$. The integral
over $k$ in the second term of Eq. (\ref{eq:firstterm2}) has a  Lorentz vector-like 
structure and can be, in general, written as
\begin{eqnarray}
\int d^{4}k\;\delta^{+}\left(k^{2}-m^{2}\right)\delta^{+}
\left[\left(q'-k\right)^{2}-m^{2}\right]\left[
\frac{k^{\mu}}{\left(k\cdot q\right)}\right] & = & I_{1}q^{\mu}+I_{2}q'^{\mu}.
\label{eq:secondintegralofthefirstterm}
\end{eqnarray}
The scalar functions $I_{1}$ and $I_{2}$ are determined by solving
the system of two equations, 
\begin{eqnarray}
\int d^{4}k\;\delta^{+}\left(k^{2}-m^{2}\right)\delta^{+}
\left[\left(q'-k\right)^{2}-m^{2}\right] & = & 
\left(q\cdot q'\right)I_{2},\nonumber \\
\int d^{4}k\;\delta^{+}\left(k^{2}-m^{2}\right)\delta^{+}
\left[\left(q'-k\right)^{2}-m^{2}\right]
\frac{\left(k\cdot q'\right)}{\left(k\cdot q\right)} & = & 
\left[\left(q\cdot q'\right)I_{1}+q'^{2}I_{2}\right],\nonumber \\
\label{eq:twoequationsioneanditwo}
\end{eqnarray}
obtained from contracting both sides of Eq. (\ref{eq:secondintegralofthefirstterm})
with $q_{\mu}$ and $q'_{\mu}$, respectively. After some algebra we find
\begin{eqnarray}
I_{2} & = & \frac{\pi}{2\left(q\cdot q'\right)}\sqrt{1-\frac{4m^{2}}{q'^{2}}},\nonumber \\
I_{1} & = & \frac{q'^{2}}{2\left(q\cdot q'\right)}\left[I_{0}-2I_{2}\right]\nonumber \\
& = & \frac{\pi q'^{2}}{4\left(q\cdot q'\right)^{2}}\left\{
\ln\left[\frac{1+\sqrt{1-4m^{2}/q'^{2}}}{1-\sqrt{1-4m^{2}/q'^{2}}}
\right]-2\sqrt{1-\frac{4m^{2}}{q'^{2}}}\right\}. 
\label{eq:itwoandione}
\end{eqnarray}
The term (\ref{eq:firstterm1}) now becomes equal to
\begin{eqnarray}
\mathcal{L}_{BH\left(1\right)}^{\mu\rho} & = & -4I_{1}q^{\mu}q^{\rho}+4\left(I_{0}-I_{2}
\right)q'^{\mu}q^{\rho}.
\label{eq:expressionintermsofinotioneitwo}
\end{eqnarray}
We repeat this procedure for all the remaining terms in the lepton tensor 
(all the integrals with the corresponding scalar functions are presented in Appendix B), 
and combine them together according to their structure. The final expression for the lepton 
tensor then reads
\begin{eqnarray}
\mathcal{L}_{BH}^{\mu\rho} & = & \frac{2\pi}{\left(q\cdot q'\right)}
\left[\mathcal{A}_{BH}q^{\mu}q^{\rho}+
\mathcal{B}_{BH}q'^{\mu}q'^{\rho}+
\mathcal{C}_{BH}\left(q^{\mu}q'^{\rho}+q'^{\mu}q^{\rho}
\right)+\mathcal{D}_{BH}g^{\mu\rho}\right],\nonumber \\
\label{eq:integratedleptonictensor}
\end{eqnarray}
where the coefficients
\begin{eqnarray}
\mathcal{A}_{BH} & = & \left[\frac{2q'^{2}}{\left(q\cdot q'\right)}-
\frac{2q'^{4}}{\left(q\cdot q'\right)^{2}}+
\frac{q'^{6}}{\left(q\cdot q'\right)^{3}}-
\frac{8m^{2}q'^{2}}{\left(q\cdot q'\right)^{2}}-\frac{4m^{4}q'^{2}}{\left(q\cdot q'\right)^{3}}+
\frac{6m^{2}q'^{4}}{\left(q\cdot q'\right)^{3}}+
\frac{4m^{2}}{\left(q\cdot q'\right)}\right]\nonumber \\
&  & \times\ln\left[\frac{1-\sqrt{1-4m^{2}/q'^{2}}}{1+\sqrt{1-4m^{2}/q'^{2}}}\right]\nonumber \\
&  & +\left[\frac{4q'^{2}}{\left(q\cdot q'\right)}-\frac{8q'^{4}}{\left(q\cdot q'\right)^{2}}+
\frac{4q'^{6}}{\left(q\cdot q'\right)^{3}}+
\frac{2m^{2}q'^{4}}{\left(q\cdot q'\right)^{3}}\right]\sqrt{1-\frac{4m^{2}}{q'^{2}}},\nonumber \\
\mathcal{B}_{BH} & = & \frac{4m^{2}}{\left(q\cdot q'\right)}\ln
\left[\frac{1-\sqrt{1-4m^{2}/q'^{2}}}{1+\sqrt{1-4m^{2}/q'^{2}}}\right]+
\frac{2q'^{2}}{\left(q\cdot q'\right)}\sqrt{1-\frac{4m^{2}}{q'^{2}}},\nonumber \\
\mathcal{C}_{BH} & = & \left[-2+\frac{2q'^{2}}{\left(q\cdot q'\right)}-
\frac{q'^{4}}{\left(q\cdot q'\right)^{2}}-
\frac{6m^{2}q'^{2}}{\left(q\cdot q'\right)^{2}}+
\frac{4m^{2}}{\left(q\cdot q'\right)}+\frac{4m^{4}}{\left(q\cdot q'\right)^{2}}
\right]\nonumber \\
&  & \times\ln\left[\frac{1-\sqrt{1-4m^{2}/q'^{2}}}{1+\sqrt{1-4m^{2}/q'^{2}}}\right]\nonumber \\
&  & +\left[-2+\frac{6q'^{2}}{\left(q\cdot q'\right)}-\frac{4q'^{4}}{\left(q\cdot q'\right)^{2}}-
\frac{2m^{2}q'^{2}}{\left(q\cdot q'\right)^{2}}\right]\sqrt{1-\frac{4m^{2}}{q'^{2}}},\nonumber \\
\mathcal{D}_{BH} & = & \left[2\left(q\cdot q'\right)-2q'^{2}+\frac{q'^{4}}{\left(q\cdot q'\right)}+
\frac{2m^{2}q'^{2}}{\left(q\cdot q'\right)}-
\frac{4m^{4}}{\left(q\cdot q'\right)}\right]\ln\left[
\frac{1-\sqrt{1-4m^{2}/q'^{2}}}{1+\sqrt{1-4m^{2}/q'^{2}}}\right]\nonumber \\
&  & +\left[2\left(q\cdot q'\right)-4q'^{2}+\frac{2q'^{4}}{\left(q\cdot q'\right)}+
\frac{2m^{2}q'^{2}}{\left(q\cdot q'\right)}
\right]\sqrt{1-\frac{4m^{2}}{q'^{2}}}.
\label{eq:bhcoefficients1}
\end{eqnarray}
are scalar functions of the invariants $m^{2}$, $q'^{2}$ and $\left(q\cdot q'\right)=\left(q'^{2}-t\right)/2$.

For the sake of completeness, it is worth noting that the same method can be applied to compute the integral 
of the lepton tensor, $L_{\mu\rho}^{C}=4\left[k_{\mu}k'_{\rho}+k_{\rho}k'_{\mu}-g_{\mu\rho}q'^{2}/2\right]$, 
given by Eq. (\ref{eq:leptonictensorcompton}) (note that $\left(k\cdot k'\right)+m^{2}=q'^{2}/2$), 
over the final lepton momenta for the Compton subprocess, namely,
\begin{eqnarray}
\mathcal{L}_{\mu\rho}^{C} & = & \int d^{4}k\int d^{4}k'\;\delta^{(4)}\left(q'-k-k'
\right)\delta^{+}\left(k^{2}-m^{2}\right)
\delta^{+}\left(k'^{2}-m^{2}\right)L_{\mu\rho}^{C}.\nonumber \\
\label{eq:leptonictensorintegralcompton1}
\end{eqnarray}
Using $\delta^{(4)}\left(q'-k-k'\right)$ to integrate over $k'$ and then writing each term separately we get
\begin{eqnarray}
\mathcal{L}_{\mu\rho}^{C} & = & 4q'_{\rho}\int d^{4}k\;\delta^{+}\left(k^{2}-m^{2}\right)
\delta^{+}\left[\left(q'-k\right)^{2}-m^{2}\right]k_{\mu}\nonumber \\
&  & +4q'_{\mu}\int d^{4}k\;\delta^{+}\left(k^{2}-m^{2}\right)\delta^{+}
\left[\left(q'-k\right)^{2}-m^{2}\right]k_{\rho}\nonumber \\
&  & -8\int d^{4}k\;\delta^{+}\left(k^{2}-m^{2}\right)\delta^{+}\left[\left(q'-k\right)^{2}-m^{2}
\right]k_{\mu}k_{\rho}\nonumber \\
&  & -2g_{\mu\rho}q'^{2}\int d^{4}k\;\delta^{+}\left(k^{2}-m^{2}\right)
\delta^{+}\left[\left(q'-k\right)^{2}-m^{2}\right].
\label{eq:leptonictensorintegralcompton2}
\end{eqnarray}
These integrals can only be constructed out of a four-vector $q'$
and a metric tensor, accompanied with the proper scalar functions of
$q'^{2}$. One can immediately discard the first two integrals. They are both of the form 
$q'_{\mu}q'_{\rho}$ and, since Eq. (\ref{eq:gaugeinvariance}) holds, they give vanishing contribution, 
when contracted with $H_{C}^{\mu\rho}$. Moreover, we write the third integral as
\begin{eqnarray}
\int d^{4}k\;\delta^{+}\left(k^{2}-m^{2}\right)\delta^{+}
\left[\left(q'-k\right)^{2}-m^{2}
\right]k_{\mu}k_{\rho} & = & N_{1}q'_{\mu}q'_{\rho}+N_{2}g_{\mu\rho}.
\label{eq:thirdterminleptonictensorintegralcompton}
\end{eqnarray}
Contracting both sides with $q'^{\mu}q'^{\rho}$ and $g^{\mu\rho}$ and evaluating the integral, we find 
the solutions for the scalar functions
\begin{eqnarray}
N_{1} & = & \frac{\pi}{2q'^{2}}\sqrt{1-\frac{4m^{2}}{q'^{2}}}
\left(q'^{2}-\frac{m^{2}}{3}\right),\nonumber \\
N_{2} & = & -\frac{\pi}{6}\sqrt{1-\frac{4m^{2}}{q'^{2}}}
\left(\frac{q'^{2}}{4}-m^{2}\right).
\label{eq:n1andn2}
\end{eqnarray}
Again, due to the gauge invariance of $H_{C}^{\mu\rho}$ only the term $N_{2}g_{\mu\rho}$ 
survives. Lastly, the final integral in Eq. (\ref{eq:leptonictensorintegralcompton2}) 
can be read off from Eqs. (\ref{eq:twoequationsioneanditwo}) and (\ref{eq:itwoandione}),
\begin{eqnarray}
\int d^{4}k\;\delta^{+}\left(k^{2}-m^{2}\right)\delta^{+}
\left[\left(q'-k\right)^{2}-m^{2}\right] & = & 
\frac{\pi}{2}\sqrt{1-\frac{4m^{2}}{q'^{2}}}.
\label{eq:fourthterminleptonictensorintegralcompton}
\end{eqnarray}
By collecting terms, $\mathcal{L}_{\mu\rho}^{C}$ reduces to
\begin{eqnarray}
\mathcal{L}_{\mu\rho}^{C} & = & \left(-8N_{2}-2q'^{2}\right)g_{\mu\rho}\nonumber \\
& = & -\frac{2\pi}{3}\left(q'^{2}+2m^{2}\right)\sqrt{1-\frac{4m^{2}}{q'^{2}}}\; g_{\mu\rho},
\label{eq:leptonictensorintegralcompton3}
\end{eqnarray}
which is identical, apart from the vanishing contribution $q'_{\mu}q'_{\rho}$,
to the expression (\ref{eq:newform}).

Next we contract Eq. (\ref{eq:integratedleptonictensor}) with the hadron 
tensor (\ref{eq:hadronictensorbh}) and follow the steps from Eq. (\ref{eq:comptoncrosssection1}) 
to Eq. (\ref{eq:comptoncrosssection4}) leading to the Bethe-Heitler differential partonic cross section, 
expressed in terms of the subprocess Mandelstam invariants
\begin{eqnarray}
\frac{d^{2}\sigma_{BH}}{dM_{pair}^{2}d\hat{t}} & = & \left(\alpha^{3}Q_{a}^{2}\right)
\frac{1}{\hat{s}^{2}\hat{t}^{2}\left(M_{pair}^{2}-\hat{t}\right)}\left[-\mathcal{E}_{BH}
\hat{u}\hat{s}+\mathcal{F}_{BH}\hat{t}\right].
\label{eq:bhsubprocesscrosssection}
\end{eqnarray}
The new coefficients are given by 
$\mathcal{E}_{BH}=\mathcal{A}_{BH}+\mathcal{B}_{BH}+2\mathcal{C}_{BH}$ 
and $\mathcal{F}_{BH}=2\mathcal{D}_{BH}$ together with the replacement
$q'^{2}\rightarrow M_{pair}^{2}$, namely,
\begin{eqnarray}
\mathcal{E}_{BH} & = & \left[-4+\frac{6M_{pair}^{2}}{\left(q\cdot q'\right)}-
\frac{4M_{pair}^{4}}{\left(q\cdot q'\right)^{2}}+\frac{M_{pair}^{6}}{\left(q\cdot q'\right)^{3}}-
\frac{20m^{2}M_{pair}^{2}}{\left(q\cdot q'\right)^{2}}-
\frac{4m^{4}M_{pair}^{2}}{\left(q\cdot q'\right)^{3}}+
\frac{6m^{2}M_{pair}^{4}}{\left(q\cdot q'\right)^{3}}\right.\nonumber \\
&  & \left.+\frac{16m^{2}}{\left(q\cdot q'\right)}+\frac{8m^{4}}{\left(q\cdot q'\right)^{2}}
\right]\times\ln\left[\frac{1-\sqrt{1-4m^{2}/M_{pair}^{2}}}{1+\sqrt{1-4m^{2}/M_{pair}^{2}}}
\right],\nonumber \\
\mathcal{F}_{BH} & = & \left[4\left(q\cdot q'\right)-4M_{pair}^{2}+
\frac{2M_{pair}^{4}}{\left(q\cdot q'\right)}+
\frac{4m^{2}M_{pair}^{2}}{\left(q\cdot q'\right)}-\frac{8m^{4}}{\left(q\cdot q'\right)}
\right]\nonumber \\
&  & \times\ln\left[\frac{1-\sqrt{1-4m^{2}/M_{pair}^{2}}}{1+\sqrt{1-4m^{2}/M_{pair}^{2}}}
\right]\nonumber \\
&  & +\left[4\left(q\cdot q'\right)-8M_{pair}^{2}+\frac{4M_{pair}^{4}}{\left(q\cdot q'\right)}+
\frac{4m^{2}M_{pair}^{2}}{\left(q\cdot q'\right)}\right]\sqrt{1-\frac{4m^{2}}{M_{pair}^{2}}}.
\label{eq:bhcoefficients2}
\end{eqnarray}
Finally, with the help of the master formula (\ref{eq:partonmodelformulaphotoproduction}), 
we obtain the Bethe-Heitler contribution to the inclusive photoproduction of lepton pairs 
in the parton model
\begin{eqnarray}
\frac{d^{3}\sigma_{BH}\left(\gamma N
\rightarrow l^{-}l^{+}X\right)}{dM_{pair}^{2}dtdx_{B}} & = & 
\alpha^{3}\frac{1}{\left(x_{B}s\right)^{2}t^{2}\left(M_{pair}^{2}-t\right)}\nonumber \\
&  & \times\left[\mathcal{E}_{BH}\left(x_{B}s+t-M_{pair}^{2}\right)x_{B}s+\mathcal{F}_{BH}
t\right]\nonumber \\
&  & \times\sum_{a}Q_{a}^{2}f_{a/N}\left(x_{B}\right).
\label{eq:finalbhprocess}
\end{eqnarray}

\section{\ \,  Interference Terms}

We turn our attention to the interference between the Compton 
and Bethe-Heitler subprocesses. Four Feynman diagrams give rise to 
eight interference terms. They all are calculated in a similar way. 
For illustration, let us consider the first two terms. 
Up to a common factor we have
\begin{eqnarray}
\overline{\left|\mathrm{T}_{1}\right|^{2}} & \sim & \frac{1}{\left(k'\cdot q\right)}\mathrm{Tr}
\left[\not\! p'
\gamma^{\mu}\left(\not\! p+\not\! q\right)\gamma^{\nu}\not\! p\gamma_{\xi}\right]\mathrm{Tr}
\left[\left(\not\! k+m\right)
\gamma_{\mu}\left(\not\! k'-m\right)\gamma_{\nu}\left(\not\! q-\not\! k'+m\right)\gamma^{\xi}
\right],\nonumber \\
\overline{\left|\mathrm{T}_{2}\right|^{2}} & \sim & \frac{1}{\left(k\cdot q\right)}\mathrm{Tr}
\left[\not\! p'
\gamma^{\mu}\left(\not\! p+\not\! q\right)\gamma^{\nu}\not\! p\gamma_{\xi}\right]\mathrm{Tr}
\left[\left(\not\! k+m\right)
\gamma_{\mu}\left(\not\! k'-m\right)\gamma^{\xi}\left(\not\! k-\not\! q+m\right)\gamma_{\nu}
\right].\nonumber \\
\label{eq:firsttwointerferenceterms}
\end{eqnarray}
After evaluating both second traces in the above expressions, we observe that they differ, apart 
from the sign, by the interchange $k\leftrightarrow k'$. Thus integrating the sum of 
$\overline{\left|\mathrm{T}_{1}\right|^{2}}$ and $\overline{\left|\mathrm{T}_{2}\right|^{2}}$ over 
the final lepton four-momenta yields zero,
\begin{eqnarray}
\int d^{4}k\int d^{4}k'\;\left(\overline{\left|\mathrm{T}_{1}\right|^{2}}+
\overline{\left|\mathrm{T}_{2}\right|^{2}}\right) & = & 0.
\label{eq:mutualcancellation}
\end{eqnarray}
By the same token, the remaining six interference terms mutually cancel each other, 
after being integrated over $k$ and $k'$.

In fact, there is a more elegant way (without cumbersome and explicit
calculation of traces) to see the cancellation of the interference
terms. Note that one needs to compare the following two double integrals
\begin{eqnarray}
\int d^{4}k\int d^{4}k'\;\delta^{\left(4\right)}\left(q'-k-k'\right)\frac{1}{\left(k'\cdot q\right)}
\mathrm{Tr}\left[\left(\not\! k+m\right)\gamma_{\mu}\left(\not\! k'-m\right)\gamma_{\nu}
\left(\not\! q-\not\! k'+m\right)
\gamma^{\xi}\right],\nonumber \\
\int d^{4}k\int d^{4}k'\;\delta^{\left(4\right)}\left(q'-k-k'\right)\frac{1}{\left(k\cdot q\right)}
\mathrm{Tr}
\left[\left(\not\! k+m\right)\gamma_{\mu}\left(\not\! k'-m\right)\gamma^{\xi}\left(\not\! k-
\not\! q+m\right)
\gamma_{\nu}\right].\nonumber \\
\label{eq:firsttwointerferencetermsastraces1}
\end{eqnarray}
Due to the presence of the delta function, one of the integrals, e.g. the integral over $k'$, 
can be trivially carried out. In addition, using the fact that the trace does not change under 
the cyclic permutation, Eq. (\ref{eq:firsttwointerferencetermsastraces1}) can be rewritten into 
\begin{eqnarray}
\int d^{4}k\;\frac{1}{\left[\left(q'-k\right)\cdot q\right]}\mathrm{Tr}\left[\gamma_{\mu}
\left(\not\! q'-\not\! k-m\right)
\gamma_{\nu}\left(\not\! q-\not\! q'+\not\! k+m\right)
\gamma^{\xi}\left(\not\! k+m\right)\right],\nonumber \\
\int d^{4}k\;\frac{1}{\left(k\cdot q\right)}\mathrm{Tr}\left[\gamma_{\mu}\left(\not\! q'-
\not\! k-m\right)
\gamma^{\xi}\left(\not\! k-\not\! q+m\right)\gamma_{\nu}\left(\not\! k+m\right)\right].
\label{eq:firsttwointerferencetermsastraces2}
\end{eqnarray}
Transposing now the trace in the first integrand, and then performing the momentum shift, 
$\tilde{k}=q'-k$, this integral turns into
\begin{equation}
\int d^{4}\tilde{k}\;\frac{1}{\left(\tilde{k}\cdot q\right)}\mathrm{Tr}\left[\left(\not\! q'-\not\!
\tilde{k}+m\right)^{\mathrm{T}}\gamma^{\xi^{\mathrm{T}}}\left(\not\! q-\not\!
\tilde{k}+m\right)^{\mathrm{T}}
\gamma_{\nu}^{\mathrm{T}}\left(\not\!\tilde{k}-m\right)^{\mathrm{T}}\gamma_{\mu}^{\mathrm{T}}\right].
\label{eq:integraloftraceovernewvariable1}
\end{equation}
Using the property $\hat{C}\gamma_{\mu}\hat{C}^{-1}=-\gamma_{\mu}^{\mathrm{T}}$,
where $\hat{C}=i\gamma^{2}\gamma^{0}$ represents the charge conjugation operator, 
and  further permuting the last two factors $\gamma_{\mu}\hat{C}^{-1}$ under the trace to 
the left side, we convert the integral finally into the form
\begin{equation}
-\int d^{4}\tilde{k}\;\frac{1}{\left(\tilde{k}\cdot q\right)}\mathrm{Tr}\left[\gamma_{\mu}
\left(\not\! q'-\not\!
\tilde{k}-m\right)\gamma^{\xi}\left(\not\!\tilde{k}-\not\! q+m\right)\gamma_{\nu}
\left(\not\!\tilde{k}+m\right)
\right].
\label{eq:integraloftraceovernewvariable2}
\end{equation}
The latter is, apart from the sign, identical to the second integral of
Eq. (\ref{eq:firsttwointerferencetermsastraces2}) and therefore, the two integrals 
exactly cancel each other. This result of cancellation is, in general, known as the Furry's theorem. 
It states that Feynman diagrams containing a closed fermion loop with an odd number of photon vertices 
can be omitted in the calculation of physical processes \cite{Greiner2}. It should be pointed out, 
however, that the cancellation of the interference terms happens only after integration over 
the momenta of the final leptons. These terms can still be accessed if one measures, for example, 
the angular distribution of the lepton pair.

\section{\ \,  Kinematics}

As the Bjorken scaling variable $x_{B}$ tends to zero, both the Compton and Bethe-Heitler 
differential cross sections, see Eqs. (\ref{eq:finalcomptonprocess})
and (\ref{eq:finalbhprocess}), become singular. Nevertheless, the following kinematical constraint
\begin{eqnarray}
s-M_{pair}^{2} & \geq\left(q-q'+P\right)^{2}\geq & M^{2}
\label{eq:kinematicalcondition1}
\end{eqnarray}
holds, where $M$ stands for the nucleon mass. Thus, in the laboratory frame, 
where the real and virtual photon momenta are $k=\left(\omega,\vec{k}\right)$ and 
$k'=\left(\omega',\vec{k}'\right)$, respectively, and the nucleon momentum is 
$P=\left(M,\vec{0}\right)$, the variable $x_{B}$ satisfies 
\begin{eqnarray}
x_{B} & \geq & \frac{1}{1-\left(2M\omega-M_{pair}^{2}\right)/t}.
\label{eq:xmin}
\end{eqnarray}
For the invariant momentum transfer \emph{t} we have
\begin{eqnarray}
t & = & M_{pair}^{2}-2\omega\left(\omega'-\sqrt{\omega'^{2}-M_{pair}^{2}}\;
\cos\theta_{\gamma\gamma}\right),
\label{eq:t}
\end{eqnarray}
where $\theta_{\gamma\gamma}$ is the scattering angle, i.e. the angle between the directions 
of the initial  and final photons.

\begin{figure}
\centerline{\epsfxsize=5in\epsffile{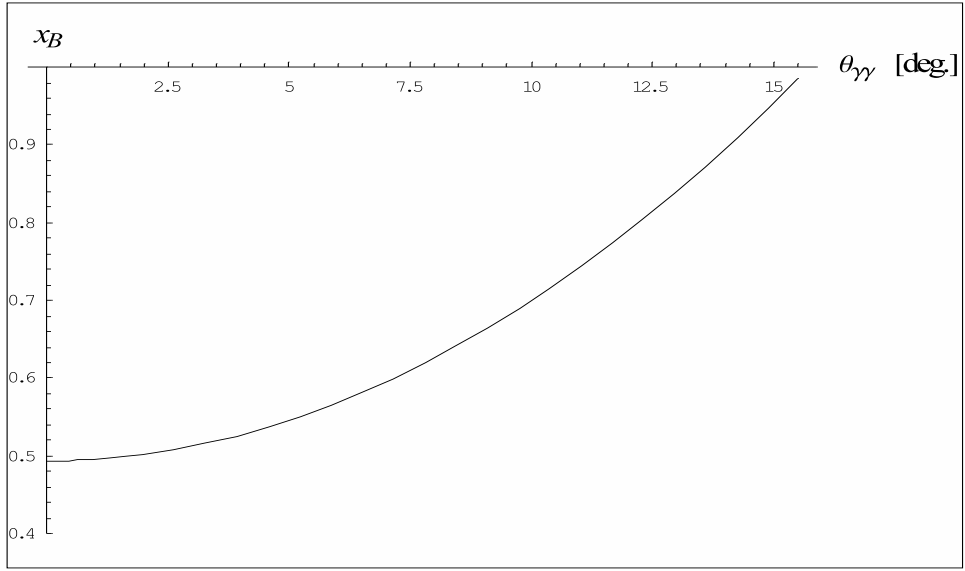}}
\caption{Bjorken scaling variable $x_{B}$ plotted as a function of the angle $\theta_{\gamma\gamma}$ 
between the incoming real and outgoing virtual photon in the target rest frame for $\omega'=10\;\mathrm{GeV}$ 
and $M_{pair}=3\;\mathrm{GeV}$ with $\omega=40\;\mathrm{GeV}$ photon beam.}
\label{inclusivebjorkenx}
\end{figure}

\begin{figure}
\centerline{\epsfxsize=5in\epsffile{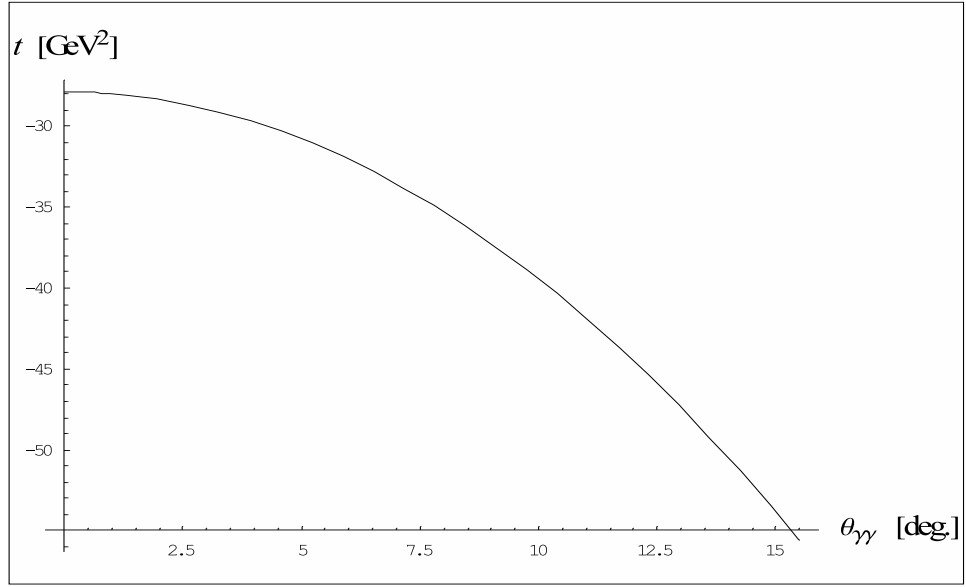}}
\caption{Invariant momentum transfer $t$ plotted as a function of the angle $\theta_{\gamma\gamma}$ 
between the incoming real and outgoing virtual photon in the target rest frame for $\omega'=10\;\mathrm{GeV}$ 
and $M_{pair}=3\;\mathrm{GeV}$ with $\omega=40\;\mathrm{GeV}$ photon beam.}
\label{inclusivet}
\end{figure}

\begin{figure}
\centerline{\epsfxsize=5in\epsffile{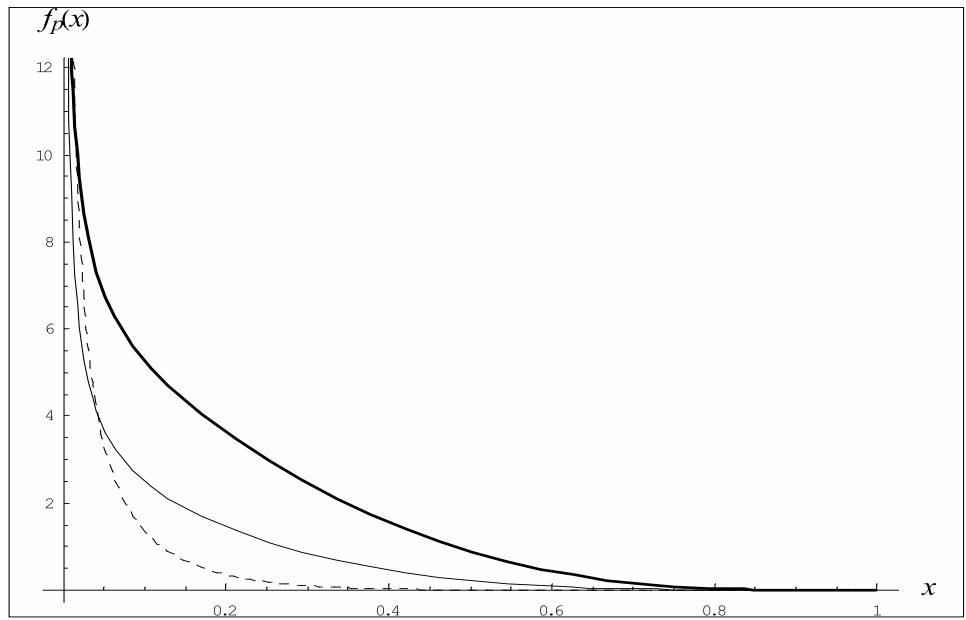}}
\caption{Unpolarized valence quark distributions for \emph{u}-flavor (bold solid line) and \emph{d}-flavor 
(solid line), and the sea quark distribution (dashed line) in the proton.}
\label{pdfsinclusive}
\end{figure}

We consider a proton target and the muons as the outgoing pair of leptons and thus we have 
$M\simeq0.94\;\mathrm{GeV}$ and $m\simeq0.106\;\mathrm{GeV}$. In Fig. \ref{inclusivebjorkenx}, the 
Bjorken scaling variable given by Eq. (\ref{eq:momentumfractionandbjorkenvariable}), is plotted 
against the angle $\theta_{\gamma\gamma}$ for fixed values of the incoming photon beam energy, 
$\omega=40\;\mathrm{GeV}$, the energy of the outgoing virtual photon, $\omega'=10\;\mathrm{GeV}$, 
and the lepton-pair mass, $M_{pair}=3\;\mathrm{GeV}$. We find $0.49 \leq x_{B} \leq 0.96$ for the 
scattering angles $0\leq\theta_{\gamma\gamma}\leq15^{0}$. In this region of $x_{B}$, the valence 
quarks play the dominant role, in particular the \emph{u}-flavor contribution, 
as shown in Fig. \ref{pdfsinclusive}. Next we plot the angular dependence of the invariant \emph{t} 
for the same kinematics, see Fig. \ref{inclusivet}. Due to high energy of the photon beam, 
rather large values of \emph{t} are expected, namely, 
$27.85\;\mathrm{GeV}^{2}\leq-t\leq53.85\;\mathrm{GeV}^{2}$. It is important to note 
that our description is not limited only to the small invariant momentum transfer, like in the case of DVCS 
kinematics discussed in Chapter IV. In contrary, it works for any kinematically allowed value of \emph{t}. 

To estimate the cross sections, we use a simplified parametrization of the unpolarized valence quark PDFs 
in the proton, given by Eq. (\ref{eq:unpolvalencedistributions}), together with the sea quark distribution, 
\begin{eqnarray}
sea\left(x\right) & = & 0.5x^{-0.75}\left(1-x\right)^{7},
\label{eq:seadistribution}
\end{eqnarray}
taken from Ref. \cite{Radyushkin:1998rt}. These are, however, the valence distributions for 
the \emph{u}- and \emph{d}-flavor components in the proton, respectively, and the sea distribution in 
the proton whereas the distributions, which appear in the cross section 
formulas (\ref{eq:finalcomptonprocess}) and (\ref{eq:finalbhprocess}) are the quark and antiquark proton PDFs. 
In fact, what one really needs is the sum of the quark and antiquark distributions,
\begin{eqnarray}
\sum_{a}f_{a/p}\left(x\right) & = & \sum_{f}\left[f_{p}\left(x\right)+
\bar{f}_{p}\left(x\right)\right].
\label{eq:sumintermsofflavors}
\end{eqnarray}
As noted in Section IV.2, sum (\ref{eq:sumintermsofflavors}) corresponds to adding the valence PDFs 
and twice the contribution from sea quarks.

\begin{figure}
\centerline{\epsfxsize=5in\epsffile{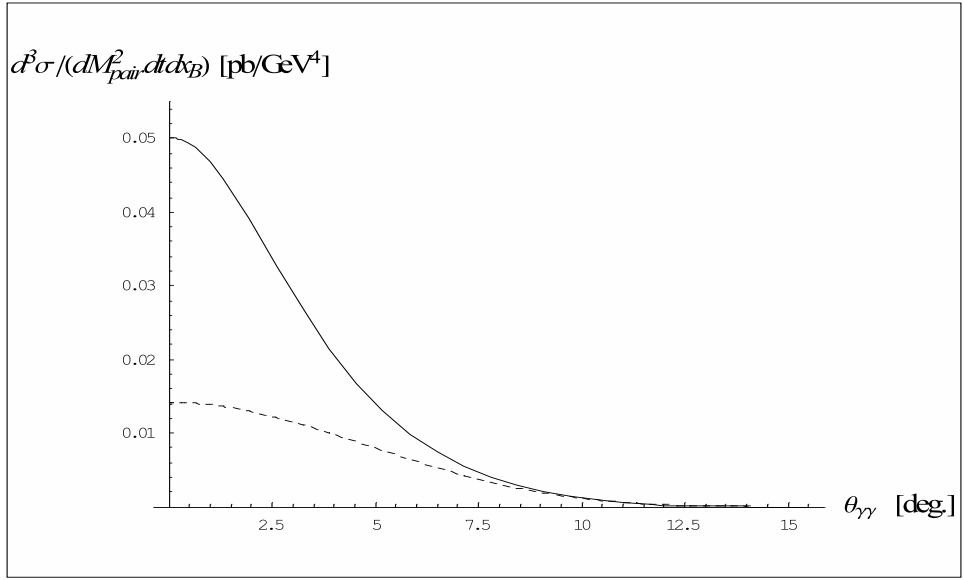}}
\caption{Compton cross section $\sigma_{C}$ (solid line) and Bethe-Heitler cross section $\sigma_{BH}$ 
(dashed line) plotted as a function of the angle $\theta_{\gamma\gamma}$ between the incoming real and 
outgoing virtual photon in the target rest frame for $\omega'=10\;\mathrm{GeV}$ and $M_{pair}=3\;\mathrm{GeV}$ 
with $\omega=40\;\mathrm{GeV}$ photon beam.}
\label{inclusivecrosssection}
\end{figure}

The unpolarized cross section results are presented in Fig. \ref{inclusivecrosssection}. 
We observe that the Compton contribution dominates by a factor $\simeq3$ in the forward direction, i.e. 
for $\theta_{\gamma\gamma}\leq3^{0}$, and falls off rapidly, even below the Bethe Heitler contribution for 
the scattering angles $\theta_{\gamma\gamma}>10{}^{0}$.

In summary, we have demonstrated the application of the QCD parton model to the high-energy photoproduction 
of lepton pairs. The Compton and Bethe-Heitler cross sections were estimated for a given simple 
parametrization of usual parton distributions in the proton. Without explicit calculation of the 
interference terms, we have shown their mutual cancellation, after being integrated over the final 
lepton momenta. 

\chapter{Exclusive Photoproduction of Lepton Pairs}

\section{\ \,  Introduction}

Deeply virtual Compton scattering is considered to be theoretically the cleanest and simplest process, 
in which GPDs can be accessed. We can now slightly modify the process, simply by interchanging the two 
photons at the subprocess level,
\begin{eqnarray}
\gamma\left(q_{1}\right)+N\left(p_{1}\right) & \longrightarrow & 
\gamma^{*}\left(q_{2}\right)+N\left(p_{2}\right).
\label{eq:TCSformula}
\end{eqnarray}
Then one arrives at time-like Compton scattering (TCS) \cite{Berger:2001xd} which is, in principle, the inverse 
process to DVCS. In a specific kinematical regime, where the virtuality of the final-state photon is 
large, $q_{2}^{2}\rightarrow\infty$, TCS becomes a handbag dominated process and accordingly, it can be 
studied in same way as DVCS was studied in Chapter IV.

\begin{figure}
\centerline{\epsfxsize=3in\epsffile{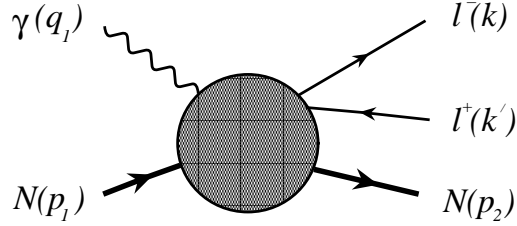}}
\caption{Exclusive photoproduction of lepton pairs.}
\label{exclusivephotoproductionfigure}
\end{figure}

\begin{figure}
\centerline{\epsfxsize=6in\epsffile{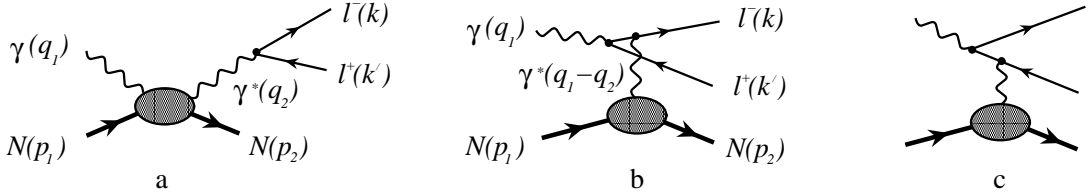}}
\caption{DVCS (a) and Bethe-Heitler (b and c) diagrams contributing to the exclusive photoproduction 
of a lepton pair.}
\label{exclusivediagrams}
\end{figure}

Time-like Compton scattering can be accessed through the physical process known as the exclusive 
photoproduction of lepton pairs. At this point, we need to draw a distinction between the exclusive and the 
inclusive photoproduction of lepton pairs. In the former case we detect, in addition to the lepton pair, also 
the scattered nucleon \emph{N} (with the four-momentum $p_{2}$), as illustrated in 
Fig. \ref{exclusivephotoproductionfigure}. Another important difference is that unlike the inclusive 
photoproduction, we are limited now to small invariant momentum transfer $t\equiv\left(p_{1}-p_{2}\right)^{2}$ 
to the nucleon. Recall that the DVCS kinematics requires small \emph{t} compared to the invariant 
$q_{2}^{2}$ (the DVCS scale $Q_{1}^{2}$ is now replaced by the virtuality $q_{2}^{2}$), and thus 
the terms $\mathcal{O}\left(-t/q_{2}^{2}\right)$ in the amplitude can be safely dropped out. 
On the other hand, the common feature of both inclusive and exclusive photoproduction processes is 
the presence of two types of diagrams. In particular, for the exclusive photoproduction of lepton pairs, 
they are shown in Fig. \ref{exclusivediagrams}. 

In Section VI.2, we define the kinematics. In the next section, we compute the leading-twist amplitude 
for the Compton part. It can be easily derived from the result for VCA established in Section IV.2. The 
relevant integrals in the Compton and Bethe-Heitler cross sections can be evaluated simply by following 
the steps of Sections V.2 and V.3. Finally, the cross sections are estimated by using the same model for 
the nucleon GPDs as in the DVCS process.

\section{\ \,  Kinematics}

In the laboratory reference frame, as the target rest frame, the photon
and nucleon four-momenta are denoted by $q_{1}=\left(\nu_{1},\vec{q}_{1}\right)$,
$q_{2}=\left(\nu_{2},\vec{q}_{2}\right)$, $p_{1}=\left(M,\vec{0}\right)$
and $p_{2}=\left(E_{2},\vec{p}_{2}\right)$ where \emph{M} stands,
as usual, for the nucleon mass. One rotates the coordinate 
system so that the \emph{z}-axis is in the direction of the incident
photon in the photon scattering plane, formed by the photon three-momenta $\vec{q}_{1}$ and $\vec{q}_{2}$. 
Similarly, the final leptons define the lepton scattering plane. We write their four-momenta as 
$k=\left(\omega,\vec{k}\right)$ and $k'=\left(\omega',\vec{k}'\right)$, respectively. 
In this reference frame, the invariant momentum transfer can be expressed as a function of the angle 
$\theta_{\gamma\gamma}$ between the incoming real and outgoing virtual photon, namely,
\begin{eqnarray}
t & \equiv\left(q_{2}-q_{1}\right)^{2}= & q_{2}^{2}-2\nu_{1}\left(\nu_{2}-
\sqrt{\nu_{2}^{2}-q_{2}^{2}}\;\cos\theta_{\gamma\gamma}\right).
\label{eq:invariantt1exclusive}
\end{eqnarray}
On the other hand, the invariant \emph{t} is also equal to
\begin{eqnarray}
t & = & 2M\left(\nu_{2}-\nu_{1}\right),
\label{eq:invariantt2exclusive}
\end{eqnarray}
which follows from the expression $t=2M^{2}-2ME_{2}$ and the energy conservation,
$M+\nu_{1}=E_{2}+\nu_{2}$. Combining Eqs. (\ref{eq:invariantt1exclusive})
and (\ref{eq:invariantt2exclusive}) gives the quadratic equation
in the energy of the virtual photon $\nu_{2}$ with the solutions
\begin{eqnarray}
\nu_{2} & = & \frac{AB\pm\sqrt{A^{2}-q_{2}^{2}\left(B^{2}-1\right)}}{B^{2}-1}.
\label{eq:nju2}
\end{eqnarray}
The coefficients are
\begin{eqnarray}
A & = & \frac{\left(1+\chi\right)M}{\cos\theta_{\gamma\gamma}}\geq M,\nonumber \\
B & = & \frac{\left(1+M/\nu_{1}\right)}{\cos\theta_{\gamma\gamma}}\geq1,
\label{eq:coefficientsAandB}
\end{eqnarray}
where the scaling variable,
\begin{eqnarray}
\chi & \equiv & \frac{q_{2}^{2}}{2\left(p_{1}\cdot q_{1}\right)}=\frac{q_{2}^{2}}{2M\nu_{1}},
\label{eq:chi}
\end{eqnarray}
was introduced. It is an analog of the Bjorken scaling variable, 
$x_{B}\equiv q_{1}^{2}/2\left(p_{1}\cdot q_{1}\right)$, in the DVCS process. Since in the DVCS kinematics 
the invariant momentum transfer \emph{t} is kept as small as possible, only the plus solution of 
Eq. (\ref{eq:nju2}) should be considered.

\begin{figure}
\centerline{\epsfxsize=5in\epsffile{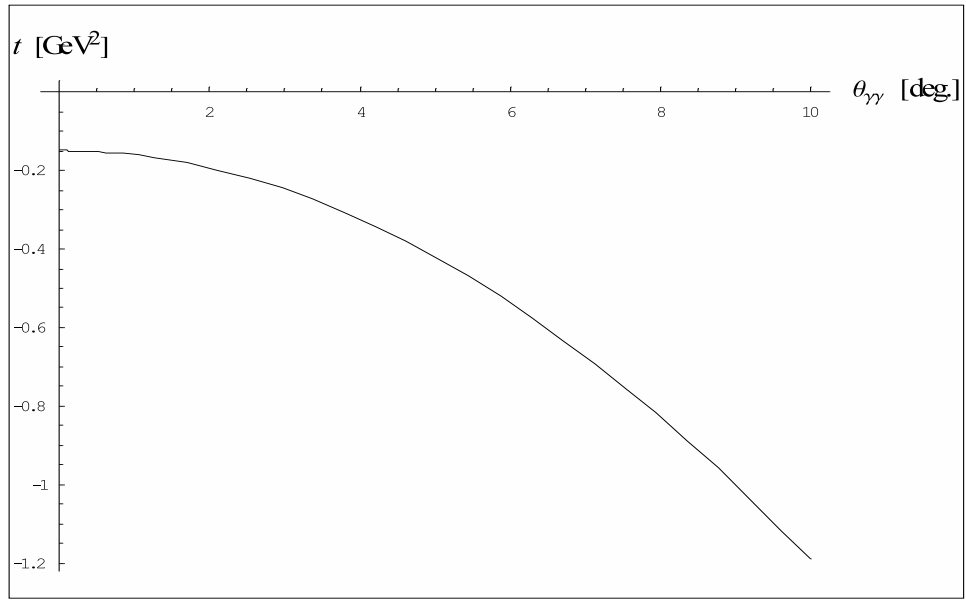}}
\caption{Invariant momentum transfer $t$ plotted as a function of the angle $\theta_{\gamma\gamma}$ 
between the incoming real and outgoing virtual photon in the target rest frame for the invariant mass of 
the lepton pair $q_{2}^{2}=3\;\mathrm{GeV}^{2}$ and $\chi=0.32$ with $\nu_{1}=5\;\mathrm{GeV}$ photon beam.}
\label{exclusivet}
\end{figure}

\begin{figure}
\centerline{\epsfxsize=5in\epsffile{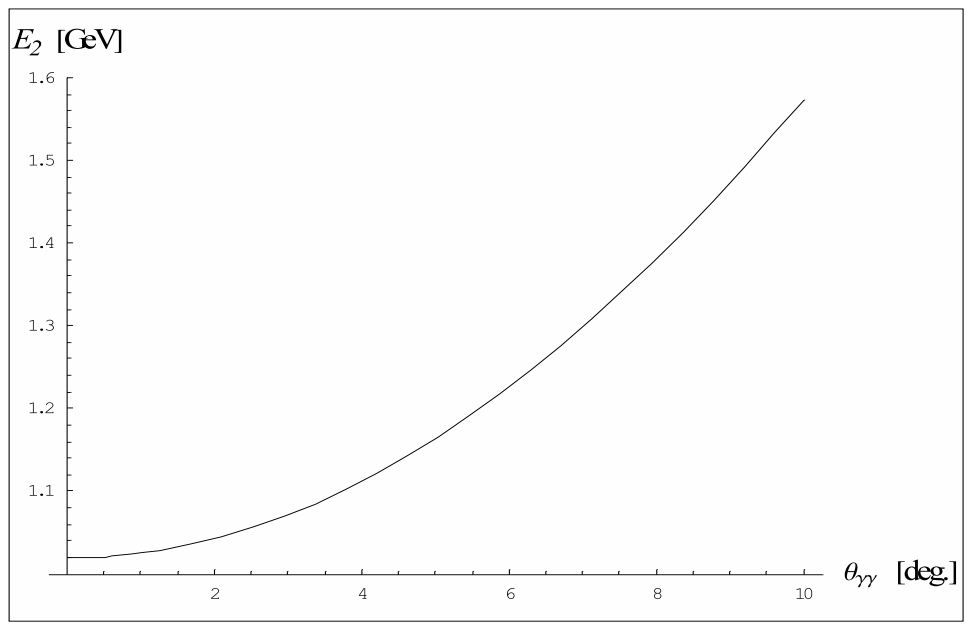}}
\caption{Energy of the scattered nucleon $E_{2}$ plotted as a function of the angle $\theta_{\gamma\gamma}$ 
between the incoming real and outgoing virtual photon in the target rest frame for the invariant mass of the 
lepton pair $q_{2}^{2}=3\;\mathrm{GeV}^{2}$ and $\chi=0.32$ with $\nu_{1}=5\;\mathrm{GeV}$ photon beam.}
\label{exclusivenucleonenergy}
\end{figure}

To illustrate the kinematics, we set the beam energy of initial photons to 
$\nu_{1}=5\;\mathrm{GeV}$, and the invariant mass of the lepton pair to $q_{2}^{2}=3\;\mathrm{GeV}^{2}$. 
Accordingly, the scaling variable (\ref{eq:chi}) is fixed at $\chi=0.32$. The invariant momentum is 
plotted as a function of the scattering angle in Fig. \ref{exclusivet}. It covers the range 
$0.15\;\mathrm{GeV}^{2}\leq-t\leq1.19\;\mathrm{GeV}^{2}$ for $\theta_{\gamma\gamma}$ up to $10^{0}$. 
For the same region in $\theta_{\gamma\gamma}$, we also plot the energy $E_{2}$ of the final-state nucleon, 
see Fig \ref{exclusivenucleonenergy}, and obtain $1.019\;\mathrm{GeV}\leq E_{2}\leq 1.574\;\mathrm{GeV}$.

\section{\ \,  Cross Section}

In general, the differential cross section for the photoproduction of lepton pairs is
\begin{eqnarray}
d\sigma & = & \frac{1}{2s}\left|\mathrm{T}\right|^{2}\frac{1}{\left(2\pi\right)^{5}}\;
\delta^{\left(4\right)}\left(p_{1}+q_{1}-p_{2}-k-k'\right)
\frac{d^{3}p_{2}}{2E_{2}}\frac{d^{3}k}{2\omega}
\frac{d^{3}k'}{2\omega'},
\label{eq:generaldiffcrosssectionexclusive}
\end{eqnarray}
where the invariant $s\equiv\left(p_{1}+q_{1}\right)^{2}=2M\nu_{1}+M^{2}$. Similarly to the DVCS case, the 
invariant matrix element consists of two contributions. Let us first investigate the Compton part.

\subsection{\ \,  Compton Process}

The T-matrix of the Compton contribution is given by the diagram (a)
in Fig. \ref{exclusivediagrams}. Using $\epsilon_{\nu}\left(q_{1}\right)$
to denote the polarization of the initial photon, we have
\begin{eqnarray}
\mathrm{T}_{C} & = & \frac{\left|e\right|^{3}}{q_{2}^{2}}\bar{u}
\left(k\right)\gamma_{\mu}v\left(k'\right)
\mathcal{T}_{TCS}^{\mu\nu}\epsilon_{\nu}\left(q_{1}\right),
\label{eq:comptonamplitudeexclusive}
\end{eqnarray}
where $\mathcal{T}_{TCS}^{\mu\nu}$ is the reduced VCA for time-like
Compton scattering, namely,
\begin{eqnarray}
\mathcal{T}_{TCS}^{\mu\nu} & = & i\int d^{4}z\; e^{i\left(q\cdot z\right)}\left
\langle N\left(p_{2},s_{2}\right)
\right|T\left\{ J_{EM}^{\mu}\left(z/2\right)J_{EM}^{\nu}
\left(-z/2\right)\right\} \left|N\left(p_{1},s_{1}
\right)\right\rangle.\nonumber \\
\label{eq:virtualcomptonamplitudeexclusive}
\end{eqnarray}
In the twist-2 approximation, this amplitude is calculated
from two (\emph{s}- and \emph{u}-channel) handbag diagrams illustrated
in Fig. \ref{handbagdiagramsDVCS}. Unlike the standard electromagnetic DVCS
process, the incoming photon is now real, $q_{1}^{2}=0$, and the
outgoing photon is time-like, $q_{2}^{2}>0$. The final result for
$\mathcal{T}_{TCS}^{\mu\nu}$ at the twist-2 level in terms of OFPDs can be derived from
the expression (\ref{eq:reducedVCAresult2}) for the reduced VCA in the DVCS process 
simply by replacing the final photon momentum with the initial one,
$q_{2}\rightarrow q_{1}$, and changing the sign in the skewness parameter 
$\eta$. Recall from Section IV.2 that the latter is defined as 
$\eta\equiv\left(r\cdot q\right)/2\left(p\cdot q\right)$, where 
$r=p_{1}-p_{2}$ is the overall momentum transfer to the nucleon, and 
$p=\left(p_{1}+p_{2}\right)/2$ and $q=\left(q_{1}+q_{2}\right)/2$ are the average nucleon 
and photon momenta, respectively. As for DVCS $\eta=x_{B}/\left(2-x_{B}\right)$ in the DVCS kinematics, 
we have for TCS a similar relation, i.e. $\eta=\chi/\left(2-\chi\right)$. Moreover, the scaling variable 
$\xi\equiv-q^{2}/2\left(p\cdot q\right)$ now coincides with $-\eta$. 

\begin{figure}
\centerline{\epsfxsize=5in\epsffile{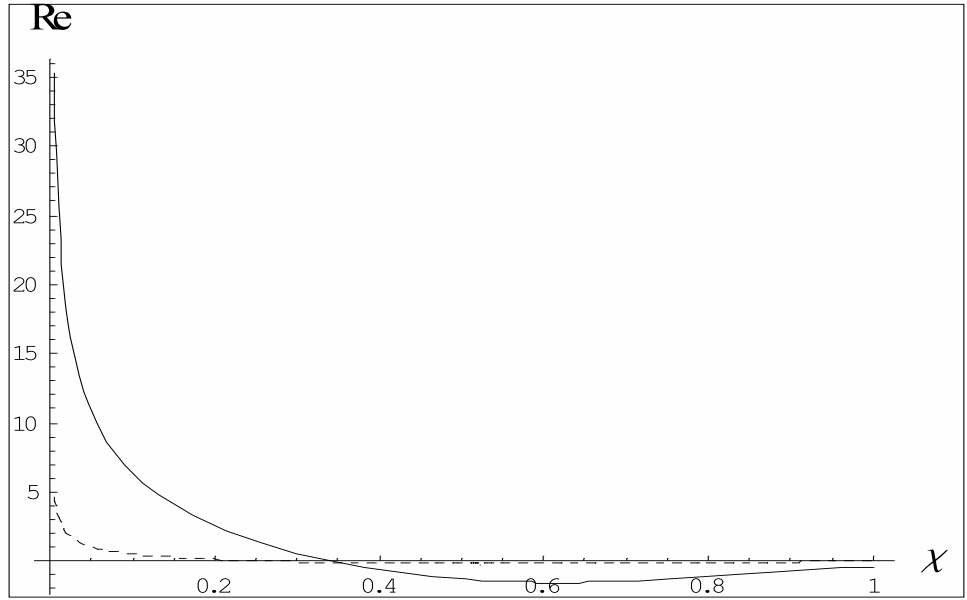}}
\caption[Contributions from \emph{u} quarks (solid line) and \emph{d} quarks (dashed line) to the 
real part of $\mathcal{H}_{TCS}^{+}$.]
{Contributions from \emph{u} quarks (solid line) and \emph{d} quarks (dashed line) to the 
real part of $\mathcal{H}_{TCS}^{+}$. 
They are divided by $F_{1u}\left(t\right)/2$ and $F_{1d}\left(t\right)$, respectively.}
\label{realHTCS}
\end{figure}

\begin{figure}
\centerline{\epsfxsize=5in\epsffile{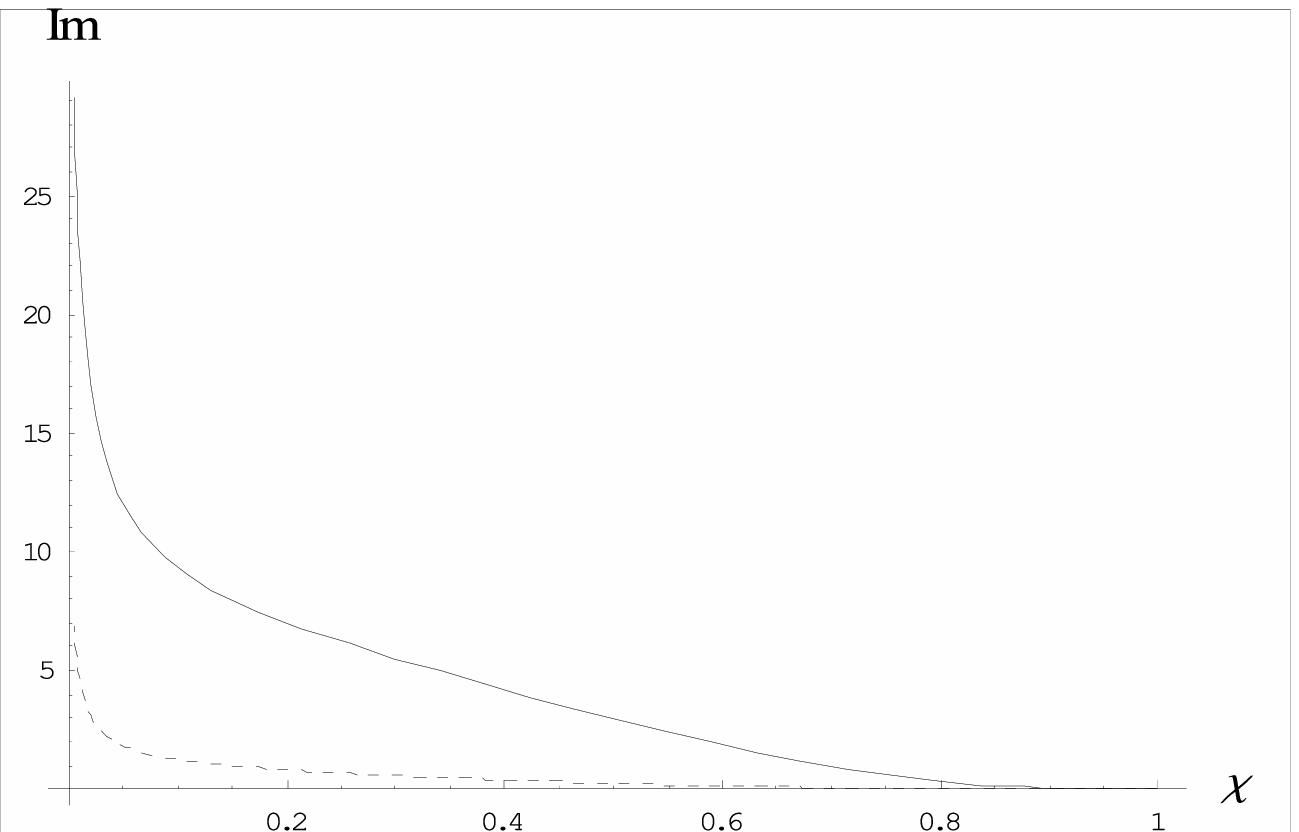}}
\caption[Contributions from \emph{u} quarks (solid line) and \emph{d} quarks (dashed line) to the 
imaginary part of $\mathcal{H}_{TCS}^{+}$.]
{Contributions from \emph{u} quarks (solid line) and \emph{d} quarks (dashed line) to the 
imaginary part of $\mathcal{H}_{TCS}^{+}$. 
They are divided by $F_{1u}\left(t\right)/2$ and $F_{1d}\left(t\right)$, respectively.}
\label{imaginaryHTCS}
\end{figure}

\begin{figure}
\centerline{\epsfxsize=5in\epsffile{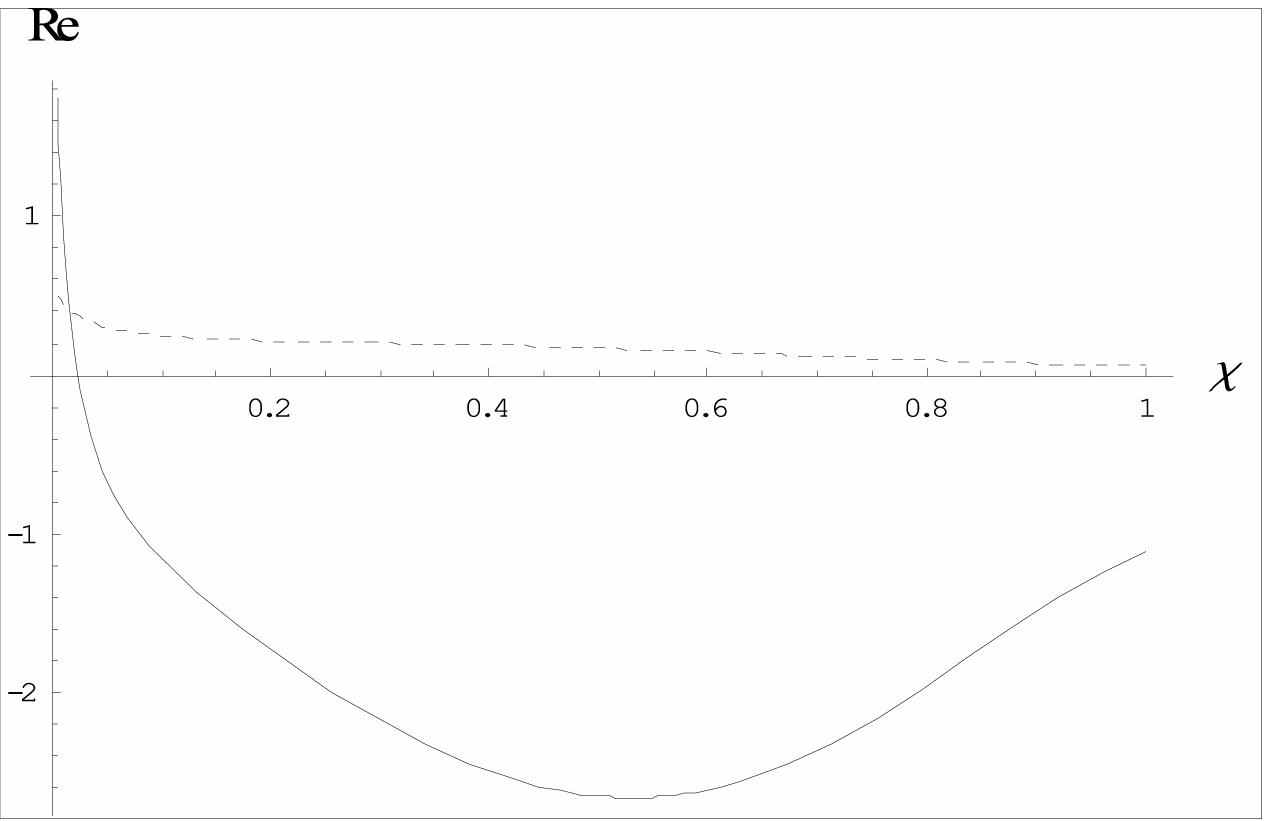}}
\caption{Contributions from \emph{u} quarks (solid line) and \emph{d} quarks (dashed line) to the 
real part of $\tilde\mathcal{H}_{TCS}^{+}$ divided by $g_{A}\left(t\right)/g_{A}\left(t=0\right)$.}
\label{realHtildaTCS}
\end{figure}

\begin{figure}
\centerline{\epsfxsize=5in\epsffile{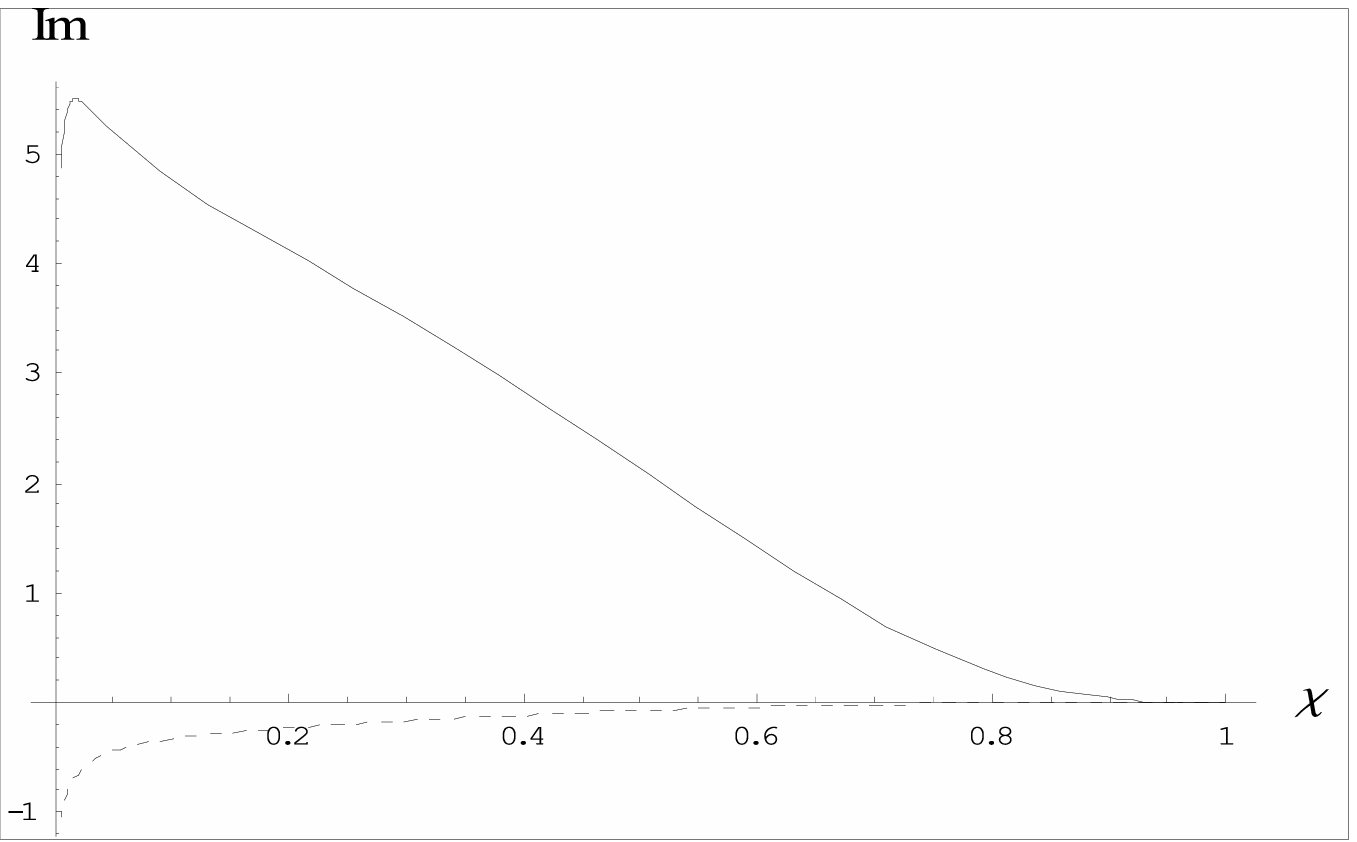}}
\caption{Contributions from \emph{u} quarks (solid line) and \emph{d} quarks (dashed line) to the 
imaginary part of $\tilde\mathcal{H}_{TCS}^{+}$ divided $g_{A}\left(t\right)/g_{A}\left(t=0\right)$.}
\label{imaginaryHtildaTCS}
\end{figure}

Substituting $x\rightarrow-x$ and using the symmetry properties of OFPDs given by 
Eq. (\ref{eq:symmetrypropertiesplusGPDs}), we obtain the reduced VCA for TCS at the leading-twist level 
in the form
\begin{eqnarray}
\mathsf{\mathcal{T}}_{TCStwist-2}^{\mu\nu}=\frac{1}{2\left(p\cdot q\right)}\sum_{f}Q_{f}^{2}
\int_{-1}^{1}\frac{dx}{\left(x-\eta-i0\right)}\nonumber \\
\times\Bigg\lbrace H_{f}^{+}\left(x,\eta,t\right)\left[\frac{1}{\left(p\cdot q_{1}\right)}
\left(p^{\mu}q_{1}^{\nu}+p^{\nu}q_{1}^{\mu}\right)-g^{\mu\nu}\right]\bar{u}\left(p_{2},s_{2}
\right)\not\! q_{1}u\left(p_{1},s_{1}\right)\nonumber \\
+E_{f}^{+}\left(x,\eta,t\right)\left[\frac{1}{\left(p\cdot q_{1}\right)}
\left(p^{\mu}q_{1}^{\nu}+p^{\nu}q_{1}^{\mu}\right)-g^{\mu\nu}\right]\bar{u}\left(p_{2},s_{2}\right)
\frac{\left(\not\! q_{1}\not\! r-\not\! r\not\! q_{1}\right)}{4M}u
\left(p_{1},s_{1}\right)\nonumber \\
+\tilde{H}_{f}^{+}\left(x,\eta,t\right)\left[\frac{1}{\left(p\cdot q_{2}\right)}i
\epsilon^{\mu\nu\rho\eta}q_{1\rho}p_{\eta}\right]\bar{u}\left(p_{2},s_{2}\right)\not\! q_{1}
\gamma_{5}u\left(p_{1},s_{1}\right)\nonumber \\
-\tilde{E}_{f}^{+}\left(x,\eta,t\right)\left[\frac{1}{\left(p\cdot q_{1}\right)}i
\epsilon^{\mu\nu\rho\eta}q_{1\rho}p_{\eta}\right]\frac{\left(q_{1}\cdot r\right)}{2M}
\bar{u}\left(p_{2},s_{2}\right)\gamma_{5}u\left(p_{1},s_{1}\right)\Bigg\rbrace.
\label{eq:VCAexclusive}
\end{eqnarray}
Since the photons in the amplitude (\ref{eq:VCAexclusive}) are interchanged compared to 
the standard VCA, the electromagnetic gauge invariance of $\mathcal{T}_{TCStwist-2}^{\mu\nu}$ 
is now satisfied with respect to the initial photon, but violated in 
$\mathcal{O}\left(r^{2}\right)$ with respect to the final one. Similarly to the standard 
DVCS process, we introduce the following integrals
\begin{eqnarray}
\mathcal{H}_{TCS}^{+}\left(\eta,t\right) & \equiv & \sum_{f}Q_{f}^{2}\int_{-1}^{1}
\frac{dx}{\left(x-\eta-i0\right)}H_{f}^{+}\left(x,\eta,t\right),\nonumber \\
\mathcal{E}_{TCS}^{+}\left(\eta,t\right) & \equiv & \sum_{f}Q_{f}^{2}\int_{-1}^{1}
\frac{dx}{\left(x-\eta-i0\right)}H_{f}^{+}\left(x,\eta,t\right),\nonumber \\
\tilde{\mathcal{H}}_{TCS}^{+}\left(\eta,t\right) & \equiv & \sum_{f}Q_{f}^{2}
\int_{-1}^{1}\frac{dx}{\left(x-\eta-i0\right)}
\tilde{H}_{f}^{+}\left(x,\eta,t\right),\nonumber \\
\tilde{\mathcal{E}}_{TCS}^{+}\left(\eta,t\right) & \equiv & \sum_{f}Q_{f}^{2}
\int_{-1}^{1}\frac{dx}{\left(x-\eta-i0\right)}\tilde{E}_{f}^{+}\left(x,\eta,t\right).
\label{eq:integralsexclusive}
\end{eqnarray}
They can be calculated with the help of Eq. (\ref{eq:principalvalueformula}). For illustration, we plot 
in Figs. \ref{realHTCS} and \ref{imaginaryHTCS} the real and imaginary parts of the convolution integral 
$\mathcal{H}_{TCS}^{+}\left(\eta,t\right)=\mathcal{H}_{TCS\left(u\right)}^{+}\left(\eta,t\right)+
\mathcal{H}_{TCS\left(d\right)}^{+}\left(\eta,t\right)$. We present the \emph{u}- and \emph{d}-quark 
contributions separately, using a GPD model from Section IV.4. To get rid of the \emph{t}-dependence, we 
divide by appropriate factors, $F_{1u}\left(t\right)/2$ and $F_{1d}\left(t\right)$. Analogous plots for 
$\tilde{\mathcal{H}}_{TCS}^{+}\left(\eta,t\right)=\tilde{\mathcal{H}}_{TCS\left(u\right)}^{+}
\left(\eta,t\right)+\tilde{\mathcal{H}}_{TCS\left(d\right)}^{+}\left(\eta,t\right)$ (here we divide by the 
factor $g_{A}\left(t\right)/g_{A}\left(t=0\right)$ to remove the dependence on \emph{t}), are given in 
Figs. \ref{realHtildaTCS} and \ref{imaginaryHtildaTCS}. All four figures clearly show that, for the proton 
target, the valence \emph{u}-quark contribution to the real and imaginary parts of the integrals 
$\mathcal{H}_{TCS}^{+}\left(\eta,t\right)$ and $\tilde{\mathcal{H}}_{TCS}^{+}\left(\eta,t\right)$ is much 
larger compared to the valence \emph{d}-quark contribution.

We will restrict ourselves to the unpolarized incoming photon beam and the unpolarized nucleon target. 
Then the spin-averaged square of the T-matrix in the factorized form is
\begin{eqnarray}
\overline{\left|\mathrm{T}_{C}\right|^{2}} & = & 
\frac{\left(4\pi\alpha\right)^{3}}{2q_{2}^{4}}L_{\mu\alpha}^{C}H_{C}^{\mu\alpha},
\label{comptonamplitudesquaredexclusive}
\end{eqnarray}
where the lepton and hadron tensors are
\begin{eqnarray}
L_{\mu\alpha}^{C} & = & 4\left[k_{\mu}k'_{\alpha}+k_{\alpha}k'_{\mu}-g_{\mu\alpha}
\left[\left(k\cdot k'
\right)+m^{2}\right]\right],\nonumber \\
H_{C}^{\mu\alpha} & = & -\frac{1}{2}\mathcal{T}_{TCS}^{\mu\nu}\left(\mathcal{T}_{\nu TCS}^{\alpha}
\right)^{*}\nonumber \\
& = & \left[\left(1-\eta^{2}\right)\left(\left|
\mathcal{H}_{TCS}^{+}\right|^{2}+\left|
\tilde{\mathcal{H}}_{TCS}^{+}\right|^{2}\right)\right.\nonumber \\
&  & -\eta^{2}\left(\mathcal{H}_{TCS}^{+*}\mathcal{E}_{TCS}^{+}+\mathcal{H}_{TCS}^{+}
\mathcal{E}_{TCS}^{+*}+
\tilde{\mathcal{H}}_{TCS}^{+*}\tilde{\mathcal{E}}_{TCS}^{+}+\tilde{\mathcal{H}}_{TCS}^{+}
\tilde{\mathcal{E}}_{TCS}^{+*}\right)\nonumber \\
&  & \left.-\left(\eta^{2}+\frac{t}{4M^{2}}\right)\left|\mathcal{E}_{TCS}^{+}\right|^{2}-
\eta^{2}\frac{t}{4M^{2}}
\left|\tilde{\mathcal{E}}_{TCS}^{+}\right|^{2}\right]\nonumber \\
&  & \times\left[\frac{1}{\left(p\cdot q_{1}\right)}
\left(p^{\mu}q_{1}^{\alpha}+p^{\alpha}q_{1}^{\mu}
\right)-\frac{M^{2}}{\left(p\cdot q_{1}\right)^{2}}\left(1-\frac{t}{4M^{2}}
\right)q_{1}^{\mu}q_{1}^{\alpha}-g^{\mu\alpha}\right],
\label{eq:exclusivetensorscompton}
\end{eqnarray}
with \emph{m} denoting the lepton mass. Note that $H_{C}^{\mu\alpha}$
has the same structure as its DVCS partner given by the expression (\ref{eq:hadronictensorDVCS}). 
To calculate the cross section, we integrate Eq. (\ref{comptonamplitudesquaredexclusive}) over the 
Lorentz-invariant phase space. This is done exactly in the same way as in Section V.2 for the case of the 
inclusive photoproduction of lepton pairs. Thus first we integrate $L_{\mu\alpha}^{C}$ over the part of 
the phase space associated with both final leptons,
\begin{eqnarray}
\int\frac{d^{3}k}{2\omega}\int\frac{d^{3}k'}{2\omega'}\;
\delta^{\left(4\right)}\left(p_{1}+q_{1}-p_{2}-k-k'
\right)L_{\mu\alpha}^{C}=\int d^{4}q_{2}\;\delta^{(4)}
\left(p_{1}+q_{1}-p_{2}-q_{2}\right)\nonumber \\
\times\left[q_{2}^{2}g_{\mu\alpha}-q_{2\mu}q_{2\alpha}
\right]\left(-\frac{2\pi}{3}
\frac{q_{2}^{2}+2m^{2}}{q_{2}^{2}}
\sqrt{1-\frac{4m^{2}}{q_{2}^{2}}}\right),\nonumber \\
\label{eq:symmetricvariablesintegralnewexclusive}
\end{eqnarray}
and then contract the tensor $\left[q_{2}^{2}g_{\mu\alpha}-q_{2\mu}q_{2\alpha}\right]$
with $H_{C}^{\mu\alpha}$. It is easy to see that, after contraction,
the first term is of the order $q_{2}^{2}$ while the second term
is $\mathcal{O}\left(M^{2}\right)$. The latter can be therefore ignored in
the limit $M^{2}/q_{2}^{2}\rightarrow0$. We are still left with two
more integrations. One of them, e.g. integration over $\vec{p}_{2}$,
is trivial due to the presence of the delta function $\delta^{(4)}\left(p_{1}+q_{1}-p_{2}-q_{2}\right)$ 
on the right-hand side of Eq. (\ref{eq:symmetricvariablesintegralnewexclusive}). The cross section 
now turns into
\begin{eqnarray}
\sigma_{C} & = & -\frac{\alpha^{3}}{3\pi}\frac{q_{2}^{2}+2m^{2}}{sq_{2}^{4}}
\sqrt{1-\frac{4m^{2}}{q_{2}^{2}}}\;
\int d^{4}q_{2}\;\delta^{+}\left[\left(p_{1}+q_{1}-q_{2}\right)^{2}-M^{2}\right]
\left[g_{\mu\alpha}H_{C}^{\mu\alpha}\right],\nonumber \\
\label{eq:comtpondifferentialcrosssection1exclusive}
\end{eqnarray}
where the contraction in the square brackets reads
\begin{eqnarray}
g_{\mu\alpha}H_{C}^{\mu\alpha} & = & -2\left[\left(1-\eta^{2}\right)\left(\left|
\mathcal{H}_{TCS}^{+}
\right|^{2}+\left|\tilde{\mathcal{H}}_{TCS}^{+}\right|^{2}\right)\right.\nonumber \\
&  & -\eta^{2}\left(\mathcal{H}_{TCS}^{+*}\mathcal{E}_{TCS}^{+}+\mathcal{H}_{TCS}^{+}
\mathcal{E}_{TCS}^{+*}+
\tilde{\mathcal{H}}_{TCS}^{+*}\tilde{\mathcal{E}}_{TCS}^{+}+\tilde{\mathcal{H}}_{TCS}^{+}
\tilde{\mathcal{E}}_{TCS}^{+*}\right)\nonumber \\
&  & \left.-\left(\eta^{2}+\frac{t}{4M^{2}}\right)\left|
\mathcal{E}_{TCS}^{+}\right|^{2}-\eta^{2}
\frac{t}{4M^{2}}\left|\tilde{\mathcal{E}}_{TCS}^{+}\right|^{2}\right].
\label{eq:comptoncontractionexclusive}
\end{eqnarray}
The remaining integral in Eq. (\ref{eq:comtpondifferentialcrosssection1exclusive}) is 
performed in the photon-nucleon center-of-mass-frame, in which the nucleon mass can be 
safely neglected. The final result for the unpolarized Compton differential cross section 
in the invariant form is
\begin{eqnarray}
\frac{d^{2}\sigma_{C}}{dM_{pair}^{2}dt} & = & -\frac{\alpha^{3}}{6}
\frac{M_{pair}^{2}+2m^{2}}{s^{2}M_{pair}^{4}}\sqrt{1-\frac{4m^{2}}{M_{pair}^{2}}}\;
\left[g_{\mu\alpha}H_{C}^{\mu\alpha}\right].
\label{eq:comtpondifferentialcrosssection2exclusive}
\end{eqnarray}

\subsection{\ \,  Bethe-Heitler Process}

The Bethe-Heitler amplitude,
\begin{eqnarray}
\mathrm{T}_{BH} & = & \frac{\left|e\right|^{3}}{t}\epsilon_{\nu}\left(q_{1}
\right)\bar{u}\left(k\right)\left[
\frac{\gamma^{\mu}\not\! q_{1}\gamma^{\nu}-2\gamma^{\mu}k'^{\nu}}{-2\left(k'\cdot q_{1}\right)}+
\frac{-\gamma^{\nu}\not\! q_{1}\gamma^{\mu}+2k^{\nu}\gamma^{\mu}}{-2\left(k\cdot q_{1}\right)}
\right]v\left(k'\right)\nonumber \\
&  & \times\left\langle N\left(p_{2},s_{2}\right)\right|J_{\mu}\left(0\right)\left|N\left(p_{1},s_{1}
\right)\right\rangle ,
\label{eq:bhamplitudeexclusive}
\end{eqnarray}
comes from the diagrams (b) and (c) in Fig. \ref{exclusivediagrams}. Dirac and Pauli nucleon form factors 
parametrize the nucleon transition current matrix element, 
\begin{eqnarray}
\left\langle N\left(p_{2},s_{2}\right)\right|J_{\mu}\left(0\right)\left|N\left(p_{1},s_{1}\right)
\right\rangle  & = & \bar{u}\left(p_{2},s_{2}\right)\left[F_{1}\left(t\right)\gamma_{\mu}-F_{2}
\left(t\right)\frac{i\sigma_{\mu\lambda}r^{\lambda}}{2M}\right]u\left(p_{1},s_{1}\right).\nonumber \\
\label{eq:transitioncurrentmatrixelementexclusive}
\end{eqnarray}
After averaging and summing $\left|\mathrm{T}_{BH}\right|^{2}$ over 
the polarizations of the initial and final particles, respectively, one has
\begin{eqnarray}
\overline{\left|\mathrm{T}_{BH}\right|^{2}} & = & 
\frac{\left(4\pi\alpha\right)^{3}}{2t^{2}}L_{BH}^{\mu\alpha}H_{\mu\alpha}^{BH},
\label{eq:bhamplitudesquaredexclusive}
\end{eqnarray}
where both tensors have already been calculated. The hadron tensor is obtained by replacing the Lorentz 
indices $\nu\rightarrow\mu$ and $\beta\rightarrow\alpha$ in the expression (\ref{eq:BHhadronictensorDVCS}),
\begin{eqnarray}
H_{\mu\alpha}^{BH} & = & t\left[g_{\mu\alpha}-\frac{r_{\mu}r_{\alpha}}{t}\right]\left[F_{1}
\left(t\right)+F_{2}\left(t\right)\right]^{2}\nonumber \\
&  & +4\left[p_{1\mu}+\frac{r_{\mu}}{2}\right]\left[p_{1\alpha}+\frac{r_{\alpha}}{2}
\right]\left[F_{1}^{2}\left(t\right)-\frac{t}{4M^{2}}F_{2}^{2}\left(t\right)\right],
\label{eq:bhhadronictensorexclusive}
\end{eqnarray}
and the lepton tensor by replacing the momenta $q\rightarrow q_{1}$ and indices 
$\rho\rightarrow\alpha$ in Eq. (\ref{eq:leptonictensorbh}),
\begin{eqnarray}
L_{BH}^{\mu\alpha} & = & 4\Bigg\lbrace\frac{1}{\left(k\cdot q_{1}\right)^{2}}
\left[\left(k\cdot q_{1}\right)
\left[k'^{\mu}q_{1}^{\alpha}+k'^{\alpha}q_{1}^{\mu}-g^{\mu\alpha}\left(k'\cdot q_{1}
\right)\right]\right.\nonumber \\
&  & -m^{2}\left[k^{\mu}k'^{\alpha}+k^{\alpha}k'^{\mu}-g^{\mu\alpha}
\left(k\cdot k'\right)\right]\nonumber \\
&  & +m^{2}\left[k'^{\mu}q_{1}^{\alpha}+k'^{\alpha}q_{1}^{\mu}-g^{\mu\alpha}\left(k'
\cdot q_{1}\right)\right]\nonumber \\
&  & \left.-m^{2}\left(k\cdot q_{1}\right)g^{\mu\alpha}+m^{4}g^{\mu\alpha}\right]\nonumber \\
&  & +\frac{1}{\left(k'\cdot q_{1}\right)^{2}}\left[\left(k'\cdot q_{1}\right)
\left[k^{\mu}q_{1}^{\alpha}+k^{\alpha}q_{1}^{\mu}-g^{\mu\alpha}\left(k\cdot q_{1}\right)\right]
\right.\nonumber \\
&  & -m^{2}\left[k^{\mu}k'^{\alpha}+k^{\alpha}k'^{\mu}-g^{\mu\alpha}
\left(k\cdot k'\right)\right]\nonumber \\
&  & +m^{2}\left[k^{\mu}q_{1}^{\alpha}+k^{\alpha}q_{1}^{\mu}-g^{\nu\alpha}
\left(k\cdot q_{1}\right)\right]\nonumber \\
&  & \left.-m^{2}\left(k'\cdot q_{1}\right)g^{\mu\alpha}+m^{4}g^{\mu\alpha}\right]\nonumber \\
&  & +\frac{1}{\left(k\cdot q_{1}\right)\left(k'\cdot q_{1}\right)}\left[2\left(k\cdot k'\right)
\left[k^{\mu}k'^{\alpha}+k^{\alpha}k'^{\mu}-g^{\mu\alpha}
\left(k\cdot k'\right)\right]\right.\nonumber \\
&  & -\left(k\cdot q_{1}\right)
\left[k^{\mu}k'^{\alpha}+k^{\alpha}k'^{\mu}-2k'^{\mu}k'^{\alpha}\right]\nonumber \\
&  & -\left(k'\cdot q_{1}\right)\left[k^{\mu}k'^{\alpha}+k^{\alpha}k'^{\mu}-2k^{\mu}k^{\alpha}
\right]\nonumber \\
&  & -\left(k\cdot k'\right)
\left[k^{\mu}q_{1}^{\alpha}+k^{\alpha}q_{1}^{\mu}+k'^{\mu}q_{1}^{\alpha}+k'^{\alpha}q_{1}^{\mu}
\right]\nonumber \\
&  & \left.+2g^{\mu\alpha}\left(k\cdot k'\right)\left[\left(k\cdot q_{1}
\right)+\left(k'\cdot q_{1}\right)\right]-m^{2}\right]-2m^{2}q_{1}^{\mu}q_{1}^{\alpha}\Bigg\rbrace.
\label{eq:bhleptonictensorexclusive}
\end{eqnarray}

The integral of $L_{BH}^{\mu\alpha}$ over the final lepton momenta is known from Section V.3, namely,
\begin{eqnarray}
\mathcal{L}_{BH}^{\mu\alpha} & \equiv & \int\frac{d^{3}k}{2\omega}\int\frac{d^{3}k'}{2\omega'}\;
\delta^{(4)}\left(q_{2}-k-k'\right)L_{BH}^{\mu\alpha}\nonumber \\
& = & \frac{2\pi}{\left(q_{1}\cdot q_{2}\right)}\left[\mathcal{A}_{BH}q_{1}^{\mu}q_{1}^{\alpha}+
\mathcal{B}_{BH}q_{2}^{\mu}q_{2}^{\alpha}+\mathcal{C}_{BH}
\left(q_{1}^{\mu}q_{2}^{\alpha}+q_{2}^{\mu}q_{1}^{\alpha}\right)+
\mathcal{D}_{BH}g^{\mu\alpha}\right],\nonumber \\
\label{eq:integratedleptonictensorexclusive}
\end{eqnarray}
where 
\begin{eqnarray}
\mathcal{A}_{BH} & = & \left[\frac{2q_{2}^{2}}{\left(q_{1}\cdot q_{2}\right)}-
\frac{2q_{2}^{4}}{\left(q_{1}\cdot q_{2}\right)^{2}}+
\frac{q_{2}^{6}}{\left(q_{1}\cdot q_{2}\right)^{3}}-
\frac{8m^{2}q_{2}^{2}}{\left(q_{1}\cdot q_{2}\right)^{2}}-
\frac{4m^{4}q_{2}^{2}}{\left(q_{1}\cdot q_{2}\right)^{3}}+
\frac{6m^{2}q_{2}^{4}}{\left(q_{1}\cdot q_{2}\right)^{3}}\right.\nonumber \\
&  & \left.+\frac{4m^{2}}{\left(q_{1}\cdot q_{2}\right)}\right]
\ln\left[\frac{1-\sqrt{1-4m^{2}/q_{2}^{2}}}{1+\sqrt{1-4m^{2}/q_{2}^{2}}}\right]\nonumber \\
&  & +\left[\frac{4q_{2}^{2}}{\left(q_{1}\cdot q_{2}\right)}-
\frac{8q_{2}^{4}}{\left(q_{1}\cdot q_{2}\right)^{2}}+
\frac{4q_{2}^{6}}{\left(q_{1}\cdot q_{2}\right)^{3}}+
\frac{2m^{2}q_{2}^{4}}{\left(q_{1}\cdot q_{2}\right)^{3}}\right]
\sqrt{1-\frac{4m^{2}}{q_{2}^{2}}},\nonumber \\
\mathcal{B}_{BH} & = & \frac{4m^{2}}{\left(q_{1}\cdot q_{2}\right)}
\ln\left[\frac{1-\sqrt{1-4m^{2}/q_{2}^{2}}}{1+\sqrt{1-4m^{2}/q_{2}^{2}}}\right]+
\frac{2q_{2}^{2}}{\left(q_{1}\cdot q_{2}\right)}\sqrt{1-\frac{4m^{2}}{q_{2}^{2}}},\nonumber \\
\mathcal{C}_{BH} & = & \left[-2+\frac{2q_{2}^{2}}{\left(q_{1}\cdot q_{2}\right)}-
\frac{q_{2}^{4}}{\left(q_{1}\cdot q_{2}\right)^{2}}-
\frac{6m^{2}q_{2}^{2}}{\left(q_{1}\cdot q_{2}\right)^{2}}+
\frac{4m^{2}}{\left(q_{1}\cdot q_{2}\right)}+
\frac{4m^{4}}{\left(q_{1}\cdot q_{2}\right)^{2}}\right]\nonumber \\
&  & \times\ln\left[\frac{1-\sqrt{1-4m^{2}/q_{2}^{2}}}{1+\sqrt{1-4m^{2}/q_{2}^{2}}}\right]\nonumber \\
&  & +\left[-2+\frac{6q_{2}^{2}}{\left(q_{1}\cdot q_{2}\right)}-
\frac{4q_{2}^{4}}{\left(q_{1}\cdot q_{2}\right)^{2}}-
\frac{2m^{2}q_{2}^{2}}{\left(q_{1}\cdot q_{2}\right)^{2}}
\right]\sqrt{1-\frac{4m^{2}}{q_{2}^{2}}},\nonumber \\
\mathcal{D}_{BH} & = & \left[2\left(q_{1}\cdot q_{2}\right)-2q_{2}^{2}+
\frac{q_{2}^{4}}{\left(q_{1}\cdot q_{2}\right)}+\frac{2m^{2}q_{2}^{2}}{\left(q_{1}\cdot q_{2}\right)}-
\frac{4m^{4}}{\left(q_{1}\cdot q_{2}\right)}\right]\nonumber \\
&  & \times\ln\left[\frac{1-\sqrt{1-4m^{2}/q_{2}^{2}}}{1+\sqrt{1-4m^{2}/q_{2}^{2}}}\right]\nonumber \\
&  & +\left[2\left(q_{1}\cdot q_{2}\right)-4q_{2}^{2}+
\frac{2q_{2}^{4}}{\left(q_{1}\cdot q_{2}\right)}+
\frac{2m^{2}q_{2}^{2}}{\left(q_{1}\cdot q_{2}\right)}\right]\sqrt{1-\frac{4m^{2}}{q_{2}^{2}}}.
\label{eq:bhcoefficientsexclusive}
\end{eqnarray}
The unpolarized Bethe-Heitler cross section is then
\begin{eqnarray}
\sigma_{BH} & = & \frac{1}{2s}\frac{1}{\left(2\pi\right)^{5}}\frac{\left(4\pi\alpha\right)^{3}}{2t^{2}}
\frac{\pi}{2}\left(\frac{s-M_{pair}^{2}}{2s}\right)\int dM_{pair}^{2}\int d\left(\cos\theta_{cm}
\right)\left[\mathcal{L}_{BH}^{\mu\alpha}H_{\mu\alpha}^{BH}\right],\nonumber \\
\label{eq:bhdifferentialcrosssection1exclusive}
\end{eqnarray}
and finally in the differential invariant form,
\begin{eqnarray}
\frac{d^{2}\sigma_{BH}}{dM_{pair}^{2}dt} & = & \frac{\alpha^{3}}{4\pi}\frac{1}{s^{2}t^{2}}
\left[\mathcal{L}_{BH}^{\mu\alpha}H_{\mu\alpha}^{BH}\right].
\label{eq:bhdifferentialcrosssection2exclusive}
\end{eqnarray}
The contraction in the square brackets of Eq. (\ref{eq:bhdifferentialcrosssection1exclusive}) gives
\begin{eqnarray}
\mathcal{L}_{BH}^{\mu\alpha}H_{\mu\alpha}^{BH} & = & \frac{4\pi}{M_{pair}^{2}-t}\left\{ \left[-
\frac{\mathcal{A}_{BH}}{4}\left(1-\frac{t}{M_{pair}^{2}}\right)^{2}+\mathcal{B}_{BH}
\left[\frac{t}{M_{pair}^{2}}-\frac{1}{4}\left(1+\frac{t}{M_{pair}^{2}}
\right)^{2}\right]\right.\right.\nonumber \\
&  & \left.+\mathcal{C}_{BH}\left[\frac{t}{M_{pair}^{2}}\left(1-\frac{t}{M_{pair}^{2}}\right)-
\frac{1}{2}\left(1-\left(\frac{t}{M_{pair}^{2}}\right)^{2}\right)\right]+3\mathcal{D}_{BH}
\frac{t}{M_{pair}^{4}}\right]\nonumber \\
&  & \times M_{pair}^{4}\left[F_{1}\left(t\right)+F_{2}\left(t\right)\right]^{2}+
\left[\mathcal{A}_{BH}\left[1-\frac{M^{2}}{s}+\frac{M_{pair}^{2}}{2s}\left(1-
\frac{t}{M_{pair}^{2}}\right)\right]^{2}\right.\nonumber \\
&  & +\mathcal{B}_{BH}\left[1+\frac{t}{s}-\frac{M^{2}}{s}+\frac{M_{pair}^{2}}{2s}\left(1+
\frac{t}{M_{pair}^{2}}\right)\right]^{2}\nonumber \\
&  & +2\mathcal{C}_{BH}\left[1-\frac{M^{2}}{s}+\frac{M_{pair}^{2}}{2s}\left(1-\frac{t}{M_{pair}^{2}}
\right)\right]\nonumber \\
&  & \times\left[1+\frac{t}{s}-\frac{M^{2}}{s}+\frac{M_{pair}^{2}}{2s}\left(1+
\frac{t}{M_{pair}^{2}}\right)\right]\nonumber \\
&  & \left.+4\mathcal{D}_{BH}\frac{M^{2}}{s^{2}}\left(1+\frac{3t}{4M^{2}}\right)
\right]s^{2}\left[F_{1}^{2}\left(t\right)-\frac{t}{4M^{2}}F_{2}^{2}\left(t\right)\right]\Bigg\rbrace.
\label{eq:bhcontractionexclusive}
\end{eqnarray}
Note that $q_{2}^{2}$ and $\left(q_{1}\cdot q_{2}\right)$, which appear in the coefficients 
(\ref{eq:bhcoefficientsexclusive}), should be replaced by $M_{pair}^{2}$ and $\left(M_{pair}^{2}-t\right)/2$, 
respectively.

\begin{figure}
\centerline{\epsfxsize=5in\epsffile{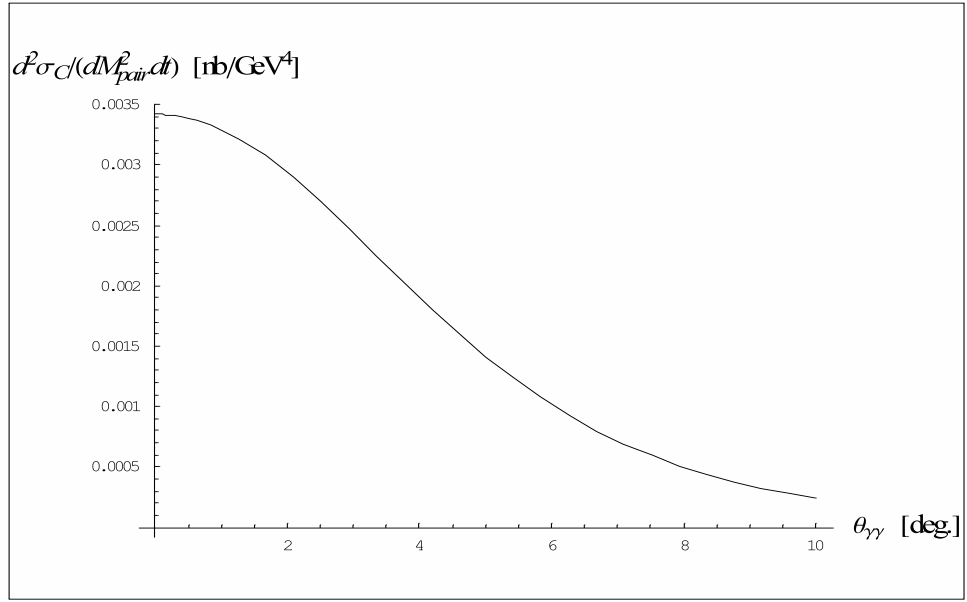}}
\caption{Compton cross section $\sigma_{C}$ plotted as a function of the angle $\theta_{\gamma\gamma}$ 
between the incoming real and outgoing virtual photon in the target rest frame for 
$M_{pair}^{2}=3\;\mathrm{GeV}^{2}$ and $\chi=0.32$ with $\nu_{1}=5\;\mathrm{GeV}$ photon beam.}
\label{exclusiveCompton}
\end{figure}

\begin{figure}
\centerline{\epsfxsize=5in\epsffile{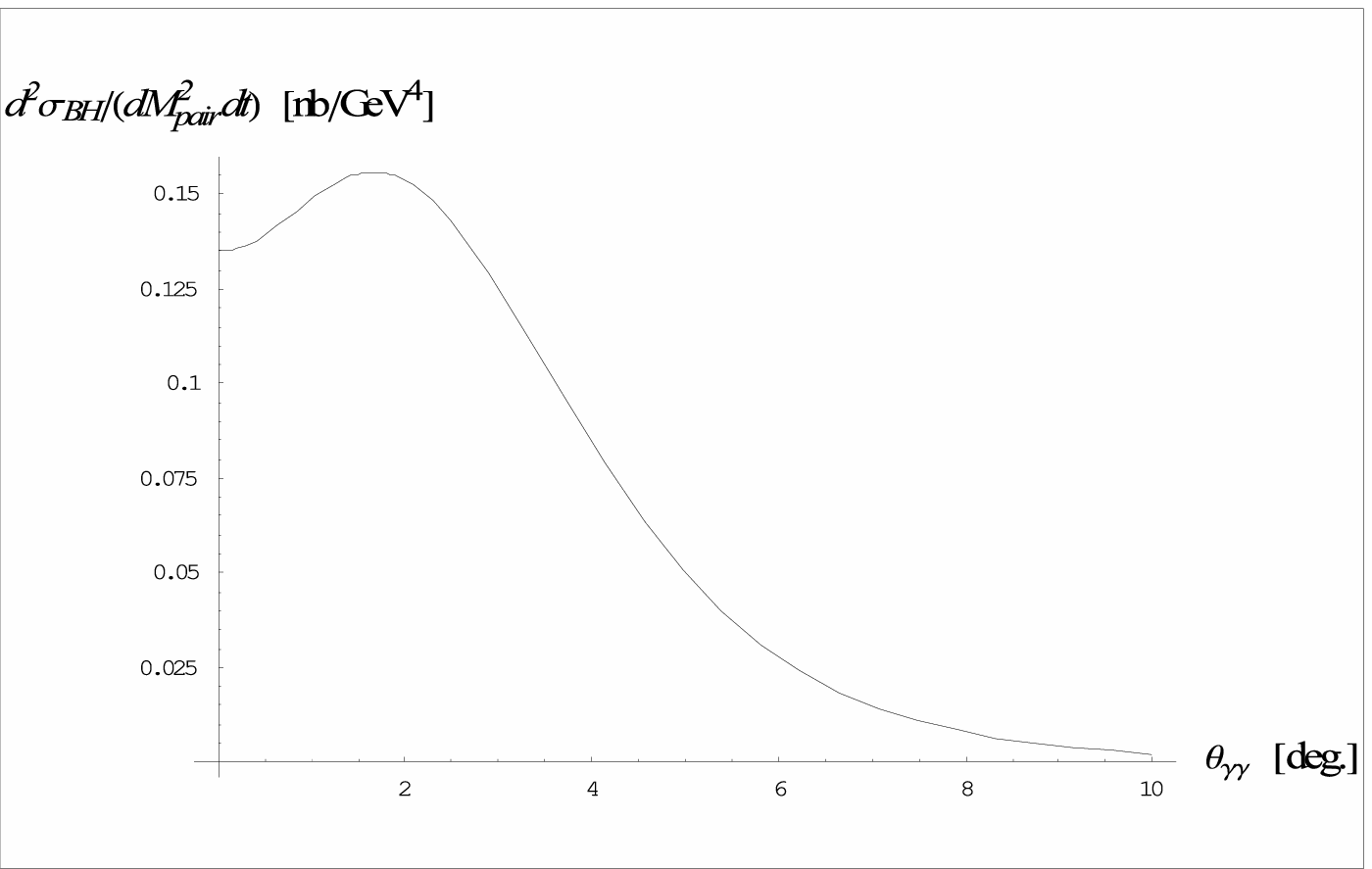}}
\caption{Bethe-Heitler cross section $\sigma_{BH}$ plotted plotted as a function of the 
angle $\theta_{\gamma\gamma}$ between the incoming real and outgoing virtual photon in the target rest 
frame for $M_{pair}^{2}=3\;\mathrm{GeV}^{2}$ and $\chi=0.32$ with $\nu_{1}=5\;\mathrm{GeV}$ photon beam.}
\label{exclusiveBH}
\end{figure}

\begin{figure}
\centerline{\epsfxsize=5in\epsffile{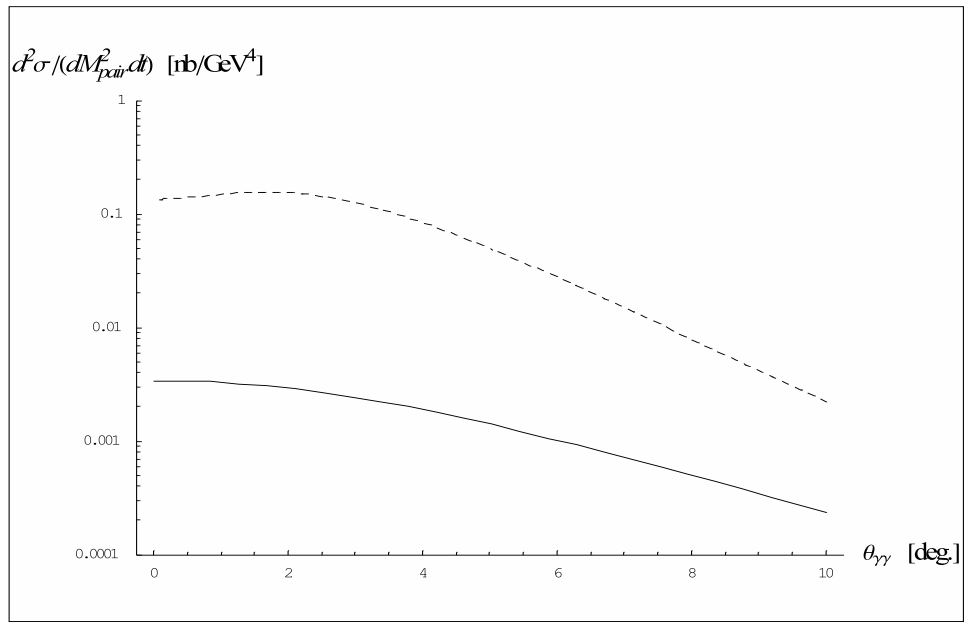}}
\caption{Compton cross section $\sigma_{C}$ (solid line) and Bethe-Heitler cross section $\sigma_{BH}$ 
(dashed line) plotted as a function of the angle $\theta_{\gamma\gamma}$ between the incoming real 
and outgoing virtual photon in the target rest frame for $M_{pair}^{2}=3\;\mathrm{GeV}^{2}$ and $\chi=0.32$ 
with $\nu_{1}=5\;\mathrm{GeV}$ photon beam.}
\label{togetherexclusive}
\end{figure}

In Figs. \ref{exclusiveCompton} and \ref{exclusiveBH}, we plot the Compton and Bethe-Heitler
contributions using a toy model for the proton valence OFPDs described in Section IV.4. Both 
cross sections are shown together on a logarithmic scale in Fig. \ref{togetherexclusive}. We notice that, 
similarly to the standard DVCS, the Bethe-Heitler process is again the dominating contribution 
(i.e. $\simeq0.95$) to the cross section.

We conclude with the remark that the cross sections presented here were integrated over the angles of the 
final-state leptons. One may expect to get additional information of GPDs, by measuring the cross sections 
that are sensitive to the angular distribution of final leptons. In particular, the interference terms 
between the Compton and Bethe-Heitler processes, together with the use of the polarized photon beam, 
may provide a new insight. This is an interesting subject by itself, and requires more attention 
in the future.

\chapter{Weak Deeply Virtual Compton Scattering}

\section{\ \,  Introduction}

Complementary to the standard electromagnetic DVCS process, different
combinations of quark flavors can be accessed by utilizing the weak
current, which couples to quarks with strengths proportional to the
quark weak charges. In analogy with DVCS, these can be studied in
neutrino-induced virtual Compton scattering,
\begin{eqnarray}
\nu\left(k\right)+N\left(p_{1}\right) & \longrightarrow & \nu\left(k'\right)+N\left(p_{2}\right)+
\gamma\left(q_{2}\right)
\label{eq:weakneutralDVCSprocess}
\end{eqnarray}
for the neutral current, and
\begin{eqnarray}
\nu\left(k\right)+N\left(p_{1}\right) & \longrightarrow & \mu^{\pm}\left(k'\right)+N'\left(p_{2}\right)+
\gamma\left(q_{2}\right)
\label{eq:weakchargedDVCSprocess}
\end{eqnarray}
for the charged current reactions (and similarly for antineutrinos).
Note that in the case of the latter the initial and final nucleons
will be different. Because of the \emph{V-A} nature of the weak interactions,
one can probe $\hat{C}$-odd combinations of GPDs as well as $\hat{C}$-even,
where $\hat{C}$ is the charge conjugation operator, and thus measure
independently both the valence and sea content of GPDs. The weak current
also allows one to study flavor nondiagonal GPDs, such as those associated 
with the neutron-to-proton transitions in charged current reactions 
in Eq. (\ref{eq:weakchargedDVCSprocess}). The use of weak currents 
can thus provide an important tool to complement the study of GPDs 
in more familiar electron-induced DVCS or exclusive meson production processes.

In this chapter, we study the weak deeply virtual Compton scattering processes given by 
Eqs. (\ref{eq:weakchargedDVCSprocess}) and (\ref{eq:weakneutralDVCSprocess}), 
and present a comprehensive account of the amplitudes and cross sections 
in the kinematics relevant to future high-intensity neutrino experiments \cite{Drakoulakos:2004gn}. 
In Section VII.2, we derive both the weak neutral and weak charged VCAs using the nonlocal 
light-cone OPE, and introduce an appropriate set of off-forward parton distributions, 
which parametrize the weak DVCS reactions. The weak DVCS processes are analyzed in Section VII.3. 
Using a simple model for nucleon OFPDs from Section IV.4 (recall that the model does not include 
the sea quark contribution), we estimate the
cross sections and compare the respective
rates in neutrino scattering with those in the standard DVCS process. 
In addition, for the sake of completeness, we also
discuss the electron-induced DVCS process associated with the exchange
of the weak boson $Z^{0}$.

Deeply virtual neutrino scattering has been discussed recently in Ref. \cite{Amore:2004ng} 
for neutral currents and in Ref. \cite{Coriano:2004bk} for charged currents. 
Also a preliminary report containing some of the formal results from Ref. \cite{Psaker1} 
appeared in Ref. \cite{Psaker:2004sf}.

\section{\ \,  Weak Virtual Compton Scattering Amplitude}

In this section, we discuss the amplitudes for the weak virtual Compton scattering process. 
Before turning to the specific amplitudes for the neutral and charged current cases, we review some 
general aspects that have been already discussed in Chapter IV.

At the subprocess level, in analogy with the photon-induced DVCS amplitude,
the weak virtual Compton scattering amplitude can be obtained by simply
replacing the incoming virtual photon with the weak boson \emph{B},
\begin{eqnarray}
B\left(q_{1}\right)+N\left(p_{1}\right) & \longrightarrow & \gamma\left(q_{2}\right)+N'
\left(p_{2}\right).
\label{eq:weakDVCSsubprocess}
\end{eqnarray}
where \emph{B} is either $Z^{0}$ or $W^{\pm}$. Note that in the
electromagnetic and weak neutral cases both the incoming and outgoing
nucleons are the same, $N=N'$. Similarly to the electromagnetic VCA
introduced in Section IV.2, for the weak process with an incoming
$W^{\pm}$ or $Z^{0}$ boson and outgoing photon, one has 
\begin{eqnarray}
T_{W}^{\mu\nu} & = & i\int d^{4}x\int d^{4}y\; e^{-i\left(q_{1}\cdot x\right)+i
\left(q_{2}\cdot y\right)}\nonumber \\
&  & \times\left\langle N'\left(p_{2},s_{2}\right)\right|T\left\{ J_{EM}^{\mu}
\left(y\right)J_{W}^{\nu}
\left(x\right)\right\} \left|N\left(p_{1},s_{1}\right)\right\rangle ,
\label{eq:weakVCA1}
\end{eqnarray}
where $J_{W}^{\nu}$ corresponds either to the weak neutral current
or the weak charged current. We will denote the currents by $J_{WN}^{\nu}$
and $J_{WC}^{\nu}$, respectively. Again, due to the current conservation
\begin{eqnarray}
T_{W}^{\mu\nu}q_{1\nu}=0 & \mathrm{and} & q_{2\mu}T_{W}^{\mu\nu}=0.
\label{eq:currentconservationweakamplitude}
\end{eqnarray}
In terms of symmetric coordinates, $X\equiv\left(x+y\right)/2$ and $z\equiv y-x$, and symmetric 
momentum variables, $q\equiv\left(q_{1}+q_{2}\right)/2$ and $p\equiv\left(p_{1}+p_{2}\right)/2$, 
the weak VCA takes the following form
\begin{eqnarray}
T_{W}^{\mu\nu} & = & \left(2\pi\right)^{4}\delta^{\left(4\right)}\left(p_{1}+q_{1}-p_{2}-q_{2}\right)
\mathsf{\mathcal{T}}_{W}^{\mu\nu},
\label{eq:weakVCA2}
\end{eqnarray}
with the expression for the reduced weak VCA
\begin{eqnarray}
\mathsf{\mathcal{T}}_{W}^{\mu\nu} & = & i\int d^{4}z\; e^{i\left(q\cdot z\right)}
\left\langle N'\left(p-r/2,s_{2}\right)
\right|T\left\{ J_{EM}^{\mu}\left(z/2\right)J_{W}^{\nu}
\left(-z/2\right)\right\} \left|N\left(p+r/2,s_{1}
\right)\right\rangle .\nonumber \\
\label{eq:reducedweakVCA}
\end{eqnarray}
Recall that, in general, the initial and final nucleons may be different,
$p_{1}^{2}=M_{1}^{2}$ and $p_{2}^{2}=M_{2}^{2}$. Accordingly, Eq.
(\ref{eq:frommassshellconditionsstandardVCA}) is now replaced by
\begin{eqnarray}
p^{2}=\frac{1}{2}\left(M_{1}^{2}+M_{2}^{2}-t/2\right) & \mathrm{and} & \left(p\cdot r\right)=
\frac{1}{2}\left(M_{1}^{2}-M_{2}^{2}\right),
\label{eq:frommassshellconditionsweak}
\end{eqnarray}
however, it is recovered by neglecting the mass difference
between the proton and neutron (we set $M_{1}=M_{2}\equiv M$).

\begin{figure}
\centerline{\epsfxsize=4in\epsffile{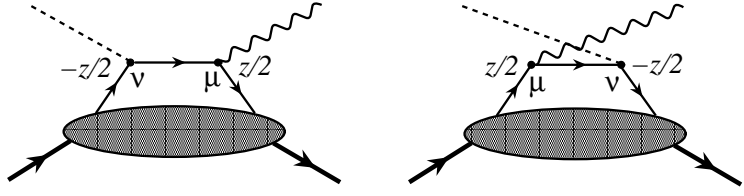}}
\caption{Handbag diagrams (\emph{s}- and \emph{u}-channel) in the weak virtual 
Compton scattering amplitude.}
\label{handbagdiagramsweakDVCS}
\end{figure}

Our starting point is the coordinate representation of the time-ordered
product of the weak and electromagnetic currents. The leading light-cone
singularity is contained in the handbag contribution illustrated in
Fig. \ref{handbagdiagramsweakDVCS}. Note that the weak current couples
to the quark fields through two types of vertices, $qqZ^{0}$ and
$qqW^{\pm}$. Therefore, the quark fields at coordinates $\pm z/2$
can carry either the same or different flavor quantum numbers. We
study these two cases separately.

\subsection{\ \,  Weak Neutral Amplitude}

We start with the expansion of the time-ordered product of the weak neutral and electromagnetic 
current. Apart from an overall factor $-\left|e\right|g/\cos\theta_{W}$ (recall that \emph{g} is the 
coupling constant of the weak interaction), we have
\begin{eqnarray}
iT\left\{ J_{EM}^{\mu}\left(z/2\right)J_{WN}^{\nu}\left(-z/2\right)\right\}  & = & i\sum_{f}Q_{f}
\left\{ \bar{\psi}_{f}\left(z/2\right)\gamma^{\mu}i\not\! S\left(z\right)\gamma^{\nu}
\left(\frac{c_{V}^{f}-\gamma_{5}c_{A}^{f}}{2}\right)\right.\nonumber \\
&  & \times\psi_{f}\left(-z/2\right)+\bar{\psi}_{f}\left(-z/2\right)\gamma^{\nu}
\left(\frac{c_{V}^{f}-\gamma_{5}c_{A}^{f}}{2}\right)\nonumber \\
&  & \times i\not\! S\left(-z\right)\gamma^{\mu}\psi_{f}\left(z/2\right)\Bigg\rbrace,
\label{eq:WNexpansion1}
\end{eqnarray}
where $c_{V}^{f}$ and $c_{A}^{f}$ are the weak vector and axial
vector charges, respectively. They are given in the standard model
by
\begin{eqnarray}
c_{V}^{u,c,t} & = & 1/2-2Q_{u,c,t}\sin^{2}\theta_{W},\nonumber \\
c_{V}^{d,s,b} & = & -1/2-2Q_{d,s,b}\sin^{2}\theta_{W},\nonumber \\
c_{A}^{u,c,t} & = & 1/2,\nonumber \\
c_{A}^{d,s,b} & = & -1/2,
\label{eq:weakneutralcharges}
\end{eqnarray}
where $\theta_{W}$ is the Weinberg angle with $\sin^{2}\theta_{W}\simeq0.23$. After 
substituting the explicit expression for the free quark propagator in the coordinate 
representation, see Eq. (\ref{eq:freequarkpropagator}), and using the formula 
(\ref{eq:gammamatrixformula}), the expansion (\ref{eq:WNexpansion1}) can be written 
in a compact form in terms of the string operators
\begin{eqnarray}
iT\left\{ J_{EM}^{\mu}\left(z/2\right)J_{WN}^{\nu}\left(-z/2\right)
\right\}  & = & \frac{z_{\rho}}{4\pi^{2}z^{4}}\sum_{f}Q_{f}\nonumber \\
&  & \times\left\{ c_{V}^{f}\left[s^{\mu\rho\nu\eta}\left[\bar{\psi}_{f}
\left(-z/2\right)\gamma_{\eta}\psi_{f}\left(z/2\right)-\left(z\rightarrow-z\right)\right]
\right.\right.\nonumber \\
&  & \left.+i\epsilon^{\mu\rho\nu\eta}\left[\bar{\psi}_{f}\left(-z/2\right)
\gamma_{\eta}\gamma_{5}\psi_{f}\left(z/2\right)+\left(z\rightarrow-z\right)\right]\right]\nonumber \\
&  & -c_{A}^{f}\left[s^{\mu\rho\nu\eta}\left[\bar{\psi}_{f}\left(-z/2\right)\gamma_{\eta}
\gamma_{5}\psi_{f}\left(z/2\right)-\left(z\rightarrow-z\right)\right]\right.\nonumber \\
&  & \left.\left.+i\epsilon^{\mu\rho\nu\eta}\left[\bar{\psi}_{f}\left(-z/2\right)
\gamma_{\eta}\psi_{f}\left(z/2\right)+\left(z\rightarrow-z\right)\right]\right]\right\}.\nonumber \\
\label{eq:WNexpansion2}
\end{eqnarray}
In contrast to the standard electromagnetic DVCS process, see Eq.
(\ref{eq:DVCSexpansion2}), we end up with two additional terms. Namely,
the presence of the axial part $\gamma_{5}c_{A}^{f}$ of the \emph{V-A}
interaction gives rise to a vector current symmetric in the Lorentz
indices $\mu$, $\nu$ and to an axial vector current antisymmetric
in $\mu$, $\nu$. The string operators here are accompanied either
with $Q_{f}c_{V}^{f}$ or $Q_{f}c_{A}^{f}$ rather than $Q_{f}^{2}$.
Also the denominator of Eq. (\ref{eq:WNexpansion2}) carries an extra
factor of 2 due to the nature of the vertex $qqZ^{0}$.

Accordingly, to obtain the twist-2 part of Eq. (\ref{eq:WNexpansion2}) we need, in addition to contracted 
string operators given by Eq. (\ref{eq:contractedoperatorsDVCS}), two extra operators. One can conveniently 
write all four contracted vector and axial vector string operators under the
same footing as
\begin{eqnarray}
\mathcal{O}^{f\pm}\left(z\right) & \equiv & \left[\bar{\psi}_{f}\left(-z/2\right)\not\! z\psi_{f}
\left(z/2\right)\pm\left(z\rightarrow-z\right)\right],\nonumber \\
\mathcal{O}_{5}^{f\pm}\left(z\right) & \equiv & \left[\bar{\psi}_{f}
\left(-z/2\right)\not\! z\gamma_{5}
\psi_{f}\left(z/2\right)\pm\left(z\rightarrow-z\right)\right].
\label{eq:contractedoperatorsWN}
\end{eqnarray}
They all should satisfy the harmonic condition (\ref{eq:d'Alembert}).
Furthermore, the parametrization of the nonforward matrix elements of these operators 
is performed by means of the relevant nonperturbative functions, namely, the off-forward parton distributions. 
A straightforward generalization of the parametrization (\ref{eq:parametrizationDVCS}) from Section IV.2 can 
be performed by introducing a new set of GPDs, i.e. the minus OFPDs. Then
\begin{eqnarray}
\left\langle N\left(p_{2},s_{2}\right)\right|\mathcal{O}^{f\pm}\left(z\right)
\left|N\left(p_{1},s_{1}\right)
\right\rangle_{z^2=0}  & = & \bar{u}\left(p_{2},s_{2}\right)\not\! zu\left(p_{1},s_{1}
\right)\nonumber \\
&  & \times\int_{-1}^{1}dx\; e^{ix\left(p\cdot z\right)}H_{f}^{\pm}\left(x,\xi,t\right)\nonumber \\
&  & +\bar{u}\left(p_{2},s_{2}\right)\frac{\left(\not\! z\not\! r-\not\! r\not\! z\right)}{4M}u
\left(p_{1},s_{1}\right)\nonumber \\
&  & \times\int_{-1}^{1}dx\; e^{ix\left(p\cdot z\right)}E_{f}^{\pm}\left(x,\xi,t\right),\nonumber \\
\left\langle N\left(p_{2},s_{2}\right)\right|\mathcal{O}_{5}^{f\pm}
\left(z\right)\left|N\left(p_{1},s_{1}
\right)\right\rangle_{z^2=0}  & = & \bar{u}
\left(p_{2},s_{2}\right)\not\! z\gamma_{5}u\left(p_{1},s_{1}\right)\nonumber \\
&  & \times\int_{-1}^{1}dx\; e^{ix\left(p\cdot z\right)}\tilde{H}_{f}^{\mp}
\left(x,\xi,t\right)\nonumber \\
&  & -\bar{u}\left(p_{2},s_{2}\right)\frac{\left(r\cdot z\right)}{2M}
\gamma_{5}u\left(p_{1},s_{1}\right)\nonumber \\
&  & \times\int_{-1}^{1}dx\; e^{ix\left(p\cdot z\right)}\tilde{E}_{f}^{\mp}\left(x,\xi,t\right).
\label{eq:parametrizationWN}
\end{eqnarray}
Note that $\mathcal{O}_{5}^{f\pm}\left(z\right)$ has a superscript
opposite in sign with respect to the corresponding tilde OFPD. While the standard DVCS process 
gives access only to the plus distributions (recall that they correspond to the sum of quark 
and antiquark distributions), scattering via the weak virtual boson exchange probes also 
the minus distributions. It can be shown that the latter correspond to the difference in 
quark and antiquark distributions, i.e.the valence configuration. Like the plus OFPDs, 
the minus distributions have also symmetry with respect to the change $x\rightarrow-x$. 
Let us summarize the symmetry properties in \emph{x} for both plus and minus OFPDs,
\begin{eqnarray}
H_{f}^{\pm}\left(x\right) & = & \mp H_{f}^{\pm}\left(-x\right),\nonumber \\
E_{f}^{\pm}\left(x\right) & = & \mp E_{f}^{\pm}\left(-x\right),\nonumber \\
\tilde{H}_{f}^{\pm}\left(x\right) & = & \pm\tilde{H}_{f}^{\pm}\left(-x\right),\nonumber \\
\tilde{E}_{f}^{\pm}\left(x\right) & = & \pm\tilde{E}_{f}^{\pm}\left(-x\right).
\label{eq:symmetrypropertiesplusminusGPDs}
\end{eqnarray}
After following the steps described in Section IV.2, we arrive at the expression for the reduced 
weak neutral VCA in the leading-twist approximation
\begin{eqnarray}
\mathsf{\mathcal{T}}_{WNtwist-2}^{\mu\nu}=\frac{1}{4\left(p\cdot q\right)}\sum_{f}Q_{f}\int_{-1}^{1}
\frac{dx}{\left(x-\xi+i0\right)}\nonumber \\
\times\Bigg\lbrace c_{V}^{f}H_{f}^{+}\left(x,\xi,t\right)\left[\frac{1}{\left(p\cdot q_{2}\right)}
\left(p^{\mu}q_{2}^{\nu}+p^{\nu}q_{2}^{\mu}\right)-g^{\mu\nu}\right]\bar{u}\left(p_{2},s_{2}
\right)\not\! q_{2}u\left(p_{1},s_{1}\right)\nonumber \\
+c_{V}^{f}E_{f}^{+}\left(x,\xi,t\right)\left[\frac{1}{\left(p\cdot q_{2}\right)}
\left(p^{\mu}q_{2}^{\nu}+p^{\nu}q_{2}^{\mu}\right)-g^{\mu\nu}\right]\bar{u}\left(p_{2},s_{2}\right)
\frac{\left(\not\! q_{2}\not\! r-\not\! r\not\! q_{2}\right)}{4M}u
\left(p_{1},s_{1}\right)\nonumber \\
-c_{V}^{f}\tilde{H}_{f}^{+}\left(x,\xi,t\right)\left[\frac{1}{\left(p\cdot q_{2}\right)}i
\epsilon^{\mu\nu\rho\eta}q_{2\rho}p_{\eta}\right]\bar{u}\left(p_{2},s_{2}\right)\not\! q_{2}
\gamma_{5}u\left(p_{1},s_{1}\right)\nonumber \\
+c_{V}^{f}\tilde{E}_{f}^{+}\left(x,\xi,t\right)\left[\frac{1}{\left(p\cdot q_{2}\right)}i
\epsilon^{\mu\nu\rho\eta}q_{2\rho}p_{\eta}\right]\frac{\left(q_{2}\cdot r\right)}{2M}\bar{u}
\left(p_{2},s_{2}\right)\gamma_{5}u\left(p_{1},s_{1}\right)\nonumber \\
-c_{A}^{f}\tilde{H}_{f}^{-}\left(x,\xi,t\right)\left[\frac{1}{\left(p\cdot q_{2}\right)}
\left(p^{\mu}q_{2}^{\nu}+p^{\nu}q_{2}^{\mu}\right)-g^{\mu\nu}\right]
\bar{u}\left(p_{2},s_{2}\right)\not\! q_{2}
\gamma_{5}u\left(p_{1},s_{1}\right)\nonumber \\
+c_{A}^{f}\tilde{E}_{f}^{-}\left(x,\xi,t\right)\left[\frac{1}{\left(p\cdot q_{2}\right)}
\left(p^{\mu}q_{2}^{\nu}+p^{\nu}q_{2}^{\mu}\right)-g^{\mu\nu}\right]
\frac{\left(q_{2}\cdot r\right)}{2M}
\bar{u}\left(p_{2},s_{2}\right)\gamma_{5}u\left(p_{1},s_{1}\right)\nonumber \\
+c_{A}^{f}H_{f}^{-}\left(x,\xi,t\right)\left[\frac{1}{\left(p\cdot q_{2}\right)}i
\epsilon^{\mu\nu\rho\eta}q_{2\rho}p_{\eta}\right]\bar{u}\left(p_{2},s_{2}\right)\not\! q_{2}u
\left(p_{1},s_{1}\right)\nonumber \\
+c_{A}^{f}E_{f}^{-}\left(x,\xi,t\right)\left[\frac{1}{\left(p\cdot q_{2}\right)}i
\epsilon^{\mu\nu\rho\eta}q_{2\rho}p_{\eta}\right]\bar{u}\left(p_{2},s_{2}\right)
\frac{\left(\not\! q_{2}\not\! r-\not\! r\not\! q_{2}\right)}{4M}u
\left(p_{1},s_{1}\right)\Bigg\rbrace.\nonumber \\
\label{eq:reducedWNVCA}
\end{eqnarray}

\subsection{\ \,  Weak Charged Amplitude}

Skipping an overall factor $-\left|e\right|g/\sqrt{2}$, the expansion of the time-ordered product of 
two currents in the weak charged sector reads
\begin{eqnarray}
iT\left\{ J_{EM}^{\mu}\left(z/2\right)J_{WC}^{\nu}\left(-z/2\right)\right\}  & = & -
\frac{z_{\rho}}{4\pi^{2}z^{4}}\sum_{f,f'}\left\{ Q_{f'}\bar{\psi}_{f'}\left(z/2\right)\gamma^{\mu}
\gamma^{\rho}\gamma^{\nu}\left(1-\gamma_{5}\right)\psi_{f}\left(-z/2\right)\right.\nonumber \\
&  & \left.-Q_{f}\bar{\psi}_{f}\left(-z/2\right)\gamma^{\nu}\left(1-\gamma_{5}\right)\gamma^{\rho}
\gamma^{\mu}\psi_{f'}\left(z/2\right)\right\} \nonumber \\
& = & \frac{z_{\rho}}{4\pi^{2}z^{4}}\sum_{f,f'}\Big\lbrace s^{\mu\rho\nu\eta}
\bigl[Q_{f}\bar{\psi}_{f}\left(-z/2\right)\gamma_{\eta}\psi_{f'}\left(z/2\right)\nonumber \\
&  & -\left(f\leftrightarrow f',z\rightarrow-z\right)\bigr]\nonumber \\
&  & +i\epsilon^{\mu\rho\nu\eta}
\bigl[Q_{f}\bar{\psi}_{f}\left(-z/2\right)\gamma_{\eta}\gamma_{5}\psi_{f'}\left(z/2\right)\nonumber \\
&  & +\left(f\leftrightarrow f',z\rightarrow-z\right)\bigr]\nonumber \\
&  & -s^{\mu\rho\nu\eta}
\bigl[Q_{f}\bar{\psi}_{f}\left(-z/2\right)\gamma_{\eta}\gamma_{5}\psi_{f'}\left(z/2\right)\nonumber \\
&  & -\left(f\leftrightarrow f',z\rightarrow-z\right)\bigr]\nonumber \\
&  & -i\epsilon^{\mu\rho\nu\eta}
\bigl[Q_{f}\bar{\psi}_{f}\left(-z/2\right)\gamma_{\eta}\psi_{f'}\left(z/2\right)\nonumber \\
&  & +\left(f\leftrightarrow f',z\rightarrow-z\right)\bigr]\Big\rbrace.
\label{eq:WCexpansion1}
\end{eqnarray}
In principle, one has to include the CKM matrix since 
for the weak charged couplings the quark mass eigenstates do not 
coincide with the weak eigenstates. However, we shall ignore all Cabibbo angles. 
Here the sum over quark flavors is subject to an extra condition,
$Q_{f}-Q_{f'}=1$ or $-1$, due the fact that the weak virtual boson
$W^{\pm}$ carries an electric charge $\pm1$ in units of $\left|e\right|$. 
For that reason, the initial and final nucleons are not the same particles anymore. 
Hence we are dealing either with the neutron-to-proton transition via the exchange of $W^{+}$, 
or with the proton-to-neutron transition via the exchange of $W^{-}$. One notices that the vector 
and axial vector string operators in the expansion (\ref{eq:WCexpansion1}) are not diagonal in quark 
flavor, i.e. they are accompanied by different quark flavors as well as by 
different electric charges. The corresponding contracted string operators, 
which appear when extracting the twist-2 part of Eq. (\ref{eq:WCexpansion1}), 
can be expressed as the linear combinations,
\begin{eqnarray}
\left[Q_{f}\bar{\psi}_{f}\left(-z/2\right)\not\! z\psi_{f'}
\left(z/2\right)\pm\left(f\leftrightarrow f',z
\rightarrow-z\right)\right] & = & Q_{\pm}\mathcal{O}^{ff'+}\left(z\right)+Q_{\mp}\mathcal{O}^{ff'-}
\left(z\right),\nonumber \\
\left[Q_{f}\bar{\psi}_{f}\left(-z/2\right)\not\! z\gamma_{5}\psi_{f'}\left(z/2\right)\pm
\left(f\leftrightarrow f',z\rightarrow-z\right)\right] & = & Q_{\pm}\mathcal{O}_{5}^{ff'+}
\left(z\right)+Q_{\mp}\mathcal{O}_{5}^{ff'-}\left(z\right),\nonumber \\
\label{eq:flavornoniagonalstringoperators}
\end{eqnarray}
of operators,
\begin{eqnarray}
\mathcal{O}^{ff'\pm}\left(z\right) & \equiv & \left[\bar{\psi}_{f}
\left(-z/2\right)\not\! z\psi_{f}\left(z/2
\right)\pm\left(f\leftrightarrow f',z\rightarrow-z\right)\right],\nonumber \\
\mathcal{O}_{5}^{ff'\pm}\left(z\right) & \equiv & \left[\bar{\psi}_{f}\left(-z/2\right)
\not\! z\gamma_{5}\psi_{f}\left(z/2\right)\pm\left(f\leftrightarrow f',z\rightarrow-z\right)\right],
\label{eq:contractedoperatorsWC}
\end{eqnarray}
with the coefficients $Q_{\pm}=\left(Q_{f}\pm Q_{f'}\right)/2$. The matrix elements of these newly 
introduced operators are parametrized in terms of the flavor nondiagonal OFPDs,
\begin{eqnarray}
\left\langle N'\left(p_{2},s_{2}\right)\right|\mathcal{O}^{ff'\pm}
\left(z\right)\left|N\left(p_{1},s_{1}
\right)\right\rangle_{z^2=0}  & = & \bar{u}\left(p_{2},s_{2}\right)\not\! zu\left(p_{1},s_{1}
\right)\nonumber \\
&  & \times\int_{-1}^{1}dx\; e^{ix\left(p\cdot z\right)}H_{ff'}^{\pm}\left(x,\xi,t\right)\nonumber \\
&  & +\bar{u}\left(p_{2},s_{2}\right)\frac{\left(\not\! z\not\! r-\not\! r\not\! z\right)}{4M}u
\left(p_{1},s_{1}\right)\nonumber \\
&  & \times\int_{-1}^{1}dx\; e^{ix\left(p\cdot z\right)}E_{ff'}^{\pm}\left(x,\xi,t\right),\nonumber \\
\left\langle N'\left(p_{2},s_{2}\right)\right|\mathcal{O}_{5}^{ff'\pm}\left(z\right)
\left|N\left(p_{1},s_{1}\right)\right\rangle_{z^2=0}  & = & \bar{u}
\left(p_{2},s_{2}\right)\not\! z\gamma_{5}u
\left(p_{1},s_{1}\right)\nonumber\\
&  & \times\int_{-1}^{1}dx\; e^{ix\left(p\cdot z\right)}
\tilde{H}_{ff'}^{\mp}\left(x,\xi,t\right)\nonumber \\
&  & -\bar{u}\left(p_{2},s_{2}\right)\frac{\left(r\cdot z\right)}{2M}
\gamma_{5}u\left(p_{1},s_{1}\right)\nonumber \\
&  & \times\int_{-1}^{1}dx\; e^{ix\left(p\cdot z\right)}\tilde{E}_{ff'}^{\mp}\left(x,\xi,t\right).
\label{eq:parametrizationWC}
\end{eqnarray}
Finally, with the help of Eqs. (\ref{eq:flavornoniagonalstringoperators})
and (\ref{eq:parametrizationWC}), the leading-twist reduced weak charged VCA can be easily obtained 
from the result for the weak neutral case with the proper replacements. We get
\begin{eqnarray}
\mathsf{\mathcal{T}}_{WCtwist-2}^{\mu\nu}=\frac{1}{4\left(p\cdot q\right)}\sum_{f,f'}\int_{-1}^{1}
\frac{dx}{\left(x-\xi+i0\right)}\nonumber \\
\times\Bigg\lbrace\left[Q_{+}H_{ff'}^{+}\left(x,\xi,t\right)+Q_{-}H_{ff'}^{-}
\left(x,\xi,t\right)\right]\nonumber \\
\times\left[\frac{1}{\left(p\cdot q_{2}\right)}\left(p^{\mu}q_{2}^{\nu}+p^{\nu}q_{2}^{\mu}
\right)-g^{\mu\nu}\right]\bar{u}\left(p_{2},s_{2}\right)
\not\! q_{2}u\left(p_{1},s_{1}\right)\nonumber \\
+\left[Q_{+}E_{ff'}^{+}\left(x,\xi,t\right)+Q_{-}E_{ff'}^{-}
\left(x,\xi,t\right)\right]\nonumber \\
\times\left[\frac{1}{\left(p\cdot q_{2}\right)}\left(p^{\mu}q_{2}^{\nu}+p^{\nu}q_{2}^{\mu}
\right)-g^{\mu\nu}
\right]\bar{u}\left(p_{2},s_{2}\right)
\frac{\left(\not\! q_{2}\not\! r-\not\! r\not\! q_{2}\right)}{4M}u
\left(p_{1},s_{1}\right)\nonumber \\
-\left[Q_{+}\tilde{H}_{ff'}^{+}\left(x,\xi,t\right)+Q_{-}\tilde{H}_{ff'}^{-}
\left(x,\xi,t\right)\right]\nonumber \\
\times\left[\frac{1}{\left(p\cdot q_{2}\right)}i\epsilon^{\mu\nu\rho\eta}q_{2\rho}p_{\eta}
\right]\bar{u}\left(p_{2},s_{2}\right)\not\! q_{2}\gamma_{5}u\left(p_{1},s_{1}\right)\nonumber \\
+\left[Q_{+}\tilde{E}_{ff'}^{+}\left(x,\xi,t\right)+Q_{-}\tilde{E}_{ff'}^{-}
\left(x,\xi,t\right)\right]\nonumber \\
\times\left[\frac{1}{\left(p\cdot q_{2}\right)}i\epsilon^{\mu\nu\rho\eta}q_{2\rho}p_{\eta}
\right]\frac{\left(q_{2}\cdot r\right)}{2M}\bar{u}\left(p_{2},s_{2}
\right)\gamma_{5}u\left(p_{1},s_{1}\right)\nonumber \\
-\left[Q_{-}\tilde{H}_{ff'}^{+}\left(x,\xi,t\right)+Q_{+}\tilde{H}_{ff'}^{-}
\left(x,\xi,t\right)\right]\nonumber \\
\times\left[\frac{1}{\left(p\cdot q_{2}\right)}\left(p^{\mu}q_{2}^{\nu}+p^{\nu}q_{2}^{\mu}
\right)-g^{\mu\nu}\right]\bar{u}\left(p_{2},s_{2}\right)\not\! q_{2}
\gamma_{5}u\left(p_{1},s_{1}\right)\nonumber \\
+\left[Q_{-}\tilde{E}_{ff'}^{+}\left(x,\xi,t\right)+Q_{+}\tilde{E}_{ff'}^{-}
\left(x,\xi,t\right)\right]\nonumber \\
\times\left[\frac{1}{\left(p\cdot q_{2}\right)}\left(p^{\mu}q_{2}^{\nu}+p^{\nu}q_{2}^{\mu}
\right)-g^{\mu\nu}\right]\frac{\left(q_{2}\cdot r\right)}{2M}\bar{u}\left(p_{2},s_{2}
\right)\gamma_{5}u\left(p_{1},s_{1}\right)\nonumber \\
+\left[Q_{-}H_{ff'}^{+}\left(x,\xi,t\right)+Q_{+}H_{ff'}^{-}\left(x,\xi,t\right)\right]\nonumber \\
\times\left[\frac{1}{\left(p\cdot q_{2}\right)}i\epsilon^{\mu\nu\rho\eta}q_{2\rho}p_{\eta}\right]
\bar{u}\left(p_{2},s_{2}\right)\not\! q_{2}u\left(p_{1},s_{1}\right)\nonumber \\
+\left[Q_{-}E_{ff'}^{+}\left(x,\xi,t\right)+Q_{+}E_{ff'}^{-}\left(x,\xi,t\right)\right]\nonumber \\
\times\left[\frac{1}{\left(p\cdot q_{2}\right)}i\epsilon^{\mu\nu\rho\eta}q_{2\rho}p_{\eta}
\right]\bar{u}\left(p_{2},s_{2}\right)
\frac{\left(\not\! q_{2}\not\! r-\not\! r\not\! q_{2}\right)}{4M}u
\left(p_{1},s_{1}\right)\Bigg\rbrace.
\label{eq:reducedWCVCA}
\end{eqnarray}

\section{\ \,  Weak DVCS Processes}

\begin{figure}
\centerline{\epsfxsize=6in\epsffile{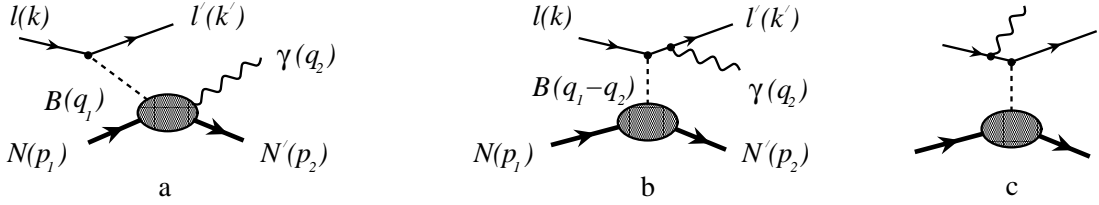}}
\caption[DVCS (a) and Bethe-Heitler (b and c) diagrams in the weak deeply virtual 
Compton scattering process.]
{DVCS (a) and Bethe-Heitler (b and c) diagrams in the weak deeply virtual 
Compton scattering process. The incoming leptons can be either neutrinos or electrons.}
\label{weakDVCSdiagrams}
\end{figure}

In the following section, we discuss three examples of the weak DVCS processes. Their kinematics has 
already been analyzed in detail in Section IV.3.  The only difference is that the incoming virtual 
photon $\gamma^{*}$ of the standard DVCS process is now being substituted by the virtual 
weak boson \emph{B}. We use the OFPD model from Section IV.4. Note that by neglecting the sea 
quark contribution, the plus OFPDs become equal to the minus ones, and hence we write 
$H_{f}^{+}=H_{f}^{-}\equiv H_{f}^{val}$, $\tilde{H}_{f}^{+}=\tilde{H}_{f}^{-}\equiv\tilde{H}_{f}^{val}$, 
$E_{f}^{+}=E_{f}^{-}\equiv E_{f}^{val}$ and 
$\tilde{E}_{f}^{+}=\tilde{E}_{f}^{-}\equiv\tilde{E}_{f}^{val}$. For the weak neutral current 
scattering process, we discuss neutrino-nucleon and electron-nucleon scattering. 
We present the calculation by taking an unpolarized proton as a target nucleon. 
In the case of an unpolarized neutron target, one can simply use the isospin (i.e. the charge) 
symmetry to express the neutron GPDs in terms of the proton ones. 
In the weak charged current interaction sector, we consider neutrino scattering off a neutron via 
the exchange of a $W^{+}$ boson producing a proton in the final state. 
All three examples are illustrated in one figure, see Fig. \ref{weakDVCSdiagrams}. 
As in Section IV.5, the relevant cross sections are plotted against the scattering angle 
$\theta_{B\gamma}$ between the directions of the incoming weak virtual boson and the outgoing 
real photon for one kinematical point with $Q_{1}^{2}=2.5\;\mathrm{GeV}^{2}$, 
$x_{B}=0.35$ and $\varphi=0$, however, for two different lepton beam energies. We take 
$\omega=5\;\mathrm{GeV}$ and $20\;\mathrm{GeV}$ rather than $\omega=5.75\;\mathrm{GeV}$ and 
$11\;\mathrm{GeV}$. Similarly to the standard DVCS process, the Bethe-Heitler cross sections are 
presented on a logarithmic scale as well as the plots with both contributions. In addition, 
for given kinematical regions, $0.28\leq x_{B}\leq0.3$ with $\omega=5\;\mathrm{GeV}$ and 
$0.26\leq x_{B}\leq0.28$ with $\omega=20\;\mathrm{GeV}$, we estimate the orders of magnitude for 
the total cross sections, coming purely from the DVCS diagrams and compare them with the Compton 
contribution to the standard electromagnetic DVCS process. They are listed in 
Table \ref{tablecomparisoncrosssection} at the end of this chapter.

\subsection{\ \,  Weak Neutral Neutrino-Proton Scattering}

\begin{figure}
\centerline{\epsfxsize=5in\epsffile{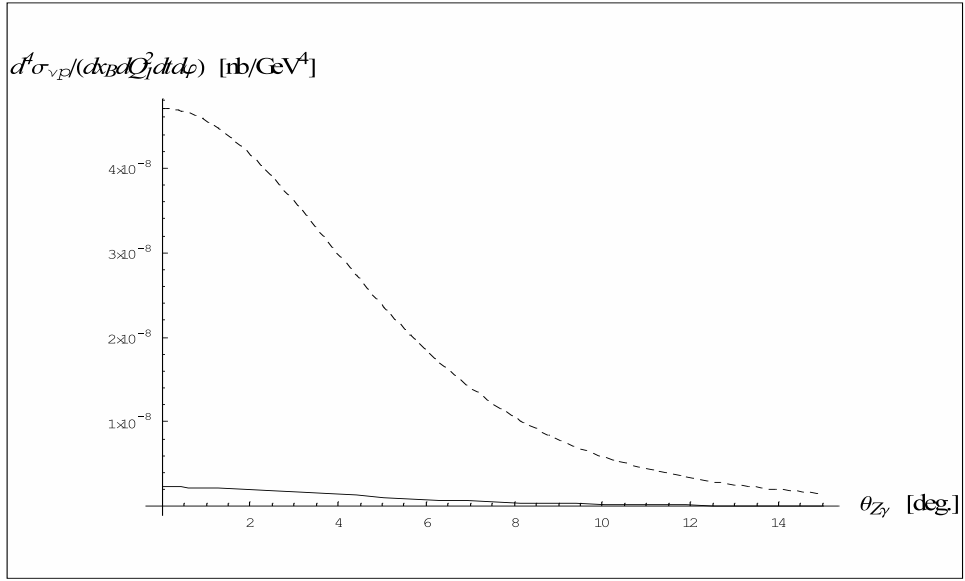}}
\caption{Weak neutral DVCS cross section $\sigma_{\nu p}$ plotted as a function of the 
angle $\theta_{B\gamma}$ between the incoming weak virtual boson and outgoing real photon in the target rest 
frame for $Q_{1}^{2}=2.5\;\mathrm{GeV}^{2}$ and $x_{B}=0.35$ with $\omega=5\;\mathrm{GeV}$ (solid line) 
and $\omega=20\;\mathrm{GeV}$ (dashed line) neutrino beam.}
\label{wndvscneutrinocrosssection}
\end{figure}

The photon cannot be emitted by the neutrino, and thus in neutrino scattering from a proton via the 
exchange of $Z^{0}$, we only measure the Compton part. The T-matrix of the process is given solely 
by the DVCS diagram, i.e. the diagram (a) in Fig. \ref{weakDVCSdiagrams},
\begin{eqnarray}
i\mathrm{T}_{\nu p} & = & \bar{u}\left(k'\right)\left(\frac{-ig}{\cos\theta_{W}}
\right)\gamma^{\lambda}\left(\frac{c_{V}^{\nu}-\gamma_{5}c_{A}^{\nu}}{2}\right)u
\left(k\right)\left[
\frac{-i\left(g_{\nu\lambda}-q_{1\nu}q_{1\lambda}/M_{Z^{0}}^{2}\right)}{q_{1}^{2}-M_{Z^{0}}^{2}}
\right]\nonumber \\
&  & \times\left(\frac{-\left|e\right|g}{\cos\theta_{W}}\right)\epsilon_{\mu}^{*}\left(q_{2}
\right)\left(-i\mathcal{T}_{WN}^{\mu\nu}\right),
\label{eq:Tmatrixneutrinoproton1}
\end{eqnarray}
with $\mathcal{T}_{WN}^{\mu\nu}$ computed in the handbag approximation
from the diagrams shown in Fig. \ref{handbagdiagramsweakDVCS}. 
The vector and axial vector couplings of the vertex $\nu\nu Z^{0}$
are $c_{V}^{\nu}=c_{A}^{\nu}=1/2$. Furthermore, since $M_{Z^{0}}^{2}\gg Q_{1}^{2}$
and $\cos\theta_{W}\equiv M_{W}/M_{Z^{0}}$ (recall that the Fermi
constant is $G_{F}\equiv g^{2}/4\sqrt{2}M_{W}^{2}$), the amplitude
turns into 
\begin{eqnarray}
\mathrm{T}_{\nu p} & = & \sqrt{2}\left|e\right|G_{F}\bar{u}\left(k'\right)\gamma_{\nu}
\left(1-\gamma_{5}\right)u(k)\epsilon_{\mu}^{*}\left(q_{2}\right)\mathcal{T}_{WN}^{\mu\nu}.
\label{eq:Tmatrixneutrinoproton2}
\end{eqnarray}
Its spin-averaged square then reads
\begin{eqnarray}
\overline{\left|\mathrm{T}_{\nu p}\right|^{2}} & = & 8
\pi\alpha G_{F}^{2}L_{\nu\beta}^{\left(\nu\right)}H_{WN}^{\nu\beta}.
\label{eq:Tmatrixsquaredneutrinoproton}
\end{eqnarray}
The neutrino tensor is simply
\begin{eqnarray}
L_{\nu\beta}^{\left(\nu\right)} & = & 8\left[k_{\nu}k'_{\beta}+k_{\beta}k'_{\nu}-g_{\nu\beta}
\left(k\cdot k'\right)+i\epsilon_{\nu\beta\sigma\tau}k^{\sigma}k'^{\tau}\right].
\label{eq:neutrinotensorweakDVCS}
\end{eqnarray}
On the other hand, the weak neutral hadron tensor has a rather complicated expression. 
Nevertheless, in the DVCS kinematics, it reduces to
\begin{eqnarray}
H_{WN}^{\nu\beta} & = & -\frac{1}{2}\mathcal{T}_{WN}^{\mu\nu}\left(\mathcal{T}_{\mu WN}^{\beta}
\right)^{*}\nonumber \\
& = & -\frac{1}{4}\left\{ \left[\left(1-\xi^{2}\right)\left(\left|\mathcal{H}_{WN}^{+}
\right|^{2}+\left|\mathcal{H}_{WN}^{-}\right|^{2}+\left|\tilde{\mathcal{H}}_{WN}^{+}
\right|^{2}+\left|\tilde{\mathcal{H}}_{WN}^{-}\right|^{2}\right)\right.\right.\nonumber \\
&  & -\left(\xi^{2}+\frac{t}{4M^{2}}\right)\left(\left|\mathcal{E}_{WN}^{+}\right|^{2}+
\left|\mathcal{E}_{WN}^{-}\right|^{2}\right)-\xi^{2}\frac{t}{4M^{2}}\left(\left|
\tilde{\mathcal{E}}_{WN}^{+}\right|^{2}+\left|
\tilde{\mathcal{E}}_{WN}^{-}\right|^{2}\right)\nonumber \\
&  & -2\xi^{2}\Re\left(\mathcal{H}_{WN}^{+*}\mathcal{E}_{WN}^{+}+\mathcal{H}_{WN}^{-*}
\mathcal{E}_{WN}^{-}+\tilde{\mathcal{H}}_{WN}^{+*}\tilde{\mathcal{E}}_{WN}^{+}+
\tilde{\mathcal{H}}_{WN}^{-*}\tilde{\mathcal{E}}_{WN}^{-}\right)\bigg]\nonumber \\
&  & \times\left[g^{\nu\beta}-\frac{1}{\left(p\cdot q_{2}\right)}
\left(p^{\nu}q_{2}^{\beta}+p^{\nu}q_{2}^{\beta}\right)+\frac{M^{2}}{\left(p\cdot q_{2}\right)^{2}}
\left(1-\frac{t}{4M^{2}}\right)q_{2}^{\nu}q_{2}^{\beta}\right]\nonumber \\
&  & +2\left[\left(1-\xi^{2}\right)\Re\left(\mathcal{H}_{WN}^{+*}\mathcal{H}_{WN}^{-}+
\tilde{\mathcal{H}}_{WN}^{+*}\tilde{\mathcal{H}}_{WN}^{-}\right)\right.\nonumber \\
&  & -\left(\xi^{2}+\frac{t}{4M^{2}}\right)\Re\left(\mathcal{E}_{WN}^{+*}\mathcal{E}_{WN}^{-}
\right)-\xi^{2}\frac{t}{4M^{2}}\Re\left(\tilde{\mathcal{E}}_{WN}^{+*}
\tilde{\mathcal{E}}_{WN}^{-}\right)\nonumber \\
&  & \left.-\xi^{2}\Re\left(\mathcal{H}_{WN}^{+*}\mathcal{E}_{WN}^{-}+\mathcal{E}_{WN}^{+*}
\mathcal{H}_{WN}^{-}+\tilde{\mathcal{H}}_{WN}^{+*}\tilde{\mathcal{E}}_{WN}^{-}+
\tilde{\mathcal{E}}_{WN}^{+*}\tilde{\mathcal{H}}_{WN}^{-}\right)\right]\nonumber \\
&  & \left.\times\frac{1}{\left(p\cdot q_{2}\right)}i
\epsilon^{\nu\beta\delta\lambda}p_{\delta}q_{2\lambda}\right\},
\label{eq:hadronictensorWN}
\end{eqnarray}
where the integrals of OFPDs are conveniently defined as
\begin{eqnarray}
\mathcal{H}_{WN}^{+\left(-\right)}\left(\xi,t\right) & \equiv & 
\sum_{f}Q_{f}c_{V\left(A\right)}^{f}\int_{-1}^{1}
\frac{dx}{\left(x-\xi+i0\right)}H_{f}^{+\left(-\right)}\left(x,\xi,t\right),\nonumber \\
\mathcal{E}_{WN}^{+\left(-\right)}\left(\xi,t\right) & \equiv & 
\sum_{f}Q_{f}c_{V\left(A\right)}^{f}\int_{-1}^{1}
\frac{dx}{\left(x-\xi+i0\right)}E_{f}^{+\left(-\right)}\left(x,\xi,t\right),\nonumber \\
\tilde{\mathcal{H}}_{WN}^{+\left(-\right)}\left(\xi,t\right) & 
\equiv & \sum_{f}Q_{f}c_{V\left(A\right)}^{f}\int_{-1}^{1}\frac{dx}{\left(x-\xi+i0\right)}
\tilde{H}_{f}^{+\left(-\right)}\left(x,\xi,t\right),\nonumber \\
\tilde{\mathcal{E}}_{WN}^{+\left(-\right)}\left(\xi,t\right) & 
\equiv & \sum_{f}Q_{f}c_{V\left(A\right)}^{f}\int_{-1}^{1}\frac{dx}{\left(x-\xi+i0\right)}
\tilde{E}_{f}^{+\left(-\right)}\left(x,\xi,t\right).
\label{eq:integralsofGPDsweakneutralcurrent}
\end{eqnarray}
After the substitution of Eq. (\ref{eq:Tmatrixsquaredneutrinoproton})
into Eq. (\ref{eq:generaldiffcrosswithQ1squared}), the unpolarized differential cross section 
for the weak neutral DVCS process on a proton target using the neutrino beam takes the following 
form
\begin{eqnarray}
\frac{d^{4}\sigma_{\nu p}}{dx_{B}dQ_{1}^{2}dtd\varphi} & = & \frac{1}{\left(2\pi\right)^{3}}
\frac{\alpha G_{F}^{2}}{16}
\frac{1+x_{B}\left(M/\omega\right)}{M^{2}\omega^{2}\left[2+\left(M/\omega\right)\right]x_{B}
\left[y+2x_{B}\left(M/\omega\right)\right]^{2}}L_{\nu\beta}^{\left(\nu\right)}H_{WN}^{\nu\beta}.\nonumber \\
\label{eq:diffcrosssectionWNneutrino}\
\end{eqnarray}
The cross section is illustrated in Fig. \ref{wndvscneutrinocrosssection}.

\subsection{\ \,  Weak Neutral Electron-Proton Scattering}

\begin{figure}
\centerline{\epsfxsize=5in\epsffile{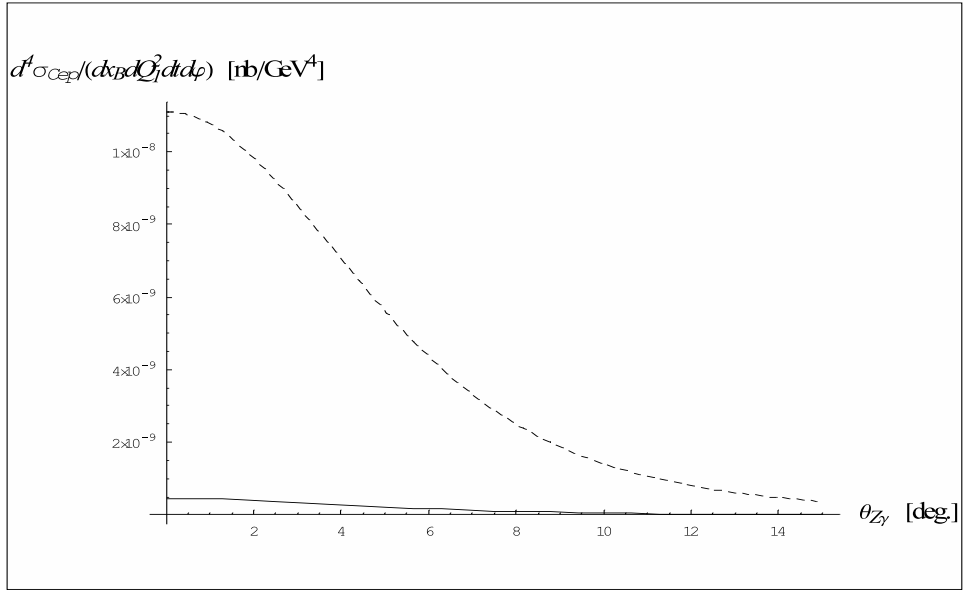}}
\caption{Compton cross section $\sigma_{Cep}$ for the weak neutral DVCS process plotted as a function of the 
angle $\theta_{B\gamma}$ between the incoming weak virtual boson and outgoing real photon in the target rest 
frame for $Q_{1}^{2}=2.5\;\mathrm{GeV}^{2}$ and $x_{B}=0.35$ with $\omega=5\;\mathrm{GeV}$ (solid line) 
and $\omega=20\;\mathrm{GeV}$ (dashed line) electron beam.}
\label{wndvsccomptonelectroncrosssection}
\end{figure}

In the electron scattering, one has to include both the Compton and the Bethe-Heitler 
contributions, $\mathrm{T}_{ep}=\mathrm{T}_{Cep}+\mathrm{T}_{BHep}$. The T-matrix 
for the Compton part can be easily obtained from Eq. (\ref{eq:Tmatrixneutrinoproton1}) 
by replacing the neutrinos with the electrons,
\begin{eqnarray}
\mathrm{T}_{Cep} & = & 2\sqrt{2}\left|e\right|G_{F}\bar{u}\left(k'\right)\gamma_{\nu}
\left(c_{V}^{e}-\gamma_{5}c_{A}^{e}\right)u(k)\epsilon_{\mu}^{*}\left(q_{2}
\right)\mathcal{T}_{WN}^{\mu\nu}.
\label{eq:Tmatrixelectronproton}
\end{eqnarray}
The couplings are now equal to $c_{V}^{e}=-1/2+2\sin^{2}\theta_{W}$ and $c_{A}^{e}=-1/2$. 
Note that the hadron tensor is the same as in the neutrino case. Due to averaging over the spin of 
incoming electrons, the electron tensor has an extra factor of $1/2$,
\begin{eqnarray}
L_{\nu\beta}^{\left(e\right)} & = & 2\left\{ \left[\left(c_{V}^{e}\right)^{2}+
\left(c_{A}^{e}\right)^{2}\right]\left[k_{\nu}k'_{\beta}+k_{\beta}k'_{\nu}-g_{\nu\beta}
\left(k\cdot k'\right)\right]-2c_{V}^{e}c_{A}^{e}i\epsilon_{\nu\beta\sigma\tau}k^{\sigma}k'^{\tau}
\right\}.\nonumber \\
\label{eq:leptonictensor1}
\end{eqnarray}
The unpolarized Compton differential cross section for the weak neutral DVCS process with a proton 
target and an electron beam reads
\begin{eqnarray}
\frac{d^{4}\sigma_{Cep}}{dx_{B}dQ_{1}^{2}dtd\varphi} & = & \frac{1}{\left(2\pi\right)^{3}}
\frac{\alpha G_{F}^{2}}{4}
\frac{1+x_{B}\left(M/\omega\right)}{M^{2}\omega^{2}\left[2+\left(M/\omega\right)\right]x_{B}
\left[y+2x_{B}\left(M/\omega\right)\right]^{2}}L_{\nu\beta}^{\left(e\right)}H_{WN}^{\nu\beta},
\nonumber \\
\label{eq:diffcrosssectionWNelectron}
\end{eqnarray}
and it is plotted in Fig. \ref{wndvsccomptonelectroncrosssection}. We notice that the cross 
section in the weak neutral current sector is larger with neutrinos than it is with electrons. 
This difference is a consequence of the structure difference in leptonic tensors given by 
Eqs.~(\ref{eq:neutrinotensorweakDVCS}) and (\ref{eq:leptonictensor1}).

\begin{figure}
\centerline{\epsfxsize=5in\epsffile{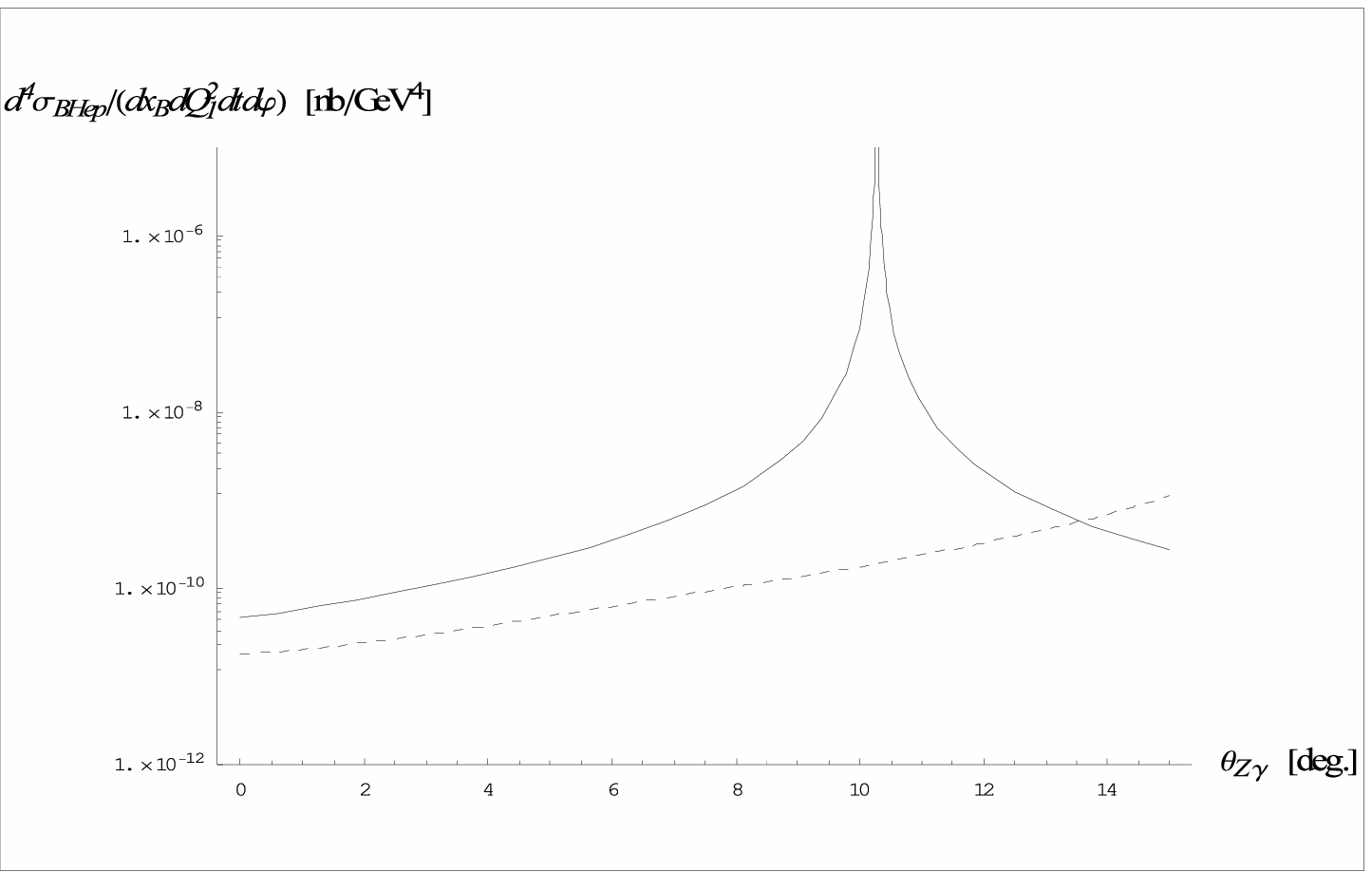}}
\caption{Bethe-Heitler cross section $\sigma_{BHep}$ for the weak neutral DVCS process plotted as a 
function of the angle $\theta_{B\gamma}$ between the incoming weak virtual boson and outgoing real photon in 
the target rest frame for $Q_{1}^{2}=2.5\;\mathrm{GeV}^{2}$ and $x_{B}=0.35$ with 
$\omega=5\;\mathrm{GeV}$ (solid line) and $\omega=20\;\mathrm{GeV}$ (dashed line) electron beam.}
\label{wndvscBH1electroncrosssection}
\end{figure}

\begin{figure}
\centerline{\epsfxsize=5in\epsffile{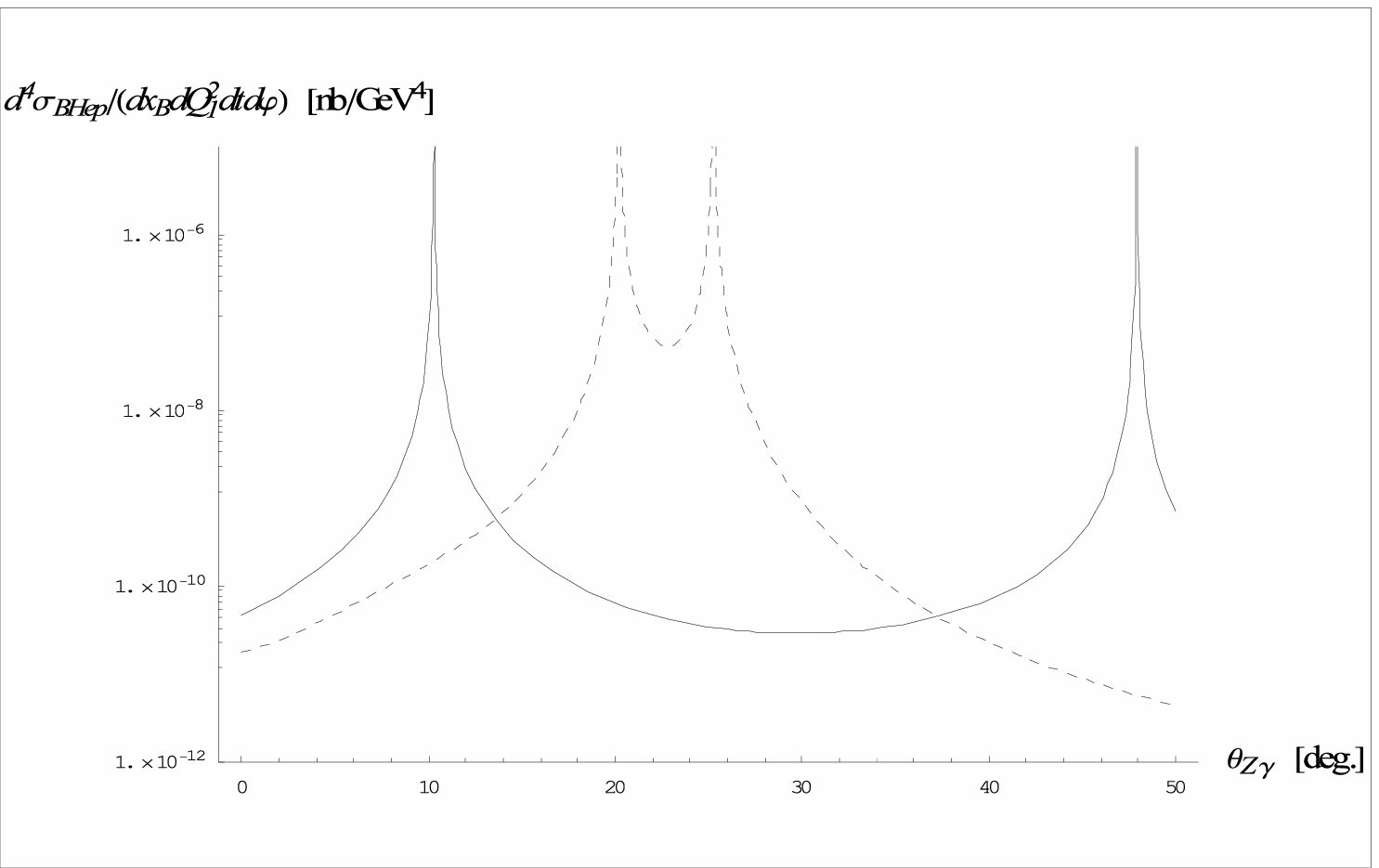}}
\caption{Bethe-Heitler cross section $\sigma_{BHep}$ for the weak neutral DVCS process plotted as a 
function of the angle $\theta_{B\gamma}$ between the incoming weak virtual boson and outgoing real photon in 
the target rest frame for $Q_{1}^{2}=2.5\;\mathrm{GeV}^{2}$ and $x_{B}=0.35$ with 
$\omega=5\;\mathrm{GeV}$ (solid line) and $\omega=20\;\mathrm{GeV}$ (dashed line) electron beam.}
\label{wndvscBH2electroncrosssection}
\end{figure}

\begin{figure}
\centerline{\epsfxsize=5in\epsffile{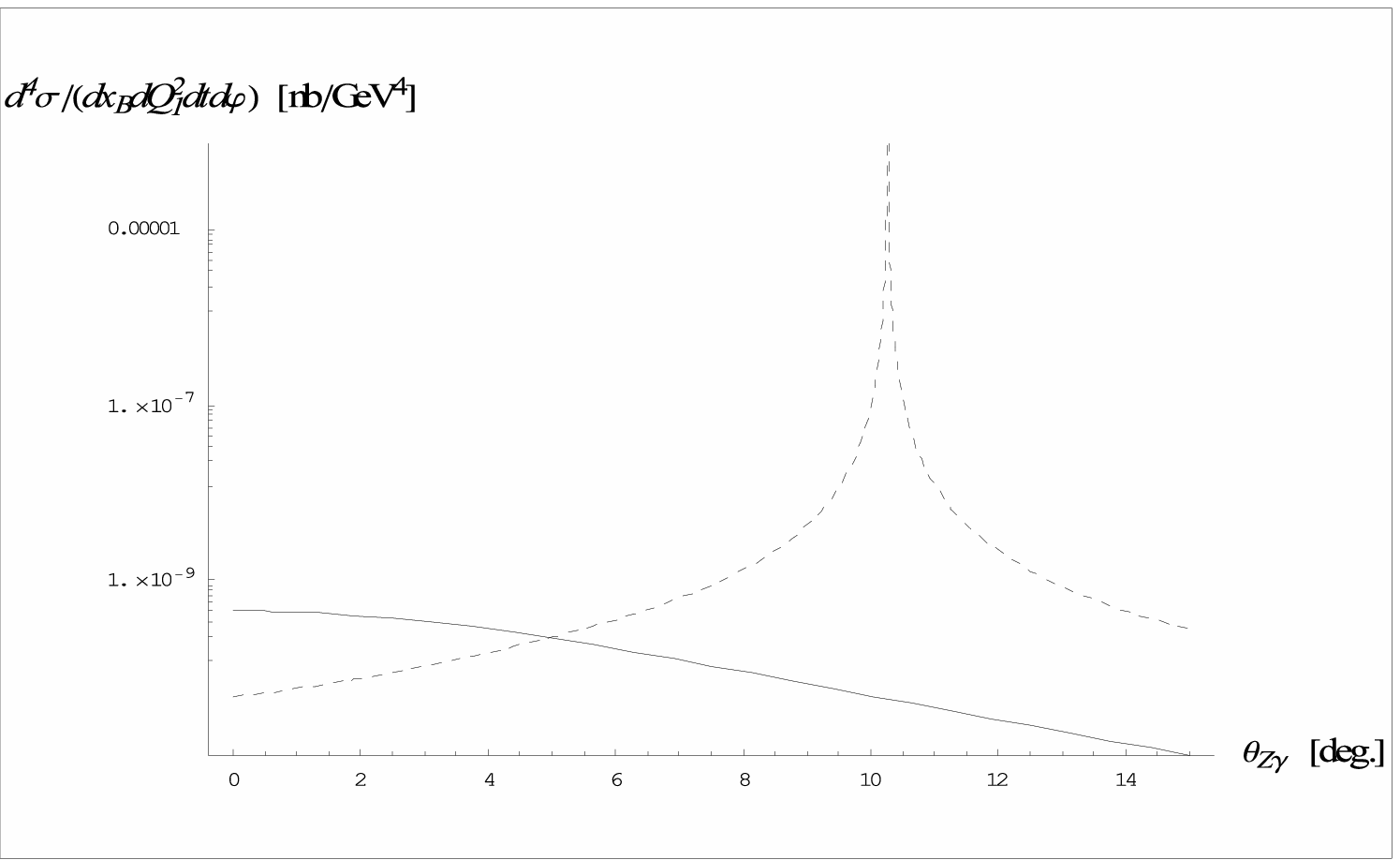}}
\caption{Compton cross section $\sigma_{Cep}$  (solid line) and Bethe-Heitler cross section $\sigma_{BHep}$ 
(dashed line) for the weak neutral DVCS process plotted as a function of the angle $\theta_{B\gamma}$ 
between the incoming weak virtual boson and outgoing real photon in the target rest frame for 
$Q_{1}^{2}=2.5\;\mathrm{GeV}^{2}$ and $x_{B}=0.35$ with $\omega=5\;\mathrm{GeV}$ electron beam.}
\label{wndvsctogetherelectroncrosssection}
\end{figure}

Next we investigate the Bethe-Heitler contribution to the process. Since incoming and scattered leptons 
are electrons, we have to include both Bethe-Heitler diagrams, see the diagrams (b) and (c) in 
Fig. \ref{weakDVCSdiagrams}. The Bethe-Heitler amplitude can be immediately written down with the help of 
the expression (\ref{eq:TmatrixBH1}) along with the following modifications:
\begin{itemize}
\item The vertex $ee\gamma$ is replaced by the vertex 
$eeZ^{0}$, $i\left|e\right|\gamma^{\nu}\rightarrow-ig\gamma^{\nu}\left(c_{V}^{e}-
\gamma_{5}c_{A}^{e}\right)/2\cos\theta_{W}$. 
\item The photon propagator is replaced by the weak boson propagator, 
$-ig_{\nu\lambda}/\left(q_{1}-q_{2}\right)^{2}\rightarrow ig_{\nu\lambda}/M_{Z^{0}}^{2}$. 
\item The coupling $\gamma p$ is replaced by the coupling $Z^{0}p$, $-i
\left|e\right|\rightarrow-ig/\cos\theta_{W}$.
\item The proton electromagnetic transition current is replaced by the proton
weak neutral transition current, 
$J_{EM}^{\lambda}\left(0\right)\rightarrow J_{NC}^{\lambda}\left(0\right)$.
\end{itemize}
Hence
\begin{eqnarray}
\mathrm{T}_{BHep} & = & 2\sqrt{2}\left|e\right|G_{F}\epsilon_{\mu}^{*}
\left(q_{2}\right)\bar{u}\left(k'\right)\left[
\frac{\gamma^{\mu}\left(\not\! k'+\not\! q_{2}\right)\gamma^{\nu}\left(c_{V}^{e}-
\gamma_{5}c_{A}^{e}\right)}{\left(k'+q_{2}\right)^{2}}\right.\nonumber \\
&  & \left.+\frac{\gamma^{\nu}\left(c_{V}^{e}-\gamma_{5}c_{A}^{e}
\right)\left(\not\! k-\not\! q_{2}\right)\gamma^{\mu}}{\left(k-q_{2}\right)^{2}}
\right]u\left(k\right)\left\langle p\left(p_{2},s_{2}\right)\right|J_{\nu}^{NC}
\left(0\right)\left|p\left(p_{1},s_{1}\right)\right\rangle,\nonumber \\
\label{eq:TmatrixwndvcsBHelectron}
\end{eqnarray}
where the transition current,
\begin{eqnarray}
J_{\nu}^{NC}\left(0\right) & = & \frac{1}{2}\left[V_{\nu}^{NC}\left(0\right)-A_{\nu}^{NC}
\left(0\right)\right]\nonumber \\
& = & \frac{1}{2}\sum_{f}\left[c_{V}^{f}\bar{\psi}_{f}\left(0\right)\gamma_{\nu}
\psi_{f}\left(0\right)-c_{A}^{f}\bar{\psi}_{f}\left(0\right)\gamma_{\nu}\gamma_{5}\psi_{f}
\left(0\right)\right],
\label{eq:hadronictransitioncurrent}
\end{eqnarray}
has both the vector $V_{\nu}^{NC}$ and axial vector $A_{\nu}^{NC}$
contributions. Their proton matrix elements are parametrized in terms
of the four form factors \cite{Thomas:2001kw}, 
\begin{eqnarray}
\left\langle p\left(p_{2},s_{2}\right)\right|V_{\nu}^{NC}\left(0\right)\left|p\left(p_{1},s_{1}
\right)\right\rangle  & = & \bar{u}\left(p_{2},s_{2}\right)\left[F_{1}^{NC}\left(t\right)
\gamma_{\nu}-F_{2}^{NC}\left(t\right)\frac{i\sigma_{\nu\lambda}r^{\lambda}}{2M}
\right]u\left(p_{1},s_{1}\right),\nonumber \\
\left\langle p\left(p_{2},s_{2}\right)\right|A_{\nu}^{NC}\left(0\right)
\left|p\left(p_{1},s_{1}\right)\right\rangle  & = & \bar{u}\left(p_{2},s_{2}\right)
\left[G_{A}^{NC}\left(t\right)\gamma_{\nu}\gamma_{5}-G_{P}^{NC}\left(t\right)
\frac{\gamma_{5}r_{\nu}}{2M}\right]u\left(p_{1},s_{1}\right).\nonumber \\
\label{eq:transitioncurrentparametrization}
\end{eqnarray}
Each of the vector form factors can be further expressed as a linear
combination of flavor triplet, octet and singlet form factors (they
correspond to the proton matrix elements of the appropriate linear
combinations of quark vector currents, $\left(\bar{u}\gamma_{\nu}u-\bar{d}\gamma_{\nu}d\right)/2$,
$\left(\bar{u}\gamma_{\nu}u+\bar{d}\gamma_{\nu}d-2\bar{s}\gamma_{\nu}s\right)$
and $\left(\bar{u}\gamma_{\nu}u+\bar{d}\gamma_{\nu}d+\bar{s}\gamma_{\nu}s\right)$,
respectively),
\begin{eqnarray}
F_{1\left(2\right)}^{NC}\left(t\right) & = & \left(1-2\sin^{2}\theta_{W}\right)
\left[F_{1\left(2\right)}^{3}\left(t\right)+\frac{1}{6}F_{1\left(2\right)}^{8}\left(t\right)
\right]-\frac{1}{6}F_{1\left(2\right)}^{0}\left(t\right).
\label{eq:F12formfactors}
\end{eqnarray}
Moreover, the axial and pseudoscalar form factors are expressed as the difference between 
the isovector and strangeness form factors (they correspond to the proton matrix elements 
of current combinations, 
$\left(\bar{u}\gamma_{\nu}\gamma_{5}u-\bar{d}\gamma_{\nu}\gamma_{5}d\right)/2$ and 
$\left(\bar{s}\gamma_{\nu}\gamma_{5}s\right)$, respectively), 
\begin{eqnarray}
G_{A\left(P\right)}^{NC}\left(t\right) & = & G_{A\left(P\right)}^{3}\left(t\right)-
\frac{1}{2}G_{A\left(P\right)}^{s}\left(t\right).
\label{eq:GAGPformfactors}
\end{eqnarray}
If the sea quarks are neglected, they all simplify to
\begin{eqnarray}
F_{1\left(2\right)}^{3} & = & \frac{1}{2}
\left[F_{1\left(2\right)u}-F_{1\left(2\right)d}\right],\nonumber \\
F_{1\left(2\right)}^{8}=F_{1\left(2\right)}^{0} & = & 
\left[F_{1\left(2\right)u}+F_{1\left(2\right)d}\right]
\label{eq:tripletoctetsinglet}
\end{eqnarray}
and
\begin{eqnarray}
G_{A}^{NC}\left(t\right) & = & \frac{g_{A}\left(t=0\right)}{2}\left(1-
\frac{t}{m_{A}^{2}}\right)^{-2},\nonumber \\
G_{P}^{NC}\left(t\right) & = & \frac{G_{A}^{NC}\left(t\right)}{2}
\frac{4M^{2}}{m_{\pi}^{2}-t}.
\label{eq:axialandpseudoscalar}
\end{eqnarray}
For the unpolarized Bethe-Heitler differential cross section we have  
\begin{eqnarray}
\frac{d^{4}\sigma_{BHep}}{dx_{B}dQ_{1}^{2}dtd\varphi} & = & \frac{1}{\left(2\pi\right)^{3}}
\frac{\alpha G_{F}^{2}}{4}
\frac{1+x_{B}\left(M/\omega\right)}{M^{2}\omega^{2}\left[2+\left(M/\omega\right)
\right]x_{B}\left[y+2x_{B}\left(M/\omega\right)
\right]^{2}}L_{BH}^{\nu\beta}H_{\nu\beta}^{BH},\nonumber \\
\label{eq:diffcrosssectionWNelectronBH}
\end{eqnarray}
where the electron tensor is
\begin{eqnarray}
L_{BH}^{\nu\beta} & = & -\frac{1}{2}\mathrm{Tr}\left\{ \not\! k'\left[
\frac{\left(\gamma^{\mu}\not\! q_{2}+2k'^{\mu}\right)\gamma^{\nu}\left(c_{V}^{e}-
\gamma_{5}c_{A}^{e}\right)}{2\left(k'\cdot q_{2}\right)}+\frac{\gamma^{\nu}\left(c_{V}^{e}-
\gamma_{5}c_{A}^{e}\right)\left(\not\! q_{2}\gamma^{\mu}-2k^{\mu}\right)}{2\left(k\cdot q_{2}
\right)}\right]\right.\nonumber \\
&  & \left.\times\not\! k\left[\frac{\gamma^{\beta}\left(c_{V}^{e}-\gamma_{5}c_{A}^{e}
\right)\left(\not\! q_{2}\gamma_{\mu}+2k'_{\mu}\right)}{2\left(k'\cdot q_{2}\right)}+
\frac{\left(\gamma_{\mu}\not\! q_{2}-2k_{\nu}\right)\gamma^{\beta}\left(c_{V}^{e}-
\gamma_{5}c_{A}^{e}\right)}{2\left(k\cdot q_{2}\right)}\right]\right\}, \nonumber \\
\label{eq:BHleptonicWNelectron}
\end{eqnarray}
and the hadron tensor is given by
\begin{eqnarray}
H_{\nu\beta}^{BH} & = & \frac{1}{2}\sum_{s_{1},s_{2}}\left\langle p\left(p_{2},s_{2}\right)
\right|J_{\nu}^{NC}\left(0\right)\left|p\left(p_{1},s_{1}\right)\right\rangle \left
\langle p\left(p_{2},s_{2}\right)\right|J_{\beta}^{NC}\left(0\right)\left|p\left(p_{1},s_{1}
\right)\right\rangle ^{*}.\nonumber \\
\label{eq:BHhadronicWNelectron}
\end{eqnarray}
The angular dependence of the cross section is shown in Figs. \ref{wndvscBH1electroncrosssection} 
and \ref{wndvscBH2electroncrosssection}. The two poles at the locations $\phi$ and $\phi'$ from 
Table \ref{tableanglesweak} are clearly seen. In Fig. \ref{wndvsctogetherelectroncrosssection}, 
both contributions are plotted together. Contrary what one might expect from the results on the 
standard DVCS, the Compton cross section in the weak neutral interaction sector is larger, 
or at least comparable, with the corresponding Bethe-Heitler cross section for the scattering 
angles $\theta_{Z\gamma}\le8^{0}$, which gives the region 
$0.15\;\mathrm{GeV}^{2}\leq-t\leq0.572\;\mathrm{GeV}^{2}$ in the invariant momentum transfer, 
see Fig. \ref{t}, and hence the ratio $-t/Q_{1}^2\le0.23$.
\begin{table}
\caption{Polar angles $\phi$ and $\phi'$ of the incoming and scattered leptons, respectively, in the target 
rest frame for $Q_{1}^{2}=2.5\;\mathrm{GeV}^{2}$ and $x_{B}=0.35$ with two different lepton beam 
energies $\omega$.}
\label{tableanglesweak}
\vspace{12pt}
\begin{center}
\begin{tabular}{lcc}\hline\hline\\
$\omega\left[\mathrm{GeV}\right]$ & 
$\phi\left[\mathrm{deg}.\right]$ & 
$\phi'\left[\mathrm{deg}.\right]$ \\\\
\hline
 & & \\
5 & 10.3 & 47.9  \\
 & & \\
20 & 20.2 & 25.3  \\\\ \hline\hline
\end{tabular}
\end{center}
\end{table}

\subsection{\ \,  Weak Charged Neutrino-Neutron Scattering}

\begin{figure}
\centerline{\epsfxsize=5in\epsffile{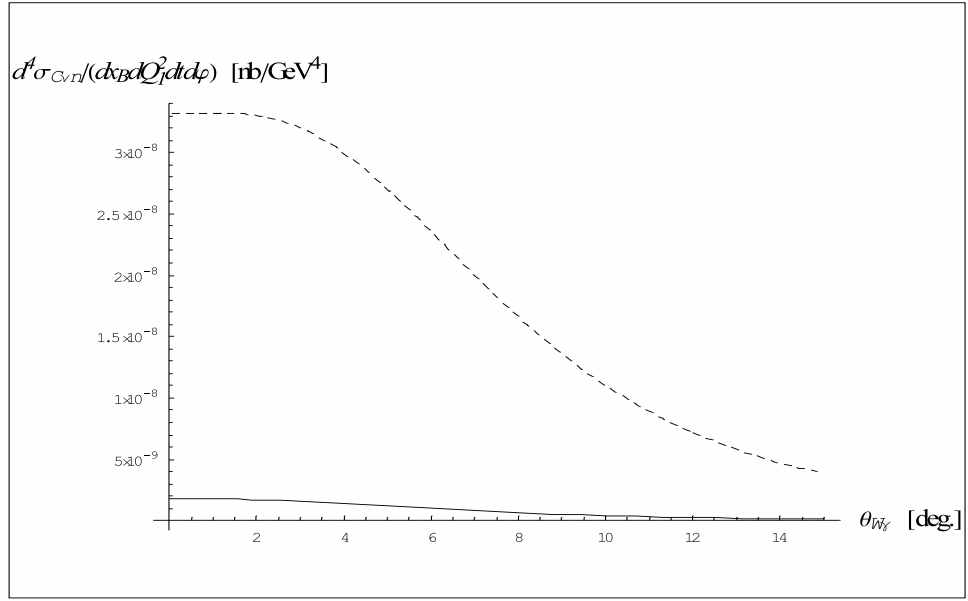}}
\caption{Compton cross section $\sigma_{C\nu n}$ for the weak charged DVCS process plotted as a function of 
the angle $\theta_{B\gamma}$ between the incoming weak virtual boson and outgoing real photon in the target 
rest frame for $Q_{1}^{2}=2.5\;\mathrm{GeV}^{2}$ and $x_{B}=0.35$ with $\omega=5\;\mathrm{GeV}$ (solid line) 
and $\omega=20\;\mathrm{GeV}$ (dashed line) neutrino beam.}
\label{comptoncharged}
\end{figure}

The invariant matrix element for the Compton contribution is
\begin{eqnarray}
\mathrm{T}_{C\nu n} & = & \sqrt{2}\left|e\right|G_{F}\bar{u}\left(k'\right)\gamma_{\mu}
\left(1-\gamma_{5}\right)u\left(k\right)\epsilon_{\mu}^{*}\left(q_{2}\right)
\mathcal{T}_{WC}^{\mu\nu},
\label{eq:TmatrixCWC}
\end{eqnarray}
where the amplitude $\mathcal{T}_{WC}^{\mu\nu}$ is, in general, given by Eq. (\ref{eq:reducedWCVCA}). 
However, in our simple model the quark flavors \emph{f} and \emph{f'} in Eqs. (\ref{eq:reducedWCVCA}) 
become \emph{d} and \emph{u}, respectively, and the coefficients are therefore equal to $Q_{+}=1/6$ 
and $Q_{-}=-1/2$. In the following, we relate the flavor nondiagonal GPDs to the flavor diagonal ones. 
This can be achieved by using the isospin symmetry. Through the isospin symmetry, 
the nucleon matrix elements, $\left\langle p\left(p_{2},s_{2}\right)\right|\mathcal{O}^{du\pm}
\left(z\right)\left|n\left(p_{1},s_{1}\right)\right\rangle$ and 
$\left\langle p\left(p_{2},s_{2}\right)\right|\mathcal{O}_{5}^{du\pm}\left(z\right)
\left|n\left(p_{1},s_{1}\right)\right\rangle$ given by Eq. (\ref{eq:contractedoperatorsWC}) 
that are nondiagonal in quark flavor, can be expressed in terms of the flavor diagonal nucleon 
matrix elements \cite{Mankiewicz:1997aa}. Namely, 
\begin{eqnarray}
\left\langle p\left(p_{2},s_{2}\right)\right|\mathcal{O}^{du\pm}\left(z\right)\left|n
\left(p_{1},s_{1}\right)\right\rangle  & = & \left\langle p\left(p_{2},s_{2}\right)
\right|\mathcal{O}^{u\pm}\left(z\right)\left|p\left(p_{1},s_{1}\right)\right\rangle \nonumber \\
&  & -\left\langle p\left(p_{2},s_{2}\right)\right|\mathcal{O}^{d\pm}
\left(z\right)\left|p\left(p_{1},s_{1}\right)\right\rangle,
\label{eq:isospinsymmetry}
\end{eqnarray}
and similarly for $\left\langle p\left(p_{2},s_{2}\right)\right|\mathcal{O}_{5}^{du\pm}
\left(z\right)\left|n\left(p_{1},s_{1}\right)\right\rangle$. Accordingly, the reduced weak charged VCA 
assumes the form
\begin{eqnarray}
\mathcal{T}_{WC}^{\mu\nu} & = & \frac{1}{4\left(p\cdot q\right)}\left\{ \left[
\frac{1}{\left(p\cdot q_{2}\right)}\left(p^{\mu}q_{2}^{\nu}+p^{\nu}q_{2}^{\mu}
\right)-g^{\mu\nu}\right]\right.\nonumber \\
&  & \times\left[\mathcal{H}_{WC}^{+}\bar{u}\left(p_{2},s_{2}\right)\not\! q_{2}u\left(p_{1},s_{1}
\right)+\mathcal{E}_{WC}^{+}\bar{u}\left(p_{2},s_{2}\right)
\frac{\left(\not\! q_{2}\not\! r-\not\! r\not\! q_{2}\right)}{4M}u\left(p_{1},s_{1}
\right)\right.\nonumber \\
&  & \left.-\tilde{\mathcal{H}}_{WC}^{-}\bar{u}\left(p_{2},s_{2}\right)\not\! q_{2}
\gamma_{5}u\left(p_{1},s_{1}\right)+\tilde{\mathcal{E}}_{WC}^{-}
\frac{\left(q_{2}\cdot r\right)}{2M}\bar{u}\left(p_{2},s_{2}\right)\gamma_{5}u\left(p_{1},s_{1}
\right)\right]\nonumber \\
&  & -\left[\frac{1}{\left(p\cdot q_{2}\right)}i\epsilon^{\mu\nu\rho\eta}q_{2\rho}p_{\eta}
\right]\nonumber \\
&  & \times\left[\tilde{\mathcal{H}}_{WC}^{+}\bar{u}\left(p_{2},s_{2}\right)\not\! q_{2}
\gamma_{5}u\left(p_{1},s_{1}\right)-\tilde{\mathcal{E}}_{WC}^{+}
\frac{\left(q_{2}\cdot r\right)}{2M}\bar{u}\left(p_{2},s_{2}\right)\gamma_{5}u\left(p_{1},s_{1}
\right)\right.\nonumber \\
&  & \left.\left.-\mathcal{H}_{WC}^{\mathit{-}}\bar{u}\left(p_{2},s_{2}
\right)\not\! q_{2}u\left(p_{1},s_{1}\right)-\mathcal{E}_{WC}^{\mathit{-}}
\bar{u}\left(p_{2},s_{2}\right)\frac{\left(\not\! q_{2}\not\! r-\not\! r\not\! q_{2}\right)}{4M}u
\left(p_{1},s_{1}\right)\right]\right\}, \nonumber \\
\label{eq:weakchargedamplitudemodel}
\end{eqnarray}
with the convolution integrals
\begin{eqnarray}
\mathcal{H}_{WC}^{+\left(-\right)}\left(\xi,t\right) & \equiv & \int_{-1}^{1}
\frac{dx}{\left(x-\xi+i0\right)}\left[Q_{+\left(-\right)}\left(H_{u}^{+}-H_{d}^{+}
\right)+Q_{-\left(+\right)}\left(H_{u}^{-}-H_{d}^{-}\right)\right],\nonumber \\
\mathcal{E}_{WC}^{+\left(-\right)}\left(\xi,t\right) & \equiv & \int_{-1}^{1}
\frac{dx}{\left(x-\xi+i0\right)}\left[Q_{+\left(-\right)}\left(E_{u}^{+}-E_{d}^{+}
\right)+Q_{-\left(+\right)}\left(E_{u}^{-}-E_{d}^{-}\right)\right],\nonumber \\
\tilde{\mathcal{H}}_{WC}^{+\left(-\right)}\left(\xi,t\right) & \equiv & \int_{-1}^{1}
\frac{dx}{\left(x-\xi+i0\right)}\left[Q_{+\left(-\right)}\left(\tilde{H}_{u}^{+}-
\tilde{H}_{d}^{+}\right)+Q_{-\left(+\right)}\left(\tilde{H}_{u}^{-}-\tilde{H}_{d}^{-}
\right)\right],\nonumber \\
\tilde{\mathcal{E}}_{WC}^{+\left(-\right)}\left(\xi,t\right) & \equiv & \int_{-1}^{1}
\frac{dx}{\left(x-\xi+i0\right)}\left[Q_{+\left(-\right)}\left(\tilde{E}_{u}^{+}-
\tilde{E}_{d}^{+}\right)+Q_{-\left(+\right)}\left(\tilde{E}_{u}^{-}-\tilde{E}_{d}^{-}
\right)\right].\nonumber \\
\label{eq:integralsofGPDsweakchargedcurrentmodel}
\end{eqnarray}
The weak charged hadron tensor,
\begin{eqnarray}
H_{WC}^{\nu\beta} & = & -\frac{1}{2}\mathcal{T}_{WC}^{\mu\nu}\left(\mathcal{T}_{\mu WC}^{\beta}
\right)^{*}\nonumber \\
& = & -\frac{1}{4}\left\{ \left[\left(1-\xi^{2}\right)\left(\left|\mathcal{H}_{WC}^{+}
\right|^{2}+\left|\mathcal{H}_{WC}^{-}\right|^{2}+\left|\tilde{\mathcal{H}}_{WC}^{+}
\right|^{2}+\left|\tilde{\mathcal{H}}_{WC}^{-}\right|^{2}\right)\right.\right.\nonumber \\
&  & -\left(\xi^{2}+\frac{t}{4M^{2}}\right)\left(\left|\mathcal{E}_{WC}^{+}\right|^{2}+
\left|\mathcal{E}_{WC}^{-}\right|^{2}\right)-\xi^{2}\frac{t}{4M^{2}}
\left(\left|\tilde{\mathcal{E}}_{WC}^{+}\right|^{2}+\left|\tilde{\mathcal{E}}_{WC}^{-}
\right|^{2}\right)\nonumber \\
&  & -2\xi^{2}\Re\left(\mathcal{H}_{WC}^{+*}\mathcal{E}_{WC}^{+}+\mathcal{H}_{WC}^{-*}
\mathcal{E}_{WC}^{-}+\tilde{\mathcal{H}}_{WC}^{+*}\tilde{\mathcal{E}}_{WC}^{+}+
\tilde{\mathcal{H}}_{WC}^{-*}\tilde{\mathcal{E}}_{WC}^{-}\right)\bigg]\nonumber \\
&  & \times\left[g^{\nu\beta}-\frac{1}{\left(p\cdot q_{2}\right)}
\left(p^{\nu}q_{2}^{\beta}+p^{\nu}q_{2}^{\beta}\right)+\frac{M^{2}}{\left(p\cdot q_{2}\right)^{2}}
\left(1-\frac{t}{4M^{2}}\right)q_{2}^{\nu}q_{2}^{\beta}\right]\nonumber \\
&  & +2\left[\left(1-\xi^{2}\right)\Re\left(\mathcal{H}_{WC}^{+*}\mathcal{H}_{WC}^{-}+
\tilde{\mathcal{H}}_{WC}^{+*}\tilde{\mathcal{H}}_{WC}^{-}\right)\right.\nonumber \\
&  & -\left(\xi^{2}+\frac{t}{4M^{2}}\right)\Re\left(\mathcal{E}_{WC}^{+*}
\mathcal{E}_{WC}^{-}\right)-\xi^{2}\frac{t}{4M^{2}}
\Re\left(\tilde{\mathcal{E}}_{WC}^{+*}\tilde{\mathcal{E}}_{WC}^{-}\right)\nonumber \\
&  & \left.-\xi^{2}\Re\left(\mathcal{H}_{WC}^{+*}\mathcal{E}_{WC}^{-}+
\mathcal{E}_{WC}^{+*}\mathcal{H}_{WC}^{-}+\tilde{\mathcal{H}}_{WC}^{+*}
\tilde{\mathcal{E}}_{WC}^{-}+\tilde{\mathcal{E}}_{WC}^{+*}\tilde{\mathcal{H}}_{WC}^{-}
\right)\right]\nonumber \\
&  & \left.\times\frac{1}{\left(p\cdot q_{2}\right)}i
\epsilon^{\nu\beta\delta\lambda}p_{\delta}q_{2\lambda}\right\},
\label{eq:hadronictensorWC}
\end{eqnarray}
has identical structure as the expression (\ref{eq:hadronictensorWN}) 
for the weak neutral hadron tensor. Moreover, by neglecting the mass of the outgoing muon, 
one can substitute the so-called muon tensor $L_{\nu\beta}^{\left(\mu\right)}$ by
the neutrino tensor $L_{\nu\beta}^{\left(\nu\right)}$, see Eq. (\ref{eq:neutrinotensorweakDVCS}).
The unpolarized Compton differential cross section for the weak charged
DVCS process on a neutron target using the neutrino beam is
\begin{eqnarray}
\frac{d^{4}\sigma_{C\nu n}}{dx_{B}dQ_{1}^{2}dtd\varphi} & = & \frac{1}{\left(2\pi\right)^{3}}
\frac{\alpha G_{F}^{2}}{16}\frac{1+x_{B}\left(M/\omega\right)}{M^{2}\omega^{2}
\left[2+\left(M/\omega\right)\right]x_{B}\left[y+2x_{B}\left(M/\omega\right)
\right]^{2}}L_{\nu\beta}^{\left(\mu\right)}H_{WC}^{\nu\beta}.\nonumber \\
\label{eq:diffcrosssectioncomptoncc}
\end{eqnarray}
The cross section is presented in Fig. \ref{comptoncharged}.

\begin{figure}
\centerline{\epsfxsize=5in\epsffile{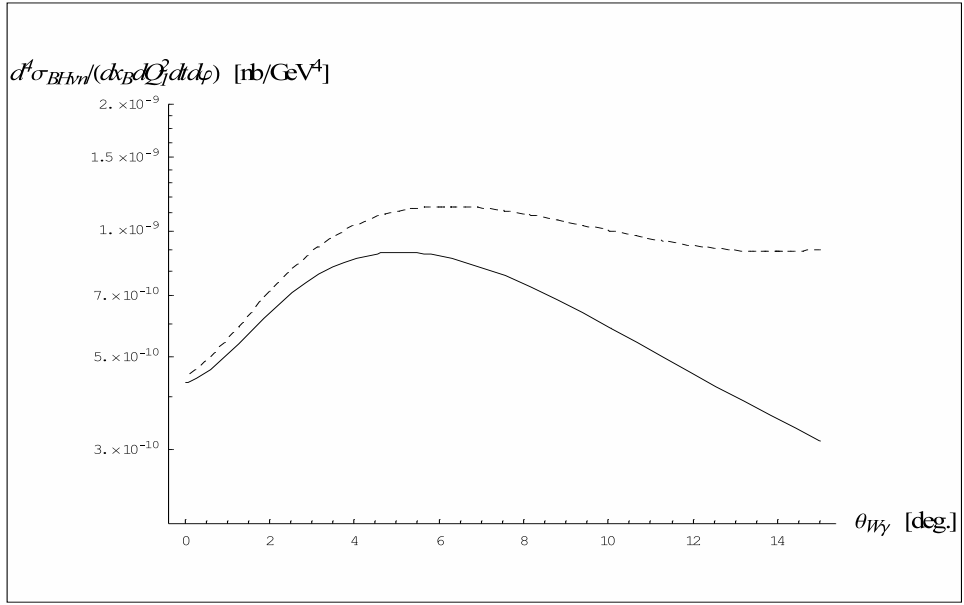}}
\caption{Bethe-Heitler cross section $\sigma_{BH\nu n}$ for the weak charged DVCS process plotted as a 
function of the angle $\theta_{B\gamma}$ between the incoming weak virtual boson and outgoing real photon 
in the target rest frame for $Q_{1}^{2}=2.5\;\mathrm{GeV}^{2}$ and $x_{B}=0.35$ with $\omega=5\;\mathrm{GeV}$ 
(solid line) and $\omega=20\;\mathrm{GeV}$ (dashed line) neutrino beam.}
\label{wcdvscBH1}
\end{figure}

\begin{figure}
\centerline{\epsfxsize=5in\epsffile{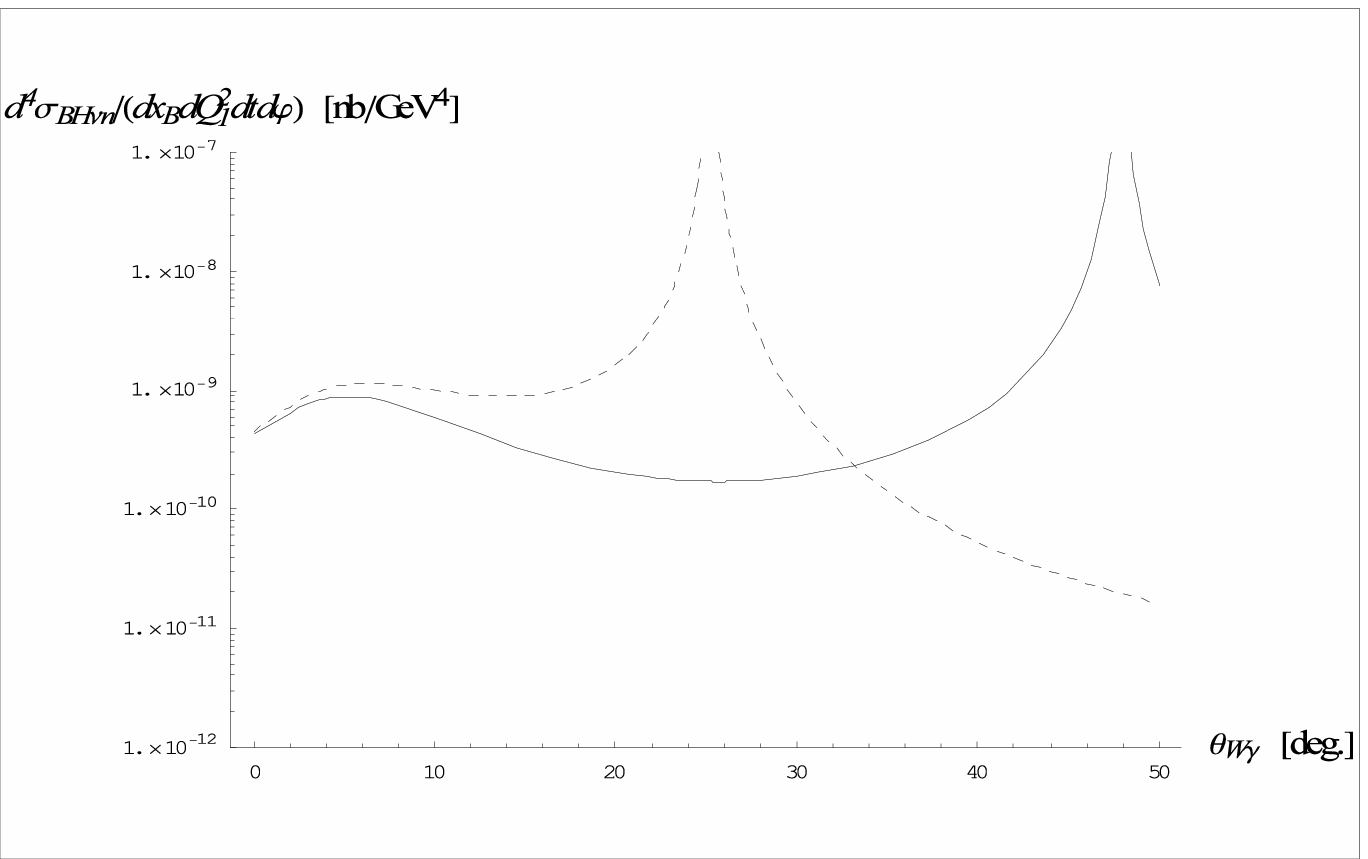}}
\caption{Bethe-Heitler cross section $\sigma_{BH\nu n}$ for the weak charged DVCS process plotted as a 
function of the angle $\theta_{B\gamma}$ between the incoming weak virtual boson and outgoing real photon 
in the target rest frame for $Q_{1}^{2}=2.5\;\mathrm{GeV}^{2}$ and $x_{B}=0.35$ with $\omega=5\;\mathrm{GeV}$ 
(solid line) and $\omega=20\;\mathrm{GeV}$ (dashed line) neutrino beam.}
\label{wcdvscBH2}
\end{figure}

\begin{figure}
\centerline{\epsfxsize=5in\epsffile{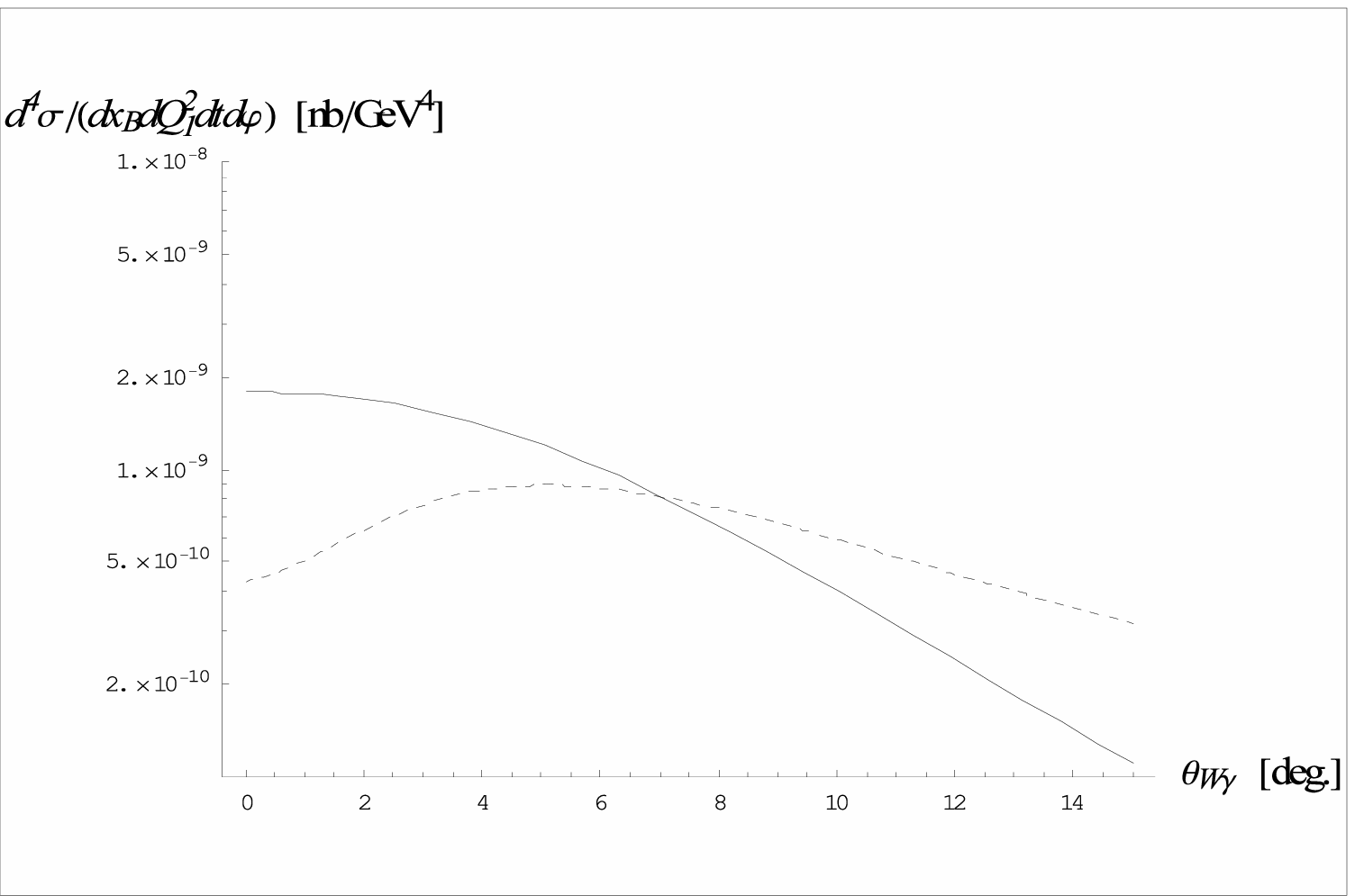}}
\caption{Compton cross section $\sigma_{C\nu n}$  (solid line) and Bethe-Heitler cross section 
$\sigma_{BH\nu n}$ (dashed line) for the weak charged DVCS process plotted as a function of the 
angle $\theta_{B\gamma}$ between the incoming weak virtual boson and outgoing real photon in the target 
rest frame for $Q_{1}^{2}=2.5\;\mathrm{GeV}^{2}$ and $x_{B}=0.35$ with $\omega=5\;\mathrm{GeV}$ neutrino beam.}
\label{wcdvsctogether}
\end{figure}

We turn our attention now to the Bethe-Heitler background. It is
given only by the diagram (b) in Fig. \ref{weakDVCSdiagrams}. At the angle $\phi'$, 
when the outgoing real photon is collinear with the scattered muon, one should expect a pole. 
The amplitude of the Bethe-Heitler contribution reads
\begin{eqnarray}
\mathrm{T}_{BH\nu n} & = & \sqrt{2}\left|e\right|G_{F}\epsilon_{\mu}^{*}\left(q_{2}\right)
\bar{u}\left(k'\right)\left[\frac{\gamma^{\mu}\left(\not\! k'+\not\! q_{2}\right)\gamma^{\nu}
\left(1-\gamma_{5}\right)}{\left(k'+q_{2}\right)^{2}}\right]u\left(k\right)\nonumber \\
&  & \times\left\langle p\left(p_{2},s_{2}\right)\right|J_{\nu}^{CC}\left(0\right)
\left|n\left(p_{1},s_{1}\right)\right\rangle.
\label{eq:TmatrixBHcharged}
\end{eqnarray}
The spin-averaged square of Eq. (\ref{eq:TmatrixBHcharged}) is
\begin{eqnarray}
\overline{\left|\mathrm{T}_{BH\nu n}\right|^{2}} & = & 8\pi
\alpha G_{F}^{2}L_{BH}^{\nu\beta}H_{\nu\beta}^{BH},
\label{eq:TmatrixsquaredBHcharged}
\end{eqnarray}
with tensors
\begin{eqnarray}
L_{BH}^{\nu\beta} & = & \frac{8}{\left(k'\cdot q_{2}\right)}
\left[k^{\nu}q_{2}^{\beta}+k^{\beta}q_{2}^{\nu}-g^{\nu\beta}\left(k\cdot q_{2}
\right)-i\epsilon^{\nu\beta\sigma\tau}k_{\sigma}q_{2\tau}\right],\nonumber \\
H_{\nu\beta}^{BH} & = & \frac{1}{2}\sum_{s_{1},s_{2}}\left\langle p
\left(p_{2},s_{2}\right)\right|J_{\nu}^{CC}\left(0\right)\left|n\left(p_{1},s_{1}
\right)\right\rangle \left\langle p\left(p_{2},s_{2}\right)\right|J_{\beta}^{CC}\left(0\right)
\left|n\left(p_{1},s_{1}\right)\right\rangle ^{*}.\nonumber \\
\label{eq:wcdvcsBHtensors}
\end{eqnarray}
The matrix element of the weak charged transition current $J_{\nu}^{CC}$
between the nucleon states is defined as 
\begin{eqnarray}
\left\langle p\left(p_{2},s_{2}\right)\right|J_{\nu}^{CC}\left(0\right)\left|n\left(p_{1},s_{1}
\right)\right\rangle  & = & \left\langle p\left(p_{2},s_{2}\right)\right|\bar{\psi}_{p}
\left(0\right)\gamma_{\nu}\frac{1}{2}\left(1-\gamma_{5}\right)\psi_{n}\left(0\right)
\left|n\left(p_{1},s_{1}\right)\right\rangle.\nonumber \\
\label{eq:transitioncurrentcc}
\end{eqnarray}
Using the same isospin symmetry relation between the flavor nondiagonal and flavor 
diagonal nucleon matrix elements, see Eq. (\ref{eq:isospinsymmetry}), we parametrize the vector current 
part of the matrix element in Eq. (\ref{eq:transitioncurrentcc}) in terms of the Dirac and Pauli 
form factors for each quark flavor,
\begin{eqnarray}
\left\langle p\left(p_{2},s_{2}\right)\right|\bar{\psi}_{p}\left(0\right)\gamma_{\nu}\psi_{n}
\left(0\right)\left|n\left(p_{1},s_{1}\right)\right\rangle =\left\langle p\left(p_{2},s_{2}
\right)\right|\bar{\psi}_{u}\left(0\right)\gamma_{\nu}\psi_{d}\left(0\right)\left|n
\left(p_{1},s_{1}\right)\right\rangle \nonumber \\
=\left\langle p\left(p_{2},s_{2}\right)\right|\bar{\psi}_{u}\left(0\right)\gamma_{\nu}
\psi_{u}\left(0\right)\left|p\left(p_{1},s_{1}\right)\right\rangle -\left\langle p\left(p_{2},s_{2}
\right)\right|\bar{\psi}_{d}\left(0\right)\gamma_{\nu}\psi_{d}\left(0\right)\left|p\left(p_{1},s_{1}
\right)\right\rangle \nonumber \\
=\bar{u}\left(p_{2},s_{2}\right)\left\{ \left[F_{1u}\left(t\right)-F_{1d}\left(t\right)
\right]\gamma_{\nu}-\left[F_{2u}\left(t\right)-F_{2d}\left(t\right)\right]
\frac{i\sigma_{\nu\lambda}r^{\lambda}}{2M}\right\} u\left(p_{1},s_{1}\right).\nonumber \\
\label{eq:vectorpart}
\end{eqnarray}
Furthermore, for the axial vector current part we have 
\begin{eqnarray}
\left\langle p\left(p_{2},s_{2}\right)\right|\bar{\psi}_{p}\left(0\right)
\gamma_{\nu}\gamma_{5}\psi_{n}\left(0\right)\left|n\left(p_{1},s_{1}\right)
\right\rangle  & = & \bar{u}\left(p_{2},s_{2}\right)
\Big[g_{A}\left(t\right)\gamma_{\nu}\gamma_{5}\nonumber \\
&  & \left.-g_{P}\left(t\right)\frac{\gamma_{5}r_{\nu}}{2M}\right]u\left(p_{1},s_{1}\right),
\label{eq:axialvectorpart}
\end{eqnarray}
where the \emph{t}-dependence of the axial and pseudoscalar from factors is given by 
\begin{eqnarray}
g_{A}\left(t\right) & = & g_{A}\left(t=0\right)\left(1-\frac{t}{m_{A}^{2}}
\right)^{-2},\nonumber \\
g_{P}\left(t\right) & = & g_{A}\left(t\right)\frac{4M^{2}}{m_{\pi}^{2}-t}.
\label{eq:axialandpseudoscalar}
\end{eqnarray}
Thus the unpolarized Bethe-Heitler differential cross section for
the weak charged DVCS process is written in the form
\begin{eqnarray}
\frac{d^{4}\sigma_{BH\nu n}}{dx_{B}dQ_{1}^{2}dtd\varphi} & = & \frac{1}{\left(2\pi\right)^{3}}
\frac{\alpha G_{F}^{2}}{16}\frac{1+x_{B}
\left(M/\omega\right)}{M^{2}\omega^{2}\left[2+\left(M/\omega\right)\right]x_{B}
\left[y+2x_{B}\left(M/\omega\right)\right]^{2}}L_{BH}^{\nu\beta}H_{\nu\beta}^{BH}.\nonumber \\
\label{eq:diffcrosssectionfinalBHcharged}
\end{eqnarray}
Its dependence on $\theta_{W\gamma}$ is plotted in Figs. \ref{wcdvscBH1} and \ref{wcdvscBH2}, 
and together with the Compton cross section in Fig. \ref{wcdvsctogether}. In particular, 
the plots in Fig. \ref{wcdvsctogether} reveal that, in the forward direction up to $4^{0}$, 
the Bethe-Heitler contamination is suppressed compared to the Compton contribution.

To summarize, we gave a comprehensive review on neutrino-induced DVCS processes. Through the weak 
interaction currents, these processes give access to two different sets of GPDs, namely, to the 
sum of quark and antiquark distributions and to their difference. Thus we can measure independently 
both the valence and sea content of GPDs. In addition, the weak charged current interaction 
probes GPDs that are nondiagonal in quark flavor, such as those associated with 
the neutron-to-proton transition. 
\begin{table}
\caption{Orders of magnitude (in nbarns) for the unpolarized total weak neutral $\sigma_{\nu p}$, 
Compton weak neutral $\sigma_{Cep}$, Compton weak charged $\sigma_{C\nu n}$ and Compton 
electromagnetic $\sigma_{Cep}$ cross sections for two different kinematical regions of the 
Bjorken scaling variable with two different lepton beam energies $\omega$.}
\label{tablecomparisoncrosssection}
\vspace{12pt}
\begin{center}
\begin{tabular}{lccccc}\hline\hline\\
kinematical region & 
$\omega\left[\mathrm{GeV}\right]$ & 
$\sigma_{\nu p}\left(Z^{0}\right)$ & 
$\sigma_{Cep}\left(Z^{0}\right)$ & 
$\sigma_{C\nu n}\left(W^{+}\right)$ & 
$\sigma_{Cep}\left(\gamma\right)$ \\\\
\hline
 & & \\
$0.28\leq x_{B}\leq0.3$ & 5 & $10^{-11}$ & $10^{-13}$ & $10^{-11}$ & $10^{-5}$  \\
 & & \\
$0.26\leq x_{B}\leq0.28$ & 20 & $10^{-9}$ & $10^{-10}$ & $10^{-9}$ & $10^{-3}$  \\\\ \hline\hline
\end{tabular}
\end{center}
\end{table}

\chapter{Wide-Angle Real Compton Scattering}

\section{\ \,  Introduction}

\begin{figure}
\centerline{\epsfxsize=6in\epsffile{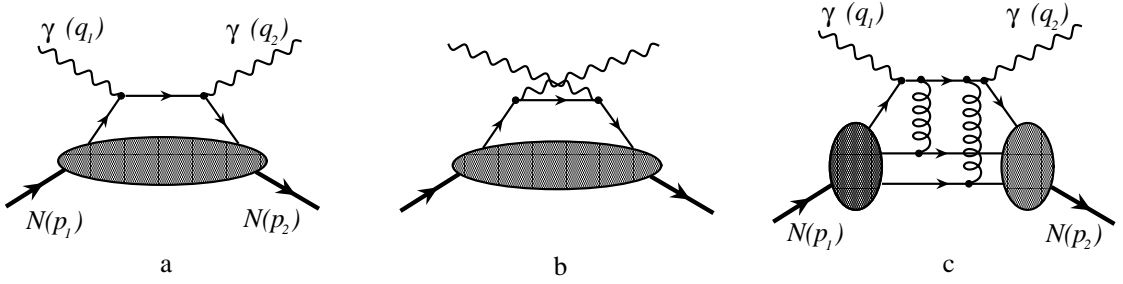}}
\caption{Handbag contribution (a and b) and the configuration (c) with the exchange of two hard gluons.}
\label{wacsdiagrams}
\end{figure}

Standard electromagnetic Compton scattering provides a unique tool
for studying hadrons. We have demonstrated in the previous chapter how the process 
can be extended into the weak interaction sector, in which the relevant Compton amplitude 
probes the hadronic structure by means of the quark electromagnetic and 
weak currents rather than two electromagnetic currents, and promise therefore 
a new insight. In the lowest QCD approximation, these currents couple to photons 
and weak bosons through point-like vertices. Moreover, in a specific kinematical 
regime, featuring the presence of a large momentum transfer, the behavior of both 
the standard and weak Compton amplitudes is dominated by the light-like distances. 
As a result the amplitudes are given in terms of the handbag diagrams.
There are several situations, where the handbag contribution plays
an essential role \cite{Radyushkin:1998rt}: 
\begin{itemize}
\item Both initial and final photons are highly virtual and have equal
space-like virtualities. This situation corresponds to the forward VCA.
Its imaginary part determines the structure functions of DIS.
\item The condition on photon virtualities may be relaxed in a sense that
the initial photon is still far off-shell but the final photon is
real and the invariant momentum transfer to the hadron is small. The
situation corresponds to the nonforward VCA. It is accessible through DVCS. 
\item The configuration, in which both photons in the initial and final states 
are real but the invariant momentum transfer is large. The physical
process corresponding to this situation is known as wide-angle real Compton 
scattering (WACS).
\end{itemize}
The Compton amplitude of WACS is a subject of the study in the present chapter, which 
contains some of the formal results from Ref. \cite{Psaker2}. 
An efficient way to study this amplitude is again to use the light-cone expansion 
of the time-ordered product of two electromagnetic currents in the coordinate 
representation. At moderately large values of Mandelstam invariants 
(\emph{s}, $-t$, $-u\leq10\;\mathrm{GeV}^{2}$), the dominant reaction mechanism corresponds to the 
handbag contribution just as in DIS and DVCS. To the leading order, the amplitude is given by 
the \emph{s-} and \emph{u}-channel handbag diagrams, see the diagrams (a) and (b) 
in Fig. \ref{wacsdiagrams}. On the other hand, for extremely large $-t\gg10\;\mathrm{GeV}^{2}$ the purely 
hard contributions involving two-gluon exchange, such as the diagram (c) in Fig. \ref{wacsdiagrams}, 
and having power-law behavior (i.e. $\sim\left(-t\right)^{N}$), become dominant. 
In particular, at large values of \emph{t}, the handbag diagrams are accompanied by the Sudakov form factor 
$S\left(t\right)\sim\exp\left[-\alpha_{s}\ln^{2}\left(-t\right)/3\pi\right]$ that 
drops faster than any power of $-t$. At moderate \emph{t}, however, the hard contributions are numerically 
suppressed by a factor $\left(\alpha_{s}/\pi\right)^{2}\simeq1/100$, and will not be considered here.

In Section VIII.2, we start from the formal light-cone expansion in 
terms of QCD string operators (with gauge links along the straight line between the fields which, 
for brevity, we do not write explicitly). In addition to the twist-2 string operators, we also include now 
the operators that are algebraically of twist-3 but given by the total derivatives of twist-2 operators. 
One refers to these as the kinematical twist-3 contributions, in order to distinguish them from the 
so-called dynamical twist-3 contributions involving quark-gluon operators. The latter cannot be reduced 
to total derivatives \cite{Radyushkin:2000ap}. Furthermore, for simplicity, we consider a spin zero 
(i.e. pion) target. For that reason, only the matrix element of a vector-type string operator is needed. 
Its spectral representation is given in terms of double distributions. It is worth noting at this point 
that even though we restrict ourselves to the pion, our approach can be used in more realistic situation 
involving the nucleon target \cite{Psaker3} as well as can be applied to the time-like region, in particular, 
to study the exclusive production of two pions in the two-photon collisions \cite{Psaker4}. 
In Section VIII.3, we compute the Compton amplitude for the pion up to the twist-3 level, 
and inspect its transversality, in other words, the electromagnetic gauge invariance with respect to both 
initial and final photons. The physical amplitudes (or alternatively the photon helicity amplitudes) of 
the WACS process are obtained by projecting the Compton amplitude on different the polarization states of 
physical photons. We estimate these amplitudes and their corresponding cross sections by taking a simple 
model for the double distribution of the pion. The results are compared with the QED calculation.

\section{\ \,  Twist Decomposition and Parametrization}

We recall from Section IV.2 that the leading light-cone singularity of 
the time-ordered product of two electromagnetic currents is contained in the 
handbag contribution,
\begin{eqnarray}
iT\left\{ J_{EM}^{\mu}\left(z/2\right)J_{EM}^{\nu}
\left(-z/2\right)\right\}  & = & \frac{z_{\rho}}{2\pi^{2}z^{4}}
\sum_{f}Q_{f}^{2}\left\{ s^{\mu\rho\nu\eta}\mathcal{O}_{\eta}^{f}
\left(z\left|0\right.\right)+i\epsilon^{\mu\rho\nu\eta}\mathcal{O}_{5\eta}^{f}
\left(z\left|0\right.\right)\right\},
\nonumber \\
\label{eq:WACSexpansion}
\end{eqnarray}
in which a highly virtual quark propagator connecting the two photon
vertices is convoluted with the blob describing the long-distance
dynamics. In Eq. (\ref{eq:WACSexpansion}) the tensor 
$s^{\mu\rho\nu\eta}\equiv g^{\mu\rho}g^{\nu\rho}+g^{\mu\eta}g^{\rho\nu}-g^{\mu\nu}g^{\rho\eta}$ 
and the QCD bilocal operators with the flavor index \emph{f} and one open Lorentz index are
\begin{eqnarray}
\mathcal{O}_{\eta}^{f}\left(z\left|0\right.\right) & = & \left[\bar{\psi}_{f}
\left(-z/2\right)\gamma_{\eta}\psi_{f}\left(z/2\right)-\bar{\psi}_{f}\left(z/2\right)
\gamma_{\eta}\psi_{f}\left(-z/2\right)\right],\nonumber \\
\mathcal{O}_{5\eta}^{f}\left(z\left|0\right.\right) & = & \left[\bar{\psi}_{f}\left(-z/2
\right)\gamma_{\eta}\gamma_{5}\psi_{f}\left(z/2\right)+\bar{\psi}_{f}\left(z/2\right)
\gamma_{\eta}\gamma_{5}\psi_{f}\left(-z/2\right)\right].
\label{eq:stringoperatorsWACS}
\end{eqnarray}
The Taylor expansion of these operators in the relative distance \emph{z}
involves covariant derivatives and yields local operators that are
not symmetric in their indices. Nevertheless, it was shown in 
Refs. \cite{Radyushkin:2000ap,Kivel:2000rb} that 
$\mathcal{O}_{\eta}^{f}\left(z\left|0\right.\right)$ and 
$\mathcal{O}_{5\eta}^{f}\left(z\left|0\right.\right)$ can be expressed in a compact form 
in terms of the contracted operators 
$\mathcal{O}^{f}\left(z\right)\equiv z^{\eta}\mathcal{O}_{\eta}^{f}\left(z\left|0\right.\right)$ 
and $\mathcal{O}_{5}^{f}\left(z\right)\equiv z^{\eta}\mathcal{O}_{5\eta}^{f}\left(z\left|0\right.\right)$ 
that produce only symmetric operators, and quark-gluon string operators. Namely, 
the fully deconstructed vector and axial vector operators take the following form
\begin{eqnarray}
\mathcal{O}_{\eta}^{f}\left(z\left|0\right.\right) & = & \int_{0}^{1}dv\;
\left\{ \cos\left[\frac{i\bar{v}\kappa}{2}\right]\partial_{\eta}+\frac{iv}{2}\sin
\left[\frac{i\bar{v}\kappa}{2}\right]
\frac{\left(z\cdot\mathcal{D}\right)\mathcal{D}_{\eta}-\mathcal{D}^{2}z_{\eta}}{\kappa}\right\} \nonumber \\
&  & \times\left[\bar{\psi}_{f}\left(-vz/2\right)\not\! z\psi_{f}\left(vz/2\right)-\bar{\psi}_{f}
\left(vz/2\right)\not\! z\psi_{f}\left(-vz/2\right)\right]\nonumber \\
&  & +\epsilon_{\eta\delta\kappa\lambda}z^{\delta}\mathcal{D}_{\kappa}
\int_{0}^{1}\frac{dv}{\kappa}\;\sin\left[\frac{i\bar{v}\kappa}{2}\right]\partial^{\lambda}\nonumber \\
&  & \times\left[\bar{\psi}_{f}\left(-vz/2\right)\not\! z\gamma_{5}\psi_{f}\left(vz/2\right)+
\bar{\psi}_{f}\left(vz/2\right)\not\! z\gamma_{5}\psi_{f}\left(-vz/2\right)\right]\nonumber \\
&  & +\left(\mathrm{quark}\!-\!\mathrm{gluon}\;\mathrm{operators}\right),\nonumber \\
\mathcal{O}_{5\eta}^{f}\left(z\left|0\right.\right) & = & \int_{0}^{1}dv\;
\left\{ \cos\left[\frac{i\bar{v}\kappa}{2}\right]\partial_{\eta}+\frac{iv}{2}\sin
\left[\frac{i\bar{v}\kappa}{2}\right]
\frac{\left(z\cdot\mathcal{D}\right)\mathcal{D}_{\eta}-\mathcal{D}^{2}z_{\eta}}{\kappa}\right\} \nonumber \\
&  & \times\left[\bar{\psi}_{f}\left(-vz/2\right)\not\! z\gamma_{5}\psi_{f}\left(vz/2\right)+
\bar{\psi}_{f}\left(vz/2\right)\not\! z\gamma_{5}\psi_{f}\left(-vz/2\right)\right]\nonumber \\
&  & +\epsilon_{\eta\delta\kappa\lambda}z^{\delta}\mathcal{D}_{\kappa}\int_{0}^{1}\frac{dv}{\kappa}\;
\sin\left[\frac{i\bar{v}\kappa}{2}\right]\partial^{\lambda}\nonumber \\
&  & \times\left[\bar{\psi}_{f}\left(-vz/2\right)\not\! z\psi_{f}\left(vz/2\right)-
\bar{\psi}_{f}\left(vz/2\right)\not\! z\psi_{f}\left(-vz/2\right)\right]\nonumber \\
&  & +\left(\mathrm{quark}\!-\!\mathrm{gluon}\;\mathrm{operators}\right),
\label{eq:decomposition1}
\end{eqnarray}
where we use the notation $\bar{v}\equiv1-v$, $\partial_{\eta}\equiv\partial/\partial z^{\eta}$ and
\begin{eqnarray}
\kappa & \equiv & \sqrt{\left(z\cdot\mathcal{D}\right)^{2}-z^{2}\mathcal{D}^{2}},
\label{eq:kappa}
\end{eqnarray}
with $\mathcal{D}\equiv\partial/\partial X$ being the total derivative.
Neglecting $z^{2}$ terms and ignoring quark-gluon operators (note
that they correspond to the dynamical twist-3 contributions, which
are not considered in our approach) the decomposition (\ref{eq:decomposition1}) gives 
\cite{Belitsky:2000vx,Radyushkin:2000ap,Kivel:2000rb}
\begin{eqnarray}
\mathcal{O}_{\eta}^{f}\left(z\left|0\right.\right) & = & \int_{0}^{1}dv\;
\left\{ \cos\left[\frac{i\bar{v}\left(z\cdot\mathcal{D}\right)}{2}\right]
\partial_{\eta}+\frac{iv}{2}\sin\left[\frac{i\bar{v}\left(z\cdot\mathcal{D}\right)}{2}
\right]\mathcal{D}_{\eta}\right\} \nonumber \\
&  & \times\left[\bar{\psi}_{f}\left(-vz/2\right)\not\! z\psi_{f}\left(vz/2\right)-
\bar{\psi}_{f}\left(vz/2\right)\not\! z\psi_{f}\left(-vz/2\right)\right]\nonumber \\
&  & +\epsilon_{\eta\delta\kappa\lambda}z^{\delta}\mathcal{D}_{\kappa}\int_{0}^{1}
\frac{dv}{\left(z\cdot\mathcal{D}\right)}\;\sin\left[\frac{i\bar{v}\left(z\cdot\mathcal{D}\right)}{2}
\right]\partial^{\lambda}\nonumber \\
&  & \times\left[\bar{\psi}_{f}\left(-vz/2\right)\not\! z\gamma_{5}\psi_{f}\left(vz/2\right)+
\bar{\psi}_{f}\left(vz/2\right)\not\! z\gamma_{5}\psi_{f}\left(-vz/2\right)\right],\nonumber \\
\mathcal{O}_{5\eta}^{f}\left(z\left|0\right.\right) & = & \int_{0}^{1}dv\;\left\{ \cos\left[
\frac{i\bar{v}\left(z\cdot\mathcal{D}\right)}{2}\right]\partial_{\eta}+\frac{iv}{2}
\sin\left[\frac{i\bar{v}\left(z\cdot\mathcal{D}\right)}{2}\right]\mathcal{D}_{\eta}\right\} \nonumber \\
&  & \times\left[\bar{\psi}_{f}\left(-vz/2\right)\not\! z\gamma_{5}\psi_{f}\left(vz/2\right)+
\bar{\psi}_{f}\left(vz/2\right)\not\! z\gamma_{5}\psi_{f}\left(-vz/2\right)\right]\nonumber \\
&  & +\epsilon_{\eta\delta\kappa\lambda}z^{\delta}\mathcal{D}_{\kappa}\int_{0}^{1}
\frac{dv}{\left(z\cdot\mathcal{D}\right)}\;\sin\left[\frac{i\bar{v}\left(z\cdot\mathcal{D}\right)}{2}
\right]\partial^{\lambda}\nonumber \\
&  & \times\left[\bar{\psi}_{f}\left(-vz/2\right)\not\! z\psi_{f}\left(vz/2\right)-\bar{\psi}_{f}
\left(vz/2\right)\not\! z\psi_{f}\left(-vz/2\right)\right].
\label{eq:decomposition2}
\end{eqnarray}
For the pion, the matrix element of the contracted axial vector operator vanishes and accordingly, 
the matrix element of the time-ordered product (\ref{eq:WACSexpansion}) reduces to
\begin{eqnarray}
\left\langle \pi\left(p_{2}\right)\right|iT\left\{ J_{EM}^{\mu}\left(z/2\right)J_{EM}^{\nu}
\left(-z/2\right)\right\} \left|\pi\left(p_{1}\right)\right\rangle =\nonumber \\
\frac{z_{\rho}}{2\pi^{2}z^{4}}\sum_{f}Q_{f}^{2}\left\{ s^{\mu\rho\nu\eta}\int_{0}^{1}dv\;
\left\{ \cos\left[\frac{\bar{v}\left(r\cdot z\right)}{2}\right]\partial_{\eta}+
\frac{v}{2}\sin\left[\frac{\bar{v}\left(r\cdot z\right)}{2}
\right]r_{\eta}\right\} \right.\nonumber \\
\left.+i\epsilon^{\mu\nu\eta\rho}\epsilon_{\eta\delta\kappa\lambda}z^{\delta}r^{\kappa}
\int_{0}^{1}\frac{dv}{\left(r\cdot z\right)}\;\sin\left[\frac{\bar{v}\left(r\cdot z\right)}{2}
\right]\partial^{\lambda}\right\} \nonumber \\
\times\left\langle \pi\left(p_{2}\right)\right|\bar{\psi}_{f}\left(-vz/2\right)\not\! z\psi_{f}
\left(vz/2\right)-\bar{\psi}_{f}\left(vz/2\right)\not\! z\psi_{f}\left(-vz/2\right)
\left|\pi\left(p_{2}\right)\right\rangle.
\label{eq:matrixelementofTproduct1}
\end{eqnarray}
We have used the fact that under the matrix element, the total derivative 
of the string operator turns into the momentum transfer, 
$i\partial/\partial X_{\kappa}\rightarrow r^{\kappa}$ 
(recall that $r\equiv p_{1}-p_{2}$ is the overall momentum transfer to the pion target).

The next step is to parametrize the nonforward matrix element of the twist-2 part of the 
contracted vector string operator in Eq. (\ref{eq:matrixelementofTproduct1}). This is performed 
in terms of the spectral functions, namely, the double distributions. Thus one needs to construct 
the parametrization for the matrix element of $\mathcal{O}^{f}\left(z\right)$ satisfying the harmonic 
condition
\begin{eqnarray}
\partial^{2}\left\langle \pi\left(p_{2}\right)\right|\mathcal{O}^{f}\left(z\right)
\left|\pi\left(p_{1}\right)\right\rangle _{twist-2} & = & 0.
\label{eq:dAlembertWACS}
\end{eqnarray}
In general, the matrix element can be regarded as a function of the
variables $\left(p\cdot z\right)$, $\left(r\cdot z\right)$ and $z^{2}$,
where $p\equiv\left(p_{1}+p_{2}\right)/2$ is the average pion momentum.
The standard way to parametrize it, is by the decomposition into plane
waves, where one explicitly separates the $\left(p\cdot z\right)$
components from the $\left(r\cdot z\right)$ components, and hence
use two spectral functions, $f_{f}\left(\beta,\alpha,t\right)$ and
$g_{f}\left(\beta,\alpha,t\right)$. Alternatively, one can write
the parametrization in the form
\begin{eqnarray}
\left\langle \pi\left(p-r/2\right)\right|\mathcal{O}^{f}\left(z\right)\left|\pi\left(p+r/2\right)
\right\rangle _{twist-2} & = & -2i\left(z^{\sigma}\partial_{\sigma}\right)
\mathcal{F}\left(z,p,r\right),
\label{eq:alternativeparametrization}
\end{eqnarray}
with
\begin{eqnarray}
\mathcal{F}\left(z,p,r\right) & = & \int_{-1}^{1}d\beta
\int_{-1+\left|\beta\right|}^{1-\left|\beta\right|}d\alpha\; e^{i\left(k\cdot z\right)}
\left[h_{f}\left(\beta,\alpha,t\right)+\mathcal{O}\left(z^{2}\right)\right],
\label{eq:functionF}
\end{eqnarray}
where the four-momentum $k=\beta p+\alpha r/2$. The latter has the same structure as that parametrizing 
the twist-2 part of the matrix element 
$\left\langle p-r/2\right|\phi\left(-z/2\right)\phi\left(z/2\right)\left|p+r/2\right\rangle $ 
in the scalar case. In particular, it satisfies $\partial^{2}\mathcal{F}\left(z,p,r\right)=0$. 
It is easy to see that for that reason
\begin{eqnarray}
\partial^{2}\left(z^{\sigma}\partial_{\sigma}\right)\mathcal{F}
\left(z,p,r\right) & = & \left(2+z^{\sigma}\partial_{\sigma}\right)
\partial^{2}\mathcal{F}\left(z,p,r\right)\nonumber \\
& = & 0.
\label{eq:relationscalarpioncase}
\end{eqnarray}
Let us note at this point that in Appendix C, we consider the scalar parametrization 
$\mathcal{F}\left(z,p,r\right)$, which includes terms up to order $\mathcal{O}\left(z^{4}\right)$. 
The inclusion of $z^{2}$ and $z^{4}$ terms in the expansion is due to the fact that the fermion propagator 
has $1/z^{4}$ singularity, and  besides the operator $\mathcal{O}^{f}\left(vz\right)$ in 
Eq. (\ref{eq:matrixelementofTproduct1}) is differentiated with respect to $z$. Now, by taking only the 
first term in the $z^{2}$ expansion in $\mathcal{F}\left(z,p,r\right)$, 
the expression (\ref{eq:alternativeparametrization}) turns into
\begin{eqnarray}
\left\langle \pi\left(p-r/2\right)\right|\mathcal{O}^{f}\left(z\right)
\left|\pi\left(p+r/2\right)\right\rangle _{twist-2} & = & 2\int_{-1}^{1}d\beta
\int_{-1+\left|\beta\right|}^{1-\left|\beta\right|}d
\alpha\; e^{i\left(k\cdot z\right)}\nonumber \\
&  & \times\left(k\cdot z\right)h_{f}\left(\beta,\alpha,t\right).
\label{eq:parametrizationWACS}
\end{eqnarray}
The representation (\ref{eq:parametrizationWACS}) is given in terms of only one double distribution 
$h_{f}\left(\beta,\alpha,t\right)$ and corresponds to the so-called single-DD 
parametrization \cite{Belitsky:2001hz}. We observe that the original distributions are 
then expressed in terms of the single one as
\begin{eqnarray}
f_{f}\left(\beta,\alpha,t\right) & = & \beta h_{f}\left(\beta,\alpha,t\right),\nonumber \\
g_{f}\left(\beta,\alpha,t\right) & = & \alpha h_{f}\left(\beta,\alpha,t\right).
\label{eq:BelitskyMuellerrelation}
\end{eqnarray}
Moreover, we establish the symmetry properties of $h_{f}\left(\beta,\alpha,t\right)$
with respect to parameters $\beta$ and $\alpha$. Since $f_{f}\left(\beta,\alpha,t\right)$ 
is odd in $\beta$ and even in $\alpha$ while for $g_{f}\left(\beta,\alpha,t\right)$
the situation is the other way around, it follows from Eq. (\ref{eq:BelitskyMuellerrelation}) 
that the double distribution $h_{f}\left(\beta,\alpha,t\right)$ is an even function in both 
variables $\beta$ and $\alpha$.

Finally, by substituting the single-DD parametrization (\ref{eq:parametrizationWACS})
into the matrix element (\ref{eq:matrixelementofTproduct1}), one arrives at
\begin{eqnarray}
\left\langle \pi\left(p_{2}\right)\right|iT\left\{ J_{EM}^{\mu}\left(z/2\right)J_{EM}^{\nu}
\left(-z/2\right)\right\} \left|\pi\left(p_{1}\right)\right\rangle =\nonumber \\
\frac{z_{\rho}}{\pi^{2}z^{4}}\sum_{f}Q_{f}^{2}\int_{-1}^{1}d\beta
\int_{-1+\left|\beta\right|}^{1-\left|\beta\right|}d\alpha\; h_{f}
\left(\beta,\alpha,t\right)\int_{0}^{1}dv\; e^{iv\left(k\cdot z\right)}\nonumber \\
\times\left\{ s^{\mu\rho\nu\eta}\left\{ \cos\left[\frac{\bar{v}\left(r\cdot z\right)}{2}
\right]\left[k_{\eta}+ivk_{\eta}\left(k\cdot z\right)\right]+\frac{v}{2}r_{\eta}
\left(k\cdot z\right)\sin\left[\frac{\bar{v}\left(r\cdot z\right)}{2}
\right]\right\} \right.\nonumber \\
\left.+i\epsilon^{\mu\nu\eta\rho}\epsilon_{\eta\delta\kappa\lambda}z^{\delta}r^{\kappa}
\frac{1}{\left(r\cdot z\right)}\sin\left[\frac{\bar{v}\left(r\cdot z\right)}{2}\right]
\left[k^{\lambda}+ivk^{\lambda}\left(k\cdot z\right)\right]\right\}.
\label{eq:matrixelementofTproduct2}
\end{eqnarray}
To isolate the leading twist-2 contribution in Eq. (\ref{eq:matrixelementofTproduct2}), 
we integrate by parts over \emph{v} the first $k_{\eta}$ term in the symmetric part 
(with respect to the Lorentz indices $\mu$ and $\nu$) of the matrix element, and the first 
$k^{\lambda}$ term in the antisymmetric part. We find
\begin{eqnarray}
\left\langle \pi\left(p_{2}\right)\right|iT\left\{ J_{EM}^{\mu}\left(z/2\right)J_{EM}^{\nu}
\left(-z/2\right)\right\} \left|\pi\left(p_{1}\right)\right\rangle =\nonumber \\
\frac{z_{\rho}}{\pi^{2}z^{4}}\sum_{f}Q_{f}^{2}\int_{-1}^{1}d\beta
\int_{-1+\left|\beta\right|}^{1-\left|\beta\right|}d\alpha\; h_{f}
\left(\beta,\alpha,t\right)\nonumber \\
\times\left\{ s^{\mu\rho\nu\eta}\left\{ k_{\eta}e^{i\left(k\cdot z\right)}+
\frac{1}{2}\left[r_{\eta}\left(k\cdot z\right)-k_{\eta}\left(r\cdot z\right)\right]
\int_{0}^{1}dv\; ve^{iv\left(k\cdot z\right)}\sin\left[\frac{\bar{v}\left(r\cdot z\right)}{2}
\right]\right\} \right.\nonumber \\
\left.+\frac{i}{2}\epsilon^{\mu\nu\eta\rho}
\epsilon_{\eta\delta\kappa\lambda}z^{\delta}r^{\kappa}k^{\lambda}
\int_{0}^{1}dv\; ve^{iv\left(k\cdot z\right)}\cos\left[\frac{\bar{v}\left(r\cdot z\right)}{2}
\right]\right\},\nonumber \\
\label{eq:matrixelementofTproduct3}
\end{eqnarray}
where only the surface term in the symmetric part corresponds to the
twist-2 contribution (note that the surface term in the antisymmetric
part vanishes). All the remaining terms are of the twist-3. We explicitly write
\begin{eqnarray}
\left\langle \pi\left(p_{2}\right)\right|iT\left\{ J_{EM}^{\mu}\left(z/2\right)J_{EM}^{\nu}
\left(-z/2\right)\right\} \left|\pi\left(p_{1}\right)\right\rangle _{twist-2}=\nonumber \\
\frac{z_{\rho}}{\pi^{2}z^{4}}\sum_{f}Q_{f}^{2}s^{\mu\rho\nu\eta}
\int_{-1}^{1}d\beta\int_{-1+\left|\beta\right|}^{1-\left|\beta\right|}d\alpha\; h_{f}
\left(\beta,\alpha,t\right)k_{\eta}e^{i\left(k\cdot z\right)}
\label{eq:twist2matrixelementWACS}
\end{eqnarray}
for the twist-2 part. Now for the twist-3 part, after expressing the sine and cosine 
functions in Eq. (\ref{eq:matrixelementofTproduct3}) in terms of the exponentials and 
combining them with the overall exponential factor $\exp\left[iv\left(k\cdot z\right)\right]$, 
we obtain
\begin{eqnarray}
\left\langle \pi\left(p_{2}\right)\right|iT\left\{ J_{EM}^{\mu}\left(z/2\right)J_{EM}^{\nu}
\left(-z/2\right)\right\} \left|\pi\left(p_{1}\right)\right\rangle _{twist-3}=\nonumber \\
\frac{iz_{\rho}}{4\pi^{2}z^{4}}\sum_{f}Q_{f}^{2}\int_{-1}^{1}d\beta
\int_{-1+\left|\beta\right|}^{1-\left|\beta\right|}d\alpha\; h_{f}
\left(\beta,\alpha,t\right)\nonumber \\
\times\left\{ s^{\mu\rho\nu\eta}\left[r_{\eta}\left(k\cdot z\right)-k_{\eta}
\left(r\cdot z\right)\right]\int_{0}^{1}dv\; v
\left[e^{iv\left(k_{1}\cdot z\right)-i\left(r\cdot z\right)/2}-e^{iv\left(k_{2}\cdot z\right)+i
\left(r\cdot z\right)/2}\right]\right.\nonumber \\
\left.+\epsilon^{\mu\nu\eta\rho}\epsilon_{\eta\delta\kappa\lambda}z^{\delta}r^{\kappa}k^{\lambda}
\int_{0}^{1}dv\; v\left[e^{iv\left(k_{1}\cdot z\right)-i
\left(r\cdot z\right)/2}+e^{iv\left(k_{2}\cdot z\right)+i\left(r\cdot z\right)/2}
\right]\right\},
\label{eq:twist3matrixelementWACS}
\end{eqnarray}
where $k_{1}=k+r/2$ and $k_{2}=k-r/2$ are the momenta of the incoming 
(with respect to the interaction with two photons) and outgoing quarks, respectively.

\section{\ \,  Compton Scattering Amplitude}

As a result of the twist decomposition of the matrix element, 
see Eqs. (\ref{eq:twist2matrixelementWACS}) and (\ref{eq:twist3matrixelementWACS}), 
the Compton scattering amplitude,
\begin{eqnarray}
\mathsf{\mathsf{\mathcal{T}}}^{\mu\nu} & = & \sum_{f}Q_{f}^{2}\int d^{4}z\; e^{i\left(q\cdot z\right)}
\left\langle \pi\left(p_{2}\right)\right|iT\left\{ J_{EM}^{\mu}\left(z/2\right)J_{EM}^{\nu}\left(-z/2
\right)\right\} \left|\pi\left(p_{1}\right)\right\rangle,\nonumber \\
\label{eq:VCAWACS}
\end{eqnarray}
where $q\equiv\left(q_{1}+q_{2}\right)/2$ is the average photon momentum, can be written in our 
approximation, as the sum of the twist-2 and twist-3 contributions,
\begin{eqnarray}
\mathsf{\mathsf{\mathcal{T}}}^{\mu\nu} & = & \mathsf{\mathsf{\mathcal{T}}}_{twist-2}^{\mu\nu}+
\mathsf{\mathsf{\mathcal{T}}}_{twist-3}^{\mu\nu}.
\label{eq:separationWACS}
\end{eqnarray}

\subsection{\ \,  Twist-2 Amplitude}

After integrating over \emph{z} with the help of Eq. (\ref{eq:integraloverzDVCS}), 
the twist-2 part of the amplitude reads
\begin{eqnarray}
\mathsf{\mathsf{\mathcal{T}}}_{twist-2}^{\mu\nu} & = & \sum_{f}Q_{f}^{2}s^{\mu\rho\nu\eta}
\int d^{4}z\; e^{i\left(k+q\right)\cdot z}\frac{z_{\rho}}{\pi^{2}z^{4}}\int_{-1}^{1}d\beta
\int_{-1+\left|\beta\right|}^{1-\left|\beta\right|}d\alpha\; h_{f}
\left(\beta,\alpha,t\right)k_{\eta}\nonumber \\
& = & 2\sum_{f}Q_{f}^{2}\int_{-1}^{1}d\beta
\int_{-1+\left|\beta\right|}^{1-\left|\beta\right|}d\alpha\; h_{f}\left(\beta,\alpha,t\right)
\frac{1}{\left(k+q\right)^{2}}\nonumber \\
&  & \times\left\{ 2k^{\mu}k^{\nu}+k^{\mu}q^{\nu}+k^{\nu}q^{\mu}-g^{\mu\nu}\left[k^{2}+
\left(k\cdot q\right)\right]\right\}.
\label{eq:twist2WACS}
\end{eqnarray}
If the pion mass $m_{\pi}$ is neglected then $k^{2}=-\left(\beta^{2}-\alpha^{2}\right)t/4$
and the denominator for the quark propagator results into
\begin{eqnarray}
\left(k+q\right)^{2} & = & \left(k_{1}+q_{1}\right)^{2}\nonumber \\
& = & \left(k_{2}+q_{2}\right)^{2}\nonumber \\
& = & s\left\{ \beta-\frac{t}{4s}\left[\left(1-\beta\right)^{2}-\alpha^{2}\right]\right\}, 
\label{eq:denominatortwist2}
\end{eqnarray}
where $s=2\left(p\cdot q\right)-t/2=2\left(p_{1}\cdot q_{1}\right)=2\left(p_{2}\cdot q_{2}\right)$ 
or, after writing $\left(1-\beta\right)^{2}=\left(1-|\beta|\right)^{2}+2\left(|\beta|-\beta\right)$,
\begin{equation}
\left(k+q\right)^{2}=\left\{ \begin{array}{cc}
\beta s-\left[\left(1-\beta\right)^{2}-\alpha^{2}\right]t/4 & \mathrm{for\;}\beta>0\\
-\beta u-\left[\left(1+\beta\right)^{2}-\alpha^{2}\right]t/4 & \mathrm{for\;}\beta<0\\
\end{array}\right.
\label{eq:hardpropagator1}
\end{equation}

\subsection{\ \,  Twist-3 Amplitude}

We compute now the twist-3 amplitude, which contains two contributions. First, the contribution from 
the $s^{\mu\rho\nu\eta}$ term in Eq. (\ref{eq:twist3matrixelementWACS}) is
\begin{eqnarray}
\mathsf{\mathsf{\mathcal{T}}}_{twist-3\left(1\right)}^{\mu\nu} & = & \frac{1}{2}
\sum_{f}Q_{f}^{2}s^{\mu\rho\nu\eta}\int_{-1}^{1}d\beta\int_{-1+\left|\beta\right|}^{1-\left|\beta\right|}d
\alpha\; h_{f}\left(\beta,\alpha,t\right)\nonumber \\
&  & \times\int_{0}^{1}dv\; v\int d^{4}z\;\left[e^{i\left(l_{1}\cdot z\right)}-e^{i\left(l_{2}\cdot z\right)}
\right]\frac{iz_{\rho}}{2\pi^{2}z^{4}}\left[r_{\eta}\left(k\cdot z\right)-k_{\eta}
\left(r\cdot z\right)\right],\nonumber \\
\label{eq:nonsurfacetermWACS}
\end{eqnarray}
where the combined exponentials, $l_{1}\equiv q_{1}+vk_{1}$ and $l_{2}\equiv q_{2}+vk_{2}$,
correspond to the effective momenta flowing through the quark propagator,
and $q_{1}=q-r/2$ and $q_{2}=q+r/2$ are the momenta of the initial
and final photons. We have two types of quark propagators here. Their
denominators are given by
\begin{eqnarray}
l_{1}^{2} & = & v\left[\left(k_{1}\cdot q_{1}\right)+vk_{1}^{2}\right]\nonumber \\
& = & v\left[\left(k+q\right)^{2}-\bar{v}k_{1}^{2}\right],\nonumber \\
l_{2}^{2} & = & v\left[\left(k_{2}\cdot q_{2}\right)+vk_{2}^{2}\right]\nonumber \\
& = & v\left[\left(k+q\right)^{2}-\bar{v}k_{2}^{2}\right],
\label{eq:denominatornonsurfacetermWACS}
\end{eqnarray}
where the incoming and outgoing quark virtualities are 
\begin{eqnarray}
k_{1}^{2} & = & \left[-\beta^{2}+\left(1+\alpha\right)^{2}\right]t/4,\nonumber \\
k_{2}^{2} & = & \left[-\beta^{2}+\left(1-\alpha\right)^{2}\right]t/4.
\label{eq:k1andk2squared}
\end{eqnarray}
After the integral over \emph{z} is carried out using of the formula
\begin{eqnarray}
\int d^{4}z\; e^{i\left(l\cdot z\right)}
\frac{iz_{\rho}z_{\sigma}}{2\pi^{2}\left(z^{2}-i0\right)^{2}} & = & 
\frac{g_{\rho\sigma}l^{2}-2l_{\rho}l_{\sigma}}{\left(l^{2}+i0\right)^{2}},
\label{eq:integralformula2}
\end{eqnarray}
the expression (\ref{eq:nonsurfacetermWACS}) turns into
\begin{eqnarray}
\mathsf{\mathsf{\mathcal{T}}}_{twist-3\left(1\right)}^{\mu\nu} & = & \sum_{f}Q_{f}^{2}\int_{-1}^{1}d
\beta\int_{-1+\left|\beta\right|}^{1-\left|\beta\right|}d\alpha\; h_{f}
\left(\beta,\alpha,t\right)\int_{0}^{1}dv\; v\nonumber \\
&  & \times\left\{ \frac{1}{l_{1}^{4}}\left[-\left(l_{1}\cdot k\right)
\left(l_{1}^{\mu}r^{\nu}+l_{1}^{\nu}r^{\mu}\right)+\left(l_{1}\cdot r\right)
\left(l_{1}^{\mu}k^{\nu}+l_{1}^{\nu}k^{\mu}\right)\right]\right.\nonumber \\
&  & \left.-\frac{1}{l_{2}^{4}}\left[-\left(l_{2}\cdot k\right)
\left(l_{2}^{\mu}r^{\nu}+l_{2}^{\nu}r^{\mu}\right)+\left(l_{2}\cdot r\right)
\left(l_{2}^{\mu}k^{\nu}+l_{2}^{\nu}k^{\mu}\right)\right]\right\}.
\label{eq:nonsurfacetermfinalWACS}
\end{eqnarray}
Next we study the second contribution to $\mathsf{\mathsf{\mathcal{T}}}_{twist-3}^{\mu\nu}$. 
It is due to the $\epsilon^{\mu\nu\eta\rho}$ term in the matrix element 
(\ref{eq:twist3matrixelementWACS}, and comes from the axial vector operator. We have
\begin{eqnarray}
\mathsf{\mathsf{\mathcal{T}}}_{twist-3\left(2\right)}^{\mu\nu} & = & \frac{1}{2}
\sum_{f}Q_{f}^{2}\int_{-1}^{1}d\beta\int_{-1+\left|\beta\right|}^{1-\left|\beta\right|}d
\alpha\; h_{f}\left(\beta,\alpha,t\right)\nonumber \\
&  & \times\epsilon^{\mu\nu\eta\rho}\epsilon_{\eta\delta\kappa\lambda}r^{\kappa}k^{\lambda}
\int_{0}^{1}dv\; v\int d^{4}z\;\left[e^{i\left(l_{1}\cdot z\right)}+e^{i\left(l_{2}\cdot z\right)}
\right]\frac{iz_{\rho}z^{\delta}}{2\pi^{2}z^{4}}\nonumber \\
& = & \sum_{f}Q_{f}^{2}\int_{-1}^{1}d\beta\int_{-1+\left|\beta\right|}^{1-\left|\beta\right|}d
\alpha\; h_{f}\left(\beta,\alpha,t\right)\int_{0}^{1}dv\; v\nonumber \\
&  & \left\{ \frac{1}{l_{1}^{4}}\left[\left(l_{1}\cdot k\right)
\left(l_{1}^{\mu}r^{\nu}-l_{1}^{\nu}r^{\mu}\right)-\left(l_{1}\cdot r\right)
\left(l_{1}^{\mu}k^{\nu}-l_{1}^{\nu}k^{\mu}\right)\right]\right.\nonumber \\
&  & \left.+\frac{1}{l_{2}^{4}}\left[\left(l_{2}\cdot k\right)
\left(l_{2}^{\mu}r^{\nu}-l_{2}^{\nu}r^{\mu}\right)-\left(l_{2}
\cdot r\right)\left(l_{2}^{\mu}k^{\nu}-l_{2}^{\nu}k^{\mu}\right)\right]\right\}.
\label{eq:antisymmetrictermfinalWACS}
\end{eqnarray}
Finally, by adding Eqs. (\ref{eq:nonsurfacetermfinalWACS}) and (\ref{eq:antisymmetrictermfinalWACS}), 
the total twist-3 result reads
\begin{eqnarray}
\mathsf{\mathsf{\mathcal{T}}}_{twist-3}^{\mu\nu} & = & 2\sum_{f}Q_{f}^{2}
\int_{-1}^{1}d\beta\int_{-1+\left|\beta\right|}^{1-\left|\beta\right|}d\alpha\; h_{f}
\left(\beta,\alpha,t\right)\int_{0}^{1}dv\; v\nonumber \\
&  & \times\left\{ \frac{1}{l_{1}^{4}}\left[-\left(l_{1}\cdot k\right)l_{1}^{\nu}r^{\mu}+
\left(l_{1}\cdot r\right)l_{1}^{\nu}k^{\mu}\right]+
\frac{1}{l_{2}^{4}}\left[\left(l_{2}\cdot k\right)l_{2}^{\mu}r^{\nu}-
\left(l_{2}\cdot r\right)l_{2}^{\mu}k^{\nu}\right]\right\} \nonumber \\
& = & 2\sum_{f}Q_{f}^{2}\int_{-1}^{1}d\beta\int_{-1+\left|\beta\right|}^{1-\left|\beta\right|}d
\alpha\; h_{f}\left(\beta,\alpha,t\right)\nonumber \\
&  & \times\left\{ \int_{0}^{1}dv\; v^{2}
\left[\frac{k_{1}^{\nu}}{l_{1}^{4}}\left[-\left(l_{1}
\cdot k\right)r^{\mu}+\left(l_{1}\cdot r\right)k^{\mu}\right]+
\frac{k_{2}^{\mu}}{l_{2}^{4}}\left[\left(l_{2}\cdot k\right)r^{\nu}-
\left(l_{2}\cdot r\right)k^{\nu}\right]\right]\right.\nonumber \\
&  & \left.+\int_{0}^{1}dv\; v\left[\frac{q_{1}^{\nu}}{l_{1}^{4}}
\left[-\left(l_{1}\cdot k\right)r^{\mu}+\left(l_{1}
\cdot r\right)k^{\mu}\right]+\frac{q_{2}^{\mu}}{l_{2}^{4}}
\left[\left(l_{2}\cdot k\right)r^{\nu}-\left(l_{2}\cdot r\right)k^{\nu}\right]
\right]\right\}, \nonumber \\
\label{eq:twist3WACS}
\end{eqnarray}
with the following scalar products
\begin{eqnarray}
\left(l_{1}\cdot k\right) & = & \left(k+q\right)^{2}/2+\left(v-1/2\right)k_{1}^{2}+
\left(\bar{v}-v\alpha\right)t/4,\nonumber \\
\left(l_{2}\cdot k\right) & = & \left(k+q\right)^{2}/2+\left(v-1/2\right)k_{2}^{2}+
\left(\bar{v}+v\alpha\right)t/4,\nonumber \\
\left(l_{1}\cdot r\right) & = & -\left(\bar{v}-v\alpha\right)t/2,\nonumber \\
\left(l_{2}\cdot r\right) & = & \left(\bar{v}+v\alpha\right)t/2.
\label{eq:scalarproductsWACS}
\end{eqnarray}
Note that the second integral over \emph{v} in the amplitude (\ref{eq:twist3WACS})
is logarithmically divergent, however, the coefficient for the divergent
integral vanishes, when $\mathsf{\mathsf{\mathcal{T}}}_{twist-3}^{\mu\nu}$
is contracted with the incoming and outgoing photon polarization vectors, 
$\epsilon_{\nu}\left(q_{1}\right)$ and $\epsilon_{\mu}\left(q_{2}\right)$.

\subsection{\ \,  Electromagnetic Gauge Invariance}

The total Compton scattering amplitude (\ref{eq:VCAWACS}) is transverse with respect to 
incoming and outgoing photon momenta
\begin{eqnarray}
\mathsf{\mathsf{\mathcal{\mathcal{T}}}}^{\mu\nu}q_{1\nu} & = & 0,\nonumber \\
q_{2\mu}\mathsf{\mathsf{\mathcal{\mathcal{T}}}}^{\mu\nu} & = & 0.
\label{eq:gaugeinvarianceWACS}
\end{eqnarray}
Let us check whether the gauge invariant condition (\ref{eq:gaugeinvarianceWACS}) also holds 
for the amplitude (\ref{eq:separationWACS}). We expand the contracted twist-2 and 
twist-3 amplitudes in terms of $t/s$ and, for simplicity, keep only the lowest (zeroth) order 
corrections. A straightforward contraction, together with imposing the symmetry properties of 
the double distribution $h_{f}\left(\beta,\alpha,t\right)$ with respect to the variables 
$\alpha$ and $\beta$, yields for the leading-twist amplitude
\begin{eqnarray}
\mathsf{\mathsf{\mathcal{\mathcal{T}}}}_{\mathit{twist-2}}^{\mu\nu}q_{1\nu} & = & \sum_{f}Q_{f}^{2}
\int_{-1}^{1}d\beta\int_{-1+\left|\beta\right|}^{1-\left|\beta\right|}d\alpha\; h_{f}
\left(\beta,\alpha,t\right)r^{\mu}/2,\nonumber \\
q_{2\mu}\mathsf{\mathsf{\mathcal{\mathcal{T}}}}_{\mathit{twist-2}}^{\mu\nu} & = & -\sum_{f}Q_{f}^{2}
\int_{-1}^{1}d\beta\int_{-1+\left|\beta\right|}^{1-\left|\beta\right|}d\alpha\; h_{f}
\left(\beta,\alpha,t\right)r^{\nu}/2
\label{eq:twist2invarianceWACS}
\end{eqnarray}
while for the contracted next-to-leading twist amplitude we get
\begin{eqnarray}
\mathsf{\mathsf{\mathcal{\mathcal{T}}}}_{\mathit{twist-3}}^{\mu\nu}q_{1\nu} & = & -\sum_{f}Q_{f}^{2}
\int_{-1}^{1}d\beta\int_{-1+\left|\beta\right|}^{1-\left|\beta\right|}d\alpha\; h_{f}
\left(\beta,\alpha,t\right)r^{\mu}/2,\nonumber \\
q_{2\mu}\mathsf{\mathsf{\mathcal{\mathcal{T}}}}_{\mathit{twist-3}}^{\mu\nu} & = & \sum_{f}Q_{f}^{2}
\int_{-1}^{1}d\beta\int_{-1+\left|\beta\right|}^{1-\left|\beta\right|}d\alpha\; h_{f}
\left(\beta,\alpha,t\right)r^{\nu}/2.
\label{eq:twist3invarianceWACS}
\end{eqnarray}
Clearly, even at the level $\mathcal{O}\left(t/s\right)^{0}$, the
amplitude $\mathsf{\mathsf{\mathcal{\mathcal{T}}}}_{\mathit{twist-2}}^{\mu\nu}$
alone is not gauge invariant. However, the electromagnetic gauge invariance at this level can be 
recovered by including the contribution from 
$\mathsf{\mathsf{\mathcal{\mathcal{T}}}}_{\mathit{twist-3}}^{\mu\nu}$. 
The latter exactly cancels Eq. (\ref{eq:twist2invarianceWACS}). 

\subsection{\ \,  Helicity Amplitudes}

We shall project the Compton scattering amplitude onto the polarization
states $\epsilon_{1\nu}\equiv\epsilon_{\nu}\left(q_{1}\right)$ and
$\epsilon_{2\mu}\equiv\epsilon_{\mu}\left(q_{2}\right)$ of the physical
photons. In the Compton-like processes, these two states can be written
as the linear combination of two basic polarization four-vectors \cite{Landau},
\begin{eqnarray}
e_{1}^{\lambda} & = & \frac{n_{1}^{\lambda}}{\sqrt{-n_{1}^{2}}},\nonumber \\
e_{2}^{\lambda} & = & \frac{n_{2}^{\lambda}}{\sqrt{-n_{2}^{2}}},
\label{eq:unitvectorsWACS}
\end{eqnarray}
where
\begin{eqnarray}
n_{1}^{\lambda} & = & p^{\lambda}-\frac{\left(p\cdot q\right)}{q^{2}}q^{\lambda}
\label{eq:n1vectorWACS}
\end{eqnarray}
is in the reaction plane and
\begin{eqnarray}
n_{2}^{\lambda} & = & \epsilon^{\lambda\mu\nu\rho}p_{\mu}r_{\nu}q_{\rho}
\label{eq:n2vectorWACS}
\end{eqnarray}
is perpendicular to it. Then, by construction 
$\left(q\cdot\epsilon_{1}\right)=\left(r\cdot\epsilon_{1}\right)=0$ 
and similarly for $\epsilon_{2}$. The contracted twist-2 amplitude
is
\begin{eqnarray}
\mathsf{\mathsf{\mathcal{\mathcal{T}}}}_{\mathit{twist-2}}^{\mu\nu}
\epsilon_{2\mu}^{*}\epsilon_{1\nu} & = & 2\sum_{f}Q_{f}^{2}\int_{-1}^{1}d\beta
\int_{-1+\left|\beta\right|}^{1-\left|\beta\right|}d\alpha\; h_{f}
\left(\beta,\alpha,t\right)\frac{1}{\left(k+q\right)^{2}}\nonumber \\
&  & \times\left\{ 2\beta^{2}\left(p\cdot\epsilon_{1}\right)\left(p\cdot\epsilon_{2}^{*}\right)-
\left(\epsilon_{1}\cdot\epsilon_{2}^{*}\right)\left[k^{2}+\left(k\cdot q\right)\right]\right\} \nonumber \\
& = & \sum_{f}Q_{f}^{2}\int_{-1}^{1}d\beta\int_{-1+\left|\beta\right|}^{1-\left|\beta\right|}d
\alpha\; h_{f}\left(\beta,\alpha,t\right)\nonumber \\
&  & \times\left\{ \left[-1+\frac{\left(\beta^{2}-\alpha^{2}-1\right)t/4}{\left(k+q\right)^{2}}
\right]\left(\epsilon_{1}\cdot\epsilon_{2}^{*}\right)+\frac{4\beta^{2}}{\left(k+q\right)^{2}}
\left(p\cdot\epsilon_{1}\right)\left(p\cdot\epsilon_{2}^{*}\right)\right\} \nonumber \\
\label{eq:contractedtwist2WACS}
\end{eqnarray}
and for the twist-3 part we find
\begin{eqnarray}
\mathsf{\mathsf{\mathcal{\mathcal{T}}}}_{\mathit{twist-3}}^{\mu\nu}\epsilon_{2\mu}^{*}
\epsilon_{1\nu} & = & -t\sum_{f}Q_{f}^{2}\int_{-1}^{1}d\beta\;\beta^{2}
\int_{-1+\left|\beta\right|}^{1-\left|\beta\right|}d\alpha\; h_{f}\left(\beta,\alpha,t\right)
\left(p\cdot\epsilon_{1}\right)\left(p\cdot\epsilon_{2}^{*}\right)\nonumber \\
&  & \times\int_{0}^{1}dv\;
\left\{ \frac{\bar{v}-v\alpha}{\left[\left(k+q\right)^{2}-\bar{v}k_{1}^{2}\right]^{2}}+
\frac{\bar{v}+v\alpha}{\left[\left(k+q\right)^{2}-\bar{v}k_{2}^{2}\right]^{2}}\right\}.
\label{eq:contractedtwist3WACS}
\end{eqnarray}

Furthermore, we introduce twist-2 and twist-3 photon helicity amplitudes by projecting the 
photon polarizations onto the basic vectors $e_{1}$ and $e_{2}$,
\begin{eqnarray}
H_{\mathit{twist-2\left(3\right)}}^{\left(1,1\right)} & = & 
\mathsf{\mathsf{\mathcal{\mathcal{T}}}}_{\mathit{twist-2}\left(3\right)}^{\mu\nu}e_{1\mu}^{*}e_{1\nu},
\nonumber \\
H_{\mathit{twist-2\left(3\right)}}^{\left(1,2\right)} & = & 
\mathsf{\mathsf{\mathcal{\mathcal{T}}}}_{\mathit{twist-2\left(3\right)}}^{\mu\nu}e_{1\mu}^{*}e_{2\nu},
\nonumber \\
H_{\mathit{twist-2\left(3\right)}}^{\left(2,1\right)} & = & 
\mathsf{\mathsf{\mathcal{\mathcal{T}}}}_{\mathit{twist-2\left(3\right)}}^{\mu\nu}e_{2\mu}^{*}e_{1\nu},
\nonumber \\
H_{\mathit{twist-2\left(3\right)}}^{\left(2,2\right)} & = & 
\mathsf{\mathsf{\mathcal{\mathcal{T}}}}_{\mathit{twist-2\left(3\right)}}^{\mu\nu}e_{2\mu}^{*}e_{2\nu}.
\label{eq:twist2and3helicityamplitudesWACS}
\end{eqnarray}
Out of eight helicity amplitudes, only three are nonzero, namely,
\begin{eqnarray}
H_{\mathit{twist-2}}^{\left(1,1\right)} & = & \sum_{f}Q_{f}^{2}\int_{-1}^{1}d\beta
\int_{-1+\left|\beta\right|}^{1-\left|\beta\right|}d
\alpha\; h_{f}\left(\beta,\alpha,t\right)\nonumber \\
&  & \times\left\{ 1-\frac{\left(\beta^{2}-\alpha^{2}-1\right)}{\left(k+q\right)^{2}}
\frac{t}{4}+\frac{4\beta^{2}}{\left(k+q\right)^{2}}\frac{us}{t}\right\} ,\nonumber \\
H_{\mathit{twist-2}}^{\left(2,2\right)} & = & \sum_{f}Q_{f}^{2}\int_{-1}^{1}d\beta
\int_{-1+\left|\beta\right|}^{1-\left|\beta\right|}d\alpha\; h_{f}
\left(\beta,\alpha,t\right)\nonumber \\
&  & \times\left\{ 1-\frac{\left(\beta^{2}-\alpha^{2}-1\right)}{\left(k+q\right)^{2}}
\frac{t}{4}\right\} ,\nonumber \\
H_{\mathit{twist-3}}^{\left(1,1\right)} & = & -us\sum_{f}Q_{f}^{2}
\int_{-1}^{1}d\beta\;\beta^{2}\int_{-1+\left|\beta\right|}^{1-\left|\beta\right|}d
\alpha\; h_{f}\left(\beta,\alpha,t\right)\nonumber \\
&  & \int_{0}^{1}dv\;
\left\{ \frac{\bar{v}-v\alpha}{\left[\left(k+q\right)^{2}-\bar{v}k_{1}^{2}\right]^{2}}+
\frac{\bar{v}+v\alpha}{\left[\left(k+q\right)^{2}-\bar{v}k_{2}^{2}\right]^{2}}
\right\} ,\nonumber \\
\label{eq:survivinghelicityamplitudesWACS}
\end{eqnarray}
where we have used 
\begin{eqnarray}
\left(p\cdot e_{1}\right)^{2} & = & -p^{2}+
\frac{\left(p\cdot q\right)^{2}}{q^{2}}\nonumber \\
& = & \frac{t}{4}-\frac{s^{2}}{t}\left(1+\frac{t}{2s}\right)^{2}\nonumber \\
& = & \frac{us}{t}.
\label{eq:scalarproductcombination}
\end{eqnarray}

\subsection{\ \,  Model}

For the pion double distribution $h_{f}\left(\beta,\alpha,t\right)$, we use a simple 
model in a factorized form \cite{Mukherjee:2002gb},
\begin{eqnarray}
h_{f}\left(\beta,\alpha,t\right) & = & \frac{1}{\left|\beta\right|}f\left(\beta\right)h
\left(\beta,\alpha\right)\mathcal{F}\left(\beta,t\right).
\label{eq:ddh}
\end{eqnarray}
The profile function is chosen to be infinitely narrow, 
\begin{eqnarray}
h\left(\beta,\alpha\right) & = & \delta\left(\alpha\right).
\label{eq:h}
\end{eqnarray}
Accordingly, the integral over $\alpha$ in the helicity 
amplitudes (\ref{eq:survivinghelicityamplitudesWACS}) is trivial. Next, for the forward distribution 
we take
\begin{eqnarray}
f\left(\beta\right) & = & \frac{3}{4}\frac{1-\left|\beta\right|}{\sqrt{\left|\beta\right|}},
\label{eq:f}
\end{eqnarray}
and the \emph{t}-dependence is described by the following Regge ansatz
\begin{eqnarray}
\mathcal{F}
\left(\beta,t\right) & = & \left|\beta\right|^{-\alpha't\left(1-\left|\beta\right|\right)},
\label{eq:tdependence}
\end{eqnarray}
where the parameter $\alpha'=1\;\mathrm{GeV}^{-2}$. The summation of quark flavors in 
Eq. (\ref{eq:survivinghelicityamplitudesWACS}) gives an overall charge factor of $5/9$. 

\begin{figure}
\centerline{\epsfxsize=5in\epsffile{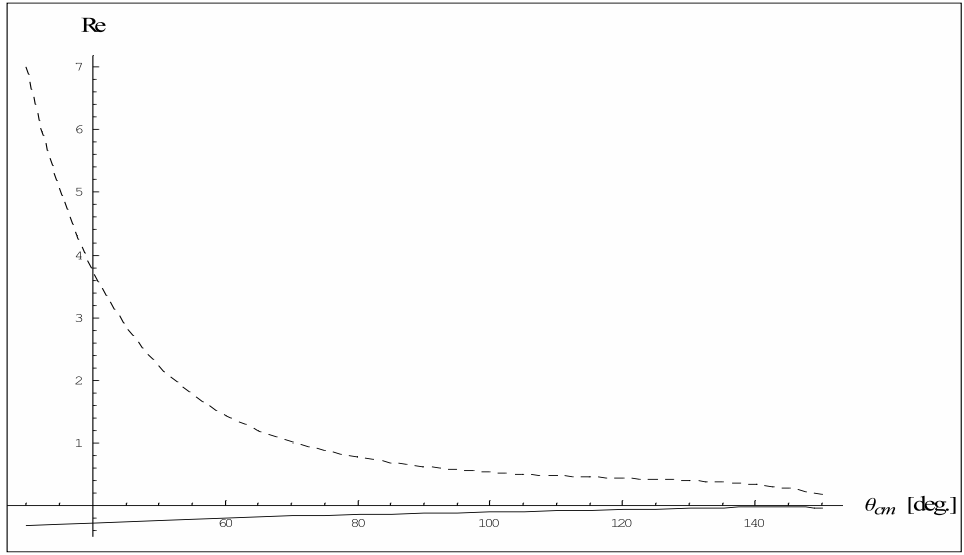}}
\caption[Real part of the helicity amplitudes $H_{\mathit{twist-2}}^{\left(1,1\right)}$ (solid line) 
and $H_{\mathit{twist-2}}^{\left(2,2\right)}$ (dashed line) plotted as a function of the center-of-mass
scattering angle $\theta_{cm}$ for the invariant $s=5\;\mathrm{GeV}^{2}$.]
{Real part of the helicity amplitudes $H_{\mathit{twist-2}}^{\left(1,1\right)}$ (solid line) and 
$H_{\mathit{twist-2}}^{\left(2,2\right)}$ (dashed line) plotted as a function of the center-of-mass 
scattering angle $\theta_{cm}$ for the invariant $s=5\;\mathrm{GeV}^{2}$. The amplitudes are divided 
by the charge factor $5/9$.}
\label{wacshelicityreal5}
\end{figure}

\begin{figure}
\centerline{\epsfxsize=5in\epsffile{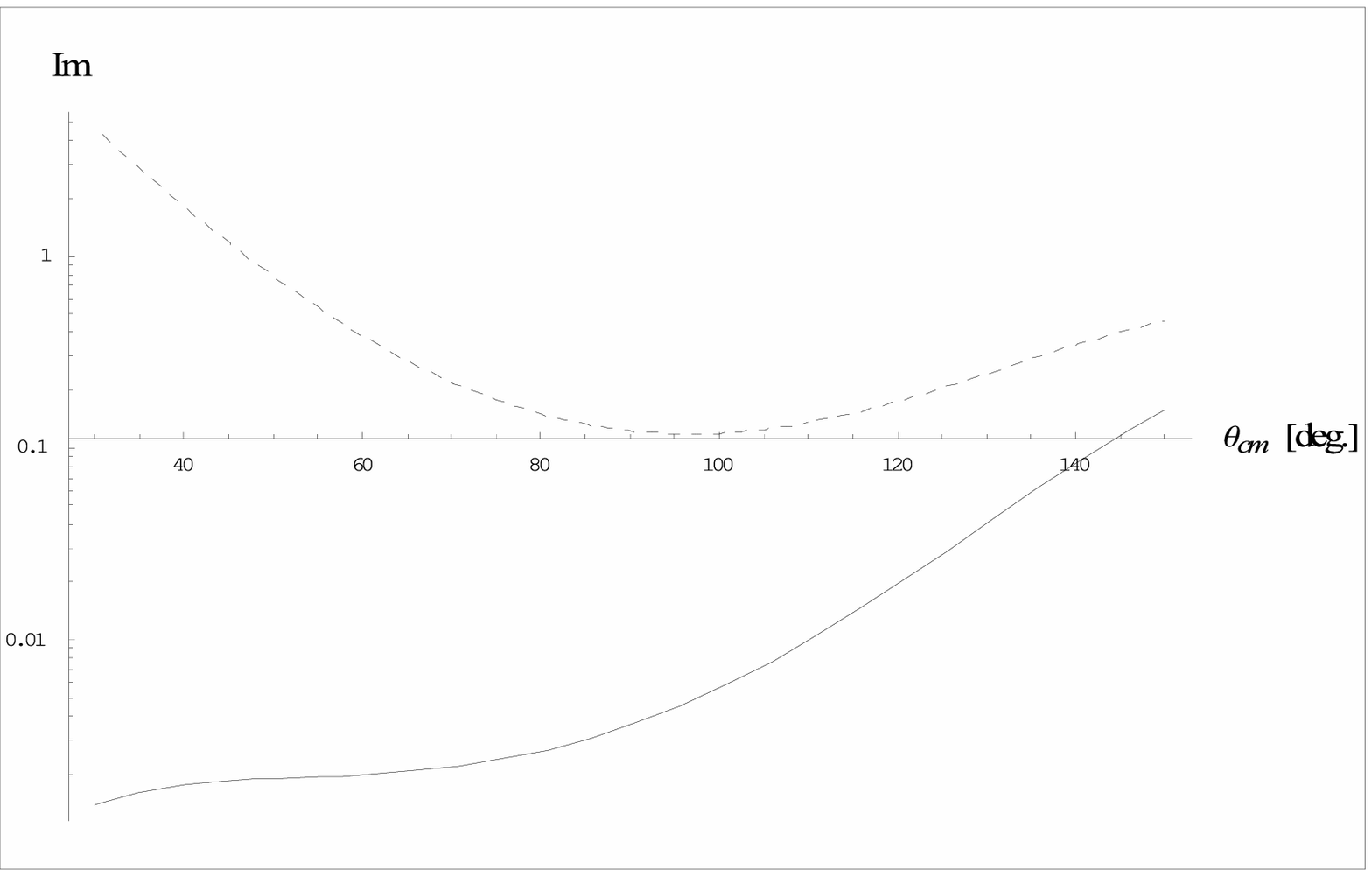}}
\caption[Imaginary part of the helicity amplitudes $H_{\mathit{twist-2}}^{\left(1,1\right)}$ (solid line) 
and $H_{\mathit{twist-2}}^{\left(2,2\right)}$ (dashed line) plotted as a function of the center-of-mass
scattering angle $\theta_{cm}$ for the invariant $s=5\;\mathrm{GeV}^{2}$.]
{Imaginary part of the helicity amplitudes $H_{\mathit{twist-2}}^{\left(1,1\right)}$ (solid line) and 
$H_{\mathit{twist-2}}^{\left(2,2\right)}$ (dashed line) plotted as a function of the center-of-mass 
scattering angle $\theta_{cm}$ for the invariant $s=5\;\mathrm{GeV}^{2}$. The amplitudes are divided by 
the charge factor $5/9$.}
\label{wacshelicityimag5}
\end{figure}

\begin{figure}
\centerline{\epsfxsize=5in\epsffile{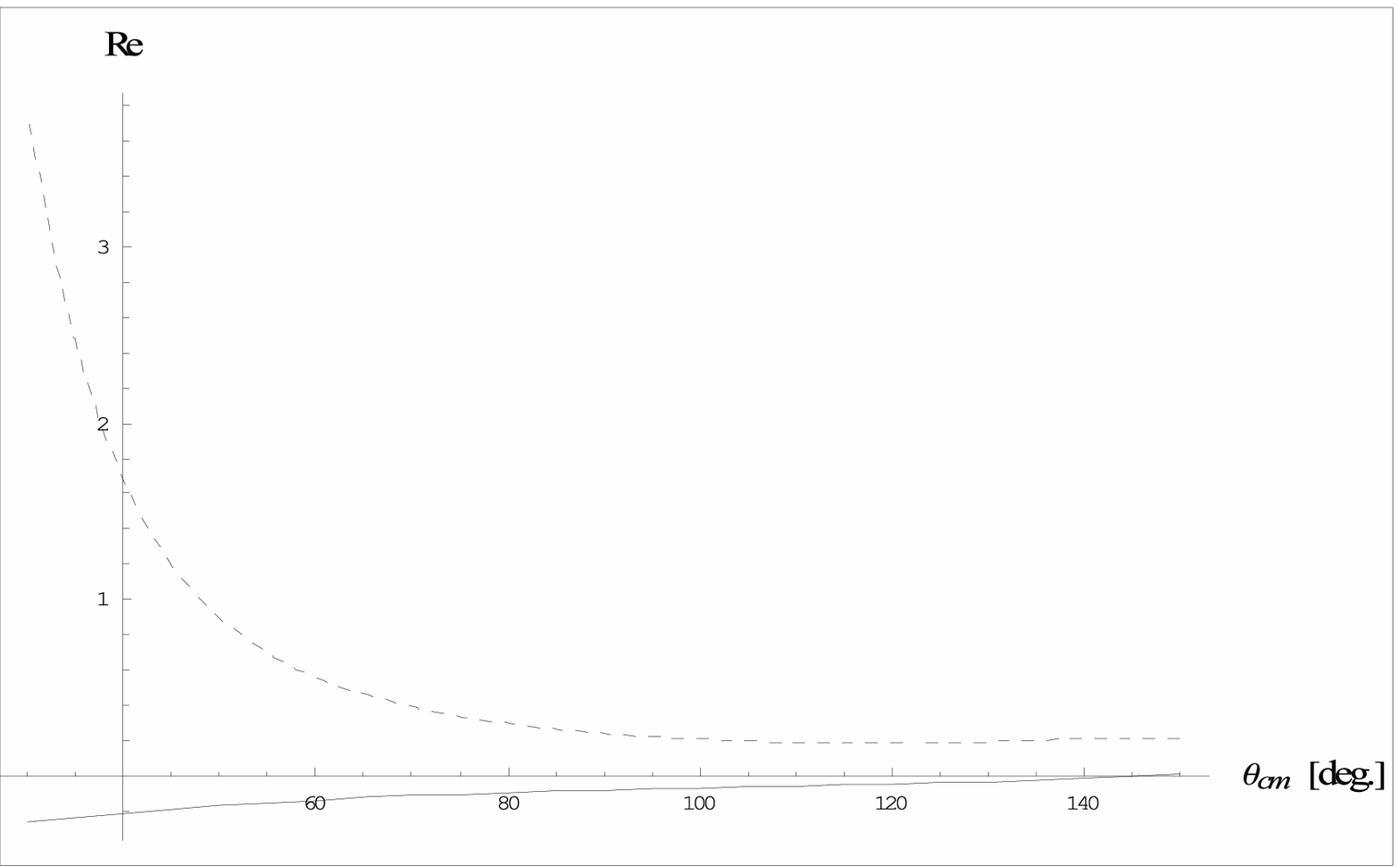}}
\caption[Real part of the helicity amplitudes $H_{\mathit{twist-2}}^{\left(1,1\right)}$ (solid line) 
and $H_{\mathit{twist-2}}^{\left(2,2\right)}$ (dashed line) plotted as a function of the center-of-mass
scattering angle $\theta_{cm}$ for the invariant $s=10\;\mathrm{GeV}^{2}$.] 
{Real part of the helicity amplitudes $H_{\mathit{twist-2}}^{\left(1,1\right)}$ (solid line) 
and $H_{\mathit{twist-2}}^{\left(2,2\right)}$ (dashed line) plotted as a function of the center-of-mass 
scattering angle $\theta_{cm}$ for the invariant $s=10\;\mathrm{GeV}^{2}$. The amplitudes are divided by 
the charge factor $5/9$.}
\label{wacshelicityreal10}
\end{figure}

\begin{figure}
\centerline{\epsfxsize=5in\epsffile{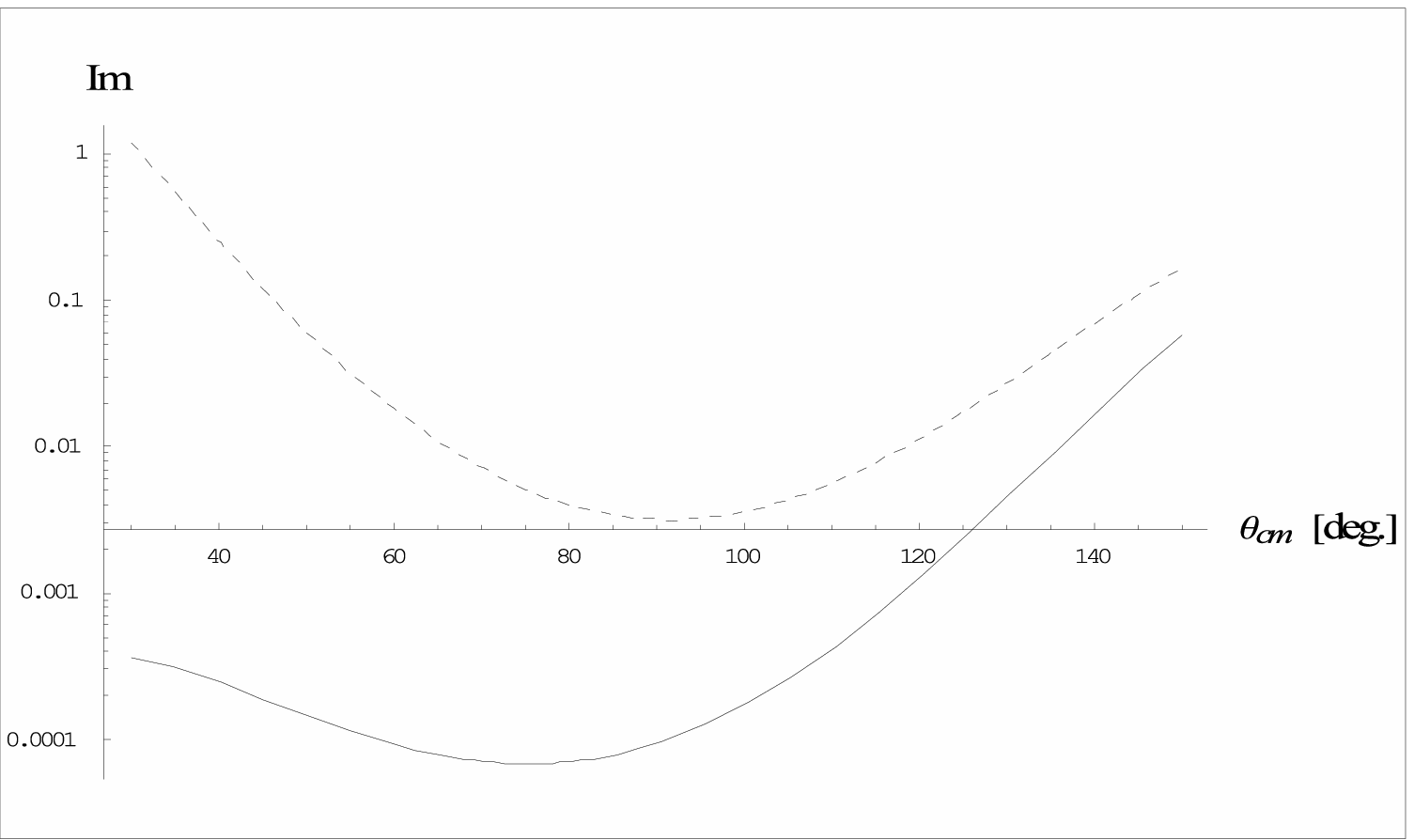}}
\caption[Imaginary part of the helicity amplitudes $H_{\mathit{twist-2}}^{\left(1,1\right)}$ (solid line) 
and $H_{\mathit{twist-2}}^{\left(2,2\right)}$ (dashed line) plotted as a function of the center-of-mass
scattering angle $\theta_{cm}$ for the invariant $s=10\;\mathrm{GeV}^{2}$.]
{Imaginary part of the helicity amplitudes $H_{\mathit{twist-2}}^{\left(1,1\right)}$ (solid line) 
and $H_{\mathit{twist-2}}^{\left(2,2\right)}$ (dashed line) plotted as a function of the center-of-mass 
scattering angle $\theta_{cm}$ for the invariant $s=10\;\mathrm{GeV}^{2}$. The amplitudes are divided by 
the charge factor $5/9$.}
\label{wacshelicityimag10}
\end{figure}

\begin{figure}
\centerline{\epsfxsize=5in\epsffile{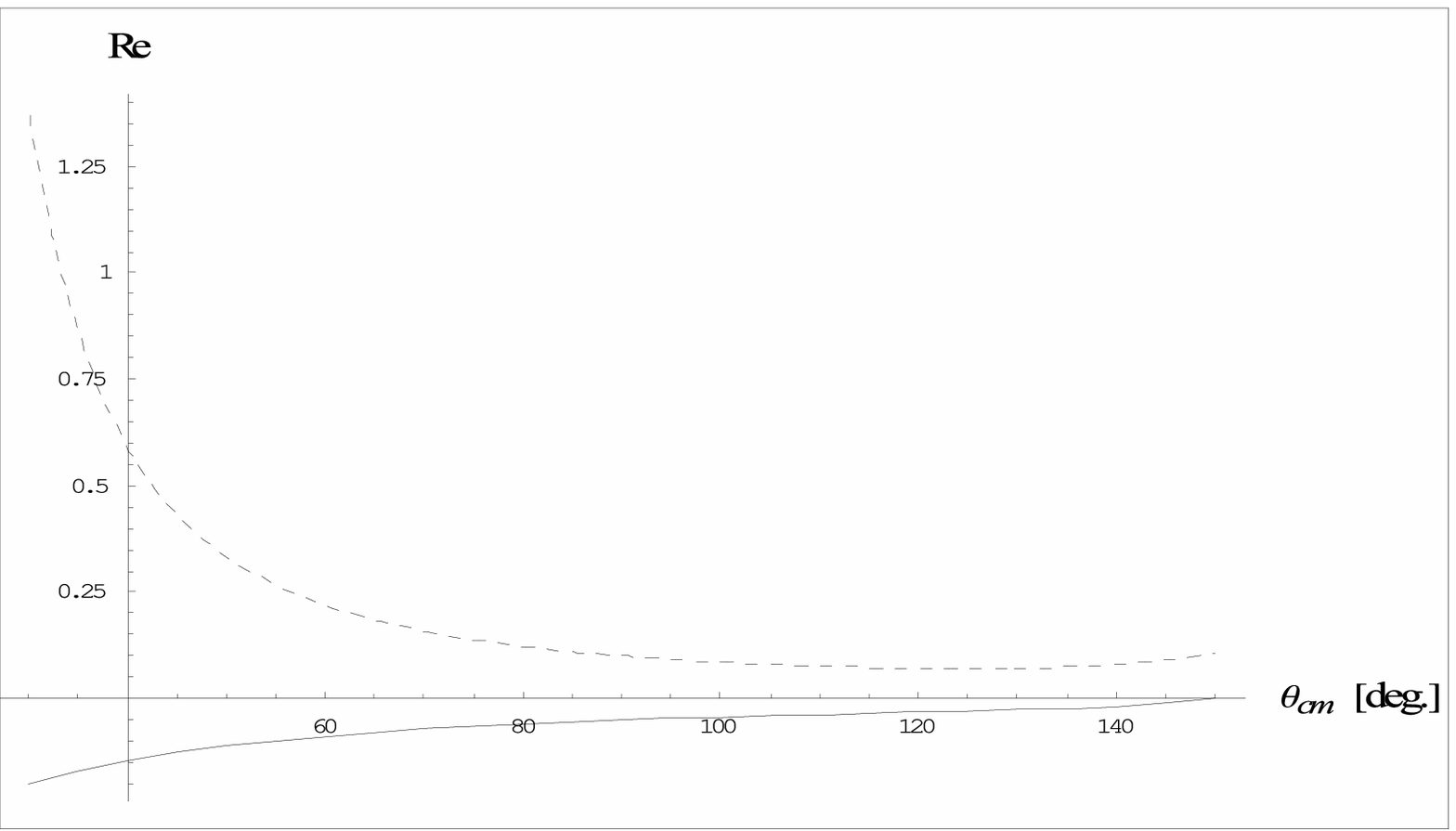}}
\caption[Real part of the helicity amplitudes $H_{\mathit{twist-2}}^{\left(1,1\right)}$ (solid line) 
and $H_{\mathit{twist-2}}^{\left(2,2\right)}$ (dashed line) plotted as a function of the center-of-mass
scattering angle $\theta_{cm}$ for the invariant $s=20\;\mathrm{GeV}^{2}$.]
{Real part of the helicity amplitudes $H_{\mathit{twist-2}}^{\left(1,1\right)}$ (solid line) and 
$H_{\mathit{twist-2}}^{\left(2,2\right)}$ (dashed line) plotted as a function of the center-of-mass 
scattering angle $\theta_{cm}$ for the invariant $s=20\;\mathrm{GeV}^{2}$. The amplitudes are divided by 
the charge factor $5/9$.}
\label{wacshelicityreal20}
\end{figure}

\begin{figure}
\centerline{\epsfxsize=5in\epsffile{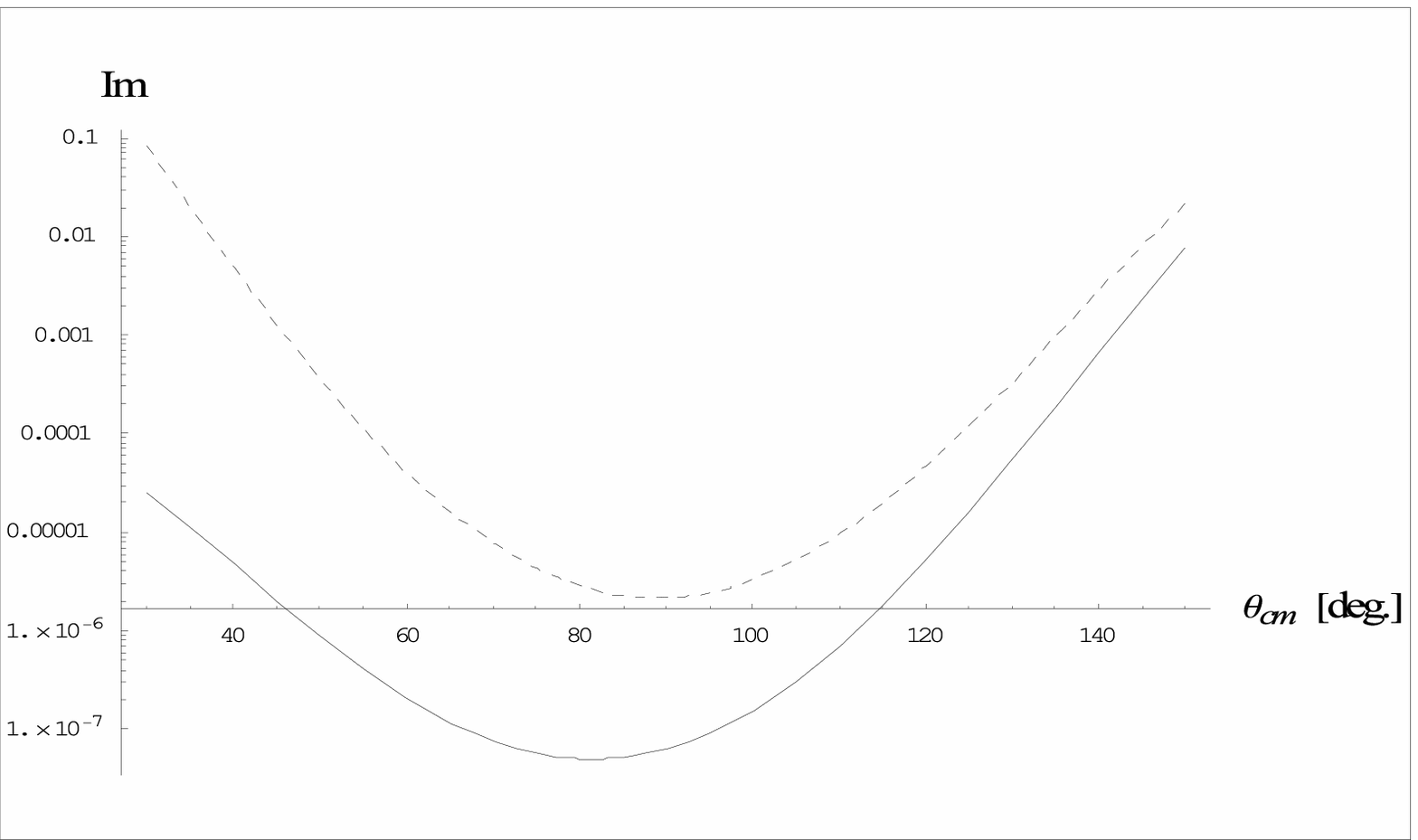}}
\caption[Imaginary part of the helicity amplitudes $H_{\mathit{twist-2}}^{\left(1,1\right)}$ (solid line) 
and $H_{\mathit{twist-2}}^{\left(2,2\right)}$ (dashed line) plotted as a function of the center-of-mass
scattering angle $\theta_{cm}$ for the invariant $s=20\;\mathrm{GeV}^{2}$.]
{Imaginary part of the helicity amplitudes $H_{\mathit{twist-2}}^{\left(1,1\right)}$ (solid line) and 
$H_{\mathit{twist-2}}^{\left(2,2\right)}$ (dashed line) plotted as a function of the center-of-mass 
scattering angle $\theta_{cm}$ for the invariant $s=20\;\mathrm{GeV}^{2}$. The amplitudes are divided by 
the charge factor $5/9$.}
\label{wacshelicityimag20}
\end{figure}

We calculate separately the \emph{s}- and \emph{u}-channel contributions in the helicity amplitudes. 
Using Eq. (\ref{eq:hardpropagator1}) and making the change in the variable $\beta\rightarrow-\beta$ 
in the \emph{u}-channel, we have for the helicity amplitude $H_{\mathit{twist-2}}^{\left(1,1\right)}$
\begin{eqnarray}
H_{\mathit{twist-2}}^{\left(1,1\right)} & = & \frac{5}{9}\int_{0}^{1}d\beta\;
\frac{1}{\beta}\frac{3}{4}\frac{\left(1-\beta\right)}{\sqrt{\beta}}
\beta^{\alpha'\left|t\right|\left(1-\beta\right)}\nonumber \\
&  & \times\left[1+\frac{\left(\beta^{2}-1\right)\left|t\right|/4}{\beta s+\left(1-\beta\right)^{2}
\left|t\right|/4}-\frac{4\beta^{2}us/\left|t\right|}{\beta s+\left(1-\beta\right)^{2}\left|t\right|/4}
\right]\nonumber \\
&  & +\frac{5}{9}\int_{0}^{1}d\beta\;\frac{1}{\beta}\frac{3}{4}
\frac{\left(1-\beta\right)}{\sqrt{\beta}}\beta^{\alpha'
\left|t\right|\left(1-\beta\right)}\nonumber \\
&  & \times\left[1+\frac{\left(\beta^{2}-1\right)\left|t
\right|/4}{\beta u+\left(1-\beta\right)^{2}\left|t\right|/4}-
\frac{4\beta^{2}us/\left|t\right|}{\beta u+\left(1-\beta\right)^{2}\left|t\right|/4}
\right],
\label{eq:H11twist2}
\end{eqnarray}
and after some manipulation
\begin{eqnarray}
H_{\mathit{twist-2}}^{\left(1,1\right)} & = & \frac{5}{9}\frac{3}{2}
\int_{0}^{1}d\beta\;\left(1-\beta\right)
\beta^{-1/2+\alpha'\left|t\right|\left(1-\beta\right)}\nonumber \\
&  & \times\left[\frac{\left(2a^{2}-1\right)\beta+a}{\beta^{2}+2a\beta+1}+
\frac{\left(2a^{2}-1\right)\beta-a}{\beta^{2}-2a\beta+1}\right],
\label{eq:H11twist2final}
\end{eqnarray}
where the parameter $a=2s/\left|t\right|-1\geq1$ depends only on the ratio $s/\left|t\right|$. 
The first term in the square brackets of Eq. (\ref{eq:H11twist2final}) corresponds to 
the \emph{s}-channel contribution. For the moderate invariant momentum transfer, the term 
has no singularities in the region $0\leq\beta\leq1$, in other words, the \emph{s}-channel 
contribution is real. The situation is, however, different with the second term in the square brackets, 
which represents the \emph{u}-channel contribution. Its denominator has the roots 
$\beta_{1,2}=a\pm\sqrt{a^{2}-1}$ (in fact, the denominator is equal to 
$\beta^{2}-2a\beta+1+i\epsilon=\left(\beta-\beta_{1}+i\epsilon\right)\left(\beta-\beta_{2}-i\epsilon\right)$) 
and, unlike the \emph{s}-channel, there is one relevant pole at $\beta=\beta_{2}+i$. The \emph{u}-channel 
contribution has therefore both the real and imaginary parts. They can be evaluated in the complex plane 
by making use of the residue theorem in the following way: one introduces a new variable, 
$\gamma=\beta-1/2$, which shifts the integration region from $\left[0,1\right]$ to $\left[-1/2,1/2\right]$. 
Then by closing the contour in the lower half-plane we find
\begin{eqnarray}
\int_{-1/2}^{1/2}d\gamma\;\left\{ ...\right\}  & = & \int_{\pi}^{2\pi}d\gamma\;
\left\{ ...\right\}, 
\label{eq:contourintegral}
\end{eqnarray}
with $\gamma=\left(1/2\right)\exp\left(i\phi\right)$. In summary, the helicity amplitudes 
$H_{\mathit{twist-2}}^{\left(1,1\right)}$ has both the real and imaginary parts, where the latter is 
generated solely by the corresponding \emph{u}-channel contribution. In the similar way, we compute 
the remaining two helicity amplitudes,
\begin{eqnarray}
H_{\mathit{twist-2}}^{\left(2,2\right)} & = & \frac{5}{9}
\frac{3}{2}\int_{0}^{1}d\beta\;\left(1-\beta\right)
\beta^{-1/2+\alpha'\left|t\right|\left(1-\beta\right)}\nonumber \\
&  & \times\left[\frac{\beta+a}{\beta^{2}+2a\beta+1}+
\frac{\beta-a}{\beta^{2}-2a\beta+1}\right]
\label{eq:H22twist2final}
\end{eqnarray}
and
\begin{eqnarray}
H_{\mathit{twist-3}}^{\left(1,1\right)} & = & \frac{5}{9}3\left(a^{2}-1\right)
\int_{0}^{1}d\beta\;\left(1-\beta\right)
\beta^{1/2+\alpha'\left|t\right|\left(1-\beta\right)}\nonumber \\
&  & \times\left[\frac{1}{\left(\beta^{2}+2a\beta+1\right)^{2}}+
\frac{1}{\left(\beta^{2}-2a\beta+1\right)^{2}}\right].
\label{eq:H11twist3final}
\end{eqnarray}
For the sake of simplicity, we have neglected terms $k_{1}^{2}$ and $k_{2}^{2}$ in the denominators 
of $H_{\mathit{twist-3}}^{\left(1,1\right)}$, see Eq. (\ref{eq:survivinghelicityamplitudesWACS}). 
Then the integral over \emph{v} simply gives $1/\left(k+q\right)^{4}$. At present, we only 
consider the twist-2 helicity amplitudes. In Figs. \ref{wacshelicityreal5} and \ref{wacshelicityimag5}, 
their real and imaginary parts are plotted against the center-of-mass scattering angle $\theta_{cm}$ 
rather than against $-t$ (recall that $t=-s\sin^{2}\left(\theta_{cm}/2\right)$) in the region of 
sufficiently large invariant $s=5\;\mathrm{GeV}^{2}$. We divide the amplitudes by $5/9$ to remove 
the overall charge factor. Analogous plots are presented in Figs. \ref{wacshelicityreal10} 
and \ref{wacshelicityimag10} for $s=10\;\mathrm{GeV}^{2}$, and in Figs. \ref{wacshelicityreal20} 
and \ref{wacshelicityimag20} for $s=20\;\mathrm{GeV}^{2}$. Due to the presence of the term 
$4\beta^{2}\left(us/t\right)/\left(k+q\right)^{2}$ in the amplitude $H_{twist-2}^{\left(1,1\right)}$, 
one might naively expect that the real part of $H_{twist-2}^{\left(1,1\right)}$ would dominate over the 
real part of $H_{twist-2}^{\left(2,2\right)}$. Both the \emph{s}-channel and the real part of 
the \emph{u}-channel contributions in $H_{twist-2}^{\left(1,1\right)}$ are indeed large, however, they 
come with the opposite signs, and as a result, the real part of the amplitude 
$H_{twist-2}^{\left(1,1\right)}$ diminishes, as oppose to the amplitude $H_{twist-2}^{\left(2,2\right)}$, 
in which both terms add up. Another observation is that the imaginary parts of the helicity amplitudes 
are suppressed with respect to their real parts.

\begin{figure}
\centerline{\epsfxsize=5in\epsffile{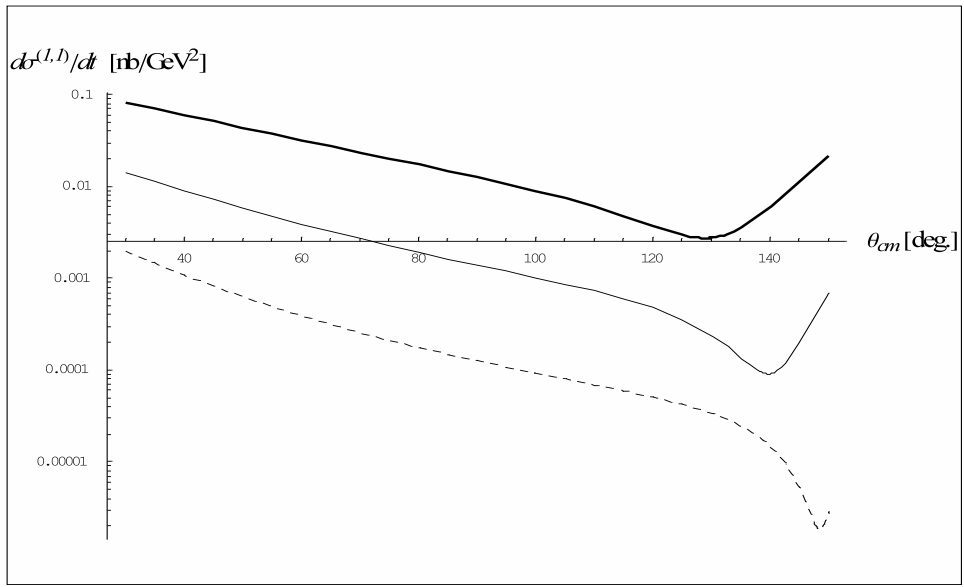}}
\caption{Twist-2 cross section $\sigma^{\left(1,1\right)}$ plotted as a function of the center-of-mass
scattering angle $\theta_{cm}$ for the invariant $s=5\;\mathrm{GeV}^{2}$ (bold solid line), 
$s=10\;\mathrm{GeV}^{2}$ (solid line) and $s=20\;\mathrm{GeV}^{2}$ (dashed line).}
\label{sigma11}
\end{figure}

\begin{figure}
\centerline{\epsfxsize=5in\epsffile{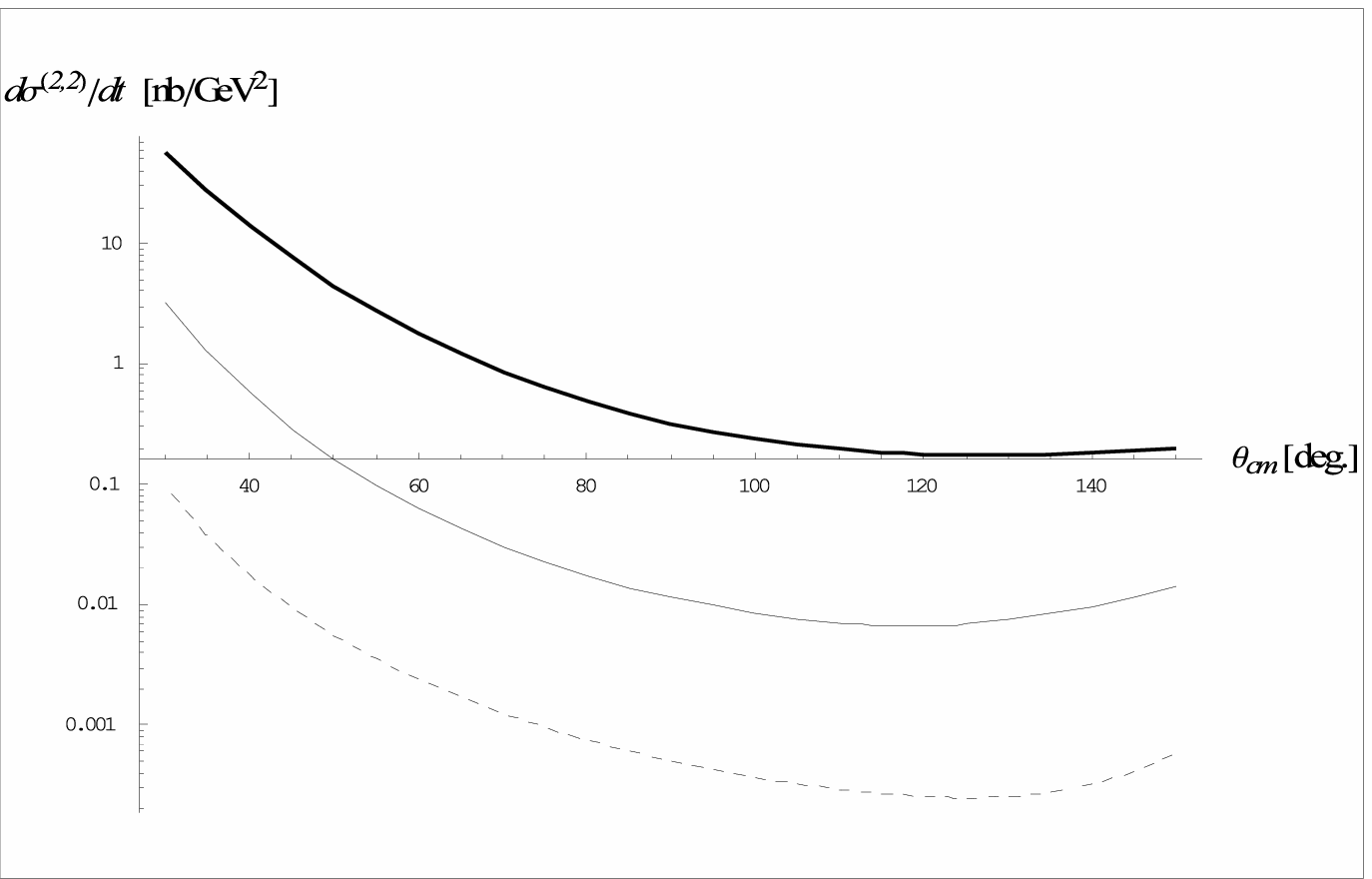}}
\caption{Twist-2 cross section $\sigma^{\left(2,2\right)}$ plotted as a function of the center-of-mass
scattering angle $\theta_{cm}$ for the invariant $s=5\;\mathrm{GeV}^{2}$ (bold solid line), 
$s=10\;\mathrm{GeV}^{2}$ (solid line) and $s=20\;\mathrm{GeV}^{2}$ (dashed line).}
\label{sigma22}
\end{figure}

\begin{figure}
\centerline{\epsfxsize=5in\epsffile{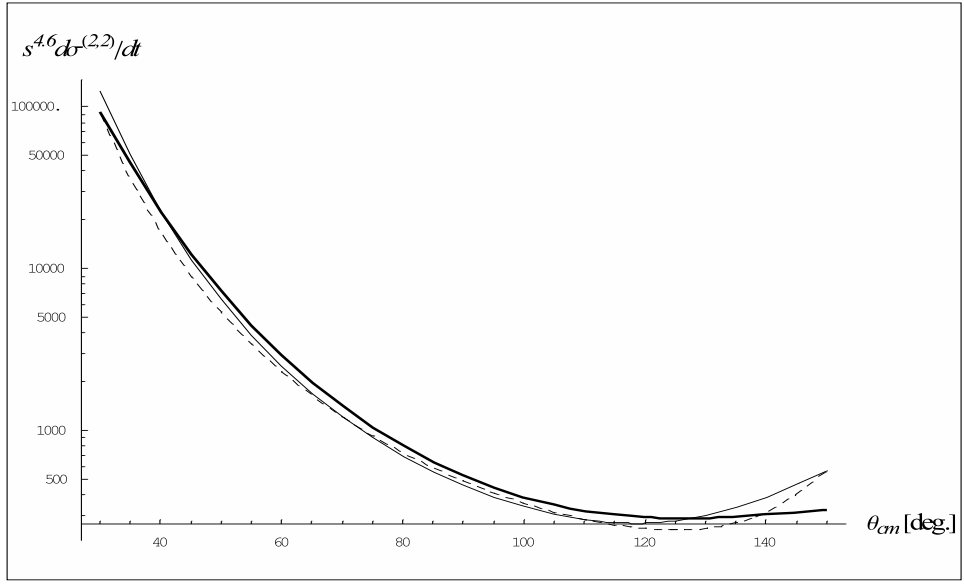}}
\caption{Combination $s^{4.6}d\sigma^{\left(2,2\right)}/dt$ plotted as a function of the center-of-mass
scattering angle $\theta_{cm}$ for the invariant $s=5\;\mathrm{GeV}^{2}$ (bold solid line), 
$s=10\;\mathrm{GeV}^{2}$ (solid line) and $s=20\;\mathrm{GeV}^{2}$ (dashed line).}
\label{wacs22scaling}
\end{figure}

We can now estimate the polarized differential cross sections 
in our simple model for the pion DD $h_{f}\left(\beta,\alpha,t\right)$. In the center-of-mass frame, 
we write
\begin{eqnarray}
d\sigma^{\left(\lambda_{1},\lambda_{2}\right)}
\left(\gamma\pi^{+}\rightarrow\gamma\pi^{+}\right) & = & 
\frac{1}{2E_{1}2\omega_{1}v_{\mathrm{rel}}}\left(2\pi\right)^{4}
\delta^{\left(4\right)}\left(p_{1}+q_{1}-p_{2}-q_{2}\right)\nonumber \\
&  & \times\frac{d\vec{p}_{2}}{\left(2\pi\right)^{3}2E_{2}}
\frac{d\vec{q}_{2}}{\left(2\pi\right)^{3}2\omega_{2}}
\left|e^{2}H^{\left(\lambda_{1},\lambda_{2}\right)}
\right|^{2}.
\label{eq:polarizedsigmaComptonQED1}
\end{eqnarray}
Note that we multiplied the helicity amplitude with $e^2$, which has been conveniently skipped in our analysis. 
Partially evaluating the integrals over the final-state momenta, e.g. integration over 
$\vec{q}_{2}$ kills $\delta^{\left(3\right)}\left(\vec{p}_{1}+\vec{q}_{1}-\vec{p}_{2}-\vec{q}_{2}\right)$ 
and sets $\vec{q}_{2}=-\vec{p}_{2}$, yields at high energies (and thus $v_{\mathrm{rel}}=2$)
\begin{eqnarray}
d\sigma^{\left(\lambda_{1},\lambda_{2}\right)}
\left(\gamma\pi^{+}\rightarrow\gamma\pi^{+}\right) & = & 
\frac{1}{2s}\frac{1}{4}\frac{1}{\left(2\pi\right)^{2}}\;\delta
\left(2E_{1}-2E_{2}\right)dE_{2}d\Omega\left|e^{2}H^{\left(\lambda_{1},\lambda_{2}\right)}
\right|^{2},\nonumber \\
\label{eq:polarizedsigmaComptonQED2}
\end{eqnarray}
where $s=E_{\mathrm{cm}}^{2}$ and $E_{1}=E_{\mathrm{cm}}/2$. The differential cross section can be 
further simplified into
\begin{eqnarray}
\frac{d^{2}\sigma^{\left(\lambda_{1},\lambda_{2}\right)}
\left(\gamma\pi^{+}\rightarrow\gamma\pi^{+}\right)}{d\Omega} & = & 
\frac{e^{4}}{64\pi^{2}s}\left|H^{\left(\lambda_{1},\lambda_{2}\right)}
\right|^{2},
\label{eq:polarizedsigmaComptonQED3}
\end{eqnarray}
which can be written in the invariant form as
\begin{eqnarray}
\frac{d\sigma^{\left(\lambda_{1},\lambda_{2}\right)}}{dt} & = & 
\frac{e^{4}}{16\pi s^{2}}\left|H^{\left(\lambda_{1},\lambda_{2}\right)}
\right|^{2}.
\label{sigmafinal}
\end{eqnarray}
The $\theta_{cm}$-dependence of the twist-2 cross sections $\sigma^{\left(1,1\right)}$ and 
$\sigma^{\left(2,2\right)}$  are illustrated on a logarithmic scale for all three values of the 
invariant \emph{s} in Figs. \ref{sigma11} and \ref{sigma22}, respectively. Similarly to the situation 
with the nucleon target (see, e.g. \cite{Radyushkin:1998rt}), the plots reveal a slight increase in 
both polarized cross sections at large scattering angles, $\theta_{cm}\geq130^{0}$. Moreover, 
by comparing their orders of magnitude, we find that the cross section $\sigma^{\left(1,1\right)}$ 
is significantly suppressed compared to $\sigma^{\left(2,2\right)}$. Next, the angular dependence of the 
combination $s^{4.6}d\sigma^{\left(2,2\right)}/dt$ is shown in Fig. \ref{wacs22scaling}. The curves 
obey scaling behavior, i.e. they basically coincide up to $\theta_{cm}\simeq100^{0}$. 

\section{\ \,  Compton Scattering on a Pion in QED}

\begin{figure}
\centerline{\epsfxsize=6in\epsffile{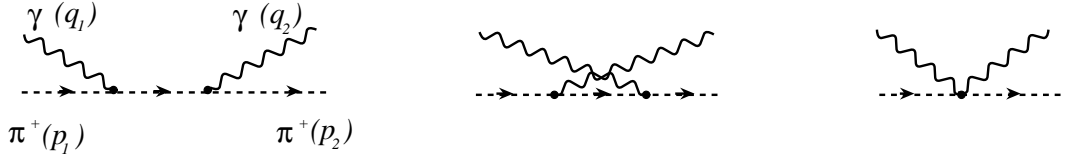}}
\caption{Feynman diagrams for Compton scattering on a pion in QED.}
\label{pionqed}
\end{figure}

We compare the QCD results of Section VIII.3 with the QED calculation. Assuming that 
$\pi^{+}$ is an elementary particle we have, in the lowest order in $\alpha_{EM}$, 
three relevant Feynman diagrams, see Fig. \ref{pionqed}, namely, 
the \emph{s}- and \emph{u}-channel tree diagrams and the four-point contact 
interaction diagram. Adding them coherently, one finds for the T-matrix
\begin{eqnarray}
i\mathrm{T} & = & \left[i\left|e\right|\left(2p_{2}+q_{2}\right)^{\mu}
\frac{i}{\left(p_{1}+q_{1}\right)^{2}-m_{\pi}^{2}}i\left|e\right|\left(2p_{1}+q_{1}
\right)^{\nu}\right.\nonumber \\
&  & +i\left|e\right|\left(2p_{2}-q_{1}\right)^{\nu}
\frac{i}{\left(p_{2}-q_{1}\right)^{2}-m_{\pi}^{2}}i\left|e\right|
\left(2p_{1}-q_{2}\right)^{\mu}\nonumber \\
&  & +2i\left|e\right|^{2}g^{\mu\nu}\Bigg]\epsilon_{1\nu}
\epsilon_{2\mu}^{*}.
\label{eq:TmatrixComptonQED1}
\end{eqnarray}
The expression in the brackets corresponds to the Compton scattering amplitude 
$\mathsf{\mathsf{\mathcal{\mathcal{T}}}}^{\mu\nu}$ in QED. We can immediately check its 
gauge invariance and find
\begin{eqnarray}
\mathsf{\mathsf{\mathcal{T}}}^{\mu\nu}q_{1\nu} & = & -ie^{2}
\left\{ \frac{\left(2p_{2}+q_{2}\right)^{\mu}\left[2\left(p_{1}\cdot q_{1}
\right)+q_{1}^{2}\right]}{2\left(p_{1}\cdot q_{1}\right)}+
\frac{\left(2p_{1}-q_{2}\right)^{\mu}\left[2\left(p_{2}\cdot q_{1}\right)-q_{1}^{2}
\right]}{-2\left(p_{1}\cdot q_{1}\right)}-2q_{1}^{\mu}\right\} \nonumber \\
& = & -2ie^{2}\left(p_{2}+q_{2}-p_{1}-q_{1}\right)^{\mu}\nonumber \\
& = & 0.
\label{eq:gaugeinvarianceqedcomptonqed}
\end{eqnarray}
Similarly, we prove that $q_{2\mu}\mathsf{\mathsf{\mathcal{\mathcal{T}}}}^{\mu\nu}=0$.

Neglecting now the pion mass in Eq. (\ref{eq:TmatrixComptonQED1}) 
(recall that $s,-t,-u\gg m_{\pi}^{2}$ and hence $s+t+u=2m_{\pi}^{2}\simeq0$), we get
\begin{eqnarray}
\mathrm{T} & = & -2e^{2}
\left[\frac{2\left(p_{1}\cdot\epsilon_{1}\right)
\left(p_{2}\cdot\epsilon_{2}^{*}\right)}{s}+
\frac{2\left(p_{1}\cdot\epsilon_{2}^{*}\right)\left(p_{2}\cdot
\epsilon_{1}\right)}{u}-
\left(\epsilon_{1}\cdot
\epsilon_{2}^{*}\right)\right].
\label{eq:TmatrixComtponQED2}
\end{eqnarray}
By writing the initial and final pion momenta as $p_{1}=p+r/2$ and $p_{2}=p-r/2$, respectively, 
and imposing $\left(r\cdot\epsilon_{1}\right)=\left(r\cdot\epsilon_{2}^{*}\right)=0$, we can 
express the T-matrix in terms of the average pion momentum, 
\begin{eqnarray}
\mathrm{T} & = & -2e^{2}\left[2\left(\frac{1}{s}+
\frac{1}{u}\right)\left(p\cdot\epsilon_{1}\right)\left(p\cdot\epsilon_{2}^{*}
\right)-\left(\epsilon_{1}\cdot\epsilon_{2}^{*}\right)\right].
\label{eq:TmatrixComtponQED4}
\end{eqnarray}
Finally, the projection of polarization vectors $\epsilon_{1}$ and $\epsilon_{2}$ onto 
$e_{1}$ and $e_{2}$ gives the following photon helicity amplitudes in QED
\begin{eqnarray}
H^{\left(1,1\right)} & = & -2e^{2}\left[2\left(\frac{1}{s}+
\frac{1}{u}\right)\frac{us}{t}+1\right]\nonumber \\
& = & 2e^{2},\nonumber \\
H^{\left(1,2\right)} & = & 0,\nonumber \\
H^{\left(2,1\right)} & = & 0,\nonumber \\
H^{\left(2,2\right)} & = & -2e^{2}.
\label{eq:helicityamplitudesComptonQED}
\end{eqnarray}
The QED helicity amplitudes $H^{\left(1,1\right)}$ and $H^{\left(2,2\right)}$ are opposite in sign 
but constant. By comparing them to the real parts of the QCD helicity amplitudes 
$H_{twist-2}^{\left(1,1\right)}$ and $H_{twist-2}^{\left(2,2\right)}$ (and, as usual, ignoring $-e^2$), 
we can establish the effective form factor, which solely describes the underlying QCD dynamics.

In summary, we presented an improved treatment of the DD formalism to the case with the pion target.
In addition to deriving the amplitude in the twist-3 approximation, the helicity properties were 
considered. We gave predictions for the polarized cross sections with the use of the simple model 
for the pion double distribution. 

\chapter{Conclusions and Outlook}

In this dissertation we have studied inclusive and exclusive Compton processes in quantum 
chromodynamics. Let us summarize our main conclusions.

First, we briefly reviewed the crown example of the hard (light-cone dominated) inclusive 
scattering process in QCD known as deeply inelastic lepton-nucleon scattering. 
We introduced the forward virtual Compton scattering amplitude and discussed its relation 
to DIS. We presented the cross sections results in terms of structure functions for both 
the electromagnetic and weak probes. The last part of Chapter II was devoted to the QCD 
parton model. This simple model, which naturally emerges from the QCD operator product 
expansion, successfully describes the scaling property of structure functions, and further 
allows to express them in terms of the parton distributions functions.

In Chapter III, we started with the systematic description of old phenomenological functions 
(form factors, parton distribution functions and distribution amplitudes) that were used 
for years in the studies of hadrons. New phenomenological functions (generalized parton 
distributions) became necessary in order to generalize the Compton scattering amplitude 
in the region of nonforward (skewed) kinematics. They were introduced in various ways and 
under different names as nonperturbative functions describing the soft (long-distance) part 
of the factorized scattering amplitudes in QCD. Generalized parton distributions accumulate 
the most complete information about the hadronic structure, and as such combine the features 
of form factors, usual parton distributions and distribution amplitudes. We presented some 
of their theoretical aspects. 

Generalized parton distributions have numerous applications to hard processes, in particular, 
to the exclusive ones. The pedagogical example in this respect is deeply virtual Compton 
scattering. In Chapter IV, we demonstrated the derivation of the amplitude at the 
leading twist-2 level in the DVCS kinematics, using the light-cone expansion in terms of QCD 
string operators in coordinate space. We investigated the electromagnetic gauge invariance of 
the amplitude. The latter is violated with respect to the initial virtual photon and therefore, 
higher twist corrections have to be taken into account. Furthermore, we discussed the kinematics 
and introduced a toy model for the nucleon off-forward parton distributions, which did not include 
the contribution from the sea quarks. We estimated the unpolarized cross sections in the 
kinematical region relevant to Jefferson Lab, and observed that unfortunately the contamination 
from the Bethe-Heitler process predominated the pure DVCS signal. The way out is to exploit the 
interference between the two processes. This approach allows to project out independently both 
the real and imaginary parts of the DVCS amplitude and probe different linear combinations of GPDs. 

In Chapter V, we studied the inclusive photoproduction of massive lepton pairs. The process was 
considered in the framework of the parton model at rather high photon beam energy. Again one 
deals with two types of subprocesses. We picked only the lowest order electromagnetic 
contributions and calculated both the Compton and Bethe cross sections. We found that in the 
forward direction the Compton contribution was slightly larger than the Bethe-Heitler 
contribution, however, it decreased much faster. In addition, we illustrated that the 
interference terms cancel in pairs, after being integrated over the momenta of final leptons. 

The inverse process to DVCS is known as time-like Compton scattering. It can be accessed 
through the photoproduction of a heavy lepton pair. In Chapter VI, we considered this 
particular reaction in the DVCS kinematics. The unpolarized cross section results revealed 
that the Compton contribution was significantly suppressed with respect to the Bethe-Heitler 
contribution. In analogy with the DVCS process, we expect that the interference 
terms between the Compton and Bethe-Heitler processes, in particular, from the experiments 
with the polarized photon beam, may play an important role in extracting new information on 
GPDs. Thus more future studies are needed in this direction.        

By utilizing the weak currents to probe the nucleon structure we are able to measure, due 
to the $V-A$ nature of interaction, a different combination of generalized parton 
distributions as well as the distributions that are nondiagonal in quark flavor. Hence 
neutrino-induced virtual Compton scattering, which has been studied in Chapter VII, provides 
an important tool to complement the study of GPDs in more familiar electron-induced DVCS 
or exclusive meson production processes. We derived the twist-2 amplitudes for the weak neutral 
and weak charged current interactions, and gave predictions for the cross sections in 
the kinematics relevant to future high-intensity neutrino experiments. At small scattering 
angles, we observed that, unlike the standard electromagnetic DVCS process, the Compton 
signal was enhanced compared to the corresponding Bethe-Heitler contribution. 
From the theoretical side, it is expected to use a more realistic model for the nucleon GPDs, 
and further check separately the contributions from the plus and minus distributions, 
in particularly, the contribution coming from the pion pole, i.e. the $\tilde{E}_{f}$ distribution. 
Moreover, one may estimate the contribution from the interference of the Compton with the 
Bethe-Heitler process, and extend the approach in order to include the twist-3 terms.

In the final chapter, the GPD formalism was applied to real Compton scattering. Again we used 
the same light-cone expansion technique as in Chapters IV and VII, however, in addition to the 
twist-2 operators we have also included a set of kinematical twist-3 operators. They appear as total 
derivatives of twist-2 operators. Unlike the previous chapters, we considered, for simplicity, the pion 
target. Instead of off-forward parton distributions, more general objects, namely, the double distributions 
were used to describe the nonperturbative stages of interaction. We obtained the expressions for the 
twist-2 and twist-3 Compton scattering amplitudes. It was shown, in the lowest order in the 
invariant momentum transfer, that the kinematical twist-3 contribution to the amplitude for pion was 
required to restore the tranversality. Next we found that a divergent part of the twist-3 amplitude had 
zero projection on the polarization vectors of initial and final photons, and calculated the photon 
helicity amplitudes, using a simple model for the pion double distribution. They were compared to 
QED helicity amplitudes. Finally, we estimated the relevant polarized cross sections, observed their 
approximate scaling behavior, and established the noncanonical powers. A straightforward generalization to 
the case with the nucleon target as well as the study of the exclusive production of two pions in two-photon 
collisions (the latter is a crossing process to Compton scattering on a pion) are in progress.

\appendix

\chapter{Glossary}

The following abbreviations are often used in a text:

\begin{eqnarray*}
\mathrm{DA} & .................... & \mathrm{distribution\; amplitude}\\
\mathrm{DD} & .................... & \mathrm{double\; distribution}\\
\mathrm{DIS} & .................... & \mathrm{deeply\; inelastic\; lepton\; scattering}\\
\mathrm{DMP} & .................... & \mathrm{deeply\; exclusive\; meson\; production}\\
\mathrm{DVCS} & .................... & \mathrm{deeply\; virtual\; Compton\; scattering}\\
\mathrm{GPD} & .................... & \mathrm{generalized\; parton\; distribution}\\
\mathrm{OPE} & .................... & \mathrm{operator\; product\; expansion}\\
\mathrm{PDF} & .................... & \mathrm{parton\; distribution\; function}\\
\mathrm{QCD} & .................... & \mathrm{quantum\; chromodynamics}\\
\mathrm{QED} & .................... & \mathrm{quantum\; electrodynamics}\\
\mathrm{TCS} & .................... & \mathrm{time\!-\!like\; Compton\; scattering}\\
\mathrm{VCA} & .................... & \mathrm{virtual\; Compton\; scattering\; amplitude}\\
\mathrm{WACS} & .................... & \mathrm{wide\!-\!angle\; real\; Compton\; scattering}\\
\end{eqnarray*}

\chapter{List of Integrals and Scalar Functions That Appear in the Bethe-Heitler Subprocess 
of the Inclusive Photoproduction of Lepton Pairs}

In Section V.3, we compute the double integral (\ref{eq:integratedleptonictensorbh2}). 
The calculation is carried out through the following set of integrals:
\begin{eqnarray}
\int d^{4}k\;\delta^{+}\left(\right)\delta^{+}\left(\right)
\frac{1}{\left(k\cdot q\right)} & = & I_{0},\nonumber \\
\int d^{4}k\;\delta^{+}\left(\right)\delta^{+}\left(\right)
\frac{k^{\mu}}{\left(k\cdot q\right)} & = & I_{1}q^{\mu}+I_{2}q'^{\mu},\nonumber \\
\int d^{4}k\;\delta^{+}\left(\right)\delta^{+}\left(\right)
\frac{k^{\mu}k^{\rho}}{
\left(k\cdot q\right)} & = & I_{3}q^{\mu}q^{\rho}+I_{4}q'^{\mu}q'^{\rho}\nonumber \\
&  & +I_{5}\left(q^{\mu}q'^{\rho}+q^{\rho}q'^{\mu}
\right)+I_{6}g^{\mu\rho},\nonumber \\
\int d^{4}k\;\delta^{+}\left(\right)\delta^{+}\left(\right)
\frac{1}{\left(q\cdot q'\right)-\left(k\cdot q\right)} & = & J_{0},\nonumber \\
\int d^{4}k\;\delta^{+}\left(\right)\delta^{+}\left(\right)
\frac{k^{\mu}}{\left(q\cdot q'\right)-
\left(k\cdot q\right)} & = & J_{1}q^{\mu}+J_{2}q'^{\mu},\nonumber \\
\int d^{4}k\;\delta^{+}\left(\right)\delta^{+}
\left(\right)\frac{k^{\mu}k^{\rho}}{\left(q\cdot q'\right)-
\left(k\cdot q\right)} & = & J_{3}q^{\mu}q^{\rho}+J_{4}q'^{\mu}q'^{\rho}\nonumber \\
&  & +J_{5}\left(q^{\mu}q'^{\rho}+q^{\rho}q'^{\mu}
\right)+J_{6}g^{\mu\rho},\nonumber \\
\int d^{4}k\;\delta^{+}\left(\right)\delta^{+}
\left(\right)\frac{1}{\left(k\cdot q\right)
\left[\left(q\cdot q'\right)-\left(k\cdot q\right)\right]} & = & 
\frac{1}{\left(q\cdot q'\right)}\left(I_{0}+J_{0}\right),\nonumber \\
\int d^{4}k\;\delta^{+}\left(\right)\delta^{+}\left(\right)
\frac{\left(k\cdot q'\right)}{\left(k\cdot q\right)
\left[\left(q\cdot q'\right)-\left(k\cdot q\right)\right]} & = & 
\frac{q'^{2}}{2\left(q\cdot q'\right)}\left(I_{0}+J_{0}\right),\nonumber \\
\int d^{4}k\;\delta^{+}\left(\right)\delta^{+}\left(\right)
\frac{\left(k\cdot q'\right)^{2}}{\left(k\cdot q\right)
\left[\left(q\cdot q'\right)-\left(k\cdot q\right)\right]} & = & 
\frac{q'^{4}}{4\left(q\cdot q'\right)}\left(I_{0}+J_{0}\right),\nonumber \\
\int d^{4}k\;\delta^{+}\left(\right)\delta^{+}\left(\right)
\frac{\left(k\cdot q\right)}{\left(q\cdot q'\right)-
\left(k\cdot q\right)} & = & \left(q\cdot q'\right)J_{2},\nonumber \\
\int d^{4}k\;\delta^{+}\left(\right)\delta^{+}\left(\right)
\frac{\left(k\cdot q'\right)k^{\mu}}{\left(k\cdot q\right)
\left[\left(q\cdot q'\right)-\left(k\cdot q\right)
\right]} & = & K_{1}q^{\mu}+K_{2}q'^{\mu},\nonumber \\
\int d^{4}k\;\delta^{+}\left(\right)\delta^{+}\left(\right)
\frac{\left(k\cdot q'\right)k^{\mu}k^{\rho}}{\left(k\cdot q\right)
\left[\left(q\cdot q'\right)-
\left(k\cdot q\right)\right]} & = & K_{3}q^{\mu}q^{\rho}+K_{4}q'^{\mu}q'^{\rho}\nonumber \\
&  & +K_{5}\left(q^{\mu}q'^{\rho}+q^{\rho}q'^{\mu}
\right)+K_{6}g^{\mu\rho},\nonumber \\
\int d^{4}k\;\delta^{+}\left(\right)\delta^{+}\left(\right)
\frac{1}{\left(k\cdot q\right)^{2}} & = & L_{0},\nonumber \\
\int d^{4}k\;\delta^{+}\left(\right)\delta^{+}\left(\right)
\frac{k^{\mu}}{\left(k\cdot q\right)^{2}} & = & L_{1}q^{\mu}+L_{2}q'^{\mu},\nonumber \\
\int d^{4}k\;\delta^{+}\left(\right)\delta^{+}
\left(\right)\frac{k^{\mu}k^{\rho}}{\left(k\cdot q
\right)^{2}} & = & L_{3}q^{\mu}q^{\rho}+L_{4}q'^{\mu}q'^{\rho}\nonumber \\
&  & +L_{5}\left(q^{\mu}q'^{\rho}+q^{\rho}q'^{\mu}
\right)+L_{6}g^{\mu\rho},\nonumber \\
\int d^{4}k\;\delta^{+}\left(\right)\delta^{+}
\left(\right)\frac{k^{\mu}}{\left(k\cdot q\right)
\left[\left(q\cdot q'\right)-\left(k\cdot q\right)
\right]} & = & M_{1}q^{\mu}+M_{2}q'^{\mu},\nonumber \\
\int d^{4}k\;\delta^{+}\left(\right)\delta^{+}\left(\right)
\frac{k^{\mu}k^{\rho}}{\left(k\cdot q\right)
\left[\left(q\cdot q'\right)-
\left(k\cdot q\right)\right]} & = & M_{3}q^{\mu}q^{\rho}+M_{4}q'^{\mu}q'^{\rho}
\nonumber \\
&  & +M_{5}\left(q^{\mu}q'^{\rho}+q^{\rho}q'^{\mu}
\right)+M_{6}g^{\mu\rho}.\nonumber\\
\label{eq:tableofintegralsappendix}
\end{eqnarray}
The abbreviation $\delta^{+}\left(\right)\delta^{+}\left(\right)
\equiv\delta^{+}\left(k^{2}-m^{2}\right)\delta^{+}\left[\left(q'-k\right)^{2}-m^{2}
\right]$ is being used together with the notation 
$\delta^{+}\left(k^{2}-m^{2}\right)\equiv\delta\left(k^{2}-m^{2}\right)
\theta\left(k_{0}\right)$,
where \emph{m} labels the lepton mass. The coefficients on the right-hand side 
of Eq. (\ref{eq:tableofintegralsappendix}) are scalar functions, which depend 
upon three invariants, i.e. $m^{2}$, $q'^{2}$ and 
$\left(q\cdot q'\right)=\left(q'^{2}-t\right)/2$, and can be
derived by contracting both sides of Eq. (\ref{eq:tableofintegralsappendix})
with the four-momenta $q_{\mu}$ and $q'_{\mu}$ of the initial real and final virtual photons, 
respectively, and with the tensors $q_{\mu}q_{\rho}$,
$q'_{\mu}q'_{\rho}$, $q'_{\mu}q_{\rho}$ and $g_{\mu\rho}$. These
functions can be further reduced to the expressions written in terms
of only two scalar functions, i.e. $I_{0}$ and $I_{2}$. After some algebra we obtain
\begin{eqnarray}
I_{0} & = & \frac{\pi}{2\left(q\cdot q'\right)}\ln
\left[\frac{1+\sqrt{1-4m^{2}/q'^{2}}}{1-\sqrt{1-4m^{2}/q'^{2}}}
\right],\nonumber \\
I_{1} & = & \frac{q'^{2}}{2\left(q\cdot q'\right)}
\left[I_{0}-2I_{2}\right],\nonumber \\
I_{2} & = & \frac{\pi}{2\left(q\cdot q'\right)}
\sqrt{1-\frac{4m^{2}}{q'^{2}}},\nonumber \\
I_{3} & = & \frac{q'^{2}}{2\left(q\cdot q'\right)^{2}}
\left[\left(\frac{q'^{2}}{2}+m^{2}\right)I_{0}-\frac{3q'^{2}}{2}I_{2}
\right],\nonumber \\
I_{4} & = & \frac{1}{2}I_{2},\nonumber \\
I_{5} & = & -\frac{1}{2\left(q\cdot q'\right)}\left[m^{2}I_{0}-
\frac{q'^{2}}{2}I_{2}\right],\nonumber \\
I_{6} & = & \frac{1}{2}\left[m^{2}I_{0}-\frac{q'^{2}}{2}I_{2}
\right],\nonumber \\
J_{0} & = & I_{0},\nonumber \\
J_{1} & = & -\frac{q'^{2}}{2\left(q\cdot q'\right)}\left[I_{0}-2I_{2}
\right],\nonumber \\
J_{2} & = & I_{0}-I_{2},\nonumber \\
J_{3} & = & \frac{q'^{2}}{2\left(q\cdot q'\right)^{2}}
\left[\left(\frac{q'^{2}}{2}+m^{2}\right)I_{0}-
\frac{3q'^{2}}{2}I_{2}\right],\nonumber \\
J_{4} & = & I_{0}-\frac{3}{2}I_{2},\nonumber \\
J_{5} & = & -\frac{1}{2\left(q\cdot q'\right)}
\left[\left(q'^{2}+m^{2}\right)I_{0}-
\frac{5q'^{2}}{2}I_{2}\right],\nonumber \\
J_{6} & = & \frac{1}{2}\left[m^{2}I_{0}-
\frac{q'^{2}}{2}I_{2}\right],\nonumber \\
K_{1} & = & 0,\nonumber \\
K_{2} & = & \frac{q'^{2}}{2\left(q\cdot q'\right)}I_{0},
\nonumber \\
K_{3} & = & \frac{q'^{4}}{2\left(q\cdot q'\right)^{3}}
\left[\left(\frac{q'^{2}}{2}+m^{2}\right)I_{0}-
\frac{3q'^{2}}{2}I_{2}\right],\nonumber \\
K_{4} & = & \frac{q'^{2}}{2\left(q\cdot q'\right)}
\left[I_{0}-I_{2}\right],\nonumber \\
K_{5} & = & -\frac{q'^{2}}{2\left(q\cdot q'\right)^{2}}
\left[\left(\frac{q'^{2}}{2}+m^{2}\right)I_{0}-
\frac{3q'^{2}}{2}I_{2}\right],\nonumber \\
K_{6} & = & \frac{q'^{2}}{2\left(q\cdot q'\right)}
\left[m^{2}I_{0}-\frac{q'^{2}}{2}I_{2}
\right],\nonumber \\
L_{0} & = & \frac{q'^{2}}{m^{2}\left(q\cdot q'\right)}I_{2},
\nonumber \\
L_{1} & = & -\frac{q'^{2}}{\left(q\cdot q'\right)^{2}}
\left[I_{0}-\frac{q'^{2}}{2m^{2}}I_{2}\right],\nonumber \\
L_{2} & = & \frac{1}{\left(q\cdot q'\right)}I_{0},\nonumber \\
L_{3} & = & -\frac{q'^{4}}{4\left(q\cdot q'\right)^{3}}
\left[6I_{0}-\left(\frac{q'^{2}}{m^{2}}+8\right)I_{2}
\right],\nonumber \\
L_{4} & = & \frac{1}{\left(q\cdot q'\right)}I_{2},
\nonumber \\
L_{5} & = & \frac{q'^{2}}{\left(q\cdot q'\right)^{2}}
\left[I_{0}-2I_{2}\right],\nonumber \\
L_{6} & = & -\frac{q'^{2}}{2\left(q\cdot q'\right)}
\left[I_{0}-2I_{2}\right],\nonumber \\
M_{1} & = & 0,\nonumber \\
M_{2} & = & \frac{1}{\left(q\cdot q'\right)}I_{0},\nonumber \\
M_{3} & = & \frac{q'^{2}}{\left(q\cdot q'\right)^{3}}
\left[\left(\frac{q'^{2}}{2}+m^{2}\right)I_{0}-
\frac{3q'^{2}}{2}I_{2}\right],\nonumber \\
M_{4} & = & \frac{1}{\left(q\cdot q'\right)}
\left[I_{0}-I_{2}\right],\nonumber \\
M_{5} & = & -\frac{1}{\left(q\cdot q'\right)^{2}}
\left[\left(\frac{q'^{2}}{2}+m^{2}\right)I_{0}-
\frac{3q'^{2}}{2}I_{2}\right],\nonumber \\
M_{6} & = & \frac{1}{\left(q\cdot q'\right)}
\left[m^{2}I_{0}-\frac{q'^{2}}{2}I_{2}\right].
\label{eq:scalarfunctionappendix}
\end{eqnarray}

\chapter{Modified Parametrization of the Nonforward Matrix Element in the Scalar Toy Model}

We modify the scalar parametrization $\mathcal{F}\left(z,p,r\right)$ of the nonforward matrix element 
$\left\langle p-r/2\right|\phi\left(-z/2\right)\phi\left(z/2\right)\left|p+r/2\right\rangle $, 
introduced in Section VIII.2, by adding the subleading $\mathcal{O}\left(z^{2}\right)$ 
and $\mathcal{O}\left(z^{4}\right)$ terms. Namely,
\begin{eqnarray}
\left\langle p-r/2\right|\phi\left(-z/2\right)\phi\left(z/2\right)
\left|p+r/2\right\rangle =\int_{-1}^{1}d\beta
\int_{-1+\left|\beta\right|}^{1-\left|\beta\right|}d
\alpha\; e^{i\left(k\cdot z\right)}\nonumber \\
\times\left[h_{f}\left(\beta,\alpha,t\right)+
\frac{z^{2}}{4}h_{2f}\left(\beta,\alpha,t\right)+\frac{z^{4}}{32}h_{4f}
\left(\beta,\alpha,t\right)\right],
\label{eq:scalarparametrizationappendix}
\end{eqnarray}
where $k=\beta p+\alpha r/2$. The function $h_{f}\left(\beta,\alpha,t\right)$
is the original double distribution or the so-called parent DD. The new functions, known as 
the daughter DDs, $h_{2f}\left(\beta,\alpha,t\right)$ and $h_{4f}\left(\beta,\alpha,t\right)$,
are not independent but rather to be determined in such a way that the matrix element 
$\left\langle p-r/2\right|\phi\left(-z/2\right)\phi\left(z/2\right)\left|p+r/2\right\rangle $ 
has only terms of the certain twist. In particular, the twist-2 part
of the matrix element should satisfy the d'Alembert equation with
respect to \emph{z},
\begin{eqnarray}
\partial^{2}\left\langle p-r/2\right|\phi\left(-z/2\right)\phi\left(z/2\right)
\left|p+r/2\right\rangle _{twist-2} & = & 0.
\label{eq:d'Alembertappendix}
\end{eqnarray}
By imposing this condition on the right-hand side of Eq. (\ref{eq:scalarparametrizationappendix}), 
one can determine both $h_{2f}\left(\beta,\alpha,t\right)$ and $h_{4f}\left(\beta,\alpha,t\right)$.
Thus in the lowest order in $z^{2}$ we find
\begin{eqnarray}
\int_{-1}^{1}d\beta\int_{-1+\left|\beta\right|}^{1-\left|\beta\right|}d
\alpha\; e^{i\left(k\cdot z\right)}\nonumber \\
\times\left[-k^{2}h_{f}\left(\beta,\alpha,t\right)+i
\left(k\cdot z\right)h_{2f}\left(\beta,\alpha,t\right)+2h_{2f}
\left(\beta,\alpha,t\right)\right] & = & 0.
\label{eq:scalarcasecondition1appendix}
\end{eqnarray}
The second term in the brackets can be written as
\begin{equation}
\int_{-1}^{1}d\beta\int_{-1+\left|\beta\right|}^{1-\left|\beta\right|}d\alpha\; h_{2f}
\left(\beta,\alpha,t\right)\left(\beta\frac{\partial}{\partial\beta}+\alpha
\frac{\partial}{\partial\alpha}\right)e^{i\left(k\cdot z\right)},
\label{eq:secondtermappendix}
\end{equation}
and further integrating it by parts and using the boundary conditions
for $h_{2f}\left(\beta,\alpha,t\right)$ (note that, like the parent
DD, both daughter distributions vanish at the boundaries of the support
region), it takes the form
\begin{equation}
-\int_{-1}^{1}d\beta\int_{-1+\left|\beta\right|}^{1-\left|\beta\right|}d
\alpha\; e^{i\left(k\cdot z\right)}\left(\beta\frac{\partial}{\partial\beta}+
\alpha\frac{\partial}{\partial\alpha}+2\right)h_{2f}\left(\beta,\alpha,t\right).
\label{eq:secondtermfinalappendix}
\end{equation}
Substituting Eq. (\ref{eq:secondtermfinalappendix}) back into 
Eq (\ref{eq:scalarcasecondition1appendix}) leads to the partial differential equation 
for $h_{2f}\left(\beta,\alpha,t\right)$,
\begin{eqnarray}
\left(\beta\frac{\partial}{\partial\beta}+\alpha\frac{\partial}{\partial\alpha}
\right)h_{2f}\left(\beta,\alpha,t\right) & = & -k^{2}h_{f}
\left(\beta,\alpha,t\right)
\label{eq:scalarcasediffequationappendix}
\end{eqnarray}
with the solution
\begin{eqnarray}
h_{2f}\left(\beta,\alpha,t\right) & = & k^{2}\int_{1}^{\infty}d\tau\;\tau h_{f}
\left(\tau\beta,\tau\alpha,t\right).
\label{eq:scalarcasesolutionappendix}
\end{eqnarray}
It is easy to see that this solution indeed satisfies 
Eq. (\ref{eq:scalarcasediffequationappendix}) since
\begin{eqnarray}
\int_{1}^{\infty}d\tau\;\tau\left(\beta\frac{\partial}{\partial\beta}+
\alpha\frac{\partial}{\partial\alpha}\right)h_{f}\left(\tau\beta,\tau\alpha,t\right) & = & 
\int_{1}^{\infty}d\tau\;\tau^{2}\frac{\partial}{\partial\tau}h_{f}
\left(\tau\beta,\tau\alpha,t\right)\nonumber \\
& = & -h_{f}\left(\beta,\alpha,t\right)\nonumber \\
&  & -2\int_{1}^{\infty}d\tau\;\tau h_{f}
\left(\tau\beta,\tau\alpha,t\right).\nonumber \\
\label{eq:scalarcasecheckappendix}
\end{eqnarray}
Similarly, in the next order in the $z^{2}$ expansion, the harmonic
condition (\ref{eq:d'Alembertappendix}) gives the partial differential
equation for the daughter DD $h_{4f}\left(\beta,\alpha,t\right)$,
\begin{eqnarray}
\left(\beta\frac{\partial}{\partial\beta}+\alpha\frac{\partial}{\partial\alpha}-1
\right)h_{4f}\left(\beta,\alpha,t\right) & = & -k^{2}h_{2f}\left(\beta,\alpha,t\right).
\label{eq:scalarcasediffequationnextorderappendix}
\end{eqnarray}
The solution is
\begin{eqnarray}
h_{4f}\left(\beta,\alpha,t\right) & = & k^{2}\int_{1}^{\infty}d\tau\; h_{2f}
\left(\tau\beta,\tau\alpha,t\right).
\label{eq:scalarcasesolutionnextorderappendix}
\end{eqnarray}

It is worth noting at this point that to evaluate the twist-2
and twist-3 contributions to the Compton scattering amplitude using
a modified parametrization (\ref{eq:scalarparametrizationappendix}), 
the inclusion of $z^{2}$ and $z^{4}$ terms requires, in addition
to Eqs. (\ref{eq:integraloverzDVCS}) and (\ref{eq:integralformula2}), 
four more integrals over \emph{z}. Here we present the list of all six integrals
\begin{eqnarray}
\int d^{4}z\; e^{i\left(l\cdot z\right)}\frac{z_{\rho}}{2\pi^{2}\left(z^{2}-i0\right)} & = & 
\frac{4l_{\rho}}{\left(l^{2}+i0\right)^{2}},\nonumber \\
\int d^{4}z\; e^{i\left(l\cdot z\right)}\frac{z_{\rho}}{2\pi^{2}\left(z^{2}-i0\right)^{2}} & = & 
\frac{l_{\rho}}{l^{2}+i0},\nonumber \\
\int d^{4}z\; e^{i\left(l\cdot z\right)}
\frac{iz_{\rho}z_{\sigma}}{2\pi^{2}\left(z^{2}-i0\right)} & = & 4
\frac{g_{\rho\sigma}l^{4}-4l_{\rho}l_{\sigma}l^{2}}{\left(l^{2}+i0\right)^{4}},
\nonumber \\
\int d^{4}z\; e^{i\left(l\cdot z\right)}\frac{iz_{\rho}z_{\sigma}}{2\pi^{2}
\left(z^{2}-i0\right)^{2}} & = & 
\frac{g_{\rho\sigma}l^{2}-2l_{\rho}l_{\sigma}}{\left(l^{2}+i0\right)^{2}},\nonumber \\
\int d^{4}z\; e^{i\left(l\cdot z\right)}
\frac{z_{\delta}z_{\rho}z_{\lambda}}{2\pi^{2}\left(z^{2}-i0\right)} & = & 16
\frac{\left(g_{\delta\rho}l_{\lambda}+g_{\delta\lambda}l_{\rho}+g_{\rho\lambda}l_{\delta}
\right)l^{2}-l_{\delta}l_{\rho}l_{\lambda}}{\left(l^{2}+i0\right)^{4}},\nonumber \\
\int d^{4}z\; e^{i\left(l\cdot z\right)}\frac{z_{\delta}z_{\rho}z_{\lambda}}{2\pi^{2}
\left(z^{2}-i0
\right)^{2}} & = & 2
\frac{\left(g_{\delta\rho}l_{\lambda}+g_{\delta\lambda}l_{\rho}+g_{\rho\lambda}l_{\delta}
\right)l^{4}-4l_{\delta}l_{\rho}l_{\lambda}l^{2}}{\left(l^{2}+i0\right)^{4}}.\nonumber \\
\label{eq:allintegralsoverz}
\end{eqnarray}

\newpage

\vita{B.S. in Physics, University of Ljubljana, March 1997\\
M.S. in Physics, Old Dominion University, May 2001}

\vitapage


\begin{thebibliography}{Bliggs}

\addtocontents{toc}{\vspace*{12pt}}

\addcontentsline{toc}{chapter}{BIBLIOGRAPHY}

\addtocontents{toc}{\vspace*{12pt}}

%
\bibitem{Zweig} G.~Zweig, Preprints CERN-TH 401 and 412 (1964).
%


\bibitem{Gell-Mann:1964nj}
  M.~Gell-Mann,
  Phys.\ Lett.\  {\bf 8}, 214 (1964).


\bibitem{Bjorken:1969ja}
  J.~D.~Bjorken and E.~A.~Paschos,
  Phys.\ Rev.\  {\bf 185}, 1975 (1969).


\bibitem{Feynman:1969ej}
  R.~P.~Feynman,
  Phys.\ Rev.\ Lett.\  {\bf 23}, 1415 (1969).


%
\bibitem{Feynman} R.~P.~Feynman, \emph{Photon-Hadron Interactions}, Reading, 
USA: W.~A.~Benjamin (1972).
%


\bibitem{Yang:1954ek}
  C.~N.~Yang and R.~L.~Mills,
  Phys.\ Rev.\  {\bf 96}, 191 (1954).


\bibitem{Gross:1973id}
  D.~J.~Gross and F.~Wilczek,
  Phys.\ Rev.\ Lett.\  {\bf 30}, 1343 (1973).


\bibitem{Politzer:1973fx}
  H.~D.~Politzer,
  Phys.\ Rev.\ Lett.\  {\bf 30}, 1346 (1973).


\bibitem{Politzer:1974fr}
  H.~D.~Politzer,
  Phys.\ Rept.\  {\bf 14}, 129 (1974).


%
\bibitem{Bogolubov} N.~N.~Bogolubov, B.~V.~Struminsky and A.~N.~Tavkhelidze, Preprint JINR D-1968, 
Dubna (1965).
%


\bibitem{Han:1965pf}
  M.~Y.~Han and Y.~Nambu,
  Phys.\ Rev.\  {\bf 139}, B1006 (1965).


%
\bibitem{Miyamoto} Y.~Miyamoto, Prog. Theor. Phys. Suppl. Extra Number (1965), 187.
%


\bibitem{Fritzsch:2002jv}
  H.~Fritzsch and M.~Gell-Mann, in:
  \emph{Proceedings of the XVI International Conference on High Energy Physics}, 
  edited by J.~D.~Jackson and A.~Roberts, Fermilab, 1972 Vol. 2, p.135, hep-ph/0208010.


\bibitem{Fritzsch:1973pi}
  H.~Fritzsch, M.~Gell-Mann and H.~Leutwyler,
  Phys.\ Lett.\ B {\bf 47}, 365 (1973).


\bibitem{Amati:1978wx}
  D.~Amati, R.~Petronzio and G.~Veneziano,
  Nucl.\ Phys.\ B {\bf 140}, 54 (1978).


\bibitem{Amati:1978by}
  D.~Amati, R.~Petronzio and G.~Veneziano,
  Nucl.\ Phys.\ B {\bf 146}, 29 (1978).


\bibitem{Libby:1978qf}
  S.~B.~Libby and G.~Sterman,
  Phys.\ Rev.\ D {\bf 18}, 3252 (1978).


\bibitem{Efremov:1978cu}
  A.~V.~Efremov and A.~V.~Radyushkin,
  Theor.\ Math.\ Phys.\  {\bf 44}, 573 (1980).


\bibitem{Efremov:1980kz}
  A.~V.~Efremov and A.~V.~Radyushkin,
  Theor.\ Math.\ Phys.\  {\bf 44}, 664 (1981).


\bibitem{Efremov:1978xm}
  A.~V.~Efremov and A.~V.~Radyushkin,
  Theor.\ Math.\ Phys.\  {\bf 44}, 774 (1981)


\bibitem{Ellis:1978ty}
  R.~K.~Ellis, H.~Georgi, M.~Machacek, H.~D.~Politzer and G.~G.~Ross,
  Nucl.\ Phys.\ B {\bf 152}, 285 (1979).


\bibitem{Mueller:1978xu}
  A.~H.~Mueller,
  Phys.\ Rev.\ D {\bf 18}, 3705 (1978).


\bibitem{Wilson:1969zs}
  K.~G.~Wilson,
  Phys.\ Rev.\  {\bf 179}, 1499 (1969).


\bibitem{Brandt:1970kg}
  R.~A.~Brandt and G.~Preparata,
  Nucl.\ Phys.\ B {\bf 27}, 541 (1972).


\bibitem{Gross:1973ju}
  D.~J.~Gross and F.~Wilczek,
  Phys.\ Rev.\ D {\bf 8}, 3633 (1973).


\bibitem{Gross:1974cs}
  D.~J.~Gross and F.~Wilczek,
  Phys.\ Rev.\ D {\bf 9}, 980 (1974).


\bibitem{Georgi:1951sr}
  H.~Georgi and H.~D.~Politzer,
  Phys.\ Rev.\ D {\bf 9}, 416 (1974).


\bibitem{Lazar:2002af}
  M.~Lazar,
  \emph{Group Theoretical Analysis of Light-Cone Dominated Hadronic Processes and Twist Decomposition of 
  Nonlocal Operators in Quantum  Chromodynamics}, Ph.D. Thesis (2002), hep-ph/0308049.


\bibitem{Muller:1998fv}
  D.~Muller, D.~Robaschik, B.~Geyer, F.~M.~Dittes and J.~Horejsi,
  Fortsch.\ Phys.\  {\bf 42}, 101 (1994), hep-ph/9812448.


\bibitem{Ji:1996ek}
  X.~D.~Ji,
  Phys.\ Rev.\ Lett.\  {\bf 78}, 610 (1997), hep-ph/9603249.


\bibitem{Ji:1996nm}
  X.~D.~Ji,
  Phys.\ Rev.\ D {\bf 55}, 7114 (1997), hep-ph/9609381.


\bibitem{Radyushkin:1996nd}
  A.~V.~Radyushkin,
  Phys.\ Lett.\ B {\bf 380}, 417 (1996), hep-ph/9604317.


\bibitem{Radyushkin:1996ru}
  A.~V.~Radyushkin,
  Phys.\ Lett.\ B {\bf 385}, 333 (1996), hep-ph/9605431.


\bibitem{Radyushkin:1997ki}
  A.~V.~Radyushkin,
  Phys.\ Rev.\ D {\bf 56}, 5524 (1997), hep-ph/9704207.
  
  
\bibitem{Vanderhaeghen:1999xj}
  M.~Vanderhaeghen, P.~A.~M.~Guichon and M.~Guidal,
  Phys.\ Rev.\ D {\bf 60}, 094017 (1999), hep-ph/9905372.


\bibitem{Goeke:2001tz}
  K.~Goeke, M.~V.~Polyakov and M.~Vanderhaeghen,
  Prog.\ Part.\ Nucl.\ Phys.\  {\bf 47}, 401 (2001), hep-ph/0106012.
  

\bibitem{Collins:1996fb}
  J.~C.~Collins, L.~Frankfurt and M.~Strikman,
  Phys.\ Rev.\ D {\bf 56}, 2982 (1997), hep-ph/9611433.


\bibitem{Radyushkin:1998rt}
  A.~V.~Radyushkin,
  Phys.\ Rev.\ D {\bf 58}, 114008 (1998), hep-ph/9803316.


\bibitem{Diehl:1998kh}
  M.~Diehl, T.~Feldmann, R.~Jakob and P.~Kroll,
  Eur.\ Phys.\ J.\ C {\bf 8}, 409 (1999), hep-ph/9811253.


\bibitem{Diehl:1999tr}
  M.~Diehl, T.~Feldmann, R.~Jakob and P.~Kroll,
  Phys.\ Lett.\ B {\bf 460}, 204 (1999), hep-ph/9903268.


%
\bibitem{Greiner1} W.~Greiner and A.~Schafer, \emph{Quantum Chromodynamics}, Berlin, Germany: 
Springer-Verlag (1994).
%


\bibitem{Thomas:2001kw}
  A.~W.~Thomas and W.~Weise,
\emph{The Structure of the Nucleon}, Berlin, Germany: Wiley-VCH (2001).


\bibitem{Jaffe:1996zw}
  R.~L.~Jaffe,
  hep-ph/9602236.


\bibitem{Bjorken:1968dy}
  J.~D.~Bjorken,
  Phys.\ Rev.\  {\bf 179}, 1547 (1969).


\bibitem{Bloom:1969kc}
  E.~D.~Bloom {\it et al.},
  Phys.\ Rev.\ Lett.\  {\bf 23}, 930 (1969).


\bibitem{Breidenbach:1969kd}
  M.~Breidenbach {\it et al.},
  Phys.\ Rev.\ Lett.\  {\bf 23}, 935 (1969).


\bibitem{Friedman:1972sy}
  J.~I.~Friedman and H.~W.~Kendall,
  Ann.\ Rev.\ Nucl.\ Part.\ Sci.\  {\bf 22}, 203 (1972).


\bibitem{Sterman:1995fz}
  G.~Sterman,
  hep-ph/9606312.


\bibitem{Peskin:1995ev}
  M.~E.~Peskin and D.~V.~Schroeder,
\emph{An Introduction to Quantum Field Theory}, Reading, USA: Addison-Wesley (1995).


\bibitem{Callan:1969uq}
  C.~G.~Callan and D.~J.~Gross,
  Phys.\ Rev.\ Lett.\  {\bf 22}, 156 (1969).


\bibitem{Muta:1987mz}
  T.~Muta,
   \emph{Foundations Of Quantum Chromodynamics: An Introduction to Perturbative Methods in Gauge Theories},
  World Sci.\ Lect.\ Notes Phys.\  {\bf 5}, 1 (1987).


\bibitem{Radyushkin:2002qt}
  A.~V.~Radyushkin,
  Nucl.\ Phys.\ A {\bf 711}, 99 (2002).


\bibitem{Radyushkin:2004sr}
  A.~V.~Radyushkin,
  hep-ph/0409215.


\bibitem{Radyushkin:2000uy}
  A.~V.~Radyushkin,
  hep-ph/0101225.


\bibitem{Diehl:2003ny}
  M.~Diehl,
  Phys.\ Rept.\  {\bf 388}, 41 (2003), hep-ph/0307382.


\bibitem{Belitsky:2005qn}
  A.~V.~Belitsky and A.~V.~Radyushkin,
  hep-ph/0504030.


\bibitem{Gribov:1972ri}
  V.~N.~Gribov and L.~N.~Lipatov,
  Sov.\ J.\ Nucl.\ Phys.\  {\bf 15}, 438 (1972)
  [Yad.\ Fiz.\  {\bf 15}, 781 (1972)].


\bibitem{Altarelli:1977zs}
  G.~Altarelli and G.~Parisi,
  Nucl.\ Phys.\ B {\bf 126}, 298 (1977).


\bibitem{Dokshitzer:1977sg}
  Y.~L.~Dokshitzer,
  Sov.\ Phys.\ JETP {\bf 46}, 641 (1977)
  [Zh.\ Eksp.\ Teor.\ Fiz.\  {\bf 73}, 1216 (1977)].
  
  
\bibitem{Radyushkin:1977gp}
  A.~V.~Radyushkin,
  hep-ph/0410276.
  
  
\bibitem{Efremov:1978rn}
  A.~V.~Efremov and A.~V.~Radyushkin,
  Theor.\ Math.\ Phys.\  {\bf 42}, 97 (1980)

  
\bibitem{Efremov:1979qk}
  A.~V.~Efremov and A.~V.~Radyushkin,
  Phys.\ Lett.\ B {\bf 94}, 245 (1980).

  
\bibitem{Lepage:1979zb}
  G.~P.~Lepage and S.~J.~Brodsky,
  Phys.\ Lett.\ B {\bf 87}, 359 (1979).


\bibitem{Lepage:1980fj}
  G.~P.~Lepage and S.~J.~Brodsky,
  Phys.\ Rev.\ D {\bf 22}, 2157 (1980).


\bibitem{Musatov:1997pu}
  I.~V.~Musatov and A.~V.~Radyushkin,
  Phys.\ Rev.\ D {\bf 56}, 2713 (1997), hep-ph/9702443.


\bibitem{Burkardt:2000za}
  M.~Burkardt,
  Phys.\ Rev.\ D {\bf 62}, 071503 (2000)
  [Erratum-ibid.\ D {\bf 66}, 119903 (2002)], hep-ph/0005108.


\bibitem{Burkardt:2002hr}
  M.~Burkardt,
  Int.\ J.\ Mod.\ Phys.\ A {\bf 18}, 173 (2003), hep-ph/0207047.


\bibitem{Ji:1998xh}
  X.~D.~Ji and J.~Osborne,
  Phys.\ Rev.\ D {\bf 58}, 094018 (1998), hep-ph/9801260.


\bibitem{Collins:1998be}
  J.~C.~Collins and A.~Freund,
  Phys.\ Rev.\ D {\bf 59}, 074009 (1999), hep-ph/9801262.


\bibitem{Lampe:1998eu}
  B.~Lampe and E.~Reya,
  Phys.\ Rept.\  {\bf 332}, 1 (2000), hep-ph/9810270.


\bibitem{Filippone:2001ux}
  B.~W.~Filippone and X.~D.~Ji,
  Adv.\ Nucl.\ Phys.\  {\bf 26}, 1 (2001), hep-ph/0101224.


\bibitem{Bass:2004xa}
  S.~D.~Bass,
  hep-ph/0411005.
  
  
\bibitem{Ji:1997gm}
  X.~D.~Ji, W.~Melnitchouk and X.~Song,
  Phys.\ Rev.\ D {\bf 56}, 5511 (1997)
  hep-ph/9702379.


\bibitem{Petrov:1998kf}
  V.~Y.~Petrov, P.~V.~Pobylitsa, M.~V.~Polyakov, I.~Bornig, K.~Goeke and C.~Weiss,
  Phys.\ Rev.\ D {\bf 57}, 4325 (1998), hep-ph/9710270.


\bibitem{Radyushkin:1998es}
  A.~V.~Radyushkin,
  Phys.\ Rev.\ D {\bf 59}, 014030 (1999), hep-ph/9805342.


\bibitem{Mankiewicz:1997uy}
  L.~Mankiewicz, G.~Piller and T.~Weigl,
  Eur.\ Phys.\ J.\ C {\bf 5}, 119 (1998), hep-ph/9711227.


\bibitem{Musatov:1999xp}
  I.~V.~Musatov and A.~V.~Radyushkin,
  Phys.\ Rev.\ D {\bf 61}, 074027 (2000), hep-ph/9905376.


\bibitem{Balitsky:1987bk}
  I.~I.~Balitsky and V.~M.~Braun,
  Nucl.\ Phys.\ B {\bf 311}, 541 (1989).


\bibitem{Anikin:2000em}
  I.~V.~Anikin, B.~Pire and O.~V.~Teryaev,
  Phys.\ Rev.\ D {\bf 62}, 071501 (2000)
  hep-ph/0003203.


\bibitem{Penttinen:2000dg}
  M.~Penttinen, M.~V.~Polyakov, A.~G.~Shuvaev and M.~Strikman,
  Phys.\ Lett.\ B {\bf 491}, 96 (2000), hep-ph/0006321.


\bibitem{Belitsky:2000vx}
  A.~V.~Belitsky and D.~Muller,
  Nucl.\ Phys.\ B {\bf 589}, 611 (2000), hep-ph/0007031.


\bibitem{Radyushkin:2000jy}
  A.~V.~Radyushkin and C.~Weiss,
  Phys.\ Lett.\ B {\bf 493}, 332 (2000), hep-ph/0008214.


\bibitem{Radyushkin:2000ap}
  A.~V.~Radyushkin and C.~Weiss,
  Phys.\ Rev.\ D {\bf 63}, 114012 (2001), hep-ph/0010296.


\bibitem{Kivel:2000cn}
  N.~Kivel and M.~V.~Polyakov,
  Nucl.\ Phys.\ B {\bf 600}, 334 (2001), hep-ph/0010150.


\bibitem{Kivel:2000fg}
  N.~Kivel, M.~V.~Polyakov and M.~Vanderhaeghen,
  Phys.\ Rev.\ D {\bf 63}, 114014 (2001), hep-ph/0012136.


\bibitem{Kivel:2000rb}
  N.~Kivel, M.~V.~Polyakov, A.~Schafer and O.~V.~Teryaev,
  Phys.\ Lett.\ B {\bf 497}, 73 (2001), hep-ph/0007315.


\bibitem{Belitsky:2000vk}
  A.~V.~Belitsky, D.~Muller, A.~Kirchner and A.~Schafer,
  Phys.\ Rev.\ D {\bf 64}, 116002 (2001), hep-ph/0011314.


\bibitem{Belitsky:2001hz}
  A.~V.~Belitsky and D.~Muller,
  Phys.\ Lett.\ B {\bf 507}, 173 (2001), hep-ph/0102224.


\bibitem{Belitsky:2001yp}
  A.~V.~Belitsky, A.~Kirchner, D.~Muller and A.~Schafer,
  Phys.\ Lett.\ B {\bf 510}, 117 (2001), hep-ph/0103343.


\bibitem{Nachtmann:1973mr}
  O.~Nachtmann,
  Nucl.\ Phys.\ B {\bf 63}, 237 (1973).


\bibitem{Georgi:1976ve}
  H.~Georgi and H.~D.~Politzer,
  Phys.\ Rev.\ D {\bf 14}, 1829 (1976).


\bibitem{Guichon:1998xv}
  P.~A.~M.~Guichon and M.~Vanderhaeghen,
  Prog.\ Part.\ Nucl.\ Phys.\  {\bf 41}, 125 (1998), hep-ph/9806305.


\bibitem{Gluck:1994uf}
  M.~Gluck, E.~Reya and A.~Vogt,
  Z.\ Phys.\ C {\bf 67}, 433 (1995).
  
  
\bibitem{Musatov:1999gv}
  I.~V.~Musatov,
  \emph{Virtual Compton Scattering Processes in Quantum  Chromodynamics}, Ph.D. Thesis (1999), 
UMI-99-49834.


\bibitem{Belitsky:2001ns}
  A.~V.~Belitsky, D.~Muller and A.~Kirchner,
  Nucl.\ Phys.\ B {\bf 629}, 323 (2002), hep-ph/0112108.


\bibitem{Goshtasbpour:1995eh}
  M.~Goshtasbpour and G.~P.~Ramsey,
  Phys.\ Rev.\ D {\bf 55}, 1244 (1997), hep-ph/9512250.


\bibitem{Penttinen:1999th}
  M.~Penttinen, M.~V.~Polyakov and K.~Goeke,
  Phys.\ Rev.\ D {\bf 62}, 014024 (2000), hep-ph/9909489.
  
  
\bibitem{Stepanyan:2001sm}
  S.~Stepanyan {\it et al.}  [CLAS Collaboration],
  Phys.\ Rev.\ Lett.\  {\bf 87}, 182002 (2001), hep-ex/0107043.

  
\bibitem{Airapetian:2001yk}
  A.~Airapetian {\it et al.}  [HERMES Collaboration],
  Phys.\ Rev.\ Lett.\  {\bf 87}, 182001 (2001), hep-ex/0106068.

   
\bibitem{Psaker:2004qf}
  A.~Psaker,
  Braz.\ J.\ Phys.\  {\bf 34}, 944 (2004), hep-ph/0404181.


\bibitem{Brodsky:1972yx}
  S.~J.~Brodsky, J.~F.~Gunion and R.~L.~Jaffe,
  Phys.\ Rev.\ D {\bf 6}, 2487 (1972).


\bibitem{Bjorken:1969uu}
  J.~D.~Bjorken and E.~A.~Paschos,
  Phys.\ Rev.\ D {\bf 1}, 1450 (1970).


%
\bibitem{Greiner2} W.~Greiner and J.~Reinhardt, \emph{Quantum Electrodynamics}, Berlin, Germany: 
Springer-Verlag (1994).
%


\bibitem{Berger:2001xd}
  E.~R.~Berger, M.~Diehl and B.~Pire,
  Eur.\ Phys.\ J.\ C {\bf 23}, 675 (2002), hep-ph/0110062.


\bibitem{Drakoulakos:2004gn}
  D.~Drakoulakos {\it et al.}  [Minerva Collaboration],
  hep-ex/0405002.


\bibitem{Amore:2004ng}
  P.~Amore, C.~Coriano and M.~Guzzi,
  JHEP {\bf 0502}, 038 (2005), hep-ph/0404121.


\bibitem{Coriano:2004bk}
  C.~Coriano and M.~Guzzi,
  Phys.\ Rev.\ D {\bf 71}, 053002 (2005), hep-ph/0411253.


\bibitem{Psaker:2004sf}
  A.~Psaker,
  hep-ph/0412321.


%
\bibitem{Psaker1} A.~Psaker, A.~V.~Radyushkin and W.~Melnitchouk, 
Weak Deeply Virtual Compton Scattering, 
work in progress.
%


\bibitem{Mankiewicz:1997aa}
  L.~Mankiewicz, G.~Piller and T.~Weigl,
  Phys.\ Rev.\ D {\bf 59}, 017501 (1999), hep-ph/9712508.


%
\bibitem{Psaker2} A.~Psaker, A.~V.~Radyushkin, 
Double Distributions, Feynman Mechanism and RCS on the Pion, 
paper under preparation.
%


%
\bibitem{Psaker3} A.~Psaker, A.~V.~Radyushkin, 
Double Distributions and Wide-Angle Real Compton Scattering on the Nucleon, 
paper under preparation.
%


%
\bibitem{Psaker4} A.~Psaker, A.~V.~Radyushkin, 
Exclusive Production of Pions in the Two-Photon Collisions in the Double-Distribution Approach, 
work in progress.
%


%
\bibitem{Landau} V.~B.~Berestetskii, E.~M.~Lifshitz and L.~P.~Pitaevskii, \emph{Quantum Electrodynamics}, 
Pergamon Press (1982). 
%


\bibitem{Mukherjee:2002gb}
  A.~Mukherjee, I.~V.~Musatov, H.~C.~Pauli and A.~V.~Radyushkin,
  Phys.\ Rev.\ D {\bf 67}, 073014 (2003), hep-ph/0205315.


\end{thebibliography}
\end{document}